\documentclass[reprint,prb,superscriptaddress,longbibliography,amsmath,amssymb]{revtex4-2}
\usepackage[utf8]{inputenc}
\usepackage[T1]{fontenc}
\usepackage{graphicx}
\usepackage{booktabs}
\usepackage{dsfont}
\usepackage[dvipsnames]{xcolor}
\usepackage{algpseudocode}
\makeatletter
\newcounter{algorithm}
\renewcommand{\thealgorithm}{\arabic{algorithm}}
\providecommand{\theHalgorithm}{}\renewcommand{\theHalgorithm}{\arabic{algorithm}}
\def\fps@algorithm{tbp}
\def\ftype@algorithm{4}
\def\ext@algorithm{loa}
\def\fnum@algorithm{Algorithm~\thealgorithm}
\newenvironment{algorithm}{\@float{algorithm}}{\end@float}
\makeatother

\usepackage{needspace}
\makeatletter
\def\fname@algorithm{Algorithm}
\algnewcommand{\LineComment}[1]{\Statex \hskip\ALG@thistlm\(\triangleright\)~\textit{#1}}
\newenvironment{breakablealgorithm}
  {\par\Needspace*{0.32\textheight}\medskip\noindent
   \refstepcounter{algorithm}%
   \hrule height.8pt depth0pt \kern2pt
   \renewcommand{\caption}[2][\relax]{{\raggedright\textbf{\fname@algorithm~\thealgorithm}\ ##2\par}\kern2pt\hrule\kern2pt}}
  {\kern2pt\hrule\relax\par\medskip}
\makeatother
\usepackage[colorlinks=true,linkcolor=blue,citecolor=blue,urlcolor=blue]{hyperref}
\graphicspath{{./}}
\renewcommand{\topfraction}{0.9}\renewcommand{\bottomfraction}{0.8}\renewcommand{\textfraction}{0.07}\renewcommand{\floatpagefraction}{0.75}

\newcommand{\kk}{\mathbf{k}}
\newcommand{\bb}{\mathbf{b}}
\newcommand{\RR}{\mathbf{R}}

\begin{document}
\title{Fully Automated Wannier Disentanglement via Projectability Selection in Gaussian and Augmented Plane-Wave Frameworks}
\author{William P. Comaskey}
\affiliation{Department of Mathematics, University of Nevada, Las Vegas, Las Vegas, NV 89154, USA}
\author{Jose L. Mendoza-Cortes}
\email{jmendoza@msu.edu}
\affiliation{Department of Chemical Engineering \& Materials Science, Michigan State University, East Lansing, Michigan 48824, USA}
\affiliation{Department of Physics and Astronomy, Michigan State University, East Lansing, Michigan 48824, USA}
\date{\today}

\begin{abstract}
Constructing the frozen manifold for entangled bands requires balancing exact reproduction of the target bands with sufficient variational freedom to build a well-localized subspace from the broader Bloch pool. We present a workflow that systematically determines the energy ceiling for a target number of selected Wannier projections, freezing every Bloch state below this threshold across the sampled $\kk$-mesh. By measuring projectability, the spectral weight of each Bloch state projected onto the target localized basis, the workflow detects missing conduction channels and quantifies discrepancies between orbital character and energy-based freezing. In comparisons where we controlled the extent of the energy ceiling, freezing all sub-ceiling states is generally favored over projectability-based frozen states. Of the seven materials we tested, freezing all sub-ceiling states lowers the band error on four from between 11~meV and 1.2~eV to 18~meV or less, and matches the projectability-based choice to within about 2~meV on the remaining three. In each of the four, projectability-based frozen states give the lower spread, showing that spread alone cannot identify the better model. With the frozen states held fixed, the workflow minimizes the gauge-invariant spread, constructs the gauge from density-matrix columns on atomic orbitals, and passes an isolated set of bands to \textsc{Wannier90} for localization. The workflow is implemented in two programs for seamless high-throughput Wannierization with WIEN2k (\textsc{wien2wannier}) and CRYSTAL (\textsc{lcao2wannier}). Using a single set of calibrated defaults, the CRYSTAL implementation with analytic overlaps produces models that reproduce the reference bands of five cells, from bulk h-BN in a joint valence-and-conduction model to a magnetic framework with 1424 orbitals. The WIEN2k implementation produces eight such models from seven calculations.

\end{abstract}
\maketitle

\section{Introduction}
\label{sec:intro}
\begin{figure*}[t!]
\centering
\includegraphics[width=\textwidth]{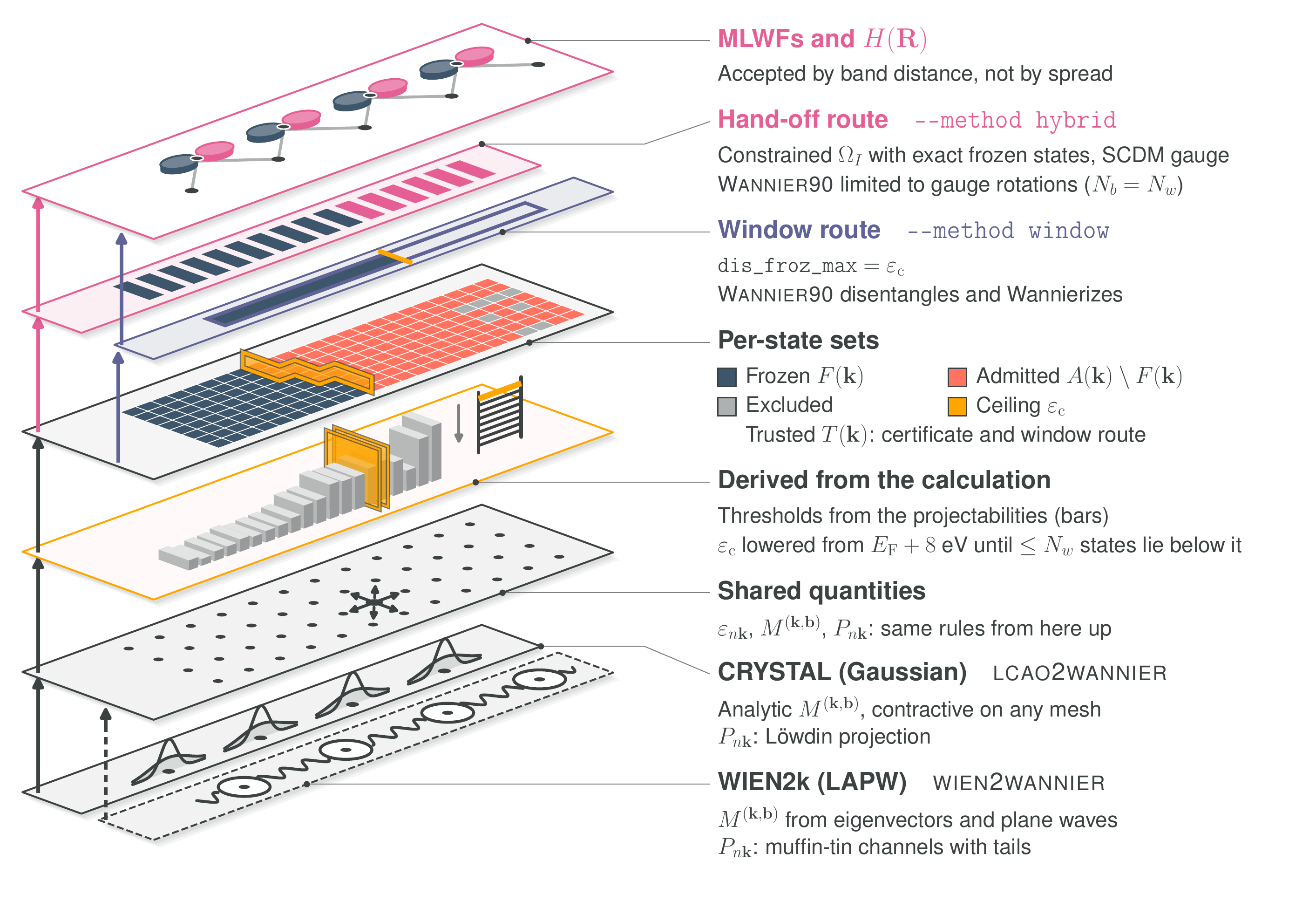}

\caption{The pipeline of both programs, read from the bottom. The two lowest sheets are the frontends. \textsc{lcao2wannier} starts from the real-space matrices $H(\RR)$ and $S(\RR)$ of a CRYSTAL calculation and computes the overlaps $M^{(\kk,\bb)}$ analytically and orthonormalizes the Bloch states in the same analytic metric, which makes the overlaps contractive on any mesh (Sec.~\ref{sec:platform}). \textsc{wien2wannier} starts from WIEN2k LAPW eigenvectors and uses projectabilities from muffin-tin channels extended by interstitial tails (Appendix~\ref{app:lapw}). From the third sheet up, both programs apply the same rules to the energies $\varepsilon_{n\kk}$, the overlaps and the projectabilities $P_{n\kk}$. On the orange sheet, two thresholds (planes) are read off the band-averaged projectabilities (bars), and the ceiling $\varepsilon_{\rm c}$ (ladder) is lowered from $E_{\mathrm{F}}+8$~eV until no $\kk$-point has more than $N_w$ states below it (Secs.~\ref{sec:masks} and \ref{sec:auto}). The next sheet shows, for h-BN on 8 of its 864 $\kk$-points and the 24 bands of its pool, the states that are frozen (slate), admitted but not frozen (coral) or excluded (gray), with the ceiling in orange. The trusted set $T(\kk)$ is used only by the certificate and, in the window route, to select the bands of the outer window. The two upper sheets are the routes to \textsc{Wannier90}. In the hand-off route (pink, \texttt{-\kern0pt-method hybrid}), $\Omega_I$ is minimized with the frozen states fixed, the gauge is built by SCDM on orbital columns, and \textsc{Wannier90} receives $N_b{=}N_w$ bands (Sec.~\ref{sec:dis}). In the window route (indigo, \texttt{-\kern0pt-method window}), the ceiling becomes \texttt{dis\_froz\_max}, and \textsc{Wannier90} also performs the disentanglement (Sec.~\ref{sec:two-routes}). Models are judged by their band distance, not by their spread. The lobes on the top sheet are schematic, and Appendix~\ref{app:constants} shows the pipeline stage by stage.}
\label{fig:pipeline}
\end{figure*}

Maximally localized Wannier functions (MLWFs)~\cite{Marzari1997,Souza2001,MarzariRMP} are the Wannier functions~\cite{Wannier1937} that minimize the total quadratic spread for a chosen band manifold, and they can be designed to decay exponentially for isolated bands~\cite{Kohn1959,Brouder2007}. They provide the foundation for tight-binding interpolation, Berry-phase calculations, and large-scale post-processing, and \textsc{Wannier90}~\cite{Mostofi2008,Pizzi2020} is the community-standard tool for constructing them. For entangled bands, the disentanglement method of Souza, Marzari, and Vanderbilt~\cite{Souza2001} uses predefined energy windows to constrain the subspace, combining an outer window of candidate states with a frozen window that preserves specific Bloch states within it exactly. A variety of automation schemes have made parameter selection increasingly systematic, ranging from SCDM-based methods~\cite{Damle2015,Damle2017,Vitale2020} and optimization protocols~\cite{Tillack2025,Tiwari2026} to projectability-disentangled Wannier functions (PDWF)~\cite{Qiao2023,Jiang2025} and the manifold-remixed Wannier functions built from them~\cite{Qiao2023b}. The PDWF thresholds can be set automatically by multi-Otsu cuts in \textsc{Wannier90}~v4.0.2~\cite{W90v402}.
Nevertheless, these approaches still rely on a single $\mathbf{k}$-uniform pair of thresholds applied globally across bands or energy ranges. As a result, large-scale high-throughput studies either fix one frozen window for all materials and correct the failures by hand (42 of 426 monolayers in Ref.~\cite{Zhou2025}) or choose one window pair per material by a scan and exclude the materials that fail their spread and band-error criteria (about 21\% in Ref.~\cite{Fang2025}). Approaches such as intrinsic Wannier functions~\cite{LiChan2026} and closest Wannier functions~\cite{Ozaki2024} dispense with both frozen manifolds and spread minimization entirely. The former weights each state by a smooth function of its atomic character (with the cutoff determined at each 
$\mathbf{k}$ by the target number of functions and the width set by hand) and projects a minimal set of free-atom orbitals onto the weighted states, following the construction of intrinsic atomic orbitals~\cite{Knizia2013}, while the latter weights states via a smooth energy window and orthonormalizes their projections onto specified localized orbitals.

Two classes of systems are particularly difficult to treat with scalar windows and thresholds. First, in a wide-gap insulator such as h-BN, a joint valence-and-conduction model must reproduce antibonding bands that cross into and out of a nearly-free-electron (NFE) continuum, whose lowest member in bulk h-BN is an interlayer state at $\Gamma$~\cite{Blase1995}. Second, in metals with strong spin--orbit coupling such as SnTe in its high-pressure B2 phase~\cite{Zhou2013}, a band changes orbital character across the zone, so that no single $\mathbf{k}$-uniform threshold can describe it. Projectability disentanglement~\cite{Qiao2023} therefore treats each state $(\mathbf{k}, n)$ individually, using its projectability to decide whether it is frozen, discarded, or retained for disentanglement, reflecting that a given band may exhibit atomic character at one $\mathbf{k}$ but free-electron behavior at another. For h-BN, using our L\"owdin projectabilities and atomic starting guesses, both hand-selected energy windows and \textsc{Wannier90}'s per-state freezing (\texttt{dis\_froz\_proj}) leave the Wannierization stalled, that is, short of its convergence criterion when the iteration budget is exhausted, at substantial band interpolation errors of 0.4--0.8~eV (Sec.~\ref{sec:baseline}).

Both approaches couple the question of how far the frozen manifold extends with that of which states to freeze. Decoupling them gives a definite rule, freezing every pool state that falls below the energy ceiling. 
With the pool, overlaps, solver, gauge, and computational budgets held fixed, this approach reproduces the bands substantially better than freezing only high-projectability states on four of the seven cells we compared, and it matches that choice to within 2.2~meV on the other three. The interpolation error drops from $52.2$ to $4.9$~meV for the h-BN conduction bands (up to $+12$~eV), and from $24.1$ to $7.0$~meV across the $[-13,+7.2]$~eV window of MgB$_2$. Where the window gives the larger error the difference is small, at most 0.6~meV, except on the GaAs conduction window (2.2~meV with the Gauss--Seidel sweep), where it lies within the dependence of that cell's Wannierization on the optimizer. 
In an all-electron LAPW bismuth bilayer, freezing only the trusted states completely loses one of the model's target bands, even though it reduces the gauge-invariant spread $\Omega_I$ (Sec.~\ref{sec:extent}). A lower spread therefore does not necessarily indicate a superior model. Setting the energy ceiling creates an inherent trade-off. Raising the ceiling shifts fidelity from the bands below the old ceiling to those above it, and once the raised ceiling encloses as many states as the Wannier count can hold, the fidelity of both ranges can degrade (Sec.~\ref{sec:extent}). We therefore align the ceiling with the target fidelity range, and lower it progressively from $E_{\mathrm{F}}+8$~eV until no $\mathbf{k}$-point contains more states below it than the target number of Wannier functions. Projectability determines the set of admitted states, selects missing conduction channels from the basis and, via an a priori certificate, quantifies how far its per-state classification departs from any uniform energy window (Appendix~\ref{app:certificate}).

The pipeline begins by minimizing $\Omega_I$ with the frozen states as exact unit columns (Sec.~\ref{sec:dis}, Fig.~\ref{fig:pipeline}, and Algorithm~\ref{alg:stage2}). The gauge is then fixed by selected columns of the density matrix (SCDM) discretized on the atomic-orbital index rather than on a real-space grid (Sec.~\ref{sec:scdm}). Finally, the pipeline passes an isolated problem with $N_b=N_w$ and no disentanglement keywords to \textsc{Wannier90}, which is limited to unitary gauge rotations.

Alternatively, the same frozen window can be passed to \textsc{Wannier90} as a standard disentanglement window (Sec.~\ref{sec:two-routes}). Once both Wannierizations are converged, the two routes yield comparable band errors on the Bi bilayer, h-BN, and GaAs. For MgB$_2$, the window route achieves lower error across the Fermi window (1.4 vs.\ 3.4~meV) but higher error across the wider window (4.4 vs.\ 3.6~meV), whereas on SnTe the hand-off reduces errors by a factor of 1.2 to 1.3. With a single fixed set of default parameters, the CRYSTAL frontend successfully models five diverse systems, ranging from a joint 16-function valence-and-conduction model of bulk h-BN (yielding errors of 0.56~meV on the valence bands and 5.2~meV over $[+5, +8]$~eV) to a magnetic two-dimensional framework with 1424 orbitals (Table~\ref{tab:defaults}). Similarly, the modified \textsc{wien2wannier} interface reproduces eight models across seven WIEN2k calculations, capturing the occupied manifolds to within 0.2--12~meV (Sec.~\ref{sec:lapw-results}).

\emph{Relation to prior work.} Per-state projectability freezing is available in \textsc{Wannier90} as \texttt{dis\_froz\_proj}~\cite{Qiao2023}, alongside a per-state admission floor via \texttt{dis\_proj\_min}. 
Using L\"owdin projectabilities rather than the pseudo-atomic projectors on which PDWF was calibrated, our controlled experiments indicate that per-state freezing stalls when applied without a band gate. In matched-extent experiments, an energy window yields substantially smaller band errors than gated per-state freezing on four of the seven materials, while matching it to within 2.2~meV on the remaining three. Gated freezing gives the lower error in only a few windows, by at most 0.6~meV, except on the GaAs conduction window, where it leads by 2.2~meV with the Gauss--Seidel sweep, a difference within the dependence of that cell's Wannierization on the optimizer.

We find that the frozen set should be an energy window below a derived ceiling, with projectability governing the admitted set and the certificate rather than state freezing. This framework is supported by three algorithmic procedures, which derive thresholds from band averages within the target fidelity range rather than from a global histogram statistic~\cite{W90v402}, check the frozen ceiling against the Wannier count, and add missing conduction channels from the basis. Once the manifold is selected, the pipeline minimizes the gauge-invariant spread $\Omega_I$ while preserving the frozen states exactly, and constructs a smooth initial gauge using an atomic-orbital column SCDM algorithm. This allows the system to be handed off to \textsc{Wannier90} as an isolated problem with $N_b = N_w$, accompanied by a diagnostic certificate that quantifies where projectability and energy ordering disagree.
 
 The atomic-orbital form of SCDM that fixes the gauge also fills a gap among existing tools. The 2024 review of the Wannier-function software ecosystem~\cite{Marrazzo2024} lists no SCDM implementation for bases of linear combinations of atomic orbitals (LCAO), although column selection on atomic orbitals has been used in related settings. CP2K's localization module includes an SCDM option for isolated states~\cite{CP2Kmanual}, and an AO-based, SCDM-type choice of initial projections over the whole $\kk$-mesh was added to its \textsc{Wannier90} exporter concurrently with this work~\cite{CP2KPR5983}. Periodic Gaussian-basis SCDM has also been run with its columns placed on real-space grids~\cite{LiChan2026}, and AO-basis SCDM has been formulated for the occupied space of molecules~\cite{Fuemmeler2023}. Our AO-SCDM differs from all of these because it starts from the $\kk$-averaged density matrix of a periodic subspace that has already been disentangled and needs neither a real-space grid nor an energy weighting. It selects the columns in the orthogonalized AO basis, as the SCDM-L variant of Ref.~\cite{Fuemmeler2023} does for molecules. The recent extension of PDWF by Jiang \emph{et al.}~\cite{Jiang2025} shares our aim of robust Wannierization for magnetic and spin--orbit-coupled systems, but it keeps per-state freezing, combined with a frozen energy window up to 2~eV above $E_{\mathrm{F}}$ or the conduction-band minimum, and our matched-extent experiments compare per-state freezing with the energy window.

The platform for this work is \textsc{lcao2wannier}, the first interface between the Gaussian-orbital code CRYSTAL~\cite{Crystal23} and \textsc{Wannier90}, which evaluates the overlap matrices $M^{(\kk,\bb)}$ analytically (Sec.~\ref{sec:platform}). Because two-center integrals between Gaussian functions have a closed form~\cite{Boys1950,Obara1986,McMurchie1978,Helgaker2000} that extends to the momentum-shifted overlaps~\cite{Saunders1983}, every $M^{(\kk,\bb)}$ is contractive on any mesh once the Bloch states are orthonormalized in the same analytic metric. Overlaps built in the midpoint-metric approximation lack this property and make the spread negative on the coarsest meshes we tested (Appendix~\ref{app:contractivity}). The CRYSTAL frontend supports two-component spin--orbit coupling~\cite{Desmarais2019a,Desmarais2020a,Desmarais2021} as well as collinear spin polarization~\cite{vonBarth1972}.

None of the existing atomic-orbital tools derives the frozen manifold from the calculation. CRYSTAL's Wannier module~\cite{ZicovichWilson2001} applies Boys localization~\cite{FosterBoys1960} to isolated composite bands, and its projective extension to entangled bands extracts the chemical character of Bloch states that the user selects with an energy window~\cite{Birkenheuer2005}. Pipek--Mezey localization with Bloch intrinsic atomic orbitals is restricted to occupied bands~\cite{Zhu2024}, and a PySCF~\cite{Sun2020} interface to \textsc{Wannier90}, which also evaluates the overlaps of AO pairs analytically~\cite{pyWannier90}, relies on hand-set windows. Intrinsic Wannier functions~\cite{LiChan2026} and, in the pseudo-atomic-orbital code OpenMX, closest Wannier functions~\cite{Ozaki2024} avoid a frozen set altogether, as described above. The rules developed here need only band projectabilities onto a labeled set of localized orbitals, which plane-wave and (L)APW workflows already compute~\cite{Qiao2023,Kunes2010}, so they are not tied to a Gaussian basis. Section~\ref{sec:lapw-results} applies them unchanged to the all-electron code WIEN2k~\cite{Blaha2020} through \textsc{wien2wannier}~\cite{Kunes2010}, which supports spin--orbit coupling and, at present, collinear magnetism, and tests them on insulating, metallic, and band-inverted cells. The Supplemental Material (SM)~\cite{SMarXiv} contains the detailed procedures, the full tables behind the figures, and the records of convergence and computational cost.

\section{Platform: analytic Gaussian overlaps and the LCAO eigenproblem}
\label{sec:platform}

A Wannierization takes two inputs from an electronic-structure calculation, namely the Bloch eigenstates on a $\kk$-mesh and the overlaps $M^{(\kk,\bb)}$ between eigenstates at neighboring mesh points. In a basis of Gaussian atomic orbitals (AOs) these overlaps can be computed exactly, and the exact overlaps have no singular value above one on any mesh, a property that keeps the gauge-invariant spread non-negative. The midpoint-metric approximation, which avoids the exact integrals, has no such bound and can make the spread negative on coarse meshes. Inside phase factors such as $e^{2\pi i\,\kk\cdot\RR}$, the wave vectors $\kk$ and $\bb$ are written in reduced coordinates of the reciprocal lattice, and $\RR$ is a lattice translation in integer coordinates. Where a wave vector multiplies a position, as in $e^{-i\bb\cdot\mathbf r}$, it denotes the Cartesian vector.

\emph{The LCAO eigenproblem.} For the $N$ AOs $\chi_\mu$ of the unit cell, CRYSTAL prints the converged Kohn--Sham matrix $H_{\mu\nu}(\RR)$, which it calls the Fock matrix, and the overlap matrix $S_{\mu\nu}(\RR)$ between AO $\mu$ in the home cell and AO $\nu$ in the cell at $\RR$~\cite{Crystal23,CrystalManual}. Their Bloch sums~\cite{DelRe1967,Andre1967,Pisani1988},
\begin{equation}
H(\kk)=\sum_{\RR} e^{2\pi i\,\kk\cdot\RR}\, H(\RR),
\qquad
S(\kk)=\sum_{\RR} e^{2\pi i\,\kk\cdot\RR}\, S(\RR),
\label{eq:blochsum}
\end{equation}
define at each $\kk$ the generalized eigenproblem of a non-orthogonal basis~\cite{Lowdin1950,Roothaan1951} in its crystal-orbital form~\cite{DelRe1967,Andre1967,Pisani1980}
\begin{equation}
H(\kk)\,C(\kk) = S(\kk)\,C(\kk)\,E(\kk),
\qquad
C^\dagger(\kk)\,S(\kk)\,C(\kk) = I .
\label{eq:lcaoeig}
\end{equation}
Column $n$ of $C(\kk)$ holds the AO coefficients of the Bloch state $\psi_{n\kk}$, and $E(\kk)$ holds the band energies $\varepsilon_{n\kk}$. Appendix~\ref{app:contractivity} describes how the triangular matrices printed by CRYSTAL are completed and how the spin--orbit case is treated.

\emph{Neighbor overlaps.} The Marzari--Vanderbilt spread~\cite{Marzari1997} is evaluated on the mesh by finite differences. Each $\kk$ is connected to neighbors $\kk+\bb$ on a few shells of vectors $\bb$, and each shell carries a finite-difference weight $w_b$, which is positive for the first-order stencil used here (Sec.~\ref{sec:dis} gives the normalization of the weights together with the spread functional). The electronic structure enters the spread only through the overlap of the cell-periodic parts $u_{n\kk}(\mathbf r)=e^{-i\kk\cdot\mathbf r}\psi_{n\kk}(\mathbf r)$ at neighboring points,
\begin{equation}
M^{(\kk,\bb)}_{mn} = \langle u_{m\kk}\,|\,u_{n,\kk+\bb}\rangle
= \langle \psi_{m\kk}\,|\,e^{-i\bb\cdot\mathbf{r}}\,|\,
\psi_{n,\kk+\bb}\rangle,
\label{eq:mmn}
\end{equation}
which \textsc{Wannier90} reads from its \texttt{.mmn} file~\cite{Pizzi2020}. In the LCAO expansion, Eq.~\eqref{eq:mmn} becomes the matrix product $C^\dagger(\kk)\,\tilde S^{(\bb)}\,C(\kk+\bb)$, in which $\tilde S^{(\bb)}$ is a Bloch sum of the momentum-shifted AO overlaps $\langle\chi_\mu|e^{-i\bb\cdot\mathbf r}|\chi_\nu\rangle$ [Eq.~\eqref{eq:mmnassembly}] that reduces to the ordinary AO overlap matrix at $\bb=0$.

\emph{Analytic evaluation.} For Gaussian AOs the momentum-shifted overlaps have a closed form. The product of two Gaussians is again a Gaussian (the Gaussian product theorem~\cite{Boys1950,Helgaker2000}), and completing the square in its exponent shows that the plane wave only multiplies the integral by a scalar and moves the center $\mathbf P$ of the product to the complex point $\mathbf P'=\mathbf P-i\bb/2p$, where $p$ is the sum of the two exponents [Eq.~\eqref{eq:complexcenter} and Ref.~\cite{Saunders1983}, Sec.~2.3]. The Obara--Saika recursion for overlaps of Cartesian Gaussians~\cite{Obara1986,Helgaker2000}, evaluated at this complex center, then gives the integrals for every angular momentum. We implement it through $\ell=4$ ($g$ shells), the highest angular momentum that CRYSTAL supports~\cite{Desmarais2018}. At $\bb=0$ the integrals reproduce the AO overlaps printed by CRYSTAL to its print precision, which tests the integral code (Appendix~\ref{app:contractivity}).

\emph{Contractivity.} The eigenvectors of Eq.~\eqref{eq:lcaoeig} are orthonormal in the metric $S(\kk)$ built from the printed overlaps, which agrees with the analytic metric only to print precision. We therefore orthonormalize them once more in the analytic metric [Eq.~\eqref{eq:gtoortho}]. The Bloch states at $\kk$ and at $\kk+\bb$ then form two orthonormal sets of functions, and $M^{(\kk,\bb)}$ is the matrix of their mutual overlaps. Such a matrix is contractive, which means that its singular values satisfy
\begin{equation}
\sigma_i\big(M^{(\kk,\bb)}\big)\le 1
\label{eq:contractive}
\end{equation}
on any mesh, with no condition on $|\bb|$ (Appendix~\ref{app:contractivity}). The bound carries over to any subspace of the pool, because projecting $M^{(\kk,\bb)}$ onto orthonormal frames at $\kk$ and at $\kk+\bb$ cannot raise its singular values, and so the squared Frobenius norm of each projected overlap is at most the number of selected bands. The gauge-invariant spread $\Omega_I$, defined in Eq.~\eqref{eq:omegai} of Sec.~\ref{sec:dis}, is a weighted sum over all links of the difference between the number of selected bands and this squared norm. With the positive weights of the first-order stencil it is therefore non-negative, whereas higher-order stencils carry negative weights on their outer shells~\cite{Lihm2026} and lose this lower bound.

The midpoint-metric approximation avoids the momentum-shifted integrals altogether. It keeps only the zeroth-order term of the expansion of $e^{-i\bb\cdot\mathbf r}$ about the midpoint of each orbital pair, because higher orders would need dipole and higher moment integrals of the AO pairs, and even the expansion through fourth order is not contractive on the coarsest h-BN mesh (Appendix~\ref{app:contractivity}). The midpoint construction evaluates the ordinary overlap at the midpoint $\kk+\bb/2$ and restores the orbital positions through a phase,
\begin{align}
M^{(\kk,\bb)}_{\text{mid}} &= C^\dagger(\kk)\,D_{\bb}\,S(\kk+\bb/2)\,D_{\bb}\,
C(\kk+\bb),
\label{eq:midpoint}\\
(D_{\bb})_{\mu\mu} &= e^{-i\bb\cdot\boldsymbol\tau_\mu/2},\nonumber
\end{align}
where $\boldsymbol\tau_\mu$ is the position of the nucleus that carries $\chi_\mu$. In a non-orthogonal basis this matrix obeys no bound of the form~\eqref{eq:contractive}. Across the mesh series of Appendix~\ref{app:contractivity}, its largest singular value exceeds one on every mesh of every cell and reaches 3.2 on bulk h-BN, and on the two coarsest h-BN meshes it makes $\Omega_I$ negative. For these reasons every result in the rest of the paper uses the analytic construction.

\section{Frozen ceiling and trust masks}
\label{sec:masks}

\begin{figure*}[t!]
\centering
\includegraphics[scale=0.67]{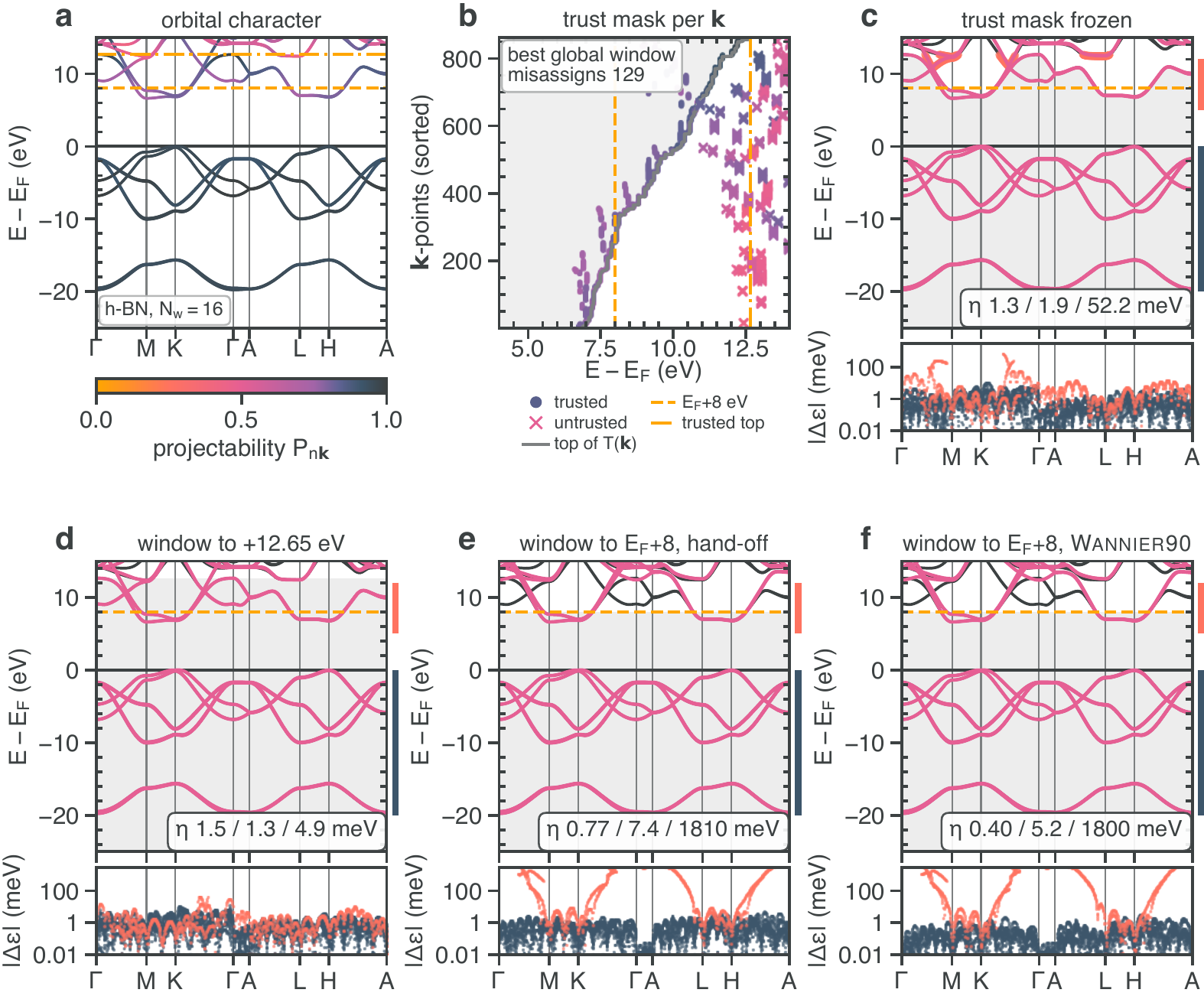}
\caption{Frozen extent and frozen membership on bulk h-BN (16 Wannier functions, $12^2{\times}6$ mesh). (a)~DFT bands along the \textsc{Wannier90} path, colored by their L\"owdin projectability $P_{n\kk}$ onto the target orbitals. The amber lines mark $E_{\mathrm{F}}+8$~eV (dashed) and the top of the trusted set at $+12.65$~eV (dash-dotted). (b)~The trust mask of Eq.~\eqref{eq:masks}. Each row is a $\kk$-point, sorted by the top of its trusted set $T(\kk)$ (gray line), and each pool state between $+4$ and $+14$~eV is drawn at its energy as a dot if trusted and a cross if not, colored by $P_{n\kk}$. Below the top of the trusted set lie 378 untrusted states, and the best single energy window misassigns 129 states. (c)--(f)~Models (pink) over the DFT bands (black), with the frozen region shaded. (c)~The trust mask is frozen ($p_{\text{froz}}=0.71$ and $p_{\text{gate}}=0.74$, from path-averaged projectabilities), and the untrusted states below the top of the trusted set stay free (coral). (d)~Every pool state below the top of the trusted set is frozen, which keeps the extent of (c) and adds those untrusted states (Sec.~\ref{sec:extent}). (e)~Every pool state below the released ceiling, $E_{\mathrm{F}}+8$~eV, is frozen. (f)~The window of (e) is disentangled and Wannierized by \textsc{Wannier90}. Panels (c)--(e) are hand-offs with the Gauss--Seidel sweep, and the conjugate-gradient default gives 0.56 and 5.2~meV for (e) (Table~\ref{tab:defaults}). The boxes give the band distance $\eta$, with bands paired by index (Table~\ref{tab:extent}), over $[-20,0]$, $[+5,+8]$ and $[+5,+12]$~eV, which the edge bars mark. The lower strips give $|\Delta\varepsilon|$ for every paired state on a log scale (slate occupied, coral empty). In (e) and (f) the band above $+8$~eV lies outside the model and appears at the top of the strip.}
\label{fig:concept}
\end{figure*}

\begin{algorithm}[tbp]
\caption{The steps of the hand-off route, in the order in which both programs run them at their defaults. Each step names the section or equation that defines it, and Algorithms~S1--S5 of the SM give the detailed procedures. A run that cannot complete a step stops and reports the cause.}
\label{alg:stage2}
\small
\begin{enumerate}\setlength{\itemsep}{1pt}\setlength{\parskip}{0pt}
\item Take the valence channels of each atom as the target orbitals, which fix the Wannier count $N_w$ [Eq.~\eqref{eq:numwann}].
\item Compute the projectability of every state onto the target.
\item If the target lacks the orbital character of the low conduction bands, or if the trust mask holds more states at some $\kk$-point than there are Wannier functions, add the missing channels from the basis. Only the CRYSTAL frontend performs this step (Sec.~\ref{sec:auto} and Algorithm~S1).
\item Read the two thresholds off the distribution of band-averaged projectabilities [Eq.~\eqref{eq:autotrust}].
\item On spin--orbit cells, average the projectability over degenerate states. Then select a contiguous pool of about $1.5N_w$ bands and build the trusted and admitted sets [Eq.~\eqref{eq:masks}], admitting every pool state in LAPW.
\item Lower the frozen ceiling progressively from $E_{\mathrm{F}}+8$~eV until no $\kk$-point has more than $N_w$ pool states below it [Eq.~\eqref{eq:walk} and Algorithm~S2].
\item Freeze every pool state below the ceiling and minimize $\Omega_I$ over the admitted states (Sec.~\ref{sec:dis} and Algorithm~S3).
\item If the subspace jumps on links that touch a band frozen at only a few $\kk$-points and of mixed character, lower the ceiling under that band and repeat step~7 (Sec.~\ref{sec:masks} and Algorithm~S2).
\item Rotate to the eigenbasis of the projected Hamiltonian [Eq.~\eqref{eq:handoff}] and fix the gauge by the AO-SCDM columns (Sec.~\ref{sec:scdm} and Algorithm~S4).
\item Check the gauge rank, the contractivity of every overlap [Eq.~\eqref{eq:handoff-acceptance}] and the finiteness of all data, then write \texttt{.eig}, \texttt{.mmn} and \texttt{.amn} with $N_b=N_w$ for \textsc{Wannier90}.
\end{enumerate}
\end{algorithm}

A model with $N_w$ Wannier functions has $N_w$ dimensions at each $\kk$-point. It can therefore hold at most $N_w$ frozen states there, and freezing all $N_w$ would leave nothing to disentangle. We take the frozen set to be the energy window below a ceiling $\varepsilon_{\rm c}$ that is derived from the Wannier count,
\begin{equation}
F(\kk) = \{n\in\text{pool} : \varepsilon_{n\kk} \le \varepsilon_{\rm c}\},
\label{eq:frozen}
\end{equation}
and we use the projectability only to decide which further states are admitted to the variational space. Section~\ref{sec:extent} tests both parts of this rule, the freezing by energy and the restriction of the projectability to admission. Freezing the states below the ceiling that lie outside the trusted set lowers the band error by factors of three to more than sixty on four of the seven cells compared, and it changes the other three by at most 2.2~meV (Table~\ref{tab:extent}). Trusted states above the ceiling are released, because raising the ceiling to include them improves the bands above $E_{\mathrm{F}}+8$~eV only at the expense of those below. Algorithm~\ref{alg:stage2} lists the steps of the hand-off route in the order in which both programs run them, and the rest of this section and the next two explain each step.

\emph{Target and Wannier count.} Let $P_{n\kk}\in[0,1]$ be the L\"owdin projectability of state $(\kk,n)$ onto the target AO set $\mathcal{T}$, that is, the weight of the state on the symmetrically orthonormalized AOs~\cite{Lowdin1950} that belong to $\mathcal{T}$. Summed over all AOs these weights give 1 for every state, because the orthonormalized AOs form an orthonormal basis of the AO span. The target set is constructed from the basis rather than chosen by hand. A table lists the valence angular-momentum channels of each element, and every radial function of those channels is marked as target, except the core functions of all-electron atoms. On Sn, Te and Bi, which carry small-core effective-core potentials (ECPs), the marked functions therefore include the outer-core $s$ and $p$ shells that these potentials leave in the basis~\cite{Metz2000,Peterson2003}, but their bands lie 85~eV or more below $E_{\mathrm{F}}$, outside the band pool. The mask is doubled for spin--orbit-coupled (SOC) spinors, and the Wannier count follows as
\begin{equation}
N_w = \sum_{\text{atoms}}\ \sum_{\ell\in\mathcal{T}} (2\ell+1),
\label{eq:numwann}
\end{equation}
doubled for SOC.

\emph{The ceiling.} Starting from the fidelity limit $E_{\text{fid}}=E_{\mathrm{F}}+8$~eV, the ceiling is lowered in steps of $\Delta E_{\rm step}=1$~eV (Algorithm~S2 of the SM) and stops at the first level at which no $\kk$-point holds more states below it than the Wannier count,
\begin{align}
\varepsilon_{\rm c} &= \max\Big\{E_j :
\max_{\kk}\big|\{n : \varepsilon_{n\kk}\le E_j\}\big| \le N_w\Big\},\nonumber\\
E_j &= E_{\text{fid}} - j\,\Delta E_{\rm step},\quad j\in\mathbb{N}_0.
\label{eq:walk}
\end{align}
If no level down to $E_{\mathrm{F}}+1$~eV satisfies this condition, the run stops and reports the exact supremum $\min_{\kk}\varepsilon_{N_w+1,\kk}$, which is the highest ceiling that would fit. Truncating the frozen set to $N_w$ states at every $\kk$ is not an option, because it would leave no free direction and the solver would report convergence at once with a zero gradient. \textsc{Wannier90} itself rejects only more than $N_w$ frozen states at a $\kk$-point. An isolated target manifold, on the other hand, legitimately freezes all $N_w$ states at every $\kk$ (the Bi bilayer freezes 16 of 16), so the search has no saturation test. The released \textsc{wien2wannier} applies the same rule (\verb|--froz-emax auto|), and the two programs therefore freeze the same set for the same eigenvalues. When the fidelity target lies above $E_{\mathrm{F}}+8$~eV, an explicit ceiling (\verb|--hybrid-froz-emax|, \verb|--froz-emax|) extends the frozen range, and the log prints the top of the trusted set as a suggested value for it.

Only the capacity test is required for feasibility, whereas the starting energy and the step are choices, and we tested each separately (SM). The starting energy sets the range over which the model is fitted, and the error of a window grows as the top of the window approaches the ceiling. Raising the start to $E_{\mathrm{F}}+10$~eV improves MgB$_2$ in both programs but makes Fe with spin--orbit coupling five to eight times worse, and lowering it to $E_{\mathrm{F}}+6$~eV leaves the h-BN conduction band outside the frozen range. The step matters only where capacity, rather than the starting energy, sets the ceiling, and there a step of 0.5~eV improves the conduction windows of the Bi bilayer and $\alpha$-Sn at the cost of a spread 19 to 32\% higher, while leaving SrVO$_3$ unchanged.

\emph{Trust and admission masks.} Two per-state sets are built from the projectability (Fig.~\ref{fig:concept}a,~b). The admitted set $A(\kk)$ bounds the variational space of the default route, while the trusted set $T(\kk)$ is never frozen and is used only by the certificate and the window route. The two sets are
\begin{align}
T(\kk) &= \big\{n : P_{n\kk}\ge p_{\text{froz}} \,\wedge\,
\bar P_n \ge p_{\text{gate}}\big\} \cup \Sigma(\kk),\nonumber\\
A(\kk) &= \big\{n : P_{n\kk}\ge p_{\text{floor}}\big\}
\cup F(\kk),
\label{eq:masks}
\end{align}
where $\bar P_n$ is the $\kk$-averaged projectability of band $n$, $\Sigma(\kk)$ is the Fermi shell defined below, and $p_{\text{floor}}=0.10$ is a projectability \emph{floor}. The floor is an admission constraint that energy windows lack, and it keeps nearly-free-electron states out of the variational space. It is the default of \textsc{lcao2wannier}, whereas the LAPW frontend admits every pool state (Appendix~\ref{app:lapw}). Where fewer than $N_w$ states are admitted at some $\kk$-point, the pool states of highest projectability below the floor are added until $N_w$ are admitted, so that the constrained problem remains feasible at every $\kk$, a repair that the \texttt{dis\_proj\_min} option of \textsc{Wannier90} does not make.

The trust mask is the input to the certificate of Appendix~\ref{app:certificate}, which counts the states on which the mask differs from the best energy window. On h-BN the top of $T(\kk)$ varies by 5.8~eV across the zone, and the best global window misassigns 129 states, compared to 108 and 74 in the two Sc-COF spin channels. In the window route the trust mask also selects the bands of the outer window (Sec.~\ref{sec:two-routes}). The band-average gate $p_{\text{gate}}$ protects the mask against isolated spikes in the projectability. Without the gate, gray-zone bands, whose band-averaged projectability lies between the continuum and the trusted tier of Sec.~\ref{sec:auto} (here $\bar P\approx0.7$--$0.8$), would enter the trusted set at the isolated $\kk$-points where $P_{n\kk}$ rises above the threshold. Freezing such a mask constrains the subspace discontinuously in $\kk$ and stalls the minimization, which is what the per-state freezing of standard \textsc{Wannier90} does in Fig.~\ref{fig:baseline}a,~c. Where most states are degenerate, as for the Kramers pairs of the spin--orbit cells (a degenerate fraction above 0.7, with $|\Delta\varepsilon|<0.02$~eV), both programs first average $P_{n\kk}$ over each degenerate group, so that the masks treat a multiplet as one unit (Appendix~\ref{app:lapw}).

\emph{Fermi shell.} The shell $\Sigma(\kk)$ adds the energy information that the projectability alone cannot supply. For metals it is
\begin{equation}
\Sigma(\kk) = \{n : \varepsilon_{n\kk}\le\varepsilon_{\text{ceil}}\},
\qquad
\varepsilon_{\text{ceil}} = \max_{\kk}\,\max_{n\in B}\,\varepsilon_{n\kk},
\label{eq:shell}
\end{equation}
where $B$ is the set of bands whose \emph{minimum} lies below $E_{\mathrm{F}}+\Delta E_{\mathrm{shell}}$, with $\Delta E_{\mathrm{shell}}=3$~eV. For insulators the shell is the occupied manifold resolved per state, that is, the set of states below mid-gap. A Fermi-crossing test decides which definition applies and treats the cell as a metal if any band has $\min_\kk\varepsilon_{n\kk}<0<\max_\kk\varepsilon_{n\kk}$. The shell matters when the trust mask itself is frozen, as in the per-state mode (\verb|--hybrid-froz-emax none|) used in the controlled experiments of Sec.~\ref{sec:results}, because nothing else then fixes the Fermi surface. Without the shell, the MgB$_2$ minimization in this mode satisfies every convergence criterion while producing a \emph{wrong} Fermi surface, with a band error of 820~meV over $[-13,+7.2]$~eV on an $8^3$ mesh, compared to 18.6~meV with the shell. Under the frozen ceiling the shell has no effect, because its states below the ceiling are frozen by Eq.~\eqref{eq:frozen} and those above it are released. The SM describes its representability check and the caps that apply in the per-state mode.

\emph{Band pool.} All of these sets are drawn from one contiguous band-index pool of at least $r_{\mathrm{pool}}N_w$ bands ($r_{\mathrm{pool}}=1.5$), placed around the best-projected bands. The pool is kept contiguous in band index so that it excludes ghost bands, which are dominated by diffuse Gaussian functions, lie at $+10$ to $+30$~eV, and nevertheless project strongly onto the compact target AOs (SM, stability lessons, rule~1).

\emph{Mixed-character pockets.} A band that dips below $\varepsilon_{\rm c}$ only on an island of $\kk$-points is frozen there and free one mesh step away. The disentangled subspace must contain the band on the island but may drop it at the island's edge, and where the band's states on the island are only partly of target character, the subspace jumps at that edge (Sec.~\ref{sec:snte}). The pocket rule protects the continuity of the subspace against this case and is applied after a first disentanglement. It is a heuristic whose four thresholds were calibrated on SnTe, which is the only one of the 14 validation hand-offs that the rule changes (Sec.~\ref{sec:snte}). The rule lowers the automatic ceiling under such a band (Algorithm~S2 of the SM) when three conditions are met. The weakest link of the hand-off, that is, the smallest singular value of $W^\dagger(\kk)M^{(\kk,\bb)}W(\kk+\bb)$, must lie below $\sigma_{\rm p}=0.5$ and touch the band, the band must lie below the ceiling at fewer than $f_{\rm p}=25\%$ of the $\kk$-points with a mean projectability of at least $p_{\rm mix}=0.5$ there, and the repeated disentanglement must raise the weakest link to at least $\sigma_{\rm p}$. These thresholds are not optimal for every cell, as the case of $\alpha$-Sn shows. There the weakest link touches a pocket of mean projectability 0.38, below $p_{\rm mix}$, so the rule leaves the pocket frozen, although lowering the ceiling under it would improve both windows there, from 2.2 and 14.2 to 2.0 and 6.3~meV (SM). Both programs apply the rule only to automatic ceilings (\verb|--hybrid-pocket-skip|, \verb|--pocket-skip|).

Every numerical choice in this section and the next two is a constant that is the same for every material, and Appendix~\ref{app:constants} lists them all. The two frontends share these constants except for three values specific to the LAPW frontend, namely the admission floor, the sweep cap and the gauge dispatch. Because the constants were calibrated on this validation set, the validation shows that one set of values covers these insulating, metallic, strong-SOC and band-inverted cells, but it does not show that the values transfer to other materials. The SM reports, for each constant, the calculations in which it affects the result.

\section{Frozen-constrained \texorpdfstring{$\Omega_I$}{Omega-I}, gauge, and hand-off}
\label{sec:dis}

The disentanglement keeps every frozen state exactly and minimizes the gauge-invariant spread over the remaining admitted states. Its result is passed to \textsc{Wannier90} as an isolated manifold, expressed in a gauge built from the density matrix, and \textsc{Wannier90} is then limited to unitary gauge rotations. In the Marzari--Vanderbilt discretization~\cite{Marzari1997}, the gauge-invariant spread of an $N_w$-dimensional subspace spanned by the semi-unitary frames $V(\kk)$ is
\begin{equation}
\Omega_I = \frac{1}{N_k}\sum_{\kk}\sum_{\bb} w_b\Big(N_w -
\big\|V^\dagger(\kk)\,M^{(\kk,\bb)}\,V(\kk+\bb)\big\|_F^2\Big),
\label{eq:omegai}
\end{equation}
where the $\bb$-shell weights $w_b$ satisfy $\sum_b w_b\,b_\alpha b_\beta = \delta_{\alpha\beta}$~\cite{Marzari1997}. Souza, Marzari, and Vanderbilt minimize this functional by a self-consistent eigenproblem in which frozen energy windows fix some columns of $V$~\cite{Souza2001}. Our minimization accepts any frozen set $F(\kk)$ per $\kk$, although the released default always passes the window of Eq.~\eqref{eq:frozen}. The frozen states enter $V(\kk)$ as exact unit columns, and the remaining $N_w-|F(\kk)|$ free columns are taken as the leading eigenvectors of a $\kk$-local operator on the admitted states that are not frozen, $D(\kk)=A(\kk)\setminus F(\kk)$,
\begin{multline}
Z_{ac}(\kk) = \sum_{\bb} w_b\,\big[M^{(\kk,\bb)}\,V(\kk+\bb)\\
\times V^\dagger(\kk+\bb)\,M^{(\kk,\bb)\dagger}\big]_{ac},
\quad a,c\in D(\kk).
\label{eq:zmat}
\end{multline}
This eigenproblem is iterated in Gauss--Seidel order, so that each $\kk$ uses the neighbors already updated within the current sweep, because the Jacobi variant was unstable in our tests. The iteration starts from the singular value decomposition (SVD) of the projections of the states in $D(\kk)$ onto the target AOs. Frozen containment, $VV^\dagger e_n = e_n$ for every frozen $(\kk,n)$, then holds exactly by construction rather than being enforced by a penalty. As a check of the implementation, the solver reproduces the $\Omega_I$ of \textsc{Wannier90}'s own disentanglement to $10^{-4}$~\AA$^2$ when both are given equivalent window constraints, and for h-BN at $6^3$ with two hand-set window pairs both codes give $\Omega_I=10.1205$ and $8.5935$~\AA$^2$.

\emph{Two solvers.} Both programs also carry a second minimizer of the same functional, conjugate gradients (CG) on the product of Grassmannians with the frozen columns held fixed (Algorithm~S3 of the SM, which also gives the gradient), and CG is the default. Direct minimization of the entangled-band problem, as an alternative to the alternating iteration, was formulated in Ref.~\cite{DamleLevittLin2019}, and we apply it here to the frozen-constrained $\Omega_I$. A CG step needs no per-$\kk$ eigendecomposition, its line search accepts only steps that lower $\Omega_I$, and its result does not depend on the number of threads. Because $\Omega_I$ is non-convex, neither solver is guaranteed to find the global minimum, which is also true of the original Souza--Marzari--Vanderbilt scheme~\cite{Souza2001}, and on entangled cells the two solvers can stop in different local minima. CG ends lower than the sweep on h-BN (by 1.8\%) and on MgB$_2$ (by 7.7\%), and there it also gives smaller band errors at the released ceiling. On the earlier 36-function SnTe target CG ends 1.25\% higher, but it still gives the usable model, because the Gauss--Seidel subspace jumps at the X points (Sec.~\ref{sec:snte}). Table~\ref{tab:defaults}, the route comparison of Sec.~\ref{sec:two-routes} and Fig.~\ref{fig:baseline} use CG. The Sc-COF rows of Table~\ref{tab:defaults}, the ceiling study of Table~\ref{tab:ceiling} and the LAPW models of Appendix~\ref{app:lapw} were made with the sweep, which was the default at the time, except where marked. Table~\ref{tab:extent} gives results for both solvers, and the SM repeats the ceiling study with CG and compares the two solvers in detail.

\emph{Hand-off.} \textsc{Wannier90} interprets the entries of the \texttt{.eig} file as the energies of the states to which the \texttt{.mmn} file refers, so the hand-off basis must diagonalize the Hamiltonian within the disentangled subspace. Since $H(\kk)$ is diagonal in the band basis of the pool, the frames are rotated to the eigenbasis of the projected Hamiltonian,
\begin{align}
V^\dagger(\kk)\,\mathrm{diag}\big(\varepsilon(\kk)\big)\,V(\kk)
&= R(\kk)\,\mathrm{diag}\big(\tilde\varepsilon(\kk)\big)\,R^\dagger(\kk),\nonumber\\
W(\kk) &= V(\kk)\,R(\kk),
\label{eq:handoff}
\end{align}
and the files written for \textsc{Wannier90} are $\texttt{.eig}=\tilde\varepsilon$, referenced to $E_{\mathrm{F}}$, $\texttt{.mmn}=W^\dagger M W$ in the analytic overlaps of Sec.~\ref{sec:platform}, and the \texttt{.amn} gauge described below. Together they define an isolated manifold with $N_b=N_w$ and \emph{no} disentanglement keywords, on which \textsc{Wannier90} minimizes only the gauge-dependent part $\tilde\Omega=\Omega_D+\Omega_{\text{OD}}$ of the spread, that is, its diagonal and off-diagonal terms in the decomposition of Ref.~\cite{Marzari1997}. The frames obey $W^\dagger(\kk)W(\kk)=I$, but the overlaps between different $\kk$-points are contractive rather than unitary, so the acceptance criterion is
\begin{equation}
\max_{\kk,\bb}\sigma_{\max}\!\left[
W^\dagger(\kk)M^{(\kk,\bb)}W(\kk+\bb)\right]\le 1+\tau_M,
\label{eq:handoff-acceptance}
\end{equation}
where $\tau_M=10^{-6}$ allows for numerical roundoff. Both frontends apply this test to every $(\kk,\bb)$ link, require every singular-value decomposition to succeed, and reject non-finite data. Overlaps are never clipped or unitarized to pass the test, and the files are validated before they replace earlier ones.

\emph{AO-SCDM gauge.}\label{sec:scdm} The disentanglement fixes the subspace, and the gauge is the basis within that subspace from which \textsc{Wannier90} starts its localization. Starting projections built from atomic-orbital trial functions can leave the Wannierization stalled, and on h-BN the window route did not converge from them within 500\,000 iterations (Sec.~\ref{sec:two-routes}). The SCDM family~\cite{Damle2015,Damle2017,Damle2018} instead derives the gauge from the density matrix of the subspace itself. Plane-wave implementations, and the periodic Gaussian-basis SCDM of Ref.~\cite{LiChan2026}, take the columns of this matrix on a real-space grid. In an LCAO basis the natural localized discretization is the AO index itself, and we call the resulting construction \emph{AO-SCDM}.

Let $B(\kk)$ hold the AO coefficients of the hand-off states, orthonormal in the analytic metric. The $\kk$-averaged density matrix $\rho = N_k^{-1}\sum_{\kk} B(\kk)B^\dagger(\kk)$ is factorized by QR with column pivoting (QRCP), which selects $N_w$ anchor AOs that are the same at every $\kk$. The trial functions are the columns of $\rho$ at these anchors, $G(\kk)=B^\dagger(\kk)\,\rho[:,\text{anchors}]$, and they lie in the span of the manifold (we write $G$ because $A(\kk)$ already denotes the admitted set). Because the anchors do not depend on $\kk$, the gauge varies with $\kk$ only through the subspace, so it is smooth wherever the subspace varies smoothly and the projection onto the anchors is nonsingular. The continuity of the subspace and the rank of the gauge are checked separately (Sec.~\ref{sec:auto}). A second anchor set, obtained by QRCP on $B^\dagger(\Gamma)$~\cite{Damle2017}, and the point-row form $B[\text{anchors}]^\dagger$ give four candidate gauges in all. The run keeps the candidate whose worst $\kk$-point is best conditioned, as measured by $\min_{\kk}\sigma_{\min}[G(\kk)]$ (Algorithm~S4 of the SM), and it stops if even this gauge is rank-deficient. The construction follows the quasi-density-matrix method of Ref.~\cite{Damle2018}, with its energy weighting (an erfc or Gaussian function of the band energy) replaced by the projectability masks of Eq.~\eqref{eq:masks} and with the primary anchors taken from the $\kk$-averaged density matrix instead of a single $\kk$-point.

\section{Automatic thresholds, channel augmentation, and checks}
\label{sec:auto}

Neither the two thresholds of the trust mask nor the composition of the target requires a value set by hand. The thresholds are read off the distribution of band-averaged projectabilities within the declared fidelity range, and channels that the target lacks are added from the basis of the calculation.

\emph{Thresholds.} No fixed pair $(p_{\text{froz}},p_{\text{gate}})$ suits every material (SM), but in all six production runs the distribution of band-averaged projectabilities separates into a trusted cluster and a continuum, with a gray plateau between them in all but the Bi bilayer, and the thresholds can be read off this distribution (Appendix~\ref{app:anatomy}). Only on the Sc-COF are the plateau and the continuum not separated by a gap. \textsc{Wannier90}~4.0.2 sets the two thresholds of its projectability disentanglement by a multi-Otsu cut of the histogram of all $P_{n\kk}$, without regard to energy and without checking whether the resulting frozen set is feasible~\cite{W90v402}. On the h-BN inputs at $6^3$ its pair $0.805/0.180$ leaves the \textsc{Wannier90} disentanglement stalled at band errors of 498 and 803~meV. Our rule instead reads band averages within the fidelity range. With the fidelity limit $E_{\text{fid}}$ of Sec.~\ref{sec:masks}, the fidelity-relevant bands $\mathcal{R}$, the gate $p_{\text{gate}}$ (the top of the gray plateau plus a margin $\delta=0.015$), the trusted bands $\mathcal{T}_{\text{tr}}$ and the state threshold $p_{\text{froz}}$ (with a margin $\delta_{\rm s}=0.02$) are
\begin{align}
\mathcal{R} &= \{n\in\text{pool} : \min_{\kk}\varepsilon_{n\kk}
\le E_{\text{fid}}\},
\nonumber\\
p_{\text{gate}} &= \max_{n\in\text{pool}\setminus\mathcal{R}} \bar P_n
+ \delta,
\nonumber\\
\mathcal{T}_{\text{tr}} &= \{n\in\mathcal{R} : \bar P_n \ge
p_{\text{gate}}\},
\nonumber\\
p_{\text{froz}} &= \operatorname{clip}\Big(
\min_{n\in\mathcal{T}_{\text{tr}}}\min_{\kk} P_{n\kk} - \delta_{\rm s},\;
p_{\mathrm{clip}}^{-},\; p_{\mathrm{clip}}^{+}\Big),
\label{eq:autotrust}
\end{align}
with the clipping bounds $(p_{\mathrm{clip}}^{-},p_{\mathrm{clip}}^{+})=(0.55,0.95)$ (Appendix~\ref{app:constants}). The tiers are separated by an energy criterion rather than by a projectability threshold. Bands that never dip below the fidelity limit cannot improve the fidelity inside the declared range, so the highest band-averaged projectability among them marks the top of the gray plateau, which the gate must exceed. When the pool lacks either tier, the rule falls back to $(p_{\text{froz}},p_{\text{gate}})=(0.95,0.85)$ and prints a warning. Under the frozen ceiling the thresholds do not change the frozen set and act only through the trust mask, which determines what the certificate tests, which bands the window route disentangles, and the top of the trusted set that the log reports as a suggested extended ceiling.

\emph{Channel augmentation.} No choice of thresholds helps when the target orbitals do not span the conduction bands. On SnTe, the coverage of the four lowest conduction bands by the chemical $sp$ valence, that is, their mean projectability onto it, is only 0.557 (Sec.~\ref{sec:snte}). No constraint or gauge can then reproduce the conduction band, whose band RMS error is 1568~meV over $[0,+6]$~eV in the channel study of the SM (scored with the earlier Fermi reference). The augmentation step therefore detects such a deficit, ranks the channels of the basis that could remove it, and adds the smallest sufficient set of channels to the target, which raises $N_w$. It is triggered when the coverage of the $N_{\mathrm{cond}}=4$ lowest conduction bands falls below $p_{\mathrm{aug}}=0.70$, or when the trust mask holds more states at some $\kk$-point than there are Wannier functions. The second condition is the test for metals, in which no single conduction band can be singled out, and of the five LCAO cells only SnTe receives added channels, through the coverage test.

Each channel $c=(\text{atom},\ell)$ of the basis outside the valence configuration, including $f$ channels, is scored by its coverage of the low conduction bands and by three further quantities. Its L\"owdin retention,
\begin{equation}
\Pi_c = \Big\langle \big|[S(\kk)^{1/2}]_{\mu\mu}\big|^2
\Big\rangle_{\mu\in c,\,\kk},
\label{eq:ceiling}
\end{equation}
is the mean on-site weight that the channel keeps after symmetric orthogonalization, which is the orthonormalization that changes the AOs least~\cite{Carlson1957}, and a channel with low retention is nearly linearly dependent on the other channels. Its semicore bands are deep occupied bands that project almost entirely onto it. Its specificity is the coverage minus the channel's projectability onto a continuum reference above the fidelity range, and it is small when the coverage only reflects the completeness of the basis. A channel is admitted if its retention is at least $\Pi_{\min}=0.40$ and it has either semicore bands or a specificity of at least $s_{\min}=0.05$. Because coverage adds over disjoint channels, the smallest set of admitted channels whose coverage lies within $\delta_C=0.02$ of that of all admitted channels consists of the channels with the highest coverage, so it can be found exactly (Algorithm~S1 of the SM). The set is added only if it raises the coverage by at least $\Delta C=0.05$, and each selected channel then adds Wannier functions shell by shell,
\begin{equation}
N_w \;\leftarrow\; N_w + \sum_{c\,\in\,\text{selected}} \bigl[g\,(2l_c+1) + n_c\bigr],
\label{eq:decoupled}
\end{equation}
where $g=2$ for spinors and 1 otherwise, and $n_c$ is the number of semicore bands of the channel inside the band pool. A channel is selected because the conduction band is made of it, so it receives one valence shell of functions, and its semicore bands in the pool form a second, deeper shell of the same $\ell$ that needs functions of its own. On SnTe this gives $10+10$ Sn-$d$ functions, the second ten for the Sn-$4d$ semicore bands, and 10 Te-$d$ functions, which raises $N_w$ from 16 to 46. The SM explains the reasons for each criterion and describes the earlier count rule, which stopped at 36 functions on SnTe.

\emph{Checks.} The pipeline checks each possible failure at the stage where it can first be assessed, and it either stops or reports the failure with the reason. The checked failures are a rank-deficient SCDM gauge, sub-ceiling states missing from the band pool, a frozen count that exhausts $N_w$, a Fermi shell that overrides the projectability threshold, a $\kk$-mesh that misses high-symmetry points, and a band distance understated by nearest-band matching. They also include a Fermi level from band filling that departs from the value of the electronic-structure code, in which case the code's value is kept for a metal, and a disentangled subspace that jumps between neighboring $\kk$-points, which is detected when the smallest singular value of $W^\dagger(\kk)M^{(\kk,\bb)}W(\kk+\bb)$ falls below $10^{-2}$ (Sec.~\ref{sec:snte}). In production these checks found real defects, for example when the $\kk$-mesh audit reported that the $8^3$ MgB$_2$ mesh of our earlier runs samples neither $K$ nor $H$. The SM lists every check and describes how each was calibrated.

\section{Results}
\label{sec:results}

The results begin with the two controlled comparisons on which the construction rests, namely frozen membership at fixed extent (Sec.~\ref{sec:extent}) and \textsc{Wannier90}'s per-state freezing on identical h-BN inputs (Sec.~\ref{sec:baseline}). They then report the released defaults on the five LCAO cells (Sec.~\ref{sec:defaults}), the SnTe cell, whose target needs added channels (Sec.~\ref{sec:snte}), the comparison of the two automatic routes (Sec.~\ref{sec:two-routes}), and the LAPW frontend (Sec.~\ref{sec:lapw-results}).

The structures and self-consistent-field settings of the LCAO cells are tabulated in the SM (structures and calculation parameters). The h-BN cell is a bulk hexagonal cell of two BN layers, with B above N and N above a hexagon center (space group $P\bar{6}m2$). This is not the AA$'$ stacking of the experimental crystal, so every comparison for this cell is made against its own reference bands, and MgB$_2$ is bulk hexagonal as well. The Bi(111) bilayer and the Sc-COF are two-dimensional slabs computed with CRYSTAL's \texttt{SLAB} option, which is periodic in two dimensions only, and the $c=500$~\AA\ axis of their \textsc{Wannier90} input stands for the non-periodic direction. SnTe is computed in a two-atom CsCl-structure (B2) cell, which is the metallic high-pressure phase that superconducts under pressure~\cite{Zhou2013}, rather than in the ambient rocksalt structure of the topological crystalline insulator~\cite{Fu2011,Hsieh2012}. With CRYSTAL's Fermi energy this cell holds 46.0 of its 46 electrons on $12^3$ and $16^3$ meshes.

All CRYSTAL calculations use hybrid functionals~\cite{Becke1993a}, which replace a fraction of the semilocal exchange by Fock exchange. The h-BN cell uses PBE0~\cite{Perdew1996b,Adamo1999,Ernzerhof1999}, that is, PBE~\cite{Perdew1996,*Perdew1997} with 25\% Fock exchange, and MgB$_2$, SnTe and the Bi bilayer use the same one-parameter form with 10\% Fock exchange [Ref.~\cite{Adamo1999}, Eq.~(5)], following Ref.~\cite{Comaskey2022}. The spin-polarized~\cite{vonBarth1972} Sc-COF uses HSE06~\cite{Heyd2003,*Heyd2006,Krukau2006}. SnTe and Bi are two-component calculations in spin-current density-functional theory~\cite{Desmarais2020a,Desmarais2020b,Bodo2022}, in which the spin--orbit operator is taken from the relativistic effective-core potentials~\cite{Ermler1981,Desmarais2019a}. B, N, Mg, H, C, O and Sc carry the all-electron pob-TZVP-rev2 basis sets~\cite{VilelaOliveira2019,Peintinger2013}, whereas Sn, Te and Bi carry Stuttgart--Cologne small-core pseudopotentials, with 28 core electrons on Sn and Te and 60 on Bi~\cite{Metz2000,Peterson2003}. Their valence basis sets are the fifth-period pob-TZVP-rev2 basis on Sn~\cite{Laun2022}, the $(10s6p6d)/[4s3p3d]$ basis of Ref.~\cite{Comaskey2022} on Bi, used with the Stuttgart--Cologne pseudopotential of Ref.~\cite{Metz2000}, and on Te the cc-pVDZ-PP basis~\cite{Peterson2003}, in which the outermost $d$ function is replaced by $d$ functions of exponents 0.6 and 0.2~bohr$^{-2}$ and an $f$ function of exponent 0.2~bohr$^{-2}$. The self-consistent densities are sampled on $\Gamma$-centered uniform nets, which CRYSTAL calls Pack--Monkhorst nets~\cite{Monkhorst1976,CrystalManual}. For the metals CRYSTAL locates $E_{\mathrm{F}}$ by integration on a second net~\cite{CrystalManual}, following Gilat and Raubenheimer~\cite{Gilat1966}, and for the insulator h-BN $E_{\mathrm{F}}$ is the top of the valence band.

The band error in every table is the band distance $\eta$ of Ref.~\cite{Vitale2020} [Eq.~(18) there], evaluated over the states inside the scoring window. It is the root-mean-square difference between model and reference energies over the states on the band path whose reference energy lies in the window, and $\eta^{\max}$ [Eq.~(19)] is the largest single difference in the same window. Both quantities pair model band $j$ with reference band $j+o$ at the offset $o$ that minimizes $\eta$, and both use \textsc{Wannier90}'s own interpolation. With this pairing, a reference band that has no partner in the model appears as a large error, whereas matching each reference state to the nearest model band can hide such a missing band (Sec.~\ref{sec:extent}). Nearest-band matching gives the same values to 0.01~meV on every row except SnTe, whose model splits the reference Kramers pairs between mesh points (Sec.~\ref{sec:snte}). Spreads are computed with \textsc{Wannier90}'s first-order finite-difference stencil. With the second-order stencil of Ref.~\cite{Lihm2026}, $\Omega_I$ is higher by 2.0\% (Bi), 3.3\% (h-BN) and 6.2\% (MgB$_2$) on these meshes, measured on the Gauss--Seidel hand-offs, while the band distances change by at most 1.5\%.

\begin{table*}[t]
\caption{Hybrid Wannier models at the automatic defaults (analytic overlaps, \texttt{-\kern0pt-method hybrid}, conjugate-gradient disentanglement except where noted, and no hand-set parameters). $p_{\text{froz}}$ and $p_{\text{gate}}$ are the derived trust-mask thresholds, $\varepsilon_{\rm c}$ is the resolved frozen ceiling above $E_{\mathrm{F}}$, $|F|/\kk$ is the range of the number of frozen states per $\kk$-point, and $\Omega_{\rm tot}/N_w$ is the total spread per Wannier function after the Wannierization. The band distance $\eta$ and the largest band error $\eta^{\max}$ are measured against the LCAO reference in each scoring window (relative to $E_{\mathrm{F}}$), with bands paired by index (Sec.~\ref{sec:results}).}
\label{tab:defaults}
\small
\begin{tabular*}{\textwidth}{@{\extracolsep{\fill}}l c r c c c c c c r r@{}}
\toprule
Cell & $\kk$ mesh & $N_w$ & $p_{\text{froz}}$ & $p_{\text{gate}}$ & $\varepsilon_{\rm c}$ & $|F|/\kk$ & $\Omega_{\rm tot}/N_w$ & Window & $\eta$ & $\eta^{\max}$ \\
 & & & & & (eV) & & (\AA$^2$) & (eV) & (meV) & (meV) \\
\midrule
h-BN & $12^2{\times}6$ & 16 & 0.706 & 0.739 & $+8$ & 8--10 & 0.74 & $[-20,0]$ & 0.56 & 3.1 \\
     &                &    &       &       &      &       &      & $[+5,+8]$ & 5.2 & 39 \\
\addlinespace
Bi bilayer & $12^2{\times}1$ & 16 & 0.855 & 0.608 & $+7$ & 16 & 2.76 & $[-15,+4]$ & 0.27 & 1.3 \\
\addlinespace
MgB$_2$ & $12^3$ & 12 & 0.766 & 0.817 & $+8$ & 5--8 & 1.28 & $[-13,+7.2]$ & 3.6 & 24 \\
        &        &    &       &       &      &      &      & $[-2,+2]$ & 3.4 & 13 \\
\addlinespace
SnTe$^{\rm b}$ & $8^3$ & 46 & 0.95$^{\rm a}$ & 0.975 & $+6.18^{\rm c}$ & 24--28 & 1.16 & $[-12,0]$ & 3.9 & 23 \\
     &       &    &      &       &      &        &      & $[0,+6]$ & 15.4 & 165 \\
\addlinespace
Sc-COF $\alpha^{\rm d}$ & $6^2{\times}1$ & 294 & 0.785 & 0.838 & $+8$ & 219--221 & 1.03 & $[-6,+3]$ & 0.11 & 0.84 \\
Sc-COF $\beta^{\rm d}$ & $6^2{\times}1$ & 294 & 0.609 & 0.837 & $+8$ & 219--221 & 1.04 & $[-6,+3]$ & 0.15 & 1.0 \\
\bottomrule
\end{tabular*}

\smallskip
\begin{minipage}{\textwidth}\footnotesize
$^{\rm a}$At the upper clip, because five pool bands above $E_{\mathrm{F}}+8$~eV project 0.955--0.96 onto the enlarged target and set the gate (Sec.~\ref{sec:snte}).\\
$^{\rm b}$Target enlarged automatically from $sp$ to $spd$, including the Sn-$4d$ semicore bands (Sec.~\ref{sec:snte}).\\
$^{\rm c}$Lowered progressively from $E_{\mathrm{F}}+8$~eV, then lowered further under a mixed-character pocket at X (Sec.~\ref{sec:masks}). At $E_{\mathrm{F}}+8$~eV the errors are 8.4 and 24.4~meV.\\
$^{\rm d}$Gauss--Seidel disentanglement, with \textsc{Wannier90} stopped at its $10^4$-iteration budget (Sec.~\ref{sec:defaults}).
\end{minipage}
\end{table*}

\subsection{Frozen extent, not frozen membership}
\label{sec:extent}

We first fix the extent and vary only the membership. At a fixed ceiling, with the pool, overlaps, solver, gauge and iteration budgets also held fixed, we compared two frozen sets, the trust mask of Eq.~\eqref{eq:masks} and the energy window below the same ceiling, which additionally freezes the untrusted states under it. The comparison was run with both solvers on the seven cells in which the two sets differ (Table~\ref{tab:extent}). The sets coincide for the LCAO Bi bilayer, for SnTe with 46 functions, whose trust mask is itself an energy window (Sec.~\ref{sec:snte}), and for LAPW MgB$_2$ and the two collinear Fe channels, while the Sc-COF channels, in which the sets do differ, were not run.

On four cells the window reduces the band error by factors of three to more than sixty. On h-BN, at the top of its own trusted set ($+12.65$~eV), freezing the 378 untrusted states below that ceiling lowers the error over the conduction band up to $+12$~eV from 52.2 to 4.9~meV, while the valence error rises slightly, from 1.3 to 1.5~meV. On MgB$_2$, at $E_{\mathrm{F}}+8$~eV, freezing the 689 untrusted states below the ceiling lowers the error over $[-13,+7.2]$~eV from 24.1 to 7.0~meV. On the LAPW frontend the comparison runs in the opposite direction, starting from the released window and releasing the untrusted states under the ceiling. On the Bi bilayer, releasing the 222 untrusted states lowers $\Omega_I$ from 80.2 to 58.4~\AA$^2$ but removes from the model the target band to which those states belong. The index-paired conduction error over $[0,+4]$~eV rises from 17.6 to 1168~meV, whereas the nearest-band metric, which matches each reference state to any model band, reports a largest deviation of only 11~meV (SM). On Fe with spin--orbit coupling, releasing the 2903 untrusted states, which are spread over every $\kk$-point, lowers $\Omega_I$ from 11.2 to 10.8~\AA$^2$ and raises the band errors from 4.5 and 5.5~meV to 174 and 301~meV. In that model the weakest link of the hand-off is 0.5 and \textsc{Wannier90} meets its criterion, so neither the spread nor the continuity check flags the degraded model (SM).

On the other three cells the two frozen sets give equivalent models. Releasing the 30 untrusted states under the ceiling of $\alpha$-Sn, or the 27 under that of SrVO$_3$, changes the band distance by at most 0.2~meV, and releasing in addition the $\alpha$-Sn shell states below $p_{\text{froz}}$, 76 states in all, lowers it by 0.1 and 2.8~meV (SM). On GaAs only a single state differs between the two sets, and the two models differ by at most 2.2~meV, which is within the dependence of this cell's Wannierization on the optimizer (SM).

The outcome does not depend on the solver, since with conjugate gradients the same four cells improve by similar factors. The trust mask then gives the lower error only on the h-BN valence band (by 0.1~meV), over the MgB$_2$ Fermi window (2.8 vs.\ 3.4~meV), and by at most 0.3~meV on GaAs, SrVO$_3$ and $\alpha$-Sn. In each of the four cells where the window is better, the trust mask nevertheless gives the lower spread, both in $\Omega_I$ and in $\Omega_{\rm tot}$, so a lower spread does not identify the better model.

Figure~\ref{fig:concept}(b)--(d) shows the h-BN case together with its certificate. At the mesh $\kk$-points the eigenvalues of a model do not depend on its Wannierization gauge, so they test the disentangled subspace alone. At these points the frozen membership already matters on h-BN, because with the trust mask frozen the subspace misses reference states over $[+5,+12]$~eV (100.6~meV RMS, 1.5~eV at most) that the window reproduces to 0.01~meV. For MgB$_2$, and for every route comparison of Table~\ref{tab:routes}, the models instead agree at the mesh points to 0.02~meV, so their differences arise between the mesh points. Rerunning the compared h-BN and MgB$_2$ models with one BLAS thread, with two MPI ranks or with guiding centers switched off moves their band errors by at most 0.1~meV (SM), with one exception, a window-route run on two MPI ranks, which is described in Sec.~\ref{sec:two-routes}.

\begin{table*}[t]
\caption{Frozen membership at fixed extent, on the seven cells where the two frozen sets differ. Each cell is run at the same ceiling $\varepsilon_{\rm c}$ once with the trust mask of Eq.~\eqref{eq:masks} frozen and once with every pool state below $\varepsilon_{\rm c}$ frozen (``window''), with the Gauss--Seidel sweep (GS) and with conjugate gradients (CG). ``Changed'' is the number of pool states $(n,\kk)$ below $\varepsilon_{\rm c}$, summed over the mesh, that the window freezes and the trust mask leaves free. The column ``Window (eV)'' gives the scoring windows relative to $E_{\mathrm{F}}$, and the columns headed ``Trust mask'' and ``Window'' give the two frozen sets. The band distance $\eta$ pairs bands by index (Sec.~\ref{sec:results}), and $\Omega_{\rm tot}$ is the total spread of the CG models after the Wannierization. Bi is the Bi bilayer, and Fe-SOC is bcc Fe with spin--orbit coupling. Three Gauss--Seidel runs stopped at the 30\,000-sweep cap with a flat tail (the Bi bilayer with both sets and the Fe-SOC trust mask, SM), and their CG counterparts converged.}
\label{tab:extent}
\small
\begin{tabular*}{\textwidth}{@{\extracolsep{\fill}}l c r c r r r r r r@{}}
\toprule
 & & & & \multicolumn{2}{c}{$\eta$, GS (meV)} & \multicolumn{2}{c}{$\eta$, CG (meV)} & \multicolumn{2}{c}{$\Omega_{\rm tot}$, CG (\AA$^2$)} \\
\cmidrule(lr){5-6}\cmidrule(lr){7-8}\cmidrule(l){9-10}
Cell & $\varepsilon_{\rm c}$ (eV) & Changed & Window (eV) & Trust mask & Window & Trust mask & Window & Trust mask & Window \\
\midrule
h-BN & $+12.65$ & 378 & $[-20,0]$ & 1.3 & 1.5 & 0.95 & 1.05 & 13.79 & 14.41 \\
 & & & $[+5,+8]$ & 1.9 & 1.3 & 1.7 & 1.1 & & \\
 & & & $[+5,+12]$ & 52.2 & 4.9 & 61.2 & 6.9 & & \\
\addlinespace
MgB$_2$ & $+8.0$ & 689 & $[-13,+7.2]$ & 24.1 & 7.0 & 11.3 & 3.6 & 15.00 & 15.40 \\
 & & & $[-2,+2]$ & 5.1 & 4.5 & 2.8 & 3.4 & & \\
\addlinespace
Bi, LAPW & $+4.0$ & 222 & $[-15,0]$ & 0.2 & 0.2 & 0.2 & 0.2 & 65.42 & 88.39 \\
 & & & $[0,+4]$ & 1167.5 & 17.6 & 1183.8 & 17.6 & & \\
\addlinespace
Fe-SOC, LAPW & $+8.0$ & 2903 & $[-10,0]$ & 174.0 & 4.5 & 155.9 & 4.5 & 11.48 & 12.73 \\
 & & & $[0,+4]$ & 301.2 & 5.5 & 226.9 & 5.7 & & \\
\addlinespace
$\alpha$-Sn, LAPW & $+6.0$ & 30 & $[-12,0]$ & 2.9 & 2.9 & 2.1 & 2.2 & 65.57 & 65.57 \\
 & & & $[0,+4]$ & 12.0 & 12.1 & 14.1 & 14.2 & & \\
\addlinespace
SrVO$_3$, LAPW & $+6.0$ & 27 & $[-8,0]$ & 0.5 & 0.7 & 0.5 & 0.7 & 20.92 & 23.47 \\
 & & & $[0,+4]$ & 1.8 & 1.6 & 1.8 & 1.6 & & \\
\addlinespace
GaAs, LAPW & $+8.0$ & 1 & $[-16,0]$ & 5.5 & 4.9 & 5.4 & 5.6 & 51.41 & 51.40 \\
 & & & $[0,+4]$ & 15.8 & 18.0 & 17.4 & 17.7 & & \\
\bottomrule
\end{tabular*}
\end{table*}

For the extent, we compared the released ceiling, lowered progressively from $E_{\mathrm{F}}+8$~eV, with a ceiling lowered progressively from the top of the trusted set (Table~\ref{tab:ceiling}). Both ceilings freeze the whole window below them, and the comparison covers every cell on which the two ceilings differ. They coincide for the LCAO Bi bilayer, LCAO MgB$_2$, the Sc-COF, SrVO$_3$ and the LAPW Bi bilayer, and for SnTe, where both searches end under the mixed-character pocket at $E_{\mathrm{F}}+6.18$~eV (Sec.~\ref{sec:masks}). Raising the ceiling moves fidelity from the energy range below $E_{\mathrm{F}}+8$~eV to the range above it. On h-BN the conduction error over $[+5,+8]$~eV falls from 5.2 to 1.1~meV and the band above $+8$~eV enters the model, while the valence error rises from 0.56 to 1.05~meV, and GaAs and the LAPW MgB$_2$ cell likewise improve in one window at the expense of the other.

Where the raised ceiling encloses as many states as the Wannier count can hold, both windows can degrade. In bcc Fe the raised ceiling freezes all nine directions at the worst $\kk$-point and fidelity is lost in both spin channels, and Fe with spin--orbit coupling, which then freezes 18 of 18 directions, degrades in the same way, from 4.5 and 5.5 to 8.4 and 6.9~meV. On $\alpha$-Sn the search from the top of the trusted set ($+9.79$~eV) ends at $+5.79$~eV, 0.2~eV below the released ceiling, only because its 1~eV steps start from a different energy. This slightly lower ceiling improves both windows, from 2.2 and 14.2 to 2.1 and 9.8~meV, which shows that a capacity-limited ceiling depends on the step of the search as well as on its starting energy. The released default ties the ceiling to the declared fidelity range, and an explicit ceiling extends it for a target above $E_{\mathrm{F}}+8$~eV. Where both Wannierizations converge, the two routes give similar models at a given extent (Fig.~\ref{fig:concept}f and Sec.~\ref{sec:two-routes}). Switching off the admission floor, finally, changes the Gauss--Seidel h-BN hand-off at $E_{\mathrm{F}}+8$~eV by at most 0.5~meV (0.79 and 7.92 vs.\ 0.77 and 7.42~meV, with bands paired by index, SM).

\begin{table}[t]
\caption{Band distance $\eta$ for two starting points of the ceiling search. The released rule starts at $E_{\mathrm{F}}+8$~eV and resolves to $+8.0$~eV in every row except $\alpha$-Sn ($+6.0$~eV), and the alternative starts at the top of the trusted set (second column). Both freeze every pool state below the resolved ceiling, and $\eta$ pairs bands by index over the scoring windows, which are relative to $E_{\mathrm{F}}$ (Sec.~\ref{sec:results}). The h-BN and $\alpha$-Sn rows use CG disentanglement, and the other LAPW rows use the Gauss--Seidel sweep, which was the default when they were made (Sec.~\ref{sec:dis}).}
\label{tab:ceiling}
\footnotesize\setlength{\tabcolsep}{3pt}
\begin{tabular*}{\columnwidth}{@{\extracolsep{\fill}}l c c r r@{}}
\toprule
 & & & \multicolumn{2}{c}{$\eta$ (meV)} \\
\cmidrule(l){4-5}
Cell & Trusted & Window & $E_{\mathrm{F}}{+}8$ & Trusted-top \\
 & top (eV) & (eV) & start & start \\
\midrule
h-BN & $+12.65$ & $[-20,0]$ & 0.56 & 1.05 \\
 & & $[+5,+8]$ & 5.2 & 1.1 \\
 & & $[+5,+12]$ & 1800 & 6.9 \\
\addlinespace
GaAs, LAPW & $+10.93$ & $[-16,0]$ & 4.9 & 7.1 \\
 & & $[0,+4]$ & 18.0 & 12.1 \\
\addlinespace
MgB$_2$, LAPW & $+13.06$ & $[-13,+7.2]$ & 11.5 & 8.0 \\
 & & $[-2,+2]$ & 4.4 & 7.9 \\
\addlinespace
$\alpha$-Sn, LAPW & $+9.79$ & $[-12,0]$ & 2.2 & 2.1 \\
 & & $[0,+4]$ & 14.2 & 9.8 \\
\addlinespace
Fe$\uparrow$, LAPW & $+14.64$ & $[-10,0]$ & 8.0 & 8.4 \\
 & & $[0,+4]$ & 15.1 & 21.3 \\
\addlinespace
Fe$\downarrow$, LAPW & $+16.32$ & $[-10,0]$ & 11.8 & 15.8 \\
 & & $[0,+4]$ & 7.0 & 11.0 \\
\addlinespace
Fe-SOC, LAPW & $+16.63$ & $[-10,0]$ & 4.5 & 8.4 \\
 & & $[0,+4]$ & 5.5 & 6.9 \\
\bottomrule
\end{tabular*}

\end{table}

\subsection{Freezing methods compared on identical h-BN inputs}
\label{sec:baseline}

The per-state mode (\texttt{-\kern0pt-hybrid-froz-emax none}) freezes the trust mask itself instead of the window. In this mode, two controlled experiments at the $6^3$ mesh show that the band-average gate decides whether the minimization converges and that the state threshold sets the conduction fidelity (SM). The released default model of this cell, which freezes the window, is shown in Fig.~\ref{fig:gallery}(d).

\begin{figure*}[tp]
\centering
\includegraphics[scale=0.67]{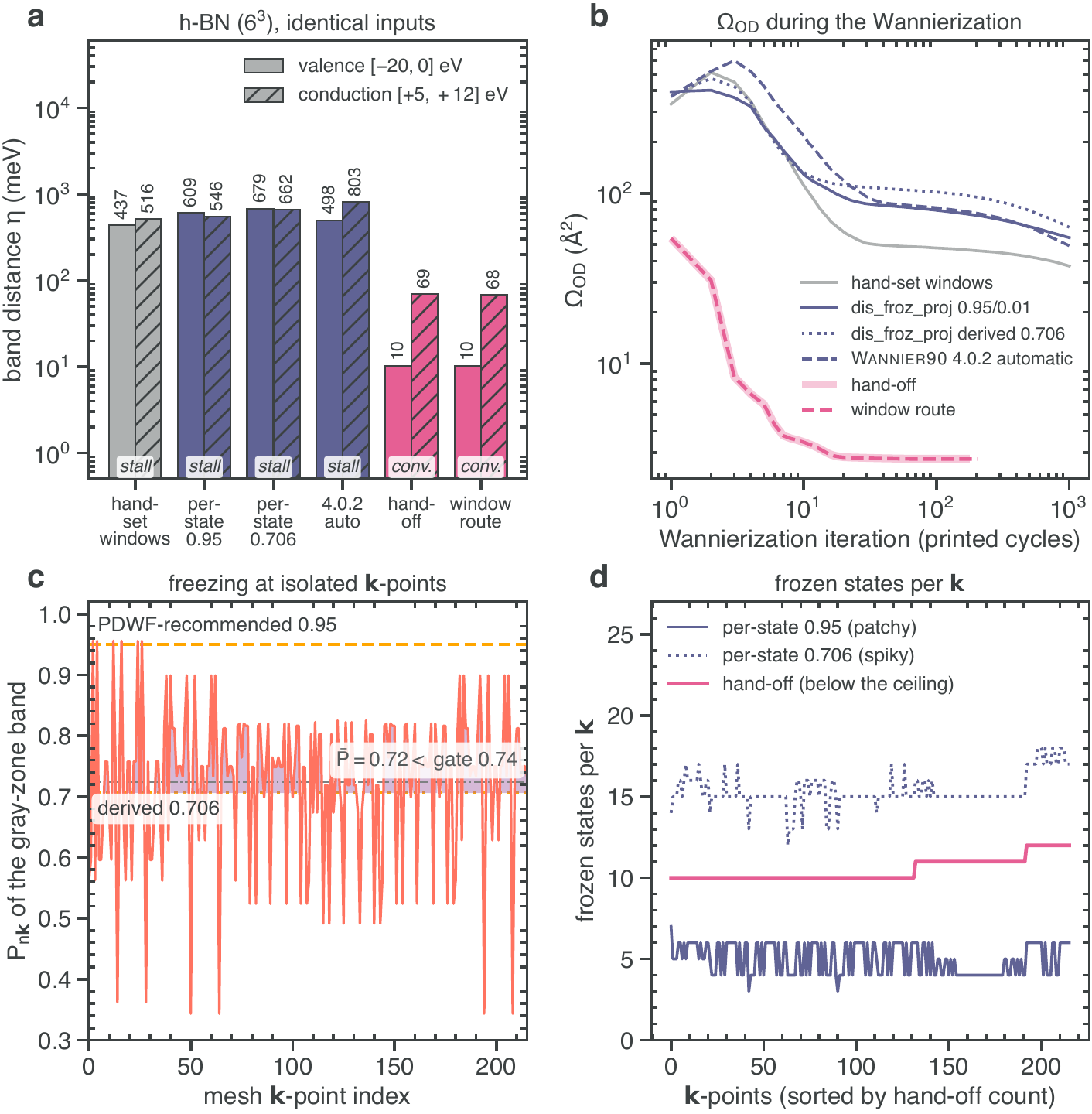}
\caption{Freezing methods on identical h-BN inputs ($6^3$ mesh, 16 functions, and the same pool, overlaps and target chemistry). (a)~Band distance over the valence window $[-20,0]$~eV and the conduction window $[+5,+12]$~eV. The four \textsc{Wannier90} settings are hand-set energy windows, per-state freezing (\texttt{dis\_froz\_proj}, without a frozen energy window) at the PDWF-recommended $0.95/0.01$ (freezing/discard) and at the derived freezing threshold $0.706$, and the automatic thresholds of \textsc{Wannier90}~4.0.2 (\texttt{dis\_proj\_auto}, $0.805/0.180$), and all four stall. The two released routes, with the ceiling at the top of the trusted set ($+12.65$~eV), converge, and the window route starts from the SCDM gauge of the hand-off. (b)~$\Omega_{\text{OD}}$ during the Wannierization, where the window route (dashed) lies on the hand-off, which is drawn as a broad pink band. (c)~Projectability $P_{n\kk}$ of the gray-zone band at each mesh $\kk$-point. Its average, 0.72, lies below the gate (0.74). A threshold on single states freezes this band at 3\% of the $\kk$-points at 0.95 and at 72\% at 0.706 (purple shading), which is the $\kk$-discontinuous freezing of Sec.~\ref{sec:masks}. (d)~Frozen states per $\kk$. Thresholds on single states give patchy (0.95) or spiky (0.706) counts, whereas the hand-off freezes every pool state below its ceiling. The \textsc{Wannier90} runs use the development build \texttt{ff955d1}, except the automatic-threshold run, which uses v4.0.2.}
\label{fig:baseline}
\end{figure*}

The controlled baseline of Fig.~\ref{fig:baseline} tests the first of these findings directly, by varying the frozen constraint, and then the starting gauge, on identical inputs. Every run used the same inputs from the same multi-zeta calculation, namely the band pool, the analytic overlaps, the L\"owdin target chemistry compressed by SVD to $N_{\text{proj}}=N_w$ (the only form that \textsc{Wannier90} accepts), and the iteration budgets. These inputs were run with four settings of standard \textsc{Wannier90} and with the two released routes, and each pair of numbers below gives the band errors over the valence window $[-20,0]$~eV and the conduction window $[+5,+12]$~eV.

The four \textsc{Wannier90} settings are hand-set energy windows, per-state freezing (\texttt{dis\_froz\_proj}) at the PDWF-recommended thresholds $0.95/0.01$~\cite{Qiao2023}, the same per-state freezing at our derived threshold $0.706$, and the automatic thresholds of \textsc{Wannier90}~4.0.2 (\texttt{dis\_proj\_auto}~\cite{W90v402}, here $0.805/0.180$). The per-state runs use no frozen energy window, whereas Ref.~\cite{Qiao2023} recommends combining the thresholds with one that reaches 2~eV above the conduction-band minimum. All four stall, at band errors of 0.44 to 0.80~eV (Fig.~\ref{fig:baseline}a, with every value listed in the SM). The $0.95$ threshold was calibrated on pseudo-atomic plane-wave projectabilities and sits at the all-electron L\"owdin retention bound $\Pi_c$ of Sec.~\ref{sec:auto}, so it freezes only 1--7 scattered states per $\kk$-point, which is exactly the $\kk$-discontinuous freezing that the band gate removes (Sec.~\ref{sec:masks}). The derived threshold freezes 8--14 states per $\kk$-point and gives even larger errors. Changing the threshold therefore does not remove the stall, which we attribute to freezing without a band gate.

The two released routes, run at the same extent with the ceiling at the top of the trusted set ($+12.65$~eV), both converge, at 10.1 and 69~meV, with the window route started from the SCDM gauge of the hand-off. The starting gauge has a separate effect, because from its own per-$\kk$ SVD starting projections the window route had not met its criterion after 500\,000 iterations, at 575 and 470~meV. Both released routes also converge when whole bands of conduction character with $P\approx0.7$--$0.8$ are frozen (Fig.~\ref{fig:baseline}a). In every configuration we tested, the stalls attributed to freezing weakly projected conduction states therefore came from the constraints and the starting gauge, that is, from hand-set windows or ungated per-state thresholds combined with atomic-orbital starting gauges. The variational problem itself is solvable, since the window derived from the automatic ceiling converges (Sec.~\ref{sec:extent}). A frozen set cannot, however, compensate for a target that does not span the bands. On SnTe the per-state channel study of the SM gives a band root-mean-square (RMS) error of 1568~meV over the conduction window for the $sp$ target, whichever states are frozen, and the conduction bands are reproduced only after channels are added to the target (Sec.~\ref{sec:snte}). That study used the earlier Fermi reference, which was 1.43~eV too high, and it was not repeated at CRYSTAL's $E_{\mathrm{F}}$.

\subsection{One fixed set of defaults on five cells}
\label{sec:defaults}

\begin{figure*}[tp]
\centering
\includegraphics[scale=0.67]{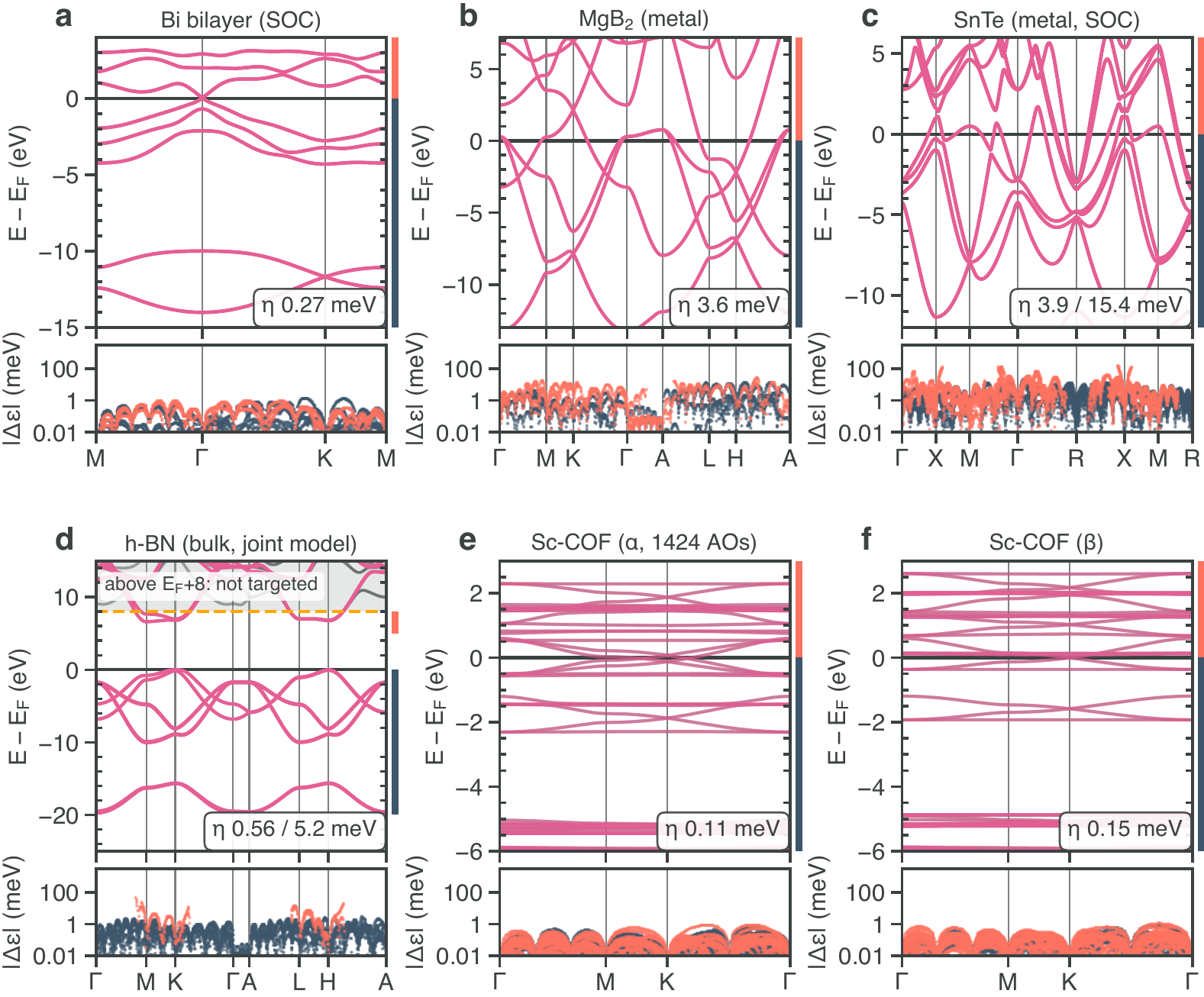}
\caption{The six production runs at the released defaults, (a)~the Bi bilayer (SOC, path M--$\Gamma$--K--M), (b)~MgB$_2$ (metal), (c)~SnTe (B2 metal, SOC, with $d$ channels added to the target automatically), (d)~bulk h-BN (joint valence-and-conduction model, drawn in gray above $E_{\mathrm{F}}+8$~eV, which is outside its target range), and (e) and (f)~the $\alpha$ and $\beta$ spin channels of the Sc-COF (1424 AOs). Each panel freezes every pool state below the automatic ceiling, which on SnTe is lowered to $E_{\mathrm{F}}+6.18$~eV under the X pocket (Fig.~\ref{fig:snte}), and draws the Wannier interpolation (pink) over the LCAO reference (thin gray). Edge bars mark the scoring windows of Table~\ref{tab:defaults} (slate below $E_{\mathrm{F}}$, coral above), and the box gives the band distance $\eta$ over them, with bands paired by index, as in the table. For SnTe and h-BN the two numbers are the occupied and the empty window, and for MgB$_2$ the box gives the wide window only. The lower strips give $|\Delta\varepsilon|$ for every paired state inside the windows on a log scale (slate occupied, coral empty).}
\label{fig:gallery}
\end{figure*}

Table~\ref{tab:defaults} and Fig.~\ref{fig:gallery} present the central validation, in which five cells are run through the identical pipeline with every threshold and the frozen ceiling derived automatically. The cells span a joint valence-and-conduction model of a wide-gap insulator, a two-dimensional bilayer with strong spin--orbit coupling, a multiband metal, a strong-SOC metal whose target needs added channels, and a magnetic framework at production scale. Each row is a \emph{single} joint Wannier model that covers the valence and low conduction bands together.

The two bulk hexagonal cells are reported on $K/H$-commensurate meshes ($3\,|\,N$ in-plane), because a mesh that misses $K$ and $H$ changes the band errors (the SM uses the earlier per-state models as an example), and the $\kk$-mesh audit of Sec.~\ref{sec:auto} reports such a mesh. Every row of Table~\ref{tab:defaults} freezes the window below the resolved ceiling (Sec.~\ref{sec:masks}), and per-state frozen sets appear only in the controlled comparisons of Secs.~\ref{sec:extent} and~\ref{sec:baseline}. The h-BN row is the model for its declared range, which extends to $E_{\mathrm{F}}+8$~eV. With the ceiling raised to the top of its trusted set ($+12.65$~eV), the same pipeline (CG) reproduces the conduction band up to $+12$~eV at 6.9~meV and gives 1.1~meV over $[+5,+8]$~eV, while the valence error rises from 0.56 to 1.05~meV (Table~\ref{tab:ceiling} and Sec.~\ref{sec:extent}).

\begin{figure*}[t]
\centering
\includegraphics[scale=0.67]{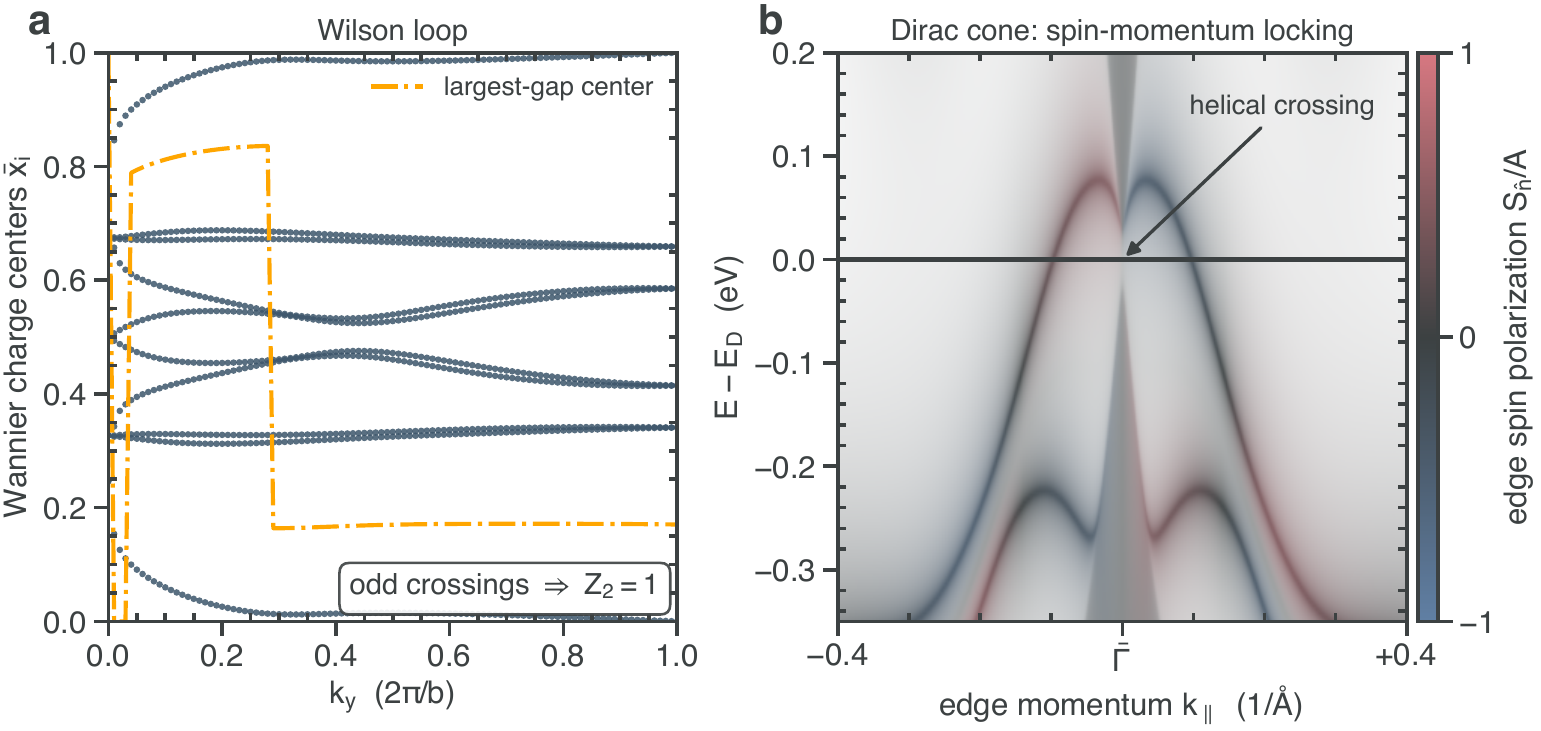}
\caption{Topology of the 16-function Bi bilayer model (band distance 0.27~meV, Table~\ref{tab:defaults}), computed with WannierTools~\cite{Wu2018}. (a)~Wilson loop of the Wannier charge centers. The largest-gap center (dash-dotted) is crossed an odd number of times, so $Z_2=1$. (b)~Spin-resolved edge spectral function near the edge Dirac point, from the surface Green's function~\cite{Sancho1985}. The intensity gives the spectral weight $A$, and the color gives the edge spin polarization $S_{\hat n}/A$, where $S_{\hat n}$ is the spectral function projected on spin along $\hat n$. Energies are relative to $E_D$, the crossing point of the two edge branches in the middle of the 52.8~meV bulk gap. The SM shows the input projectabilities, the edge Fermi points and the quantized ballistic conductance.}
\label{fig:topo}
\end{figure*}

\emph{Topology.} Figure~\ref{fig:topo} shows a topological characterization~\cite{Wu2018,Sancho1985,Soluyanov2011} of the automatically generated Bi bilayer model, whose band distance is 0.27~meV. Wilson-loop Wannier charge centers give $Z_2=1$~\cite{KaneMele2005,FuKane2006}, and the edge spectral function shows the gapless helical Dirac crossing inside the $52.8$~meV inverted bulk gap, consistent with the quantum spin Hall state predicted for Bi(111) bilayers~\cite{Murakami2006}.

The Wilson loop is computed on the Wannier model itself, so it is direct evidence for the topology of the model. Because the model reproduces the bands to sub-meV accuracy and keeps the inverted gap open, we take its invariant to be that of the underlying spin--orbit calculation. LCAO calculations of heavy-element materials with hybrid exchange--correlation functionals~\cite{Bodo2022,Comaskey2022}, whose Fock exchange is needed for a correct treatment of spin--orbit coupling~\cite{Desmarais2019b}, can therefore be analyzed with Wannier-based tools for $Z_2$ invariants, surface spectra and transport.

\emph{Scale.} The largest cell we ran is the Sc-COF, a two-dimensional scandium covalent organic framework with composition C$_{36}$H$_{24}$B$_{12}$O$_{12}$Sc$_5$ and with 89 atoms and 1424 AOs per cell. It is spin-polarized and has 294 Wannier functions (WFs) in each spin channel. Its projectability distribution is also the least favorable case for the automatic rule, because the band-averaged projectabilities of the 441 pool bands fall from 0.98 to 0.39--0.43 without any jump larger than 0.07, so that the gray plateau and the continuum are not separated by a gap (Appendix~\ref{app:anatomy}). The bands are nevertheless interpolated at 0.11 and 0.15~meV ($\alpha$ and $\beta$) over $[-6,+3]$~eV, with one command per spin channel (SM). The frozen ceiling resolves to $E_{\mathrm{F}}+8$~eV in both channels, with 221 of 294 directions frozen at the worst $\kk$-point. The constrained $\Omega_I$ minimization used the Gauss--Seidel sweep, which was the default when these models were made, and it converged in 14\,806 and 13\,881 sweeps within the cap of 20\,000, and the models were not rerun with CG. Their \textsc{Wannier90} runs stopped at the budget of 10\,000 iterations, with per-iteration changes of $|\Delta\Omega|=2.7\times10^{-9}$ and $7.1\times10^{-6}$~\AA$^2$ ($\alpha$ and $\beta$), so these are the only models in the paper that were not run to the convergence criterion, and they are scored at that point. They improve on the earlier per-state model (1.23 and 0.25~meV), whose disentanglement had stopped at its sweep budget, but extent, frozen membership and convergence all changed between those runs, so the improvement cannot be attributed to any one of them.

\subsection{SnTe: channel augmentation, Wannier count and subspace continuity}
\label{sec:snte}

SnTe is computed in the B2 cell, a metal with strong spin--orbit coupling (Sec.~\ref{sec:results}). Its chemical $sp$ target cannot represent the conduction manifold, because the four lowest conduction bands have a coverage of only 0.557 on it. The automatic augmentation adds the Sn-$d$ and Te-$d$ channels (coverage 0.884) and rejects Te-$f$, whose specificity of $-0.24$ shows that its coverage gain only reflects the completeness of the basis. The count rule of Eq.~\eqref{eq:decoupled} then gives $N_w=46$, made up of the 16 valence functions, a $d$ shell on each atom, and 10 functions for the Sn-$4d$ semicore bands inside the pool. The channel-set study behind these rules (SM) was run in the per-state mode with the earlier Fermi reference, and the lower block of that table compares the 36- and 46-function counts at the released defaults.

\begin{figure*}[tp]
\centering
\includegraphics[scale=0.67]{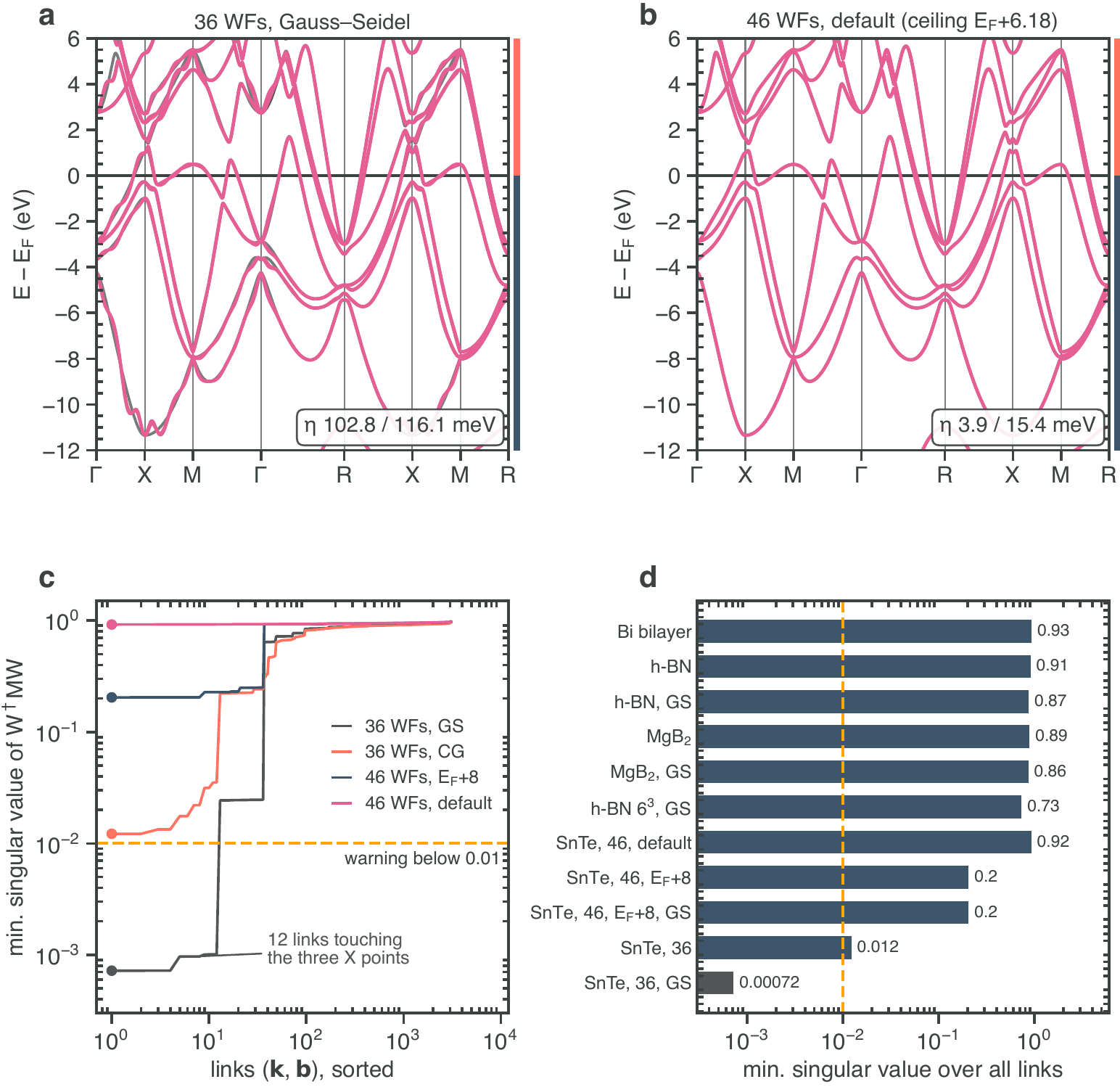}
\caption{SnTe in the B2 cell (metal, SOC, $8^3$ mesh, CRYSTAL's $E_{\mathrm{F}}$). (a)~The earlier count of 36 WFs with Gauss--Seidel disentanglement and the ceiling at $E_{\mathrm{F}}+8$~eV, for which \textsc{Wannier90} had not converged after 300\,000 iterations. (b)~The default of 46 WFs with CG disentanglement and the ceiling lowered to $E_{\mathrm{F}}+6.18$~eV under the X pocket (Sec.~\ref{sec:masks}), for which \textsc{Wannier90} converged in 5\,180 iterations. The model (pink) is drawn over the LCAO reference (gray), and the boxes give the band distance $\eta$ over $[-12,0]$ and $[0,+6]$~eV (edge bars), with bands paired by index. (c)~The smallest singular value of $W^\dagger(\kk)M^{(\kk,\bb)}W(\kk+\bb)$ on every link, sorted, for four SnTe hand-offs. The 12 lowest links of the 36-function Gauss--Seidel hand-off touch the three X points of the simple-cubic zone, and the dashed line marks the warning threshold of the continuity check, $10^{-2}$. (d)~The smallest value over all links for each hand-off named on the axis, including the models of Table~\ref{tab:defaults}. The 46-function hand-offs at $E_{\mathrm{F}}+8$~eV with CG and with the Gauss--Seidel sweep agree to $4\times10^{-5}$.}
\label{fig:snte}
\end{figure*}

With the ceiling at $E_{\mathrm{F}}+8$~eV, 32 of the 46 directions are frozen at the worst $\kk$-point. The CG and Gauss--Seidel disentanglers reach the same minimum and the same subspace (SM), in which the smallest singular value of $W^\dagger(\kk)M^{(\kk,\bb)}W(\kk+\bb)$ is 0.20 for both, on links that touch the X points, while every other link lies above 0.93 (Fig.~\ref{fig:snte}c,d). \textsc{Wannier90} converges from both hand-offs, at 8.4 and 24.4~meV over $[-12,0]$ and $[0,+6]$~eV from the CG one and at 8.5 and 24.0~meV from the Gauss--Seidel one. The eigenvalues at the mesh $\kk$-points are exact to 0.01~meV, so the remaining error comes from the interpolation between them. It is largest over $[+4,+6]$~eV (29.8~meV), on path segments that touch X. There, a conduction pocket at $E_{\mathrm{F}}+6.28$~eV is frozen at 17\% of the $\kk$-points and its states are only partly of target character ($\langle P\rangle=0.72$), while the pair of bands above it is frozen only at the three X points and has $\langle P\rangle=0.59$. These are the conditions of the pocket skip (Sec.~\ref{sec:masks}), which lowers the ceiling to $E_{\mathrm{F}}+6.18$~eV. The weakest link then rises to 0.92, and \textsc{Wannier90} converges at 3.9 and 15.4~meV (Fig.~\ref{fig:snte}b), which is the row of Table~\ref{tab:defaults}. Freezing the pocket's states had also reduced the fidelity below the pocket, since with the lower ceiling the error falls from 29.8 to 22.9~meV over $[+4,+6]$~eV and from 20.3 to 7.2~meV over $[0,+4]$~eV. Adding Te-$f$ to the earlier 36-function target (50 functions) made both windows worse with the earlier Fermi reference (21.7 and 52.7~meV), which supports the rejection of Te-$f$.

Matching each reference state to the nearest model band gives lower values for this cell, 3.2 and 13.0~meV, because of the way the Kramers pairs are treated. The reference bands are Kramers pairs, doubly degenerate at every $\kk$ by inversion combined with time reversal, but their median splitting is 0.05~meV because CRYSTAL's two-component self-consistent field (SCF) uses neither symmetry~\cite{CrystalManual}. The model splits these pairs between mesh points (median 1.5~meV over $[-12,+6]$~eV), and nearest-band matching then lets both states of a pair match the nearer model state.

The earlier count rule (Sec.~\ref{sec:auto}) stopped at 36 functions, which leaves only 4 free directions per $\kk$-point. With these, the Gauss--Seidel disentanglement converges ($\Omega_I=49.20$~\AA$^2$ after 15\,889 sweeps), yet the subspace jumps between neighboring $\kk$-points. At the X points three Kramers pairs between $E_{\mathrm{F}}+4$ and $+11$~eV are frozen, but one mesh step away only two are, and the smallest singular value drops to $7\times10^{-4}$ on the 12 links that touch the three X points (Fig.~\ref{fig:snte}a,c). No change of gauge can remove such a jump, because rotating the frames at $\kk$ and at $\kk+\bb$ by unitary matrices leaves the singular values of $W^\dagger(\kk)M^{(\kk,\bb)}W(\kk+\bb)$ unchanged, and none of the gauges we tried on this subspace gives a usable model. The four SCDM candidates give 47--94~meV over the occupied window, \textsc{Wannier90}'s conjugate-gradient restarts (\texttt{num\_cg\_steps}~$=200$) give 102~meV, a gauge-continuous first stage (SM) gives 88.8~meV, and \textsc{Wannier90} does not converge in 300\,000 iterations (102.8 and 116.1~meV, bands paired by index). On the same 36-function target the CG disentangler stops at a different minimum, 1.25\% higher in $\Omega_I$ (49.82~\AA$^2$), whose weakest link is $1.2\times10^{-2}$, and its model converges in 42\,130 iterations at 21.8 and 47.1~meV. The additional Sn-$d$ shell of the 46-function count therefore removes both the jump and the dependence on the solver, lowers the band errors by factors of 2.6 and 1.9, and cuts the Wannierization from 42\,130 to 1\,460 iterations. The pipeline now reports the continuity of the subspace and warns below $10^{-2}$ before \textsc{Wannier90} runs (Sec.~\ref{sec:auto}).

The larger pool also changes the derived trust thresholds. With 46 functions the pool holds 69 bands, and five of the bands above $E_{\mathrm{F}}+8$~eV project 0.955--0.96 onto the enlarged target. These are bands of $d$ character that the target represents, rather than the gray plateau that the rule of Eq.~\eqref{eq:autotrust} expects outside the fidelity range, so the highest of them sets the gate (0.975) and $p_{\text{froz}}$ reaches its upper clip (0.95). Because every pool state below the ceiling is frozen whatever the thresholds are, the model is unchanged, but the trust mask then becomes an energy window at every $\kk$-point, with its top at $E_{\mathrm{F}}+7.64$~eV (Appendix~\ref{app:certificate}).

\subsection{Two automatic routes}
\label{sec:two-routes}

Both released programs also contain a second automatic method, the window route (\verb|--method window|), which passes energy windows to \textsc{Wannier90} instead of an isolated problem. It classifies the bands as in the projectability-disentangled method of Ref.~\cite{Qiao2023}, but it takes the two thresholds from the same distribution as the hybrid rule. The upper threshold is the band gate, and the lower one, which separates the continuum tier below the gate, is set at the largest jump in the sorted band-averaged projectabilities, or at half the gate when no jump exceeds 0.05. The frozen window ends at the automatic ceiling of the hybrid route (Eq.~\eqref{eq:walk}), every band with a state below that ceiling is exported, the outer window is chosen by the gauge-invariant spread, and \textsc{Wannier90} disentangles with $N_b>N_w$. Table~\ref{tab:routes} compares the two routes on the same inputs and at equal ceilings, and it lists the number of \textsc{Wannier90} iterations of every Wannierization. Because the ceiling and the inputs are common to both routes, they differ only in the parts fixed by their construction, namely the bands exported, the outer window and the states admitted to it, and the program that performs the disentanglement.

When both Wannierizations are run to convergence, the two routes give models of similar quality. The window route gives 0.27~meV on the Bi bilayer, the same as the hand-off, 0.4 and 5.2~meV on h-BN compared to 0.56 and 5.2~meV for the hand-off, and 4.4 and 1.4~meV on MgB$_2$ compared to 3.6 and 3.4~meV. On MgB$_2$ it also reaches a lower total spread for the 12 functions (13.03 vs.\ 15.40~\AA$^2$), and inside the same h-BN window the disentanglement of \textsc{Wannier90} reaches a lower $\Omega_I$ than the constrained minimization (9.52 vs.\ 9.79~\AA$^2$), which is a different local minimum of the non-convex functional.

The number of iterations that \textsc{Wannier90} needs differs between the routes, and which route needs more depends on the cell. On h-BN the hand-off reaches its criterion in 1\,400 iterations, whereas the window route is still far from its minimum after 10\,000 (258~meV over the valence band) and needs 18\,260 in one uninterrupted run with two threads. With one thread or with two MPI ranks the same Wannierization converges in 2\,800 and 4\,230 iterations (SM), so this count depends on the execution path. On MgB$_2$ the order is reversed, with 3\,040 iterations for the hand-off and 1\,690 for the window route. The window route also depends on the starting projections it is given. On h-BN at $6^3$, started from the per-$\kk$ SVD of its target projections, it had not met its criterion after 500\,000 iterations, whereas started from the SCDM gauge of the hand-off, with the same \texttt{.win}, \texttt{.mmn} and \texttt{.eig} files, it converges in 2\,260 iterations with the ceiling at $E_{\mathrm{F}}+8$~eV and in 1\,870 with the ceiling at the top of the trusted set. On MgB$_2$ with two MPI ranks the SVD start stalls (7.7 and 6.5~meV after 200\,000 iterations), while the SCDM start converges in 1\,140 iterations, to 4.5 and 1.8~meV.

On SnTe, at CRYSTAL's $E_{\mathrm{F}}$ with the ceiling at $E_{\mathrm{F}}+8$~eV and 46 functions, the two routes give different models. The hand-off (CG) converges in 1\,460 iterations at 8.4 and 24.4~meV. The window route exports 69 bands ($1.5N_w$), and its disentanglement converges in 10\,197 iterations to a lower $\Omega_I$ than that of the hand-off (45.14 vs.\ 46.52~\AA$^2$). It reaches the lower value by admitting every state in the outer window, including the states below the projectability floor of the hand-off, and its subspace jumps between neighboring $\kk$-points at the three X points, where the smallest singular value of $V^\dagger(\kk)M^{(\kk,\bb)}V(\kk+\bb)$ is 0.031, compared to 0.20 for the hand-off. Restarted from the converged disentanglement, the default optimizer of \textsc{Wannier90} meets the criterion after 29\,950 iterations, at a lower total spread than the hand-off (53.66 vs.\ 56.41~\AA$^2$) but at larger band errors of 10.0 and 31.3~meV. Its eigenvalues at the mesh $\kk$-points are exact to 0.01~meV, so the larger error comes from the interpolation between them. The two-stage optimizer of the SM ends at 12.2 and 31.9~meV from the per-$\kk$ SVD starting projections and at 14.4 and 34.3~meV from the SCDM gauge. Removing the floor from the hand-off reproduces both the lower $\Omega_I$ (45.14~\AA$^2$) and the jump at X and raises its errors to 16.2 and 34.4~meV, so on this cell it is the admission floor that keeps the subspace of the hand-off continuous at X.

On the LAPW frontend, the window route of the GaAs cell, at the same $E_{\mathrm{F}}+8$~eV ceiling and within the same budget of 50\,000 iterations, gives 5.5 and 17.1~meV compared to 4.9 and 18.0~meV for the hand-off, at $\Omega_{\rm tot}/N_w=2.78$ compared to 2.92~\AA$^2$. When each route is run to its criterion in a single run, the hand-off stays at 4.9 and 18.0~meV (59\,500 iterations) and the window route gives 5.6 and 17.1~meV (66\,500 iterations, $\Omega_{\rm tot}/N_w=2.78$). Over the five cells, band fidelity therefore favors neither route on the Bi bilayer, h-BN and GaAs, favors the window route over the Fermi window of MgB$_2$, and favors the hand-off on SnTe. The hand-off also provides an isolated, already disentangled $N_b{=}N_w$ problem for tools that need one, together with a gauge built from the density matrix, and it reaches a converged model at lower cost (Table~\ref{tab:routes}). The window route is preferable when the disentanglement should be performed by \textsc{Wannier90} and its tools themselves.

The last two columns of Table~\ref{tab:routes} give the wall time of each route to a converged model on one machine, and the SM breaks it down by stage. On the entangled LCAO cells the window route costs 6.6 (MgB$_2$), 8.1 (h-BN) and 21 (SnTe) times as much as the hand-off, and both stages contribute to the difference. The disentanglement in \textsc{Wannier90} takes 1\,398 to 10\,197 iterations at 0.28 to 0.70~s each, whereas the constrained minimization takes 149 to 655 CG gradient evaluations at 0.05 to 0.20~s each. The Wannierization of the window route also needs more iterations than that of the hand-off on h-BN and SnTe, although fewer on MgB$_2$. The h-BN ratio depends on the execution path, because with the 2\,800 or 4\,230 Wannierization iterations of the one-thread and two-rank runs, at the measured time per iteration, the window route would cost 3.0 or 3.5 times as much as the hand-off. On the isolated Bi bilayer there is nothing to disentangle, and the two routes cost the same. On GaAs the Wannierization dominates both routes, and the window route costs 1.2 times as much. Each time is a single wall-clock measurement on a machine that the runs of different cells shared (SM), so the ratios carry the corresponding uncertainty.

\begin{table*}[t]
\caption{The two automatic routes at equal frozen ceilings ($E_{\mathrm{F}}+8$~eV, and $E_{\mathrm{F}}+7$~eV for the Bi bilayer). The table gives the band distance $\eta$ over the scoring windows of the second column (relative to $E_{\mathrm{F}}$), with bands paired by index (Sec.~\ref{sec:results}), and the number of Wannierization iterations in \textsc{Wannier90}. The window route was first run within an iteration budget and, where the budget stopped it (h-BN, SnTe and GaAs), continued to its convergence criterion, so the criterion columns are empty for the Bi bilayer and MgB$_2$, which converged within the budget. Every disentanglement converged (SM), the LCAO hand-offs use CG disentanglement, and the GaAs hand-off is the Gauss--Seidel model of Table~\ref{tab:lapw}. The time is the wall time from the CRYSTAL output or the \textsc{wien2wannier} files to the converged model, on one machine (Apple M5 Pro, two threads per LCAO run and one per LAPW \textsc{Wannier90} run), and the SM breaks it down by stage.}
\label{tab:routes}
\small
\begin{tabular*}{\textwidth}{@{\extracolsep{\fill}}l c r r r r r r r r@{}}
\toprule
 & & \multicolumn{2}{c}{Hand-off} & \multicolumn{2}{c}{Window route, budget} & \multicolumn{2}{c}{Window route, criterion} & \multicolumn{2}{c}{Time (s)} \\
\cmidrule(lr){3-4}\cmidrule(lr){5-6}\cmidrule(lr){7-8}\cmidrule(l){9-10}
Cell & Window (eV) & $\eta$ (meV) & Iter. & $\eta$ (meV) & Iter. & $\eta$ (meV) & Iter. & Hand-off & Window \\
\midrule
Bi bilayer & $[-15,+4]$ & 0.27 & 2\,230 & 0.27 & 2\,170 & & & 52 & 55 \\
\addlinespace
h-BN & $[-20,0]$ & 0.56 & 1\,400 & 258 & $10^4$\,$^{\rm a}$ & 0.4 & 18\,260 & 312 & 2\,519$^{\rm f}$ \\
 & $[+5,+8]$ & 5.2 & & 11.7 & & 5.2 & & & \\
\addlinespace
MgB$_2$ & $[-13,+7.2]$ & 3.6 & 3\,040 & 4.4 & 1\,690 & & & 462 & 3\,040 \\
 & $[-2,+2]$ & 3.4 & & 1.4 & & & & & \\
\addlinespace
SnTe$^{\rm b}$ & $[-12,0]$ & 8.4 & 1\,460 & -- & $10^4$\,$^{\rm c}$ & 10.0 & 29\,950$^{\rm d}$ & 1\,338 & 28\,281$^{\rm e}$ \\
 & $[0,+6]$ & 24.4 & & -- & & 31.3 & & & \\
\addlinespace
GaAs, LAPW & $[-16,0]$ & 4.9 & 59\,500 & 5.5 & $5\times10^4$\,$^{\rm a}$ & 5.6 & 66\,500 & 3\,546 & 4\,094 \\
 & $[0,+4]$ & 18.0 & & 17.1 & & 17.1 & & & \\
\bottomrule
\end{tabular*}

\smallskip
\begin{minipage}{\textwidth}\footnotesize
$^{\rm a}$Stopped at the budget, where the per-iteration change $|\Delta\Omega|$ was $6.7\times10^{-4}$ (h-BN) and
$6\times10^{-7}$~\AA$^2$ (GaAs).\\
$^{\rm b}$Both routes use the ceiling $E_{\mathrm{F}}+8$~eV. Because the window route has no pocket skip, the hand-off (46 functions) is compared at
the automatic ceiling before the pocket skip, which the default model of Table~\ref{tab:defaults} applies.\\
$^{\rm c}$Stopped at the budget (per-iteration $|\Delta\Omega|\approx5\times10^{-8}$~\AA$^2$). Because \textsc{Wannier90} writes no checkpoint during the Wannierization, there is no model to score at the budget.\\
$^{\rm d}$Run with 69 bands and restarted from the converged disentanglement (10\,197 iterations) with the default optimizer, a restart that keeps the gauge because $N_b>N_w$. The two-stage optimizer (SM) gives 12.2 and 31.9~meV from the per-$\kk$ SVD start and 14.4 and 34.3~meV from the SCDM gauge.\\
$^{\rm e}$The Wannierization was timed in a separate run, restarted from the disentanglement checkpoint while other jobs ran on the machine.\\
$^{\rm f}$One run with two threads (18\,260 Wannierization iterations). With one thread or with two MPI ranks the same Wannierization converges in 2\,800 and 4\,230 iterations (SM), which at the measured time per iteration would take about 930 and 1\,080~s.
\end{minipage}
\end{table*}

\subsection{Portability: the LAPW frontend}
\label{sec:lapw-results}

The frozen-ceiling rule, the trust masks and the hand-off need only the projectabilities of the bands onto a labeled set of localized orbitals, so they do not depend on the LCAO basis. The procedure of Algorithm~\ref{alg:stage2} therefore runs unchanged on the all-electron LAPW code WIEN2k~\cite{Blaha2020} through \textsc{wien2wannier}~\cite{Kunes2010}, in our modified release v3.0 (see Data and code availability), with every LCAO-specific input replaced by its LAPW analogue. The radial projectors are continued past the muffin-tin radius into the interstitial region (Appendix~\ref{app:lapw}). The constraint and the hand-off are also unchanged, except that the LAPW models admit every pool state and that the gauge construction keeps the SCDM columns only at well-conditioned $\kk$-points and transports the gauge to the others (Appendix~\ref{app:lapw}). Table~\ref{tab:lapw} lists eight models from seven WIEN2k calculations, with the thresholds and the frozen ceiling derived as in the LCAO frontend (the ceiling by Eq.~\eqref{eq:walk}), on meshes that contain the high-symmetry points of each cell, and with the disentanglement settings of Appendix~\ref{app:lapw}. The cells cover an insulator (GaAs), a metal (MgB$_2$), a band-inverted semimetal with spin--orbit coupling ($\alpha$-Sn, 20 functions), a two-dimensional bilayer with spin--orbit coupling (Bi), a ferromagnet (bcc Fe) treated both in separate spin channels and with spin--orbit coupling, and a correlated oxide (SrVO$_3$, with V~$t_{2g}$, O~$p$ and Sr~$d$ functions). Every occupied window is reproduced to 12~meV or better and every low conduction window to 18~meV or better, with one command per cell and no threshold or window set by hand. The Wannierization of the Bi bilayer uses guiding centers, and the bands of all eight models are shown in Appendix~\ref{app:lapw}.

MgB$_2$ was run in both frontends with the same window, Wannier count and frozen ceiling ($E_{\mathrm{F}}+8$~eV) and on $K/H$-commensurate meshes, but with different functionals, PBE with 10\% Fock exchange in CRYSTAL and PBE in WIEN2k. Over $[-13,+7.2]$~eV the LCAO model has an error of 7.0~meV with the Gauss--Seidel sweep (3.6~meV with CG, Table~\ref{tab:defaults}) and the LAPW model one of 11.5~meV. The LAPW value depends on the order in which the sweep visits the $\kk$-points. The LAPW Gauss--Seidel models were made with the colored order (Appendix~\ref{app:lapw}), and with the serial order of the LCAO sweep the LAPW hand-off reaches a lower minimum of $\Omega_I$ and gives 7.0~meV (SM). Because each model is scored against its own reference bands on its own mesh, these numbers do not compare the two basis sets.

\begin{table}[t]
\caption{WIEN2k (LAPW) models with the thresholds and ceiling of the LAPW release (Appendix~\ref{app:lapw}). All models use the L\"owdin full-cell projectability, freeze every pool state below the automatic ceiling $\varepsilon_{\rm c}$, and use the SCDM gauge and Gauss--Seidel disentanglement in the colored order with over-relaxation, except $\alpha$-Sn, which uses CG. $\Omega_{\rm tot}/N_w$ is the total spread per Wannier function after the Wannierization, and the band distance $\eta$ is measured against the \texttt{lapw1}($+$\texttt{so}) eigenvalues on the band paths in each scoring window (relative to $E_{\mathrm{F}}$), with bands paired by index (Sec.~\ref{sec:results}). Bi is the Bi bilayer, and the two spin channels of collinear Fe are separate rows.}
\label{tab:lapw}
\footnotesize\setlength{\tabcolsep}{3pt}
\begin{tabular*}{\columnwidth}{@{\extracolsep{\fill}}l c r c c c r@{}}
\toprule
Cell & $\kk$ mesh & $N_w$ & $\varepsilon_{\rm c}$ & $\Omega_{\rm tot}/N_w$ & Window & $\eta$ \\
 & & & (eV) & (\AA$^2$) & (eV) & (meV) \\
\midrule
GaAs & $8^3$ & 18 & $+8.0$ & 2.92 & $[-16,0]$ & 4.9 \\
 & & & & & $[0,+4]$ & 18.0 \\
\addlinespace
MgB$_2$ & $12^2{\times}8$ & 12 & $+8.0$ & 1.51 & $[-13,+7.2]$ & 11.5 \\
 & & & & & $[-2,+2]$ & 4.4 \\
\addlinespace
$\alpha$-Sn (SOC) & $8^3$ & 20 & $+6.0$ & 3.28 & $[-12,0]$ & 2.2 \\
 & & & & & $[0,+4]$ & 14.2 \\
\addlinespace
Bi (SOC) & $12^2{\times}1$ & 36 & $+4.0$ & 2.45 & $[-15,0]$ & 0.2 \\
 & & & & & $[0,+4]$ & 17.6 \\
\addlinespace
bcc Fe$\uparrow$ & $8^3$ & 9 & $+8.0$ & 0.83 & $[-10,0]$ & 8.0 \\
 & & & & & $[0,+4]$ & 15.1 \\
\addlinespace
bcc Fe$\downarrow$ & $8^3$ & 9 & $+8.0$ & 0.84 & $[-10,0]$ & 11.8 \\
 & & & & & $[0,+4]$ & 7.0 \\
\addlinespace
bcc Fe (SOC) & $8^3$ & 18 & $+8.0$ & 0.71 & $[-10,0]$ & 4.5 \\
 & & & & & $[0,+4]$ & 5.5 \\
\addlinespace
SrVO$_3$ & $8^3$ & 19 & $+6.0$ & 1.24 & $[-8,0]$ & 0.7 \\
 & & & & & $[0,+4]$ & 1.6 \\
\bottomrule
\end{tabular*}
\end{table}

The LAPW runs also test the certificate and the projectors. Because the released frozen sets are energy windows by construction, the certificate of the frozen set is zero, but applied to the trust mask the certificate still measures, before any minimization, how far the classification of states by projectability departs from that window. Trust masks built from the projectability inside the muffin tins alone misassign 613 (GaAs), 21 (MgB$_2$), and 450 ($\alpha$-Sn) states against the best window, and at 330 of the 512 $\kk$-points the GaAs trusted set is not an energy interval (the earlier porting runs, SM). Releasing the states under the ceiling that lie outside the trust mask nevertheless does not improve the model (Sec.~\ref{sec:extent}). The interstitial tails of the projectors change the model only when the frozen set is taken from the projectability. At the automatic ceiling, the muffin-tin and the full-cell columns give the same masks and the same models on GaAs, MgB$_2$, and $\alpha$-Sn, because every state below the ceiling is frozen and the ceiling is set by the Wannier count. Without a ceiling, the muffin-tin projectability, which omits the interstitial weight of the conduction states of free-electron character, raises the error over the GaAs conduction window to 52.7~meV, compared to 30.5~meV with the full-cell columns and 18.0~meV at the release protocol (Appendix~\ref{app:lapw}). Both programs write the masks with every hand-off (\verb|--dump-masks|, \verb|--hybrid-report-json|, to \texttt{<out>.masks} and \texttt{<out>.anatomy}), and the shipped scripts draw Fig.~\ref{fig:concept}(b) and the figures of Appendices~\ref{app:certificate} and~\ref{app:anatomy} for any run. The dictionary of LCAO$\to$LAPW replacements, the released implementation, the run details and the design choices tested on both frontends are given in Appendix~\ref{app:lapw}, and the porting runs that preceded the release are described in the SM.

\section{Discussion}
\label{sec:discussion}

Of the parts of the method, only the analytic overlaps and the atomic-orbital discretization of SCDM are specific to LCAO, whereas the selection rule and the hand-off are not, and the WIEN2k runs test this directly. With atomic-channel projectabilities from WIEN2k as input, the same rule and hand-off, with the release thresholds and ceiling, reproduce eight models from seven calculations, and the released LAPW frozen sets are energy windows below the same automatic ceiling. Releasing the untrusted states under that ceiling loses a target band of the Bi bilayer and raises the band errors of Fe with spin--orbit coupling more than thirtyfold, just as it lowers the fidelity of the LCAO models (Sec.~\ref{sec:extent}).

PDWF~\cite{Qiao2023} and its extension to magnetic and spin--orbit-coupled systems~\cite{Jiang2025} freeze states by projectability inside \textsc{Wannier90} (\texttt{dis\_froz\_proj}), in their recommended setting together with a frozen energy window up to 2~eV above $E_{\mathrm{F}}$ or the conduction-band minimum, and our experiments at matched extent bear directly on that design. At equal extent, the energy window below the ceiling gives far smaller band errors than the set selected by projectability on four of the seven cells compared, and it agrees with that set to within 2.2~meV on the other three (Table~\ref{tab:extent}). The two approaches also take the target from different sources. PDWF takes its projectors from the pseudo-atomic orbitals of the pseudopotential~\cite{Qiao2023}, and Ref.~\cite{Jiang2025} adds analytic hydrogenic orbitals for every orbital of a fixed per-element list that the pseudopotential lacks. Our procedure instead starts from the chemical channels and adds channels from the basis of the calculation only when the low conduction bands are poorly covered or the trust mask holds more states than there are Wannier functions, choosing the smallest set of channels that restores the coverage (Sec.~\ref{sec:auto}). One per-state criterion remains in our default route, the admission floor, and its effect depends on the cell. A floor of 0.10 is the LCAO default, but on the LAPW cells the same floor breaks the continuity of three subspaces at the automatic ceiling, and the pocket skip restores it only on SrVO$_3$, by lowering the ceiling to $+4.62$~eV, so the LAPW frontend admits every pool state (Appendix~\ref{app:lapw} and SM).

Reference~\cite{Tiwari2026} automates the choice of inputs by search, in a meta-optimization loop that repeats the Wannierization to minimize a loss combining the band-interpolation error and the spread, scored against separately computed reference bands. For silicon the search covered the inner and outer window bounds and 285 types of initial projections, computed on a $6{\times}6{\times}6$ mesh. Our ceiling and thresholds are instead computed once from the calculation and need no reference bands, so the two approaches can be combined, with the computed configuration as the starting point of the search. Meta-optimization works from coarse meshes ($6^3$ and $4^3$ for silicon in Ref.~\cite{Tiwari2026}, with plane-wave overlaps), and such meshes need care in a non-orthogonal LCAO basis. There, overlaps in the midpoint-metric approximation are not contractive on any mesh tested and make the h-BN spread negative at $3^3$ and $4^3$ (Appendix~\ref{app:contractivity}), whereas the analytic $M^{(\kk,\bb)}$ stays contractive on these meshes, although such meshes can still undersample the zone.

The comparisons reported here hold under two conditions, namely that the interpolated bands are read from the \texttt{band.dat} file of \textsc{Wannier90}, which includes the Wigner--Seitz corrections, and that they are scored on Wannierizations run to their criterion in one uninterrupted run. The second condition is needed because, in the \textsc{Wannier90} builds we used, restarting a run without disentanglement from its checkpoint discards the gauge while the spread still appears converged, and the interpolated bands are then wrong (SM). Every $N_b{=}N_w$ hand-off is a run of this kind, because it has no disentanglement step. The two MgB$_2$ models, finally, differ in functional and are each scored against their own reference bands, so they do not compare the basis sets (Sec.~\ref{sec:lapw-results}).

Now that the method has been ported to a muffin-tin code (Appendix~\ref{app:lapw}), other frontends need different amounts of work. A second Gaussian-basis frontend such as CP2K would require only a parser, whereas a code with numeric atomic orbitals such as FHI-aims would also need its overlaps evaluated numerically. A plane-wave frontend remains future work, since pseudo-atomic projectabilities in plane-wave codes are the one class of projectors not yet tested. On the methodological side, the ceiling divides a fixed number of variational directions between the energy range below it and the range above, and a variational formulation of that division would replace the declared fidelity range by one obtained from the calculation.

\section{Conclusion}

At equal extent, freezing every state below a derived ceiling lowered the band error by factors of three to more than sixty on four of the seven cells compared, and it matched freezing only the states of the trusted set to within 2.2~meV on the other three, with either solver. On the four cells where the window gave the better bands, the selective set had nevertheless reached the lower spread, so the spread alone does not identify the better frozen set. On this evidence the frozen set should be an energy window, even though hybridization varies from state to state, and the projectability is better used to define the admitted set, select the target channels, and certify, before any minimization, where the per-state classification disagrees with the energy ordering. With the ceiling derived from the Wannier count, the window can be passed to \textsc{Wannier90} either as a standard disentanglement window or, after a constrained $\Omega_I$ minimization and with an atomic-orbital SCDM gauge, as an isolated $N_b{=}N_w$ problem. In this way, joint valence-and-conduction models that stalled under hand-set windows were obtained from LCAO input without manual parameters, with analytic overlaps and at production scale, and the same rule gave eight models on an all-electron LAPW frontend (Appendix~\ref{app:lapw}).

\begin{acknowledgments}
We thank Alessandro Erba for discussions and assistance with CRYSTAL, and Manuel Antonio Garc\'{\i}a Bl\'azquez for extending, at our request, the CRYSTAL properties module to support the two-component (spin--orbit) implementation of \textsc{lcao2wannier}. This work received no external funding. Computational resources were provided by the Mendoza-Cortes group at Michigan State University. This work was supported in part through computational resources and services provided by the Institute for Cyber-Enabled Research at Michigan State University.
\end{acknowledgments}

\section*{Data and code availability}
A public archive~\cite{dataZenodo} will hold the CRYSTAL outputs, the generated \textsc{Wannier90} inputs and outputs, the band-comparison caches and the data files of each figure for all production runs, the overlap mesh series, the baseline runs, and the matched-extent and frozen-ceiling runs together with their intact \textsc{Wannier90} records and checkpoints (SM). It will also hold the LAPW validation archives, namely the WIEN2k and \textsc{wien2wannier} cases of the seven calculations of Table~\ref{tab:lapw}, with the command lines, masks and per-state projectability files of each cell, and the three cells of the earlier porting runs on muffin-tin projectability described in the SM. Every Wannierization used the development build \texttt{ff955d1} of \textsc{Wannier90}, whose results v4.0.2 reproduces bitwise, with local patches to its command-line handling and none to its algorithms (SM).

The method is released in two programs, \textsc{lcao2wannier}~v1.0 (GPLv2 or later)~\cite{lcao2wannierZenodo}, which contains the full pipeline of this paper with an end-to-end tutorial, and a modified \textsc{wien2wannier}, released as v3.0 (GPLv3 or later, upstream v2.0)~\cite{wien2wannierZenodo}, which contains the hybrid pre-Wannierization of Appendix~\ref{app:lapw} with its user guide and a worked example. Both programs contain the two disentanglers of Sec.~\ref{sec:dis} (\texttt{-\kern0pt-hybrid-dis-solver cg|gs}, with CG as the default and the colored sweep as an option), the subspace-continuity check of Sec.~\ref{sec:auto}, the graded classification of runs that reach the sweep cap (Appendix~\ref{app:constants}), and the two automatic routes of Sec.~\ref{sec:two-routes} (\verb|--method hybrid| and \verb|--method window|), which share the frozen ceiling (\verb|--hybrid-froz-emax| and \verb|--froz-emax|). Both write the masks and the per-state projectabilities behind every hand-off, ship the scripts that draw Fig.~\ref{fig:concept}(b) and the figures of Appendices~\ref{app:certificate} and~\ref{app:anatomy} from these files, and document one command per material (SM).

\appendix
\section{Analytic overlaps and contractivity}
\label{app:contractivity}

This appendix gives the details behind Sec.~\ref{sec:platform}. It derives the LCAO
form of the neighbor overlaps $M^{(\kk,\bb)}$, evaluates the
momentum-shifted Gaussian integrals they require, describes the two approximate
constructions used for comparison, proves that the analytic construction
is contractive, and compares the three constructions on a series of
meshes. All integrals are in atomic units (bohr), and, as in Sec.~\ref{sec:platform}, $\kk$, $\bb$ and the lattice vectors $\mathbf g$ are in reduced coordinates inside phase factors $e^{2\pi i\,\kk\cdot\mathbf g}$, whereas their Cartesian forms, used where they multiply a position, carry the subscript ``cart''.

\emph{Matrices from CRYSTAL.} CRYSTAL prints the lower triangle of each real-space matrix. The upper triangle follows from the partner cell $-\RR$, because translational invariance and Hermiticity of the Hamiltonian give $H(-\RR)=H(\RR)^\dagger$, the relation $F^{\mathbf g}_{12}=F^{-\mathbf g}_{21}$ by which CRYSTAL reduces the stored Fock matrices~\cite{Dovesi1986,CrystalManual}. Writing $H_{\text{raw}}(\RR)$ for the printed lower triangle, the full matrix is \begin{equation} H(\RR)=\operatorname{tril}\big(H_{\text{raw}}(\RR)\big) + \operatorname{tril}_{<}\big(H_{\text{raw}}(-\RR)\big)^{\dagger}, \label{eq:trilfill} \end{equation} where $\operatorname{tril}$ and $\operatorname{tril}_{<}$ take the lower-triangular and strictly lower-triangular parts, and $S(\RR)$ is completed in the same way. Under two-component spin--orbit coupling (SOC)~\cite{Desmarais2019a,Desmarais2020a,Desmarais2021} the rule is applied to the $2N\times2N$ matrix assembled from the four spin blocks. The SOC Wannier count (Eq.~\eqref{eq:numwann}, doubled) is even, which allows Kramers pairs but does not enforce them. The SnTe model, for example, splits the reference pairs between mesh points (Sec.~\ref{sec:snte}). Our implementation follows CRYSTAL's conventions for Bloch phases~\cite{Pisani1988}, basis-function normalization~\cite{Desmarais2018,CrystalManual} and orbital order~\cite{CrystalManual}, so it applies directly to the coefficients CRYSTAL prints. The PySCF interface to \textsc{Wannier90} also builds $M^{(\kk,\bb)}$ from analytic Fourier transforms of Gaussian AO pairs~\cite{pyWannier90}.

At $\bb=0$ the integrals reduce to the AO overlaps that CRYSTAL also prints, and comparing the two tests the integral code. Against the eight-significant-digit matrix dump of the 1424-orbital Sc-COF ($f$ shells included, $4.7\times10^5$ AO pairs), the analytic overlaps match CRYSTAL's at the print precision for most pairs (90th percentile of the deviation $3\times10^{-8}$). A small systematic tail, reaching $4\times10^{-4}$ on pairs of diffuse functions, comes from the three-decimal precision of the printed atomic coordinates from which the integrals are rebuilt (changing the coordinates by half of the last printed digit reproduces it), not from the integral evaluation.

\emph{Assembly.} CRYSTAL builds its Bloch functions from AOs $\chi_\mu$
centered on the nuclei and attaches the phase factor to the lattice vector
of the cell, not to the position of the orbital~\cite{Andre1967,Pisani1980,Pisani1988,CrystalManual}, the tight-binding convention in which $H(\kk)$ is periodic in the reciprocal lattice~\cite{Bena2009},
\begin{align}
\phi_{\mu\kk}(\mathbf r) &= \sum_{\mathbf g} e^{2\pi i\,\kk\cdot\mathbf g}\,
\chi_\mu(\mathbf r - \boldsymbol\tau_\mu - \mathbf g_{\text{cart}}),\nonumber\\
\psi_{n\kk} &= \sum_\mu C_{\mu n}(\kk)\,\phi_{\mu\kk},
\label{eq:blochfn}
\end{align}
where $\boldsymbol\tau_\mu$ is the position, in the home cell, of the
nucleus that carries $\chi_\mu$, and the sum runs over all lattice vectors
$\mathbf g$. Because the phases are attached to lattice vectors,
$\phi_{\mu,\kk+\mathbf G}=\phi_{\mu\kk}$ for every reciprocal-lattice vector
$\mathbf G$, so a neighbor $\kk+\bb$ outside the first Brillouin zone uses the
coefficients of its image inside the zone, with no additional phase.
Inserting Eq.~\eqref{eq:blochfn} into Eq.~\eqref{eq:mmn} and using the
lattice periodicity gives
\begin{align}
M^{(\kk,\bb)} &= C^\dagger(\kk)\,\tilde S^{(\bb)}(\kk+\bb)\,C(\kk+\bb),\nonumber\\
\tilde S^{(\bb)}(\kk+\bb) &= \sum_{\mathbf g}
e^{2\pi i\,(\kk+\bb)\cdot\mathbf g}\, S^{(\bb)}(\mathbf g),
\label{eq:mmnassembly}
\end{align}
in which $S^{(\bb)}(\mathbf g)$ is the matrix of momentum-shifted overlaps
between AO $\mu$ in the home cell and AO $\nu$ in cell $\mathbf g$,
\begin{equation}
S^{(\bb)}_{\mu\nu}(\mathbf g)=\int\chi^*_\mu(\mathbf r-\boldsymbol\tau_\mu)\,
e^{-i\bb_{\text{cart}}\cdot\mathbf r}\,
\chi_\nu(\mathbf r-\boldsymbol\tau_\nu-\mathbf g_{\text{cart}})\,d^3r .
\label{eq:sbint}
\end{equation}
At $\bb=0$, $S^{(0)}(\mathbf g)$ is the ordinary AO overlap matrix, the
analytic counterpart of the printed $S(\RR)$.

The eigenvectors of Eq.~\eqref{eq:lcaoeig} are orthonormal in the metric
$S(\kk)$ built from the printed overlaps, not in the analytic metric
$\tilde S^{(0)}(\kk)$. The contractivity proof below needs the analytic
metric, so the eigenvectors are orthonormalized once more by the symmetric
(L\"owdin) transformation~\cite{Lowdin1950},
\begin{equation}
O(\kk) = C^\dagger(\kk)\,\tilde S^{(0)}(\kk)\,C(\kk),
\qquad
C(\kk) \leftarrow C(\kk)\,O(\kk)^{-1/2}.
\label{eq:gtoortho}
\end{equation}
Here $O(\kk)$ is the overlap matrix of the Bloch states in the analytic metric, and
it would be the identity if the printed and analytic overlaps agreed exactly, and after the transformation $M^{(\kk,0)}=I$ holds exactly. For
two-component spinors the coefficient matrix splits into the rows of the
two spin components, $C_\alpha$ and $C_\beta$ [Ref.~\cite{Desmarais2020a}, Eq.~(5)]. The Gaussian overlap acts on
the spatial coordinates only and is diagonal in spin, so the two components
are summed,
\begin{equation}
M^{(\kk,\bb)} = C_\alpha^\dagger(\kk)\, \tilde S^{(\bb)}\, C_\alpha(\kk+\bb)
+ C_\beta^\dagger(\kk)\, \tilde S^{(\bb)}\, C_\beta(\kk+\bb),
\label{eq:spinormmn}
\end{equation}
with $\tilde S^{(\bb)}=\tilde S^{(\bb)}(\kk+\bb)$, and $O(\kk)$ in
Eq.~\eqref{eq:gtoortho} is formed in the same way.

\emph{Integrals.} The AOs are contractions of primitive Cartesian Gaussians
\begin{equation}
g_a(\mathbf r)=(x{-}A_x)^{l_a}\,(y{-}A_y)^{m_a}\,(z{-}A_z)^{n_a}\,
e^{-\alpha|\mathbf r-\mathbf A|^2},
\label{eq:cartgauss}
\end{equation}
with exponent $\alpha$, center $\mathbf A$ and Cartesian powers
$(l_a,m_a,n_a)$. The function $g_b$ is defined in the same way with $\beta$, $\mathbf B$
and $(l_b,m_b,n_b)$. The exponential factors combine into one
Gaussian~\cite{Boys1950},
\begin{gather}
e^{-\alpha|\mathbf r-\mathbf A|^2}\,e^{-\beta|\mathbf r-\mathbf B|^2}
= K_{AB}\,e^{-p|\mathbf r-\mathbf P|^2},\nonumber\\
p=\alpha+\beta,\quad
\mathbf P=\frac{\alpha\mathbf A+\beta\mathbf B}{p},\quad
K_{AB}=e^{-\alpha\beta|\mathbf A-\mathbf B|^2/p}.
\label{eq:gpt}
\end{gather}
Completing the square in the exponent of the product with the plane wave
($\bb$ is Cartesian here and in the rest of this paragraph) gives
\begin{align}
e^{-p|\mathbf r-\mathbf P|^2}\, e^{-i\bb\cdot\mathbf r}
&= e^{-i\bb\cdot\mathbf P}\, e^{-|\bb|^2/4p}\,
e^{-p(\mathbf r-\mathbf P')\cdot(\mathbf r-\mathbf P')},\nonumber\\
\mathbf P' &= \mathbf P - \frac{i\,\bb}{2p}.
\label{eq:complexcenter}
\end{align}
The integrand is therefore again a Gaussian, centered at the complex point
$\mathbf P'$, and the integral factorizes into one-dimensional factors,
\begin{equation}
\langle g_a|e^{-i\bb\cdot\mathbf r}|g_b\rangle
= K_{AB}\, e^{-i\bb\cdot\mathbf P}\, e^{-|\bb|^2/4p}\;
S^x_{l_a l_b}\,S^y_{m_a m_b}\,S^z_{n_a n_b},
\label{eq:primfactor}
\end{equation}
with
$S^x_{l_a l_b}=\int(x-A_x)^{l_a}(x-B_x)^{l_b}\,e^{-p(x-P'_x)^2}\,dx$ and
likewise for $y$ and $z$. The one-dimensional factors follow from the
Obara--Saika recursion~\cite{Obara1986,Helgaker2000},
\begin{equation}
\begin{aligned}
S^x_{l_a+1,\,l_b} &= X'_{PA}\, S^x_{l_a l_b}
+ \tfrac{1}{2p}\big(l_a S^x_{l_a-1,\,l_b} + l_b S^x_{l_a,\,l_b-1}\big),\\
S^x_{l_a,\,l_b+1} &= X'_{PB}\, S^x_{l_a l_b}
+ \tfrac{1}{2p}\big(l_a S^x_{l_a-1,\,l_b} + l_b S^x_{l_a,\,l_b-1}\big),
\\
S^x_{00} &= \sqrt{\pi/p},
\end{aligned}
\label{eq:hermite}
\end{equation}
with the complex distances $X'_{PA}=P'_x-A_x$ and $X'_{PB}=P'_x-B_x$. The
recursion is usually stated for a real center, but integral formulas for Gaussians with real centers carry over to complex centers~\cite{Saunders1983}. Both sides of
Eq.~\eqref{eq:hermite} are analytic functions of $P'_x$ (the Gaussian
decays along any line parallel to the real axis), and they agree for real
$P'_x$, so they agree for complex $P'_x$ as well. Equivalently,
$S^x_{l_a l_b}$ is proportional to the $t=0$ Hermite expansion coefficient
of the McMurchie--Davidson scheme~\cite{McMurchie1978,Helgaker2000}.

CRYSTAL's AOs are contractions of real solid-harmonic Gaussians~\cite{Desmarais2018}. Each real solid harmonic of degree $\ell$ is a fixed linear combination of the Cartesian monomials of degree $\ell$~\cite{Schlegel1995,Helgaker2000}, so its integrals are the same combinations of the Cartesian ones. The combinations are tabulated per
$\ell$ in CRYSTAL's internal storage order~\cite{CrystalManual} ($s$, then $p$ in the order
$x,y,z$, then $d$ in the order $2z^2{-}x^2{-}y^2$, $xz$, $yz$, $x^2{-}y^2$, $xy$, then the seven
$f$ and the nine $g$ functions). CRYSTAL's input contraction coefficients refer to normalized primitives, and the coefficients it prints differ from them~\cite{CrystalManual}, because the printed coefficients multiply unnormalized primitives. Each contracted AO is normalized to unit self-overlap, as in CRYSTAL~\cite{Desmarais2018,CrystalManual}. With these conventions the
$\bb=0$ integrals reproduce the printed $S(\RR)$ to the precision stated in
Sec.~\ref{sec:platform}, and this comparison tests the integral code. The lattice sum in Eq.~\eqref{eq:mmnassembly} runs over the cells for
which CRYSTAL prints $S(\RR)$, and AO pairs whose centers are more than
16~bohr apart are skipped. The truncation lies outside the contractivity argument of this appendix and is checked numerically instead, since every link of every run passes the acceptance test of Eq.~\eqref{eq:handoff-acceptance}. The integrals for different pairs
$(\bb,\mathbf g)$ are independent and are computed in parallel.

\emph{The two approximate constructions.} Both approximate constructions use only the printed matrices $S(\RR)$, and they differ in how they treat the position dependence of the plane-wave factor. The midpoint construction, Eq.~\eqref{eq:midpoint},
replaces each momentum-shifted overlap by the zeroth-order term of its expansion about the pair midpoint,
\begin{equation}
S^{(\bb)}_{\mu\nu}(\mathbf g)\approx
e^{-i\bb_{\text{cart}}\cdot(\boldsymbol\tau_\mu+\boldsymbol\tau_\nu
+\mathbf g_{\text{cart}})/2}\,S_{\mu\nu}(\mathbf g),
\label{eq:midpointapprox}
\end{equation}
in which the plane wave is evaluated at the midpoint of the two orbital
centers. Inserted into Eq.~\eqref{eq:mmnassembly}, the $\mathbf g$-dependent
part of the phase moves the Bloch sum to
$\kk_{\text{mid}}=\kk+\tfrac12\bb$, and the $\boldsymbol\tau$-dependent part
gives the matrices $D_{\bb}$ of Eq.~\eqref{eq:midpoint}. Here $\bb$ is the
neighbor vector before reduction to the first zone (the reciprocal-lattice
shift in the neighbor list is added back), because $S(\kk)$ is periodic in
$\kk$ but $\tfrac12\mathbf G$ is not a reciprocal-lattice vector.
Equation~\eqref{eq:midpointapprox} places the dipole matrix element
$\langle\chi_\mu|\mathbf r|\chi_\nu\rangle$ at the midpoint of the two
centers, so it misses dipoles between different orbitals on the same atom
(an $s$ and a $p$ function, for example), and its error is first order in
$|\bb|$. Keeping the expansion of $e^{-i\bb\cdot(\mathbf r-\mathbf c)}$ about the pair midpoint $\mathbf c$ to higher order, with the moment integrals $\langle\chi_\mu|[\bb\cdot(\mathbf r-\mathbf c)]^j|\chi_\nu\rangle$, removes the negative $\Omega_I(\mathds{1})$ but does not make $M^{(\kk,\bb)}$ contractive. Through fourth order, $\Omega_I(\mathds{1})$ is positive on every mesh tested and within 0.23\% of the analytic value, and $\sigma_{\max}$ exceeds one only on h-BN at $3^3$ (1.00037). Through first order, $\Omega_I(\mathds{1})$ is negative on every mesh, because $\Omega_I$ is second order in $|\bb|$ (SM). No finite order is contractive by construction, and with the printed $S(\RR)$ as the zeroth-order term, the relative distance to the analytic $M$ stops falling at $(0.8$--$3.5)\times10^{-4}$, the size of the difference between the printed overlaps and those rebuilt from the printed coordinates (up to $4\times10^{-4}$, above). On MgB$_2$ the relative Frobenius distance between the midpoint
and the analytic $M^{(\kk,\bb)}$ falls from 0.26 at $6^3$ to 0.13 at
$12^3$, linearly in the mesh spacing. Without the phase matrices $D_{\bb}$
the distance doubles (0.55 at $6^3$, 0.26 at $12^3$), and with the
opposite sign of the phase it roughly quadruples. The L\"owdin construction, which we built for this
comparison, orthonormalizes the coefficients symmetrically,
$\tilde C(\kk)=S(\kk)^{1/2}C(\kk)$, and takes the position operator to be
diagonal in the orthonormalized AO basis, with each orbital at its
nucleus,
\begin{equation}
M^{(\kk,\bb)}_{mn} = \sum_\mu \tilde C^{*}_{\mu m}(\kk)\,
e^{-i\,\bb\cdot\boldsymbol\tau_\mu}\,\tilde C_{\mu n}(\kk+\bb).
\label{eq:mmnlowdin}
\end{equation}
Since $\tilde C(\kk)$ and $\tilde C(\kk+\bb)$ have orthonormal columns and
the phase matrix is unitary, this $M$ is contractive by construction, but it relies on the diagonal position approximation (Table~\ref{tab:mmn}).

\emph{Contractivity.} Let $\{\psi_m(\kk)\}$ and $\{\psi_n(\kk+\bb)\}$ be
the Bloch states after the orthonormalization of Eq.~\eqref{eq:gtoortho}.
Their cell-periodic parts $\{u_{m\kk}\}$ and $\{u_{n,\kk+\bb}\}$ are then two
orthonormal families in the Hilbert space $L^2$ of square-integrable
functions on the unit cell, and $M^{(\kk,\bb)}_{mn}=\langle
u_{m\kk}|u_{n,\kk+\bb}\rangle$ is the matrix of overlaps between the two
families. If each family is extended to an orthonormal basis of $L^2$, the overlaps between the two bases form a unitary matrix whose upper-left block is $M^{(\kk,\bb)}$. A block of a unitary matrix has all singular values at
most one, so Eq.~\eqref{eq:contractive} holds on any mesh, with no
assumption that $|\bb|$ is small. The midpoint construction replaces the
momentum-shifted metric by the approximation~\eqref{eq:midpointapprox}
instead of evaluating it, so its $M$ is not an overlap matrix of two
orthonormal families, and nothing bounds its singular values by one.

\begin{figure*}[t]
\centering
\includegraphics[scale=0.67]{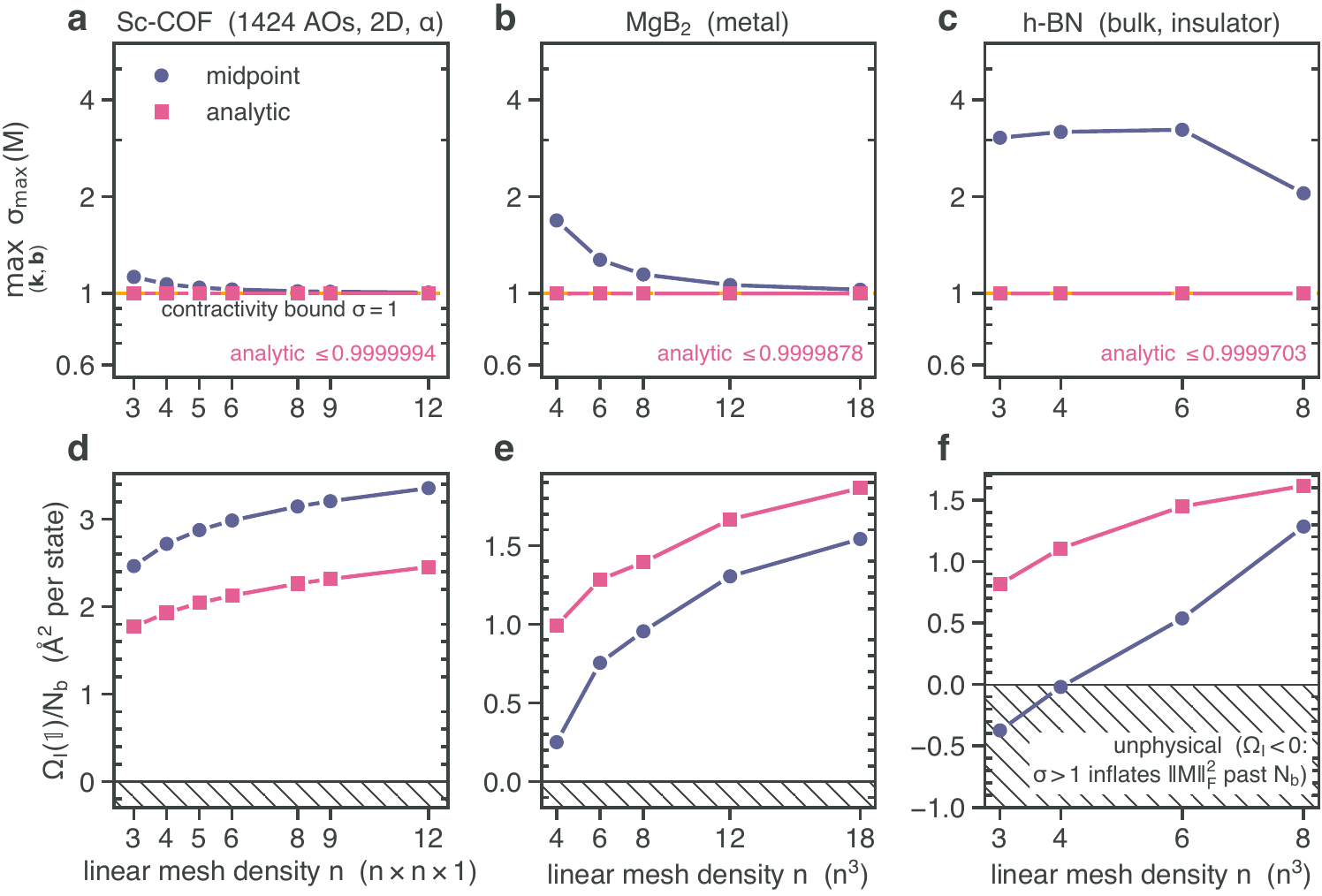}
\caption{Contractivity and stability of the overlaps across a mesh series, computed with the projectability pipeline at each mesh for the Sc-COF, MgB$_2$ and bulk h-BN (columns). (a)--(c)~Largest singular value of $M^{(\kk,\bb)}$ over all blocks. The midpoint construction, Eq.~\eqref{eq:midpoint}, exceeds the bound $\sigma=1$ on every mesh (up to 3.2, on h-BN at $6^3$), while the analytic construction stays below it on every mesh (largest value 0.9999994). (d)--(f)~Spread of the emitted pool per state, $\Omega_I(\mathds{1})/N_b$. The midpoint values turn negative, which is unphysical (hatched), on the two coarsest h-BN meshes and differ from the analytic values on every mesh, lying below them on MgB$_2$ and h-BN and above them on the Sc-COF. The analytic values are positive and rise smoothly with the mesh density.}
\label{fig:stability}
\end{figure*}

\emph{The three constructions across a mesh series.}
Figure~\ref{fig:stability} and Table~\ref{tab:mmn} compare the
constructions on the Sc-COF, MgB$_2$ and bulk h-BN, using the
projectability pipeline at each mesh. The quantity compared is the spread
of the whole emitted band pool, $\Omega_I(\mathds{1})$, which is Eq.~\eqref{eq:omegai}
with $V(\kk)$ the identity on the $N_b$ pool bands, so that each term is
$N_b-\|M^{(\kk,\bb)}\|_F^2$. The largest singular value of the midpoint $M$
exceeds one on every mesh of every cell, with values of 1.03--1.07 on the Sc-COF,
1.03--1.69 on MgB$_2$ and 2.0--3.2 on bulk h-BN. On the two coarsest h-BN
meshes ($3^3$ and $4^3$) this makes $\Omega_I(\mathds{1})$ negative, which
is unphysical. The analytic construction stays below one on the same
meshes (largest value $0.9999994$, on the Sc-COF), and its pool spread is
positive and rises smoothly with the mesh density. The L\"owdin
construction is contractive but carries the diagonal position approximation, and its pool spread also differs from the analytic one. On the densest MgB$_2$ mesh ($18^3$) the midpoint pool spread lies 17\% below the analytic value and the L\"owdin value 12\% above it, whereas on the Sc-COF at $6\times6\times1$ the midpoint value lies 40\% above the analytic one and the L\"owdin value 6\% below it. The coarse meshes that expose the negative spread
($3^3$, $4^3$) miss the hexagonal zone-corner point H, and $4^3$ also misses K (SM). They are comparison points for the overlap
constructions, not production meshes, and the good behavior of the analytic construction on them does not make them adequate samplings of the zone. The Sc-COF
series uses the 318-band pool of the projectability method, whereas the hybrid
models of the Sc-COF use $1.5N_w=441$ bands (SM).

\begin{table*}[t]
\caption{The three constructions of the overlap matrices $M^{(\kk,\bb)}$ between neighboring $\kk$-points (Appendix~\ref{app:contractivity}). The column $\sigma_{\max}$ gives the largest singular value over all
$(\kk,\bb)$ blocks, so a construction is contractive if $\sigma_{\max}\le1$, and $\Omega_I(\mathds{1})/N_b$ gives the spread of the emitted pool per band in \AA$^2$. The coarse mesh is the Sc-COF at
$6\times6\times1$ (1424 AOs, the 318-band pool of the projectability method), and the dense mesh is MgB$_2$ at $18^3$ (14 bands).}
\label{tab:mmn}
\small
\begin{tabular*}{\textwidth}{@{\extracolsep{\fill}}l l c c c c c@{}}
\toprule
 & & & \multicolumn{2}{c}{Sc-COF, $6{\times}6{\times}1$} & \multicolumn{2}{c}{MgB$_2$, $18^3$} \\
\cmidrule(lr){4-5}\cmidrule(l){6-7}
Construction & Approximation & Contractive & $\sigma_{\max}$ & $\Omega_I(\mathds{1})/N_b$ & $\sigma_{\max}$ & $\Omega_I(\mathds{1})/N_b$ \\
\midrule
Midpoint & dipoles at pair midpoints & no & 1.030 & 2.99 & 1.028 & 1.54 \\
L\"owdin & diagonal position & yes & 1.000000 & 2.01 & 0.9999998 & 2.10 \\
Analytic & none & yes (exact) & 0.9999994 & 2.13 & 0.9999877 & 1.87 \\
\bottomrule
\end{tabular*}
\end{table*}

\section{Certificate: how far the trust mask departs from a window}
\label{app:certificate}
We call a family of state sets $\{T(\kk)\}$ \emph{window-representable} if scalar bounds $a\le b$ exist with $T(\kk)=\{n:\varepsilon_{n\kk}\in[a,b]\}$ at every $\kk$, which is exactly what the \texttt{dis\_froz\_min/max} parameters of \textsc{Wannier90} can express. Because the bands are ordered in energy, this requires (i) that each $T(\kk)$ be an energy interval and (ii) that the interval be \emph{the same} at every $\kk$. The certificate is evaluated on the per-state trust masks of Eq.~\eqref{eq:masks} (\texttt{frozen\_trust\_mask} in the \verb|--dump-masks| export, at the meshes of Table~\ref{tab:defaults}). Since the frozen set used by the disentanglement is an energy window by construction (Sec.~\ref{sec:masks}), the certificate measures how far the per-state classification by projectability departs from the energy ordering.

\begin{figure*}[t]
\centering
\includegraphics[scale=0.67]{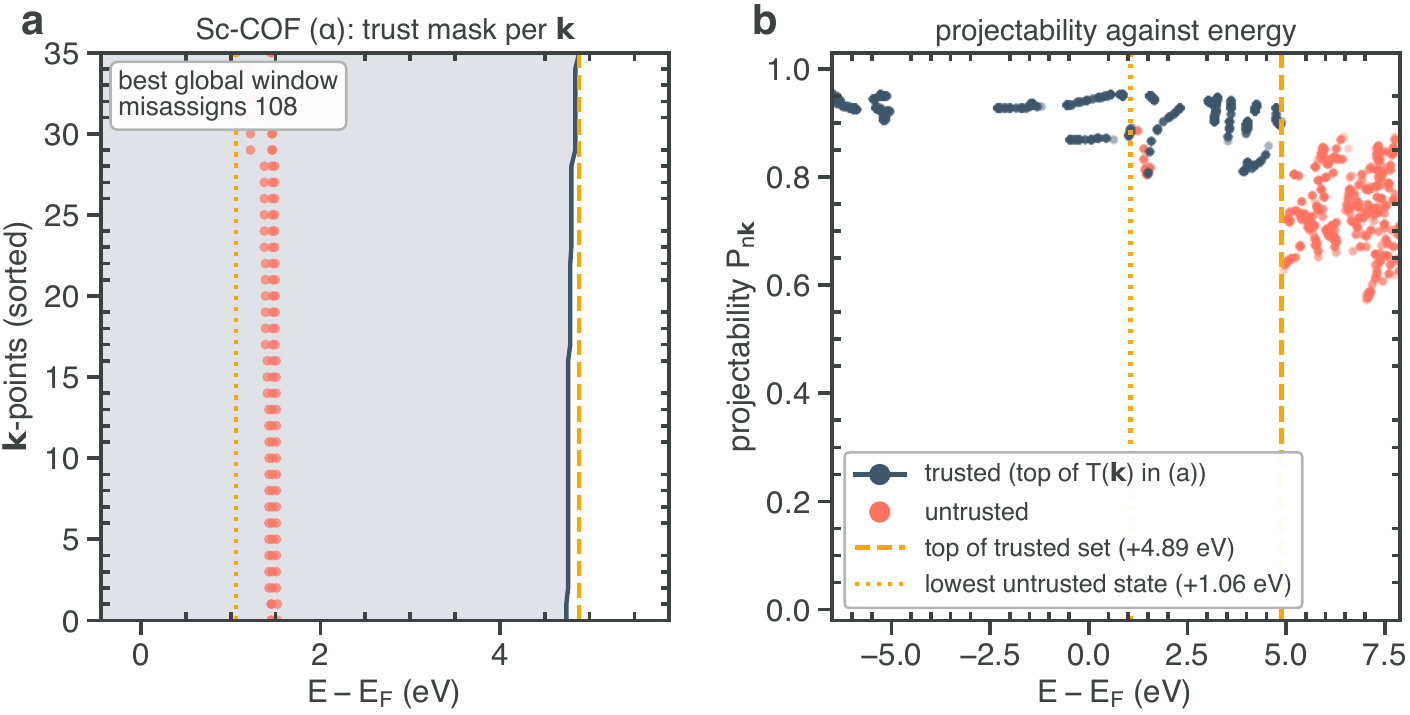}
\caption{Certificate of the Sc-COF ($\alpha$) trust mask, the counterpart of Fig.~\ref{fig:concept}(b) for h-BN. (a)~The top of the trusted set $T(\kk)$ at each of the 36 $\kk$-points (slate line, $\kk$ sorted by this top, trusted span shaded) and the 108 untrusted states that lie inside the trusted interval (coral). The best global window ends between the two amber lines. A window ending at the top of the trusted set (dashed, $+4.89$~eV) freezes all of these untrusted states, and a window ending just below the lowest untrusted state (dotted, $+1.06$~eV) freezes none of them but leaves out 936 trusted states. The best global window misassigns 108 states. (b)~Projectability $P_{n\kk}$ against energy for the trusted (slate) and untrusted (coral) pool states. The 108 untrusted states inside the trusted interval ($P=0.81$--$0.89$) overlap in projectability with the trusted states between $+0.5$ and $+2.5$~eV (0.81--0.95), but they belong to bands whose average lies below the gate (0.838). The mask therefore follows bands, not energy, and no energy window reproduces it. The disentanglement freezes the window below the automatic ceiling instead (Sec.~\ref{sec:masks}), and Sec.~\ref{sec:extent} compares the two.}

\label{fig:certificate}
\end{figure*}

Of the LCAO cells at the defaults, only h-BN and the Sc-COF have masks that are not window-representable, and the two cells violate different conditions. On h-BN, condition (i) holds at every $\kk$ and only condition (ii) fails, because the interval selected by the trust mask moves across the zone, with its top ranging over $[+6.85,+12.65]$~eV (a change of 5.8~eV). Any global window that reaches the top of the whole trusted set ($b=+12.65$~eV) must then also freeze the 378 untrusted pool states beneath it (at 318 of 864 $\kk$-points), and the largest window free of untrusted states ($b=+10.93$~eV) leaves out 185 trusted states (Fig.~\ref{fig:concept}b). On the Sc-COF, condition (i) fails at every $\kk$ (Fig.~\ref{fig:certificate}). In the $\alpha$ channel the top of the trusted set hardly moves ($[+4.74,+4.89]$~eV), but at each of the 36 $\kk$-points untrusted states lie inside the trusted interval, 108 in all, between $+1.06$ and $+1.52$~eV. The largest window free of untrusted states ($+1.06$~eV) leaves out 936 of the 6480 trusted states, and the best global window misassigns 108 states (74 in the $\beta$ channel, computed by Algorithm~S5 of the SM). These counts are exact for the masks of each run, which are archived with it. The masks of Bi and MgB$_2$, in contrast, are window-representable at the automatic thresholds, and so is that of SnTe with 46 functions, whose thresholds sit at the upper clip (Sec.~\ref{sec:snte}). With the earlier 36 functions, the top of the SnTe trusted set was at $+19.40$~eV and the best global window misassigned 756 states.

The same obstruction occurs on the LAPW frontend when the masks are built from muffin-tin projectability, for which the construction gives 613 (GaAs), 21 (MgB$_2$), and 450 ($\alpha$-Sn) misassigned states (the porting runs of the SM), and on GaAs condition (i) itself fails, since the trusted set is not an interval at 330 of 512 $\kk$-points. With the full-cell projectability used by the released \textsc{wien2wannier}, the trust masks themselves are window-representable (Appendix~\ref{app:lapw}), and the released frozen sets are energy windows by construction.

The certificate does not show which set gives the better model, a question that the matched-extent comparison of Sec.~\ref{sec:extent} answers. Freezing the whole span below the ceiling, including the 378 h-BN states outside the trust mask, improves the conduction fidelity, from 1.9 to 1.3~meV over $[+5,+8]$~eV and from 52.2 to 4.9~meV over $[+5,+12]$~eV at an unchanged ceiling of $+12.65$~eV. On the LAPW frontend, releasing the states outside the trust mask under the ceiling loses a target band of the Bi bilayer and raises the band errors of Fe with spin--orbit coupling more than thirtyfold (Sec.~\ref{sec:extent}). A nonzero misassignment count therefore identifies the states on which projectability and energy ordering disagree, and under the frozen ceiling those states are assigned by their energy.

\section{Projectability distributions and derived thresholds}
\label{app:anatomy}
\begin{figure*}[tp]
\centering
\includegraphics[scale=0.67]{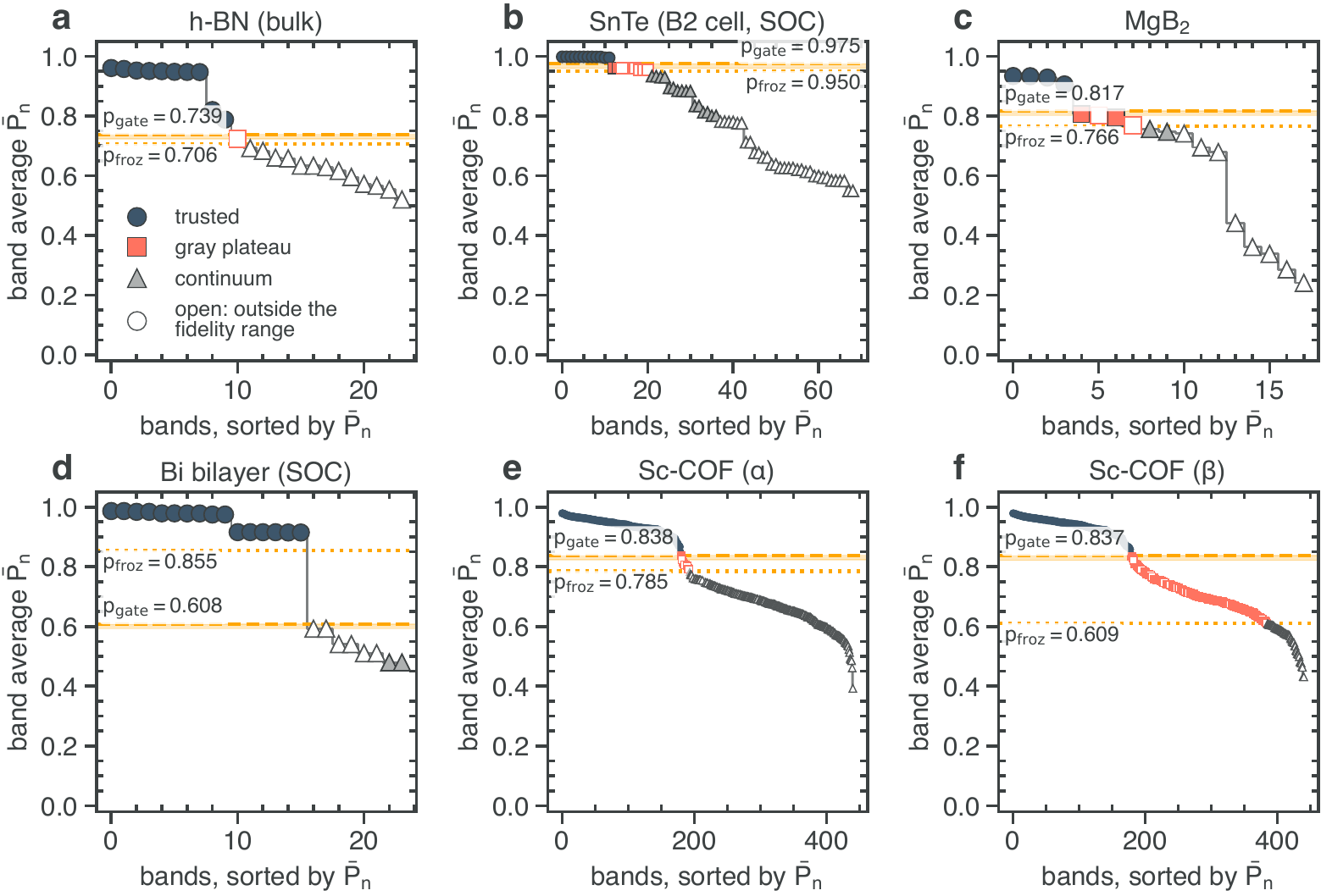}
\caption{Automatic trust thresholds read off the projectability distribution of each production run. Each panel shows the sorted band-averaged projectability $\bar P_n$ of the pool bands for (a)~bulk h-BN, (b)~SnTe (B2 cell, SOC), (c)~MgB$_2$, (d)~the Bi bilayer (SOC), and (e) and (f)~the $\alpha$ and $\beta$ channels of the Sc-COF. Bands are marked as trusted (slate circles), in the gray plateau of intermediate projectability (coral squares) or in the continuum (gray triangles), and open markers are bands that never enter the fidelity range. The Bi bilayer has no band in the gray plateau. The amber lines are the derived gate $p_{\text{gate}}$ (dashed, the highest band average outside the fidelity range plus $\delta$) and the state threshold $p_{\text{froz}}$ (dotted, the worst-$\kk$ projectability of the trusted bands minus $\delta_{\rm s}$, clipped to $[p_{\mathrm{clip}}^{-},p_{\mathrm{clip}}^{+}]$), labeled with the values of Table~\ref{tab:defaults}. On SnTe, five well-projected bands outside the fidelity range (open squares) raise the gate to 0.975, so that $p_{\text{froz}}$ reaches its upper clip of 0.95 (Sec.~\ref{sec:snte}). The data are the \texttt{-\kern0pt-hybrid-report-json} exports of these runs (\texttt{figures\_src/report\_to\_f3.py}).}
\label{fig:anatomy}
\end{figure*}

Figure~\ref{fig:anatomy} shows the sorted band-averaged projectability distributions, with the derived thresholds, for all six production runs, counting the two spin channels of the Sc-COF separately. All six distributions show a trusted cluster of well-projected bands and a continuum of weakly projected bands, and all but the Bi bilayer show a gray plateau of intermediate projectability between the two.

\section{Algorithm parameters and reported values}
\label{app:constants}
\begin{figure*}[tp]
\centering
\includegraphics[width=\textwidth]{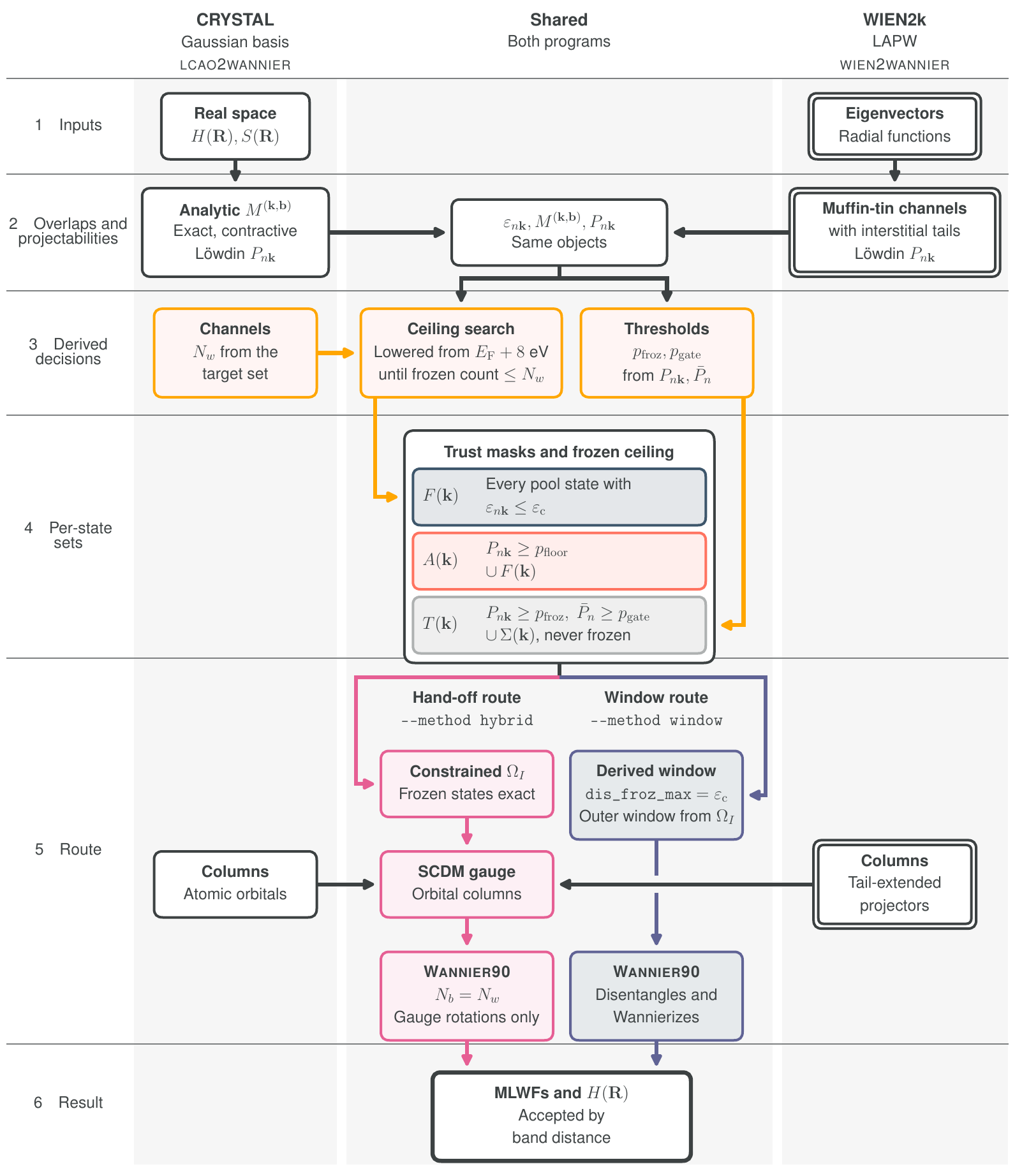}
\caption{The pipeline of Fig.~\ref{fig:pipeline} stage by stage, from top to bottom, for the CRYSTAL frontend (\textsc{lcao2wannier}, left), the steps shared by both programs (center) and the WIEN2k frontend (\textsc{wien2wannier}, right, double-ruled boxes). The hand-off is drawn in pink (\texttt{-\kern0pt-method hybrid}) and the window route in indigo (\texttt{-\kern0pt-method window}), and the per-state sets are those of Eqs.~\eqref{eq:masks} and \eqref{eq:frozen}. After stage 2 only CRYSTAL adds channels (stage 3) and applies the admission floor, and the SCDM gauge columns are atomic orbitals for CRYSTAL and tail-extended projectors for WIEN2k (Appendix~\ref{app:constants}). $\Sigma(\kk)$ is the Fermi shell (Sec.~\ref{sec:masks}), and the amber arrows mark the set that each derived decision (Sec.~\ref{sec:auto}) defines.}
\label{fig:lanes}
\end{figure*}

Figure~\ref{fig:lanes} lays out the pipeline of Fig.~\ref{fig:pipeline} stage by stage for both programs, and the parameters listed in this appendix belong to its stages 3 and 4.

The symbols below denote the parameters of the algorithm, given with the values used in the reported reference pipeline. These values are choices, not universal constants, and the material-dependent thresholds are outputs of the rule for the given parameters. Benchmark overrides and the choices specific to the LAPW frontend are stated separately.

\emph{Masks} (Sec.~\ref{sec:masks}). The admission floor is $p_{\text{floor}}=0.10$ in \textsc{lcao2wannier}, whereas the LAPW frontend has none (see below). The pool is contiguous, has a minimum size of $r_{\mathrm{pool}}N_w$ ($r_{\mathrm{pool}}=1.5$), is grown only to cover the seed span, and is seeded within $[E_{\mathrm{F}}+\Delta E_{\mathrm{seed}}^{-},\allowbreak E_{\mathrm{F}}+\Delta E_{\mathrm{seed}}^{+}]$ with $(\Delta E_{\mathrm{seed}}^{-},\Delta E_{\mathrm{seed}}^{+})=(-30,20)$~eV. In the per-state mode the selection caps the trusted counts at $N_w$ per $\kk$, taking the shell states first, and under the frozen ceiling the ceiling search enforces the same limit. Admitted sets with fewer than $N_w$ states are filled up to $N_w$ with the states of highest $P$.
The LAPW models (Table~\ref{tab:lapw}) admit every pool state, which is the default of the released \textsc{wien2wannier}, because on that frontend a floor of 0.10 breaks the subspace continuity of the Bi bilayer and of Fe with spin--orbit coupling. The Wannierizations of these two cells then do not meet their criterion within 300\,000 iterations, and their band errors become 7 to 28 times larger. The floor also breaks the continuity of the SrVO$_3$ subspace at the automatic ceiling, where the pocket skip restores it by lowering the ceiling to $+4.62$~eV. The same floor raises the band errors of $\alpha$-Sn, although it lowers those of MgB$_2$ (from 11.5 and 4.4 to 4.8 and 1.5~meV), and a floor that excludes only the states of exactly zero projectability breaks the continuity of Fe with spin--orbit coupling as well (Appendix~\ref{app:lapw} and SM).
\emph{Frozen ceiling} [Eq.~\eqref{eq:walk}]. The search starts at $E_{\text{fid}}$ and lowers the ceiling in steps of $\Delta E_{\rm step}=1$~eV, two values that are set rather than derived and whose effect is measured in the SM. It stops as soon as no $\kk$-point has more than $N_w$ states below the ceiling, and its lowest level is $E_{\mathrm{F}}+1$~eV. The frozen set is every pool state at or below the ceiling, and \textsc{wien2wannier}~v3.0 uses the same levels, starting energy and stopping condition.
\emph{Pocket skip} (Sec.~\ref{sec:masks}). The weak-link threshold is $\sigma_{\rm p}=0.5$, the largest pocket fraction is $f_{\rm p}=0.25$ of the
$\kk$-points, the mixed-character threshold is $p_{\rm mix}=0.5$, and the margin below the pocket band is $\delta_{\rm p}=0.1$~eV. The ceiling is lowered at most three
times and never below $E_{\mathrm{F}}+1$~eV (\texttt{POCKET\_*} in \texttt{hybrid.py} and \texttt{hybrid\_prep.F}).
\emph{Fermi shell}. For metals $\Delta E_{\mathrm{shell}}=3$~eV and the shell-floor exclusion budget is $f_{\mathrm{shell}}=0.01$, whereas for insulators the shell is the per-state occupied manifold below mid-gap.

\emph{Trust rule} [Eq.~\eqref{eq:autotrust}]. The fidelity offset is
$\Delta E_{\mathrm{fid}}=8$~eV, the margins are $\delta=0.015$ and $\delta_{\rm s}=0.02$, the
clipping bounds are $(p_{\mathrm{clip}}^{-},p_{\mathrm{clip}}^{+})=(0.55,0.95)$, and the
fallback pair is $(p_{\mathrm{froz}}^{\mathrm{fb}},p_{\mathrm{gate}}^{\mathrm{fb}})
=(0.95,0.85)$.

\emph{Augmentation} (Sec.~\ref{sec:auto}). The coverage is assessed on the $N_{\mathrm{cond}}=4$ lowest conduction bands, channels are added when it falls below $p_{\mathrm{aug}}=0.70$, and a channel is retained if $\Pi_c\ge\Pi_{\min}=0.40$. The criteria that identify semicore bands use $\Delta E_{\mathrm{sc}}=2$~eV and $p_{\mathrm{sc}}=0.85$. The continuum reference contains the $N_{\mathrm{ref}}=12$ lowest bands that lie entirely above $E_{\text{fid}}$ and have a mean energy below $E_{\mathrm{F}}+\Delta E_{\mathrm{ref}}$, where $\Delta E_{\mathrm{ref}}=40$~eV. A channel is admitted if it carries semicore bands or has a specificity $s_c\ge s_{\min}=0.05$. Of the admitted channels, the smallest set whose coverage lies within $\delta_C=0.02$ of that of all admitted channels is retained, and channels are added only if the coverage gain is at least $\Delta C=0.05$ (\texttt{set\_tol} and \texttt{min\_gain} in \texttt{channel\_augment.py}). All projectability, retention, specificity and coverage parameters in this list are dimensionless.

\emph{Disentanglement} (Sec.~\ref{sec:dis}). Grassmann CG is the default and stops when the projected gradient falls to $\max(10^{-10},10^{-6}|g_0|)$, and \texttt{-\kern0pt-hybrid-dis-solver gs} selects the serial Gauss--Seidel sweep. The sweep cap is $N_{\mathrm{sweep}}^{\max}=20\,000$ in \textsc{lcao2wannier} and 30\,000 in \textsc{wien2wannier}. A Gauss--Seidel run that stops at the cap is graded by comparing $\Omega_I$ with its value one tenth of the sweeps earlier, as `flat' below a relative change of $10^{-5}$ and otherwise as `still descending', and the grade is written into the \texttt{.win}. The absolute spread tolerance is
$\tau_{\Omega}=10^{-9}$~\AA$^2$, the relative tolerance is
$\tau_{\mathrm{rel}}=10^{-10}$, and the convergence-check interval is
$N_{\mathrm{check}}=1$ sweep. The \textsc{lcao2wannier} sweep uses neither over-relaxation nor the colored order, whereas the LAPW Gauss--Seidel models used both, with convergence checks at adaptive intervals starting from every 10 sweeps (Appendix~\ref{app:lapw}).
\emph{Gauge and continuity}. Gauge acceptance (Sec.~\ref{sec:scdm}) uses a condition-number limit
$\kappa_{\max}=10^{10}$ and no absolute $\sigma_{\min}$ floor, and the
measured $\sigma_{\min}$ spans $0.39$ (Bi) to $3.7\times10^{-3}$ (SnTe) for the models of Table~\ref{tab:defaults}. The subspace-continuity check (Sec.~\ref{sec:auto}) warns below $\min\sigma[W^\dagger(\kk)M^{(\kk,\bb)}W(\kk+\bb)]=10^{-2}$.
\emph{Other parameters}. The hand-off acceptance test [Eq.~\eqref{eq:handoff-acceptance}] uses the contractivity tolerance $\tau_M=10^{-6}$. The analytic overlaps (Appendix~\ref{app:contractivity}) skip AO pairs more than 16~bohr apart. The LAPW frontend (Appendix~\ref{app:lapw}) uses a degeneracy tolerance $\Delta E_{\mathrm{deg}}=0.02$~eV and transports the SCDM gauge from the best of up to 16 candidate seed $\kk$-points.
The hand-off conventions, $N_b=N_w$, energies relative to $E_{\mathrm{F}}$, and
the absence of \texttt{dis\_*} keywords, are fixed and are not tunable
parameters.

\section{Validation on an all-electron LAPW frontend}
\label{app:lapw}

\begin{figure*}[t]
\centering
\includegraphics[scale=0.67]{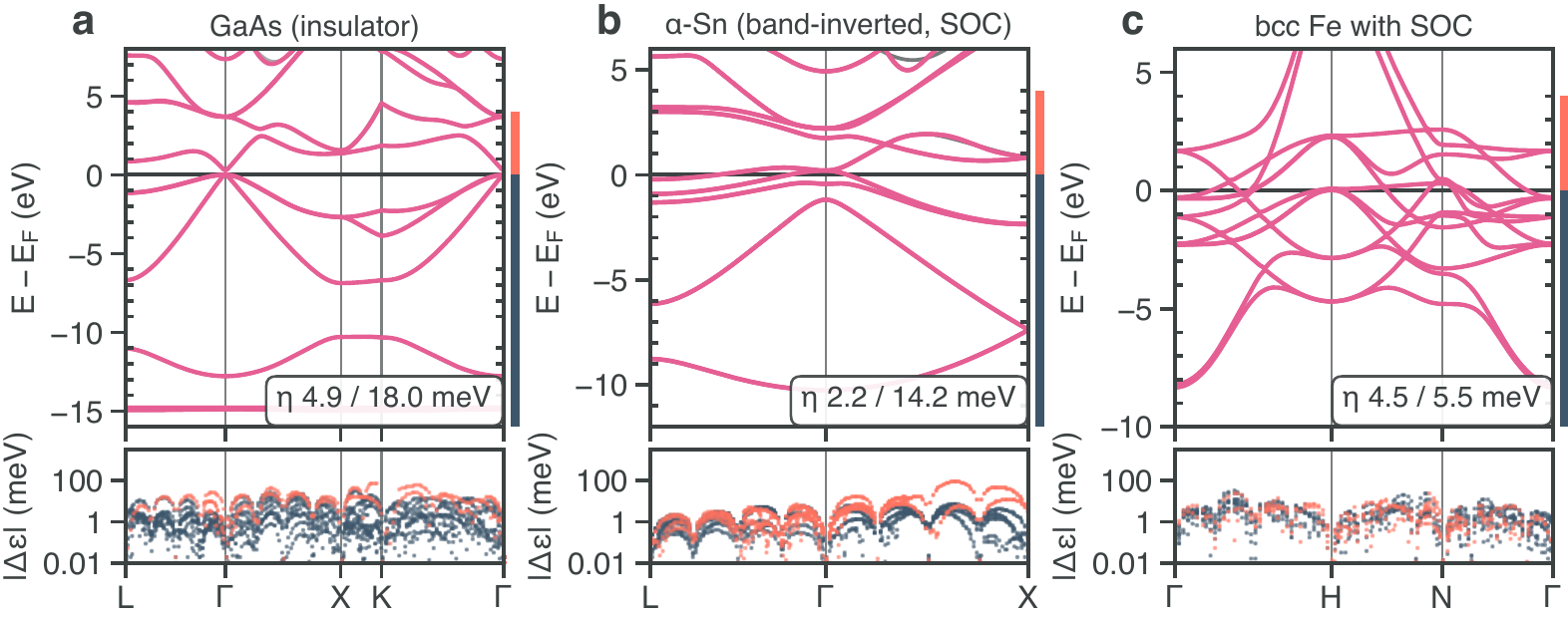}
\caption{LAPW band validation for (a)~GaAs (insulator, 18 WFs), (b)~$\alpha$-Sn (band-inverted, SOC, 20 WFs) and (c)~bcc Fe with SOC (18 WFs). The WIEN2k \texttt{lapw1}($+$\texttt{so}) reference is drawn thin and gray, and the Wannier interpolation, from the \texttt{\_hr.dat} with Wigner--Seitz corrections, thick and pink. Edge bars mark the two scoring windows of Table~\ref{tab:lapw}, in slate below $E_{\mathrm{F}}$ and in coral above it, and the box gives the band distance $\eta$ over them, with bands paired by index. The lower strips give $|\Delta\varepsilon|$ for every paired state inside the windows on a log scale, in slate for occupied and in coral for empty states. The paths follow the shipped \texttt{band\_compare.py} specification, with L--$\Gamma$--X only for $\alpha$-Sn.}
\label{fig:lapwbands}
\end{figure*}

\begin{figure*}[tp]
\centering
\includegraphics[scale=0.67]{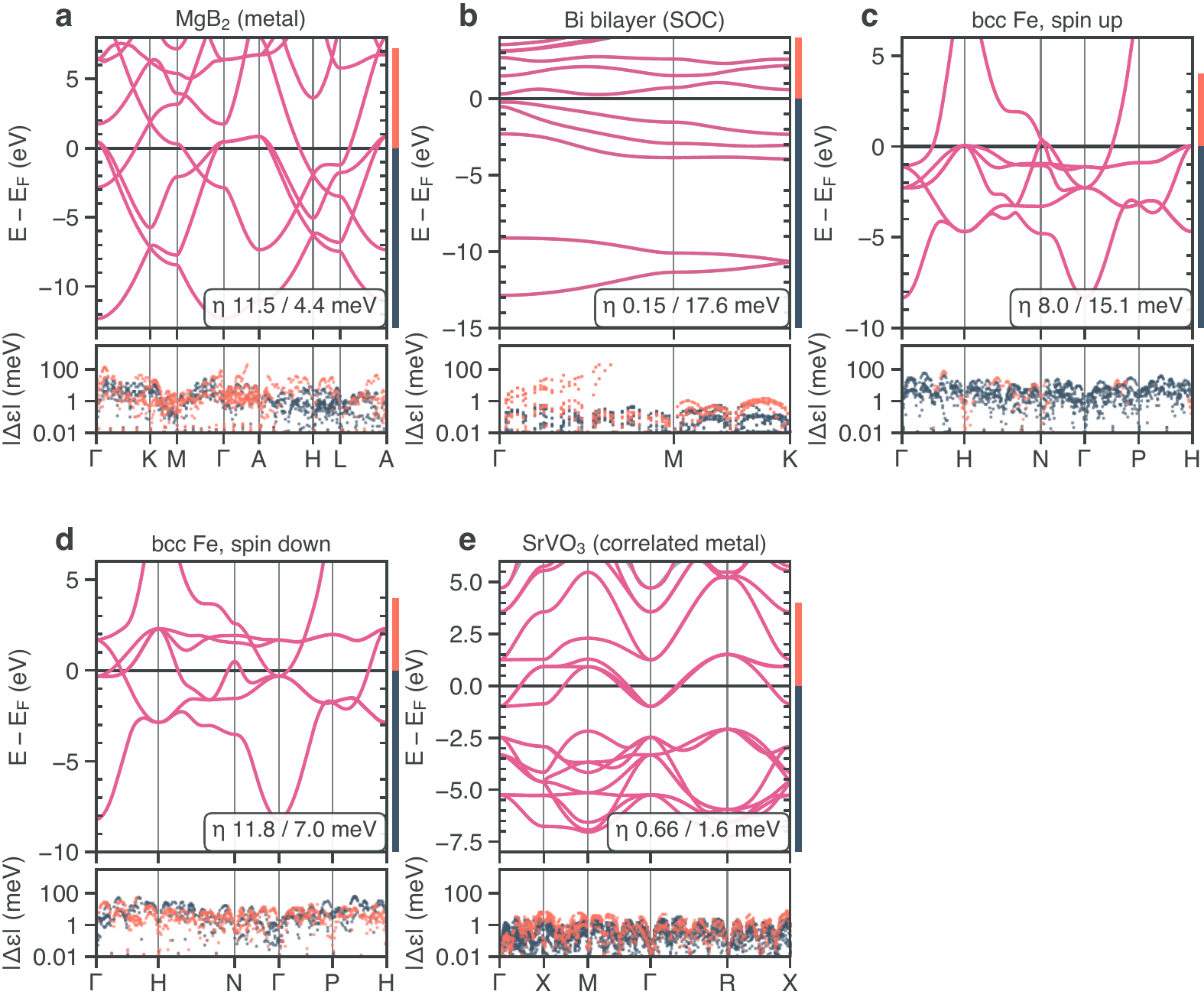}
\caption{The other five LAPW models of Table~\ref{tab:lapw}, drawn as in Fig.~\ref{fig:lapwbands}. (a)~MgB$_2$ (metal, 12 WFs), (b)~Bi bilayer (SOC, 36 WFs, path $\Gamma$--M--K only), (c)~bcc Fe spin up and (d)~spin down (9 WFs each), (e)~SrVO$_3$ (19 WFs).}
\label{fig:lapwbands2}
\end{figure*}

The method needs only the projectabilities of the bands onto a labeled set of localized orbitals, and the functions $u_\ell Y_{\ell m}$ inside the muffin-tin spheres of (L)APW codes~\cite{Kunes2010} provide these just as Gaussian AOs do. To test this, the procedure of Algorithm~\ref{alg:stage2}, with its trust masks, frozen ceiling, frozen-constrained $\Omega_I$ minimization, SCDM density-column gauge, $N_b{=}N_w$ hand-off and automatic thresholds, was run from the all-electron LAPW code WIEN2k~\cite{Blaha2020} through its \textsc{wien2wannier} interface~\cite{Kunes2010}. Every LCAO-specific input was replaced by its muffin-tin analogue, and the constraint, gauge and hand-off steps were left unchanged. The main results, namely the eight models from seven calculations, the MgB$_2$ runs in both frontends, the certificate for both projectabilities and the cost, are given in Sec.~\ref{sec:lapw-results}. This appendix gives the dictionary of replacements, the released implementation, the detailed certificate analysis and the tests of individual design choices.

\emph{Calculations.} The seven calculations use WIEN2k~24.1~\cite{Blaha2020}, with the augmented plane waves of Slater~\cite{Slater1937} linearized~\cite{Andersen1975,Koelling1975} in the APW+lo form~\cite{Sjostedt2000,Madsen2001} and local orbitals for the semicore states~\cite{Singh1991}. GaAs and $\alpha$-Sn also carry a $p$ local orbital built from the second energy derivative of the radial function~\cite{Karsai2017,Blaha2020}. The spheres use $\ell_{\max}=10$, and $R_{\rm MT}^{\min}K_{\max}$ is 7.5 (GaAs, $\alpha$-Sn), 6.04 (MgB$_2$), 7.0 (Bi, SrVO$_3$) and 8.0 (Fe). GaAs is computed in the local-density approximation~\cite{KohnSham1965,PerdewWang1992,*PerdewWang2018} and the other cells with PBE~\cite{Perdew1996,*Perdew1997}, and bcc Fe is spin-polarized~\cite{vonBarth1972}. Spin--orbit coupling enters $\alpha$-Sn, the Bi bilayer and Fe by second variation~\cite{MacDonald1980} on the scalar-relativistic states~\cite{Koelling1977}, self-consistently for $\alpha$-Sn and the Bi bilayer and in one step on the converged scalar-relativistic potential for Fe. $E_{\mathrm{F}}$ comes from the tetrahedron method~\cite{Blochl1994} for GaAs, MgB$_2$, $\alpha$-Sn, Fe and SrVO$_3$, and from a Fermi--Dirac smearing of 0.002~Ry for the Bi bilayer. Structures, muffin-tin radii and the settings of each self-consistent field are in the SM (structures and calculation parameters).

\emph{Dictionary.} The L\"owdin projectability is replaced by the muffin-tin channel projectability
$P_{n\kk} = \sum_{\mu,m} |\langle u_\ell Y_{\ell m}\,|\,
\psi_{n\kk}\rangle|^2$ in the projector convention of the $A_{mn}$ of the interface ($\dot u$ excluded)~\cite{Kunes2010}, so that the masks, the thresholds and the emitted \texttt{.amn} share one set of projectors. The muffin-tin projectability differs from the L\"owdin one in two respects, and the construction handles both. (i)~Inside the muffin-tin spheres alone, $P_{n\kk}$ is not a partition of unity, because part of the charge lies in the interstitial region outside the spheres. The released implementation removes this deficiency at its source by extending each projector past the sphere with a matched interstitial tail, so that the interstitial region enters the projectability itself (\emph{The consolidated LAPW implementation}, below, and the SM for the percentile rescaling used in the earlier porting runs). (ii)~Under strong SOC, sorting by energy splits the orbital character of a Kramers pair arbitrarily between band indices, so $P$ is symmetrized over degenerate groups ($|\Delta\varepsilon| < \Delta E_{\mathrm{deg}}$, $\Delta E_{\mathrm{deg}}=0.02$~eV). When the fraction of degenerate states is large ($\approx$1 with SOC vs.\ $<$0.25 otherwise), the per-state mode freezes whole bands ($\kk$-independent freezing) instead of single states, because in band-inverted SOC systems per-state freezing selects \emph{different} bands at different $\kk$, and the resulting $\kk$-discontinuous subspace stalls the minimization. This is the same failure that the band-average gate suppresses in the LCAO setting. In the SCDM discretization the role of the AO index is taken by the (atom, $\ell$, $m$) channel index, and the QRCP anchor selection then chooses the $N_w$ channels that best represent the disentangled subspace, which amounts to an automatic selection of projections.

\emph{The consolidated LAPW implementation.} The released \textsc{wien2wannier} runs the whole of Stage~2 itself and treats the LAPW-specific representation at the level of the projectors. Each radial projector $u_\ell$ is continued past the muffin-tin radius by a decaying exponential matched to its logarithmic derivative, the overlap of this tail with the interstitial plane-wave expansion is added to the muffin-tin matching, and the column is then renormalized. The per-state projectability is the L\"owdin form~\cite{Lowdin1950} $\operatorname{diag}(A S^{-1}A^\dagger)$ over these full-cell columns, which lies in $[0,1]$ and includes the interstitial charge, so that no percentile correction is needed. Channels are ranked by their sphere-summed weight, while the masks use the full-cell L\"owdin $P$. Because the per-$\kk$ SCDM gauge is fragile at ill-conditioned $\kk$-points, the gauge construction depends on the conditioning. Well-conditioned $\kk$-points keep their SCDM columns, while the others receive the seed subspace by orthogonal-Procrustes~\cite{Green1952,Schoenemann1966} parallel transport from a seed $\kk$-point, chosen as the best of up to 16 candidates, after which sweeps smooth the gauge. As an option, the SCDM pivots can be closed over whole $(\text{atom},\ell)$ shells, so that the anchor set forms a basis of a representation. The implementation contains both minimizers of Sec.~\ref{sec:dis}, the Gauss--Seidel sweep and the Grassmann-CG solver, which is the default. The sweep can visit the $\kk$-points in a colored order, which allows a parallel sweep but can change the minimum it reaches, and it has options for scheduled over-relaxation~\cite{Frankel1950,Young1954} and a safeguarded Anderson acceleration with a user-set history depth~\cite{Anderson1965,Walker2011}. The Gauss--Seidel models of Table~\ref{tab:lapw} used the colored order and over-relaxation (run details below). The numbers of Table~\ref{tab:lapw} are those of the native release run (Sec.~\ref{sec:lapw-results}), and the porting runs that preceded it are described in the SM.

\emph{Run details of Table~\ref{tab:lapw}.} The Gauss--Seidel hand-offs were made with three sweep options that are off by default, namely the colored update order, scheduled over-relaxation starting at a factor of 1.9, and convergence checks at adaptive intervals (\texttt{-\kern0pt-colored}, \texttt{-\kern0pt-relax -1.9}, \texttt{-\kern0pt-adaptive}). Over-relaxation and the adaptive checks change the number of sweeps but not the result, whereas the order of the sweep can change the result. Below the frozen ceiling the serial order gives the same subspace on every cell, and above it the serial order gives the same subspace on GaAs and both Fe spin channels and, to within 0.9~meV, on SrVO$_3$. On Fe with spin--orbit coupling the two orders reach the same $\Omega_I$ to $8\times10^{-6}$~\AA$^2$ but different subspaces above $+8.2$~eV, just above the ceiling of $+8.0$~eV, where a subspace eigenvalue changes by up to 1.1~eV, and the serial-order hand-off of this cell was not Wannierized. On MgB$_2$, however, it reaches a lower minimum of $\Omega_I$, whose model has 7.0 and 1.5~meV instead of 11.5 and 4.4~meV (SM, sweep options of the LAPW models).
Each hand-off was passed to \textsc{Wannier90} with its SCDM \texttt{.amn}, without guiding centers except on the Bi bilayer, and was Wannierized to the criterion in one uninterrupted run, at the end of which $\Omega_{\rm tot}$ is taken, and the SM lists the sweep and iteration counts. On $\alpha$-Sn, computed on the coarsest fcc mesh that contains $K$ and $U$ ($8^3$), the table reports the conjugate-gradient model, whose $\Omega_I$ lies $1.0\times10^{-3}$~\AA$^2$ below that of the colored sweep, and in the serial order the sweep does not converge within 30\,000 sweeps. On the Bi bilayer the sweep stopped at its cap of 30\,000 sweeps ($|\Delta\Omega_I|/\Omega_I=1.3\times 10^{-9}$), the SCDM anchors were ill-conditioned ($\sigma_{\min}=1.4\times 10^{-5}$), and the Wannierization oscillated from a random start, so the model uses guiding centers on Bi $s$, $p$, $d$ and then reproduces the occupied manifold to 0.2~meV. The subspace continuity of the GaAs hand-off, $5\times10^{-4}$, lies below the warning level of $10^{-2}$, yet the converged model interpolates the bands at 4.9 and 18.0~meV (SM), so a warning does not by itself imply a poor model.

\emph{Where the interstitial enters.} Both projectabilities are available in the release, which allows a direct comparison. At the automatic ceiling the muffin-tin and the full-cell columns give the same masks and the same model on GaAs, MgB$_2$, and $\alpha$-Sn ($\Omega_I$ identical to every printed digit), because every state below the ceiling is frozen and the ceiling is set by the Wannier count. The projectability affects the frozen set only when the per-state rule runs without a ceiling, and in that case the muffin-tin projectability, which lacks the interstitial weight of the conduction states of free-electron character, places fewer of these states in the trusted set. On GaAs it freezes 9.1 states per $\kk$-point compared to 12.3 with the full-cell columns and leaves 2582 unfrozen states inside its own frozen span, compared to 395 with the full-cell columns. Its band gate falls to 0.27, compared to 0.36 with the full-cell columns, and with the admission mask active the error over the conduction window rises to 52.7~meV, compared to 30.5~meV with the full-cell columns and 18.0~meV at the release protocol. On $\alpha$-Sn both variants driven by the projectability are worse than the release over the occupied window (5.1~meV with the muffin-tin columns and 6.6~meV with the full-cell columns, compared to 2.2~meV). Over the conduction window the better of them lies within 0.5~meV of the release (17.6 and 13.7 vs.\ 14.2~meV), so there the loss comes from the absent ceiling and not from the missing interstitial weight. A selection driven by the projectability therefore needs the interstitial tails to classify the conduction states of free-electron character correctly, whereas the released model does not depend on that classification, because the automatic ceiling fixes the frozen set. For the gauge, the full-cell columns help in a different way, since seeding the SCDM anchors from them improves the conditioning (on MgB$_2$, $\sigma_{\min}$ rises from $8\times10^{-5}$ to $1\times10^{-2}$) without changing the localized model ($\Omega_{\rm tot}$ identical to four digits). On the Bi bilayer, however, the better conditioning does not stop the oscillation of a Wannierization from a random start, whereas the guiding centers do.

\emph{Interpolation fidelity (Figs.~\ref{fig:lapwbands} and~\ref{fig:lapwbands2}).} Each of the eight hybrid Wannier models reproduces the WIEN2k LAPW bands over the stated windows to the $\eta$ of Table~\ref{tab:lapw}. The residual strips give the error of every paired state, and the gray LAPW bands are visible only above the fidelity window, which the model does not target.

\emph{Earlier porting runs.} The SM describes the certificate analysis of the muffin-tin masks of these runs, the joint GaAs model, and the cost comparison of the two routes in that setup.

\emph{Design choices tested on both frontends.} Three design choices of the LCAO frontend were tested again on the LAPW frontend. (i)~Guiding centers on the hand-off, placed automatically at the host atom of each SCDM anchor column (one card per anchor column), change nothing where the Wannier functions are centered on atoms (on MgB$_2$ $\Omega$ is identical to seven digits, and on $\alpha$-Sn in the porting runs $\Omega$ is 72.15 compared to 72.11~\AA$^2$, with $\eta$ within 0.2~meV), but they are harmful where the model contains diffuse functions centered off the atoms. On the joint GaAs model $\Omega/N_w$ rises from 2.7 to 3.9~\AA$^2$ and $\eta$ doubles, because guiding centers on atomic sites pull the functions that carry the states of free-electron character toward those sites, and the Wannierization ends in a minimum of higher spread. On the LCAO hand-offs guiding changes nothing on h-BN and MgB$_2$ and improved the occupied window of SnTe, so the LCAO hand-offs are guided. In LAPW the harm follows the diffuse content of the Wannier functions and is unrelated to spin--orbit coupling or to whether the cell is a metal. The LAPW hand-off is therefore unguided by default, since the SCDM gauge already provides a start that needs no guiding, whereas the window route depends on guided real projections, as in the LCAO setting [SM, stability lessons, rule~(7)]. (ii)~The unweighted density matrix gives better bands than erfc energy weighting (Sec.~\ref{sec:scdm}), and the band distance of MgB$_2$ improves by about 20\% when the weighting is removed, because the frozen and admitted sets already select the states. (iii)~Guiding centers taken from the SCDM anchors give no improvement over the standard rule, which uses guiding centers only with real projections, in the cases where guiding does no harm.

\emph{Admission floor.} The admission floor is the one LCAO default that does not carry over to the LAPW frontend. With the LCAO floor ($p_{\text{floor}}=0.10$), the full-cell projectability leaves 26 to 7\,292 non-frozen states per cell out of the admitted set, and the subspaces of the Bi bilayer and of Fe with spin--orbit coupling lose their continuity (weakest links $1\times10^{-3}$ and $3\times10^{-5}$), so that their Wannierizations do not meet the criterion within 300\,000 iterations and their band errors grow 7 to 28 times. The SrVO$_3$ subspace also loses its continuity at the automatic ceiling (weakest link $4.8\times10^{-19}$), and there the pocket skip restores it by lowering the ceiling to $+4.62$~eV, where the weakest link is 0.90 and the band errors are 0.5 and 1.7~meV, compared to 0.7 and 1.6~meV with every state admitted. On $\alpha$-Sn the floor excludes 557 states and raises the band errors from 2.2 and 14.2 to 4.2 and 17.5~meV without breaking the continuity of the subspace. The floor improves MgB$_2$ (from 11.5 and 4.4 to 4.8 and 1.5~meV) and the minority-spin channel of Fe, and it worsens the majority-spin channel and GaAs. Excluding only the states of exactly zero projectability breaks the continuity of the Fe-SOC subspace as well, through a zero at H on a band whose states at the neighboring $\kk$-points stay admitted. For these reasons the LAPW models admit every pool state (SM).

\emph{Scope.} At the release protocol, the window below the automatic ceiling gives all eight models through the hand-off, and the GaAs model also through the disentanglement of \textsc{Wannier90} itself. Releasing the untrusted states under the ceiling removes a target band from the Bi bilayer model, raises the band errors of Fe with spin--orbit coupling more than thirtyfold, and leaves GaAs, SrVO$_3$ and $\alpha$-Sn unchanged to within 2.2~meV (Sec.~\ref{sec:extent}). The WIEN2k models thus repeat the outcome of the h-BN and MgB$_2$ matched-extent experiments of Sec.~\ref{sec:extent} with a different code, basis set and type of projector. In the earlier porting runs hand-set windows fail on GaAs and give comparable errors on MgB$_2$ and $\alpha$-Sn, and the fraction of $\kk$-points at which the trust set is not an interval predicts which cells fail (SM). The data are the \textsc{wien2wannier} test archives, which contain the seven calculations of Table~\ref{tab:lapw} with the command line, masks and per-state projectability files of each cell, and the GaAs, MgB$_2$ and $\alpha$-Sn porting runs with their command lines. The \textsc{Wannier90} development build \texttt{ff955d1} was used throughout (SM), and the certificates and the values of $\eta$ were computed with the shipped scripts.

\nocite{AiidaW90,Li2023,StengelSpaldin2006}
\bibliography{references,arxiv_extra}

\clearpage
\onecolumngrid
\makeatletter
\setcounter{section}{0}\setcounter{subsection}{0}\setcounter{figure}{0}\setcounter{table}{0}
\setcounter{equation}{0}\setcounter{algorithm}{0}
\@removefromreset{equation}{section}
\setcounter{secnumdepth}{3}
\renewcommand{\thesection}{S\arabic{section}}
\renewcommand{\thesubsection}{\thesection.\arabic{subsection}}
\renewcommand{\thesubsubsection}{\thesubsection.\arabic{subsubsection}}
\renewcommand{\p@subsection}{}\renewcommand{\p@subsubsection}{}
\renewcommand{\thefigure}{S\arabic{figure}}\renewcommand{\theHfigure}{S\arabic{figure}}
\renewcommand{\thetable}{S\arabic{table}}\renewcommand{\theHtable}{S\arabic{table}}
\renewcommand{\theequation}{S\arabic{equation}}\renewcommand{\theHequation}{S\arabic{equation}}
\renewcommand{\thealgorithm}{S\arabic{algorithm}}\renewcommand{\theHalgorithm}{S\arabic{algorithm}}
\renewcommand{\theHsection}{S\arabic{section}}
\def\@hangfrom@section#1#2#3{\@hangfrom{#1#2}\MakeTextUppercase{#3}}
\let\@sectioncntformat\relax
\makeatother
\algrenewcommand\algorithmicindent{1.3em}
\renewcommand{\topfraction}{0.9}\renewcommand{\bottomfraction}{0.8}\renewcommand{\textfraction}{0.07}\renewcommand{\floatpagefraction}{0.85}
\begin{center}
{\large\bfseries Supplemental Material for\\[0.3em] Fully Automated Wannier Disentanglement via Projectability Selection in Gaussian and Augmented Plane-Wave Frameworks}
\end{center}
\vspace{1em}
This document collects the controlled comparisons, convergence records and certificates to which the main text refers, together with material moved out of the main text. Every number is read from a run log, a \textsc{Wannier90}~\cite{Pizzi2020} output file or a mask export named in the text. Runs are identified by their directory in the two calculation trees of the data archive~\cite{dataZenodo}, the LCAO tree (\texttt{calculations/}) and the \textsc{wien2wannier} test tree, both of which will be deposited with the paper. Unless stated otherwise, band errors are given as the band distance $\eta$ of the main text, with bands paired by index, in meV.

\section{Frozen extent: the ceiling study}
\label{sm:ceiling}

Section~\ref{sec:extent} of the main text separates the frozen extent from the frozen membership, and this section lists every run behind that comparison. All LCAO runs use the same pool, analytic overlaps, automatically derived thresholds, SCDM gauge~\cite{Damle2015,Damle2017} and \textsc{Wannier90} settings as Table~\ref{tab:defaults}, with the \textsc{Wannier90} development build \texttt{ff955d1}, whose results v4.0.2~\cite{W90v402} reproduces bitwise. Their disentanglement minimizes the constrained $\Omega_I$~\cite{Souza2001} by the Gauss--Seidel sweep, which was the default when they were made, and converged to a relative change of $10^{-10}$ within the cap of 20\,000 sweeps. The exceptions are the h-BN rows of Table~\ref{tab:sm-ceiling}, which are conjugate-gradient models and of which the $E_{\mathrm{F}}+8$~eV row is the model of Table~\ref{tab:defaults}. SnTe, at the Fermi energy of CRYSTAL~\cite{Crystal23}, is discussed below the table. The LCAO runs are in \texttt{calculations/ceiling\_study}, \texttt{calculations/final\_default}, \texttt{calculations/table1\_cg} and \texttt{calculations/table1\_cg46}, and the LAPW runs are in the test tree of our \textsc{wien2wannier}~\cite{Kunes2010} release v3.0 (\texttt{tests/regress\_out/ceiling\_study\_lapw.sh}).

\paragraph{Membership at fixed extent (LCAO).} On h-BN the two frozen sets are compared at the top of the trusted set, $+12.65$~eV. With the trust mask frozen, $\Omega_I=11.37$ and $\Omega_{\rm tot}=14.01$~\AA$^2$, and the band errors are 1.3, 1.9 and 52.2~meV over $[-20,0]$, $[+5,+8]$ and $[+5,+12]$~eV. When every pool state below $+12.65$~eV is frozen, which adds 378 states to the frozen set, the minimization reaches $\Omega_I=11.89$~\AA$^2$ in 316 sweeps, the Wannierization meets its criterion at 1900 iterations with $\Omega_{\rm tot}=14.60$~\AA$^2$, and the errors over the same windows are 1.5, 1.3 and 4.9~meV. On MgB$_2$ at $E_{\mathrm{F}}+8$~eV the window adds 689 states and gives $\Omega_I=12.87$~\AA$^2$ in 1510 sweeps and $\Omega_{\rm tot}=16.68$~\AA$^2$, with the criterion met at 5690 iterations. Its errors over $[-13,+7.2]$ and $[-2,+2]$~eV are 7.0 and 4.5~meV, compared to 24.1 and 5.1~meV with the trust mask frozen.

\paragraph{Extent (both frontends).} Table~\ref{tab:sm-ceiling} lists the runs of main-text Table~\ref{tab:ceiling} with their spreads and convergence. The top of the trusted set is computed from the trust mask of each run, for the LAPW cells from the release masks with the L\"owdin~\cite{Lowdin1950} full-cell projectability.

\begin{table}[htbp]
\caption{Frozen-ceiling runs. The rule ``fid'' is the released rule, which lowers the ceiling progressively from $E_{\mathrm{F}}+8$~eV, and the rule ``top'' lowers it progressively from the top of the trusted set. ``Frozen'' is the number of frozen states at the worst $\kk$-point, to be compared with $N_w$, and $\Omega_{\rm tot}$ is the total spread after the Wannierization. Band errors are in meV, with bands paired by index, over the scoring windows of main-text Table~\ref{tab:ceiling} (relative to $E_{\mathrm{F}}$) in its order ($[-20,0]$, $[+5,+8]$ and $[+5,+12]$~eV for h-BN, and the two windows of that table for the other cells). ``Dis.'' gives the disentanglement steps (sw for Gauss--Seidel sweeps and ev.\ for CG gradient evaluations), and ``W90'' gives the Wannierization iterations, or conv.\ when the criterion was met.}
\label{tab:sm-ceiling}
\small
\begin{tabular*}{\textwidth}{@{\extracolsep{\fill}}l l c r c l l r@{}}
\toprule
Cell ($N_w$) & Rule & $\varepsilon_{\rm c}$ & Frozen & $\Omega_{\rm tot}$ & Band error & Dis. & W90 \\
 & & (eV) & & (\AA$^2$) & (meV) & & \\
\midrule
h-BN (16) & fid & $+8.00$ & 10 & 11.88 & 0.56; 5.2; 1800 & 149 ev. & 1\,400 \\
 & top & $+12.65$ & 12 & 14.41 & 1.05; 1.1; 6.9 & 144 ev. & 8\,160 \\
GaAs (18) & fid & $+8.0$ & 14 & 52.49 & 4.9; 18.0 & 120 sw & 59\,500 \\
 & top & $+10.93$ & 17 & 53.18 & 7.1; 12.1 & 121 sw & 7\,000 \\
MgB$_2$, LAPW (12) & fid & $+8.0$ & 8 & 18.15 & 11.5; 4.4 & 120 sw & conv. \\
 & top & $+13.06$ & 12 & 23.23 & 8.0; 7.9 & 674 sw & 12\,000 \\
$\alpha$-Sn (20) & fid & $+6.0$ & 18 & 65.57 & 2.2; 14.2 & 2\,883 ev. & 30\,500 \\
 & top & $+5.79$ & 18 & 61.73 & 2.1; 9.8 & 75 ev. & 500 \\
Fe$\uparrow$ (9) & fid & $+8.0$ & 7 & 7.48 & 8.0; 15.1 & 110 sw & conv. \\
 & top & $+14.64$ & 9 & 8.61 & 8.4; 21.3 & 110 sw & 500 \\
Fe$\downarrow$ (9) & fid & $+8.0$ & 7 & 7.58 & 11.8; 7.0 & 110 sw & conv. \\
 & top & $+16.32$ & 9 & 9.10 & 15.8; 11.0 & 129 sw & 2\,000 \\
Fe-SOC (18) & fid & $+8.0$ & 14 & 12.73 & 4.5; 5.5 & 8049 sw & 64\,500 \\
 & top & $+16.63$ & 18 & 18.70 & 8.4; 6.9 & 2558 sw & 49\,500 \\
\bottomrule
\end{tabular*}

\end{table}

SnTe has no row, because with 46 functions both searches end at $E_{\mathrm{F}}+6.18$~eV under the pocket skip (main-text Sec.~\ref{sec:masks}). An explicit ceiling at the top of its trusted set ($E_{\mathrm{F}}+7.64$~eV), which the pocket skip does not change, gives 11.4 and 24.1~meV (5\,560 iterations), compared to 8.4 and 24.4~meV at an explicit $E_{\mathrm{F}}+8$~eV. The following run is a historical diagnostic and not a released result. With the earlier count of 36 functions and the earlier Fermi reference, and frozen to the top of its trusted set (Gauss--Seidel, with all 36 directions frozen at the worst $\kk$), it did not meet the criterion in 400\,000 iterations with the default settings, since $\Omega_{\rm tot}$ held at 108.478~\AA$^2$ to three decimals from 175\,000 iterations on, with recurring excursions to 109.0~\AA$^2$. The run converged with \texttt{precond = .true.} (76\,530 iterations) or \texttt{num\_cg\_steps = 200} (15\,440), at 38 and 45--49~meV.

A condition on spare capacity, which extends the ceiling only while the worst $\kk$-point keeps a fraction of the $N_w$ directions free, was tested on all cells before any run and rejected. The resolved ceilings of MgB$_2$ (LAPW), Fe$\downarrow$ and Fe-SOC change between $+8$ and $+12$~eV as the fraction varies from 10 to 25\%, so the rule would depend on a constant tuned to this set of cells.

\section{The extent and ceiling studies with the conjugate-gradient solver}
\label{sm:cg-studies}

Main-text Table~\ref{tab:ceiling} reports Gauss--Seidel models, the default when they were made, except where marked, and main-text Table~\ref{tab:extent} gives both solvers. Table~\ref{tab:sm-cg-studies} repeats the two LCAO cells of the extent comparison and the Gauss--Seidel rows of the ceiling comparison with the conjugate-gradient solver and every other input unchanged. The unchanged inputs are the code that made each model (for LCAO the code of main-text Table~\ref{tab:defaults}, Sec.~\ref{app:perf}, and for LAPW release v3.0), the pool, the thresholds, the ceiling, the gauge and the \textsc{Wannier90} input file, so that within each pair of columns only the solver of the constrained minimization differs. In what follows, a frozen set or a ceiling search is called better when its index-paired $\eta$ is lower, and the comparisons are between frozen sets or between ceiling searches and do not rank the solvers. Every Wannierization met its criterion in one run, within 25\,500 iterations, and the LAPW cells of the extent comparison are listed with both solvers in Sec.~\ref{sm:mixed}.

In the ceiling comparison every cell keeps the order of Table~\ref{tab:ceiling} in every window. The search from $E_{\mathrm{F}}+8$~eV is better in both windows of Fe$\uparrow$, Fe$\downarrow$ and Fe with spin--orbit coupling, over the occupied window of GaAs and over the Fermi window of MgB$_2$, whereas the search from the top of the trusted set is better over the conduction window of GaAs and the wide window of MgB$_2$. The band errors move by at most 0.8~meV, except over the GaAs conduction window of the search from the trusted top (12.1 to 16.9~meV). In the extent comparison the conjugate-gradient solver lowers the errors of both frozen sets on MgB$_2$ (trust mask from 24.1 and 5.1 to 11.3 and 2.8~meV, window from 7.0 and 4.5 to 3.6 and 3.4~meV) and on the h-BN valence band, and it raises both over $[+5,+12]$~eV on h-BN (trust mask from 52.2 to 61.2~meV, window from 4.9 to 6.9~meV). The order changes in one window, the MgB$_2$ Fermi window, where the trust mask now gives the lower error (2.8 vs.\ 3.4~meV). In the other four windows the order is as in Table~\ref{tab:extent}, with the window giving the lower error everywhere except over the h-BN valence band (trust mask 0.95 vs.\ 1.05~meV).

\begin{table}[htbp]
\caption{The two LCAO cells of main-text Table~\ref{tab:extent} (extent) and the Gauss--Seidel rows of main-text Table~\ref{tab:ceiling} (ceiling)
with the conjugate-gradient solver. GS denotes the Gauss--Seidel sweep, as published, and CG conjugate gradients. The band distance $\eta$ is in meV, with bands paired by index, over the scoring windows of the main-text tables, which the column ``Window (eV)'' gives relative to $E_{\mathrm{F}}$. In the upper block the columns headed ``Trust mask'' and ``Window'' give the two frozen sets.}
\label{tab:sm-cg-studies}
\small
\begin{tabular*}{\textwidth}{@{\extracolsep{\fill}}l l r r r r@{}}
\toprule
\multicolumn{6}{l}{Extent (Table~\ref{tab:extent}), frozen set at the same ceiling $\varepsilon_{\rm c}$} \\
\midrule
Cell, $\varepsilon_{\rm c}$ & Window (eV) & \multicolumn{2}{c}{Trust mask} & \multicolumn{2}{c}{Window} \\
\cmidrule(lr){3-4}\cmidrule(l){5-6}
 & & GS & CG & GS & CG \\
\midrule
h-BN, $+12.65$~eV & $[-20,0]$ & 1.3 & 0.95 & 1.5 & 1.05 \\
 & $[+5,+8]$ & 1.9 & 1.7 & 1.3 & 1.1 \\
 & $[+5,+12]$ & 52.2 & 61.2 & 4.9 & 6.9 \\
MgB$_2$, $+8.0$~eV & $[-13,+7.2]$ & 24.1 & 11.3 & 7.0 & 3.6 \\
 & $[-2,+2]$ & 5.1 & 2.8 & 4.5 & 3.4 \\
\midrule
\multicolumn{6}{l}{Ceiling (Table~\ref{tab:ceiling}), LAPW cells, with the ceiling lowered progressively from} \\
\midrule
Cell & Window (eV) & \multicolumn{2}{c}{$E_{\mathrm{F}}+8$~eV} & \multicolumn{2}{c}{Trusted-set top} \\
\cmidrule(lr){3-4}\cmidrule(l){5-6}
 & & GS & CG & GS & CG \\
\midrule
GaAs & $[-16,0]$ & 4.9 & 5.6 & 7.1 & 6.3 \\
 & $[0,+4]$ & 18.0 & 17.7 & 12.1 & 16.9 \\
MgB$_2$ & $[-13,+7.2]$ & 11.5 & 11.9 & 8.0 & 7.8 \\
 & $[-2,+2]$ & 4.4 & 4.6 & 7.9 & 7.4 \\
Fe$\uparrow$ & $[-10,0]$ & 8.0 & 8.0 & 8.4 & 8.4 \\
 & $[0,+4]$ & 15.1 & 15.2 & 21.3 & 21.3 \\
Fe$\downarrow$ & $[-10,0]$ & 11.8 & 11.5 & 15.8 & 15.0 \\
 & $[0,+4]$ & 7.0 & 6.9 & 11.0 & 10.7 \\
Fe-SOC & $[-10,0]$ & 4.5 & 4.5 & 8.4 & 8.4 \\
 & $[0,+4]$ & 5.5 & 5.7 & 6.9 & 6.7 \\
\bottomrule
\end{tabular*}
\end{table}

\section{The two constants of the ceiling search}
\label{sm:walkconst}

The ceiling search of main-text Eq.~\eqref{eq:walk} has two constants that are set rather than derived, the starting energy $E_{\text{fid}}=E_{\mathrm{F}}+8$~eV and the step $\Delta E_{\rm step}=1$~eV. Each was changed separately, with every other input kept as published. The CRYSTAL cells were run with the protocol of main-text Table~\ref{tab:defaults} and \texttt{-\kern0pt-hybrid-fid-emax}, and the LAPW cells with builds of the released \textsc{wien2wannier} in which either the step or the two compile-time values that hold 8~eV (the starting energy of the search and the fidelity limit of the automatic thresholds) are changed. The reference builds reproduce the published hand-offs of $\alpha$-Sn, SrVO$_3$ and the Bi bilayer byte for byte, and \textsc{Wannier90} runs every changed model to its criterion in one run (runs in \texttt{calculations/r2\_sensitivity} and \texttt{calculations/fid10}).

\paragraph{Fidelity limit.} On h-BN and MgB$_2$ the search stops at its first level for every starting energy, so the ceiling equals $E_{\text{fid}}$ (Table~\ref{tab:sm-efid}), and the model is not fitted above it. At $E_{\mathrm{F}}+6$~eV the h-BN ceiling lies below the conduction band and freezes only the eight valence bands. Below the ceiling, the error over a window grows as the top of the window approaches the ceiling. Over $[-13,+6]$~eV MgB$_2$ gives 7.7, 3.2 and 2.9~meV with the window top 0, 2 and 4~eV below the ceiling, and over $[+5,+8]$~eV h-BN gives 5.2 and 1.7~meV with the window top 0 and 2~eV below it. Windows that end at least 2~eV below both ceilings change by less than 0.5~meV, and each 2~eV added to $E_{\text{fid}}$ raises $\Omega_{\rm tot}/N_w$ by 3.6 to 10.6\%.

\begin{table}[htbp]
\caption{h-BN and MgB$_2$ (spin channel $\alpha$ of the open-shell calculation with zero moment, Table~\ref{tab:sm-crystal}) at three fidelity limits, with the protocol of main-text
Table~\ref{tab:defaults} otherwise (conjugate-gradient hand-off, \textsc{Wannier90} to its criterion). The band distance $\eta$ is in meV with bands paired by index, the scoring windows are relative to $E_{\mathrm{F}}$, and the ceiling equals $E_{\text{fid}}$ in every column.}
\label{tab:sm-efid}
\centering\small
\begin{tabular*}{\textwidth}{@{\extracolsep{\fill}}l l r r r@{}}
\toprule
Cell & & $E_{\mathrm{F}}+6$~eV & $E_{\mathrm{F}}+8$~eV & $E_{\mathrm{F}}+10$~eV \\
\midrule
h-BN & $\eta$ over $[-20,0]$ & 0.52 & 0.56 & 0.88 \\
 & $\eta$ over $[+5,+8]$ & 420.5 & 5.2 & 1.7 \\
 & $\eta$ over $[+5,+10]$ & 1710 & 1547 & 21.5 \\
 & $\Omega_{\rm tot}/N_w$ (\AA$^2$) & 0.717 & 0.743 & 0.821 \\
 & \textsc{Wannier90} iterations & 1\,580 & 1\,400 & 1\,870 \\
\addlinespace
MgB$_2$ & $\eta$ over $[-13,+6]$ & 7.7 & 3.2 & 2.9 \\
 & $\eta$ over $[-13,+7.2]$ & 130.5 & 3.6 & 3.0 \\
 & $\eta$ over $[-2,+2]$ & 3.0 & 3.4 & 2.9 \\
 & $\eta$ over $[-13,+10]$ & 261.3 & 63.5 & 5.3 \\
 & $\Omega_{\rm tot}/N_w$ (\AA$^2$) & 1.181 & 1.284 & 1.365 \\
 & \textsc{Wannier90} iterations & 2\,520 & 3\,040 & 2\,630 \\
\bottomrule
\end{tabular*}
\end{table}

\paragraph{$E_{\mathrm{F}}+10$~eV as the default.} The starting energy $E_{\text{fid}}=E_{\mathrm{F}}+10$~eV was also run on h-BN, MgB$_2$ and the LAPW models other than the Bi bilayer (Table~\ref{tab:sm-efid10}). It improves MgB$_2$ in both programs, most on the LAPW frontend (from 11.5 and 4.4 to 2.9 and 1.0~meV), and improves the two Fe spin channels slightly. It changes GaAs by at most 0.6~meV, and it leaves unchanged the models whose ceiling is set by the Wannier count (the CRYSTAL Bi bilayer at $+7$~eV and LAPW $\alpha$-Sn and SrVO$_3$ at $+6$~eV), whose hand-offs are identical byte for byte. It makes Fe with spin--orbit coupling five to eight times worse. With the ceiling at $+10$~eV, 436 more states are frozen, $\Omega_I$ rises from 11.2 to 12.5~\AA$^2$, and the Wannierization ends at $\Omega_{\rm tot}=16.09$ compared to 12.73~\AA$^2$. A convergence window of 100 iterations gives the same value as the published window of 5, so this value is a minimum and not an early stop. For this reason the default starting energy stays at $E_{\mathrm{F}}+8$~eV.

\begin{table}[htbp]
\caption{$E_{\text{fid}}=E_{\mathrm{F}}+8$~eV (published) compared to $E_{\mathrm{F}}+10$~eV on the models whose ceiling the change moves.
The columns $\varepsilon_{\rm c}$ and ``Window (eV)'' give the resolved ceiling and the scoring windows relative to $E_{\mathrm{F}}$. The band distance $\eta$ is in meV with bands paired by index, and $\Omega_{\rm tot}/N_w$ is the total spread per Wannier function after the Wannierization. The CRYSTAL Bi bilayer ($+7$~eV) and LAPW $\alpha$-Sn and SrVO$_3$ ($+6$~eV) are unchanged, with hand-offs identical byte for
byte, because the Wannier count sets their ceilings. On GaAs the pocket skip acts at $+10$~eV, but its result is rejected because the weakest link stays below 0.5, so the ceiling stays at $+10$~eV.}
\label{tab:sm-efid10}
\centering\small
\begin{tabular*}{\textwidth}{@{\extracolsep{\fill}}l c l r r r r@{}}
\toprule
 & & & \multicolumn{2}{c}{$\eta$ (meV)} & \multicolumn{2}{c}{$\Omega_{\rm tot}/N_w$ (\AA$^2$)} \\
\cmidrule(lr){4-5}\cmidrule(l){6-7}
Cell & $\varepsilon_{\rm c}$ (eV) & Window (eV) & $E_{\mathrm{F}}+8$ & $E_{\mathrm{F}}+10$ & $E_{\mathrm{F}}+8$ & $E_{\mathrm{F}}+10$ \\
\midrule
h-BN & $+8\to+10$ & $[-20,0]$ & 0.56 & 0.88 & 0.74 & 0.82 \\
 & & $[+5,+8]$ & 5.2 & 1.7 & & \\
MgB$_2$ & $+8\to+10$ & $[-13,+7.2]$ & 3.6 & 3.0 & 1.28 & 1.36 \\
 & & $[-2,+2]$ & 3.4 & 2.9 & & \\
\addlinespace
GaAs, LAPW & $+8\to+10$ & $[-16,0]$ & 4.9 & 5.2 & 2.92 & 2.82 \\
 & & $[0,+4]$ & 18.0 & 17.4 & & \\
MgB$_2$, LAPW & $+8\to+10$ & $[-13,+7.2]$ & 11.5 & 2.9 & 1.51 & 1.46 \\
 & & $[-2,+2]$ & 4.4 & 1.0 & & \\
Fe$\uparrow$, LAPW & $+8\to+10$ & $[-10,0]$ & 8.0 & 6.6 & 0.83 & 0.86 \\
 & & $[0,+4]$ & 15.1 & 14.2 & & \\
Fe$\downarrow$, LAPW & $+8\to+10$ & $[-10,0]$ & 11.8 & 10.6 & 0.84 & 0.87 \\
 & & $[0,+4]$ & 7.0 & 6.3 & & \\
Fe-SOC, LAPW & $+8\to+10$ & $[-10,0]$ & 4.5 & 23.9 & 0.71 & 0.89 \\
 & & $[0,+4]$ & 5.5 & 41.7 & & \\
\bottomrule
\end{tabular*}
\end{table}

\paragraph{Step size.} The step matters only where the Wannier count sets the ceiling, that is, on the LAPW cells $\alpha$-Sn, SrVO$_3$ and the Bi bilayer, which were run here, and on the CRYSTAL Bi bilayer ($+7$~eV), which was not rerun, since on every other model the search stops at its first level. A step of 0.5~eV raises each of the three LAPW ceilings by 0.5~eV (Table~\ref{tab:sm-step}). The conduction window improves on the Bi bilayer (from 17.6 to 2.5~meV), whose window top lay at the ceiling, and on $\alpha$-Sn (from 14.2 to 7.4~meV), whereas SrVO$_3$ does not change (0.6 and 1.6 vs.\ 0.7 and 1.6~meV). In exchange, $\Omega_{\rm tot}$ rises by 19 to 32\%, the weakest link of the hand-off overlaps drops on all three cells, the occupied windows of $\alpha$-Sn and the Bi bilayer rise by 0.5 and 0.1~meV, and SrVO$_3$ needs three times as many \textsc{Wannier90} iterations. The released programs therefore keep the step of 1~eV.

\begin{table}[htbp]
\caption{Ceiling search with steps of 1~eV (published) and 0.5~eV on the cells whose ceiling is set by the Wannier count. The weakest link is the smallest singular value of the hand-off overlaps (main-text Sec.~\ref{sec:masks}), $\Omega_I$ is the gauge-invariant spread of the disentangled subspace, and the column \textsc{Wannier90} gives the Wannierization iterations. The band distance $\eta$ is in meV with bands paired by index, over the two scoring windows of main-text Table~\ref{tab:lapw} (relative to $E_{\mathrm{F}}$).}
\label{tab:sm-step}
\centering\small
\begin{tabular*}{\textwidth}{@{\extracolsep{\fill}}l l r r r r r r@{}}
\toprule
Cell & Step & $\varepsilon_{\rm c}$ (eV) & $\Omega_I$ (\AA$^2$) & Weakest link & \textsc{Wannier90} & Occupied & Conduction \\
\midrule
$\alpha$-Sn & 1~eV & $+6.0$ & 55.94 & 0.496 & 30\,500, met & 2.2 & 14.2 \\
 & 0.5~eV & $+6.5$ & 64.21 & 0.262 & 25\,500, met & 2.7 & 7.4 \\
\addlinespace
Bi bilayer & 1~eV & $+4.0$ & 80.19 & 0.609 & 58\,500, met & 0.2 & 17.6 \\
 & 0.5~eV & $+4.5$ & 101.16 & 0.442 & 84\,500, met & 0.3 & 2.5 \\
\addlinespace
SrVO$_3$ & 1~eV & $+6.0$ & 21.96 & 0.863 & 45\,500, met & 0.7 & 1.6 \\
 & 0.5~eV & $+6.5$ & 24.73 & 0.810 & 142\,000, met & 0.6 & 1.6 \\
\bottomrule
\end{tabular*}
\end{table}

\section{Releasing states under the ceiling on the LAPW frontend}
\label{sm:mixed}

The released \textsc{wien2wannier} freezes every pool state below the automatic ceiling (main-text Eq.~\eqref{eq:frozen}). The implementation reaches this in two steps, in which the ceiling removes the trusted states above it and a per-$\kk$ freezing floor adds the states below it that the projectability rule left out of the trusted set. This freezing floor is distinct from the admission floor $p_{\rm floor}$ of Sec.~\ref{sm:floor}. We tested whether releasing these states below the ceiling improves the model by disabling the freezing floor, with every other input identical (L\"owdin full-cell projectability, the resolved ceiling, the auto8 SCDM gauge, admission of every pool state, the same solver and the same \textsc{Wannier90} input). Variant x1 disables the freezing floor, and variant x2 also raises the projectability threshold for Fermi-shell states from 0.15 to $p_{\rm froz}$, so that the shell states must pass the same test as every other state and those below $p_{\rm froz}$ are released as well. Both variants were built from the release source with \texttt{tests/regress\_out/mixed\_window\_flags.patch} applied, and the released code does not contain the two options.

\begin{table}[htbp]
\caption{Releasing states under the ceiling on two LAPW cells. ``release'' is the published frozen set, which freezes every pool state below the ceiling, x1 disables the per-$\kk$ freezing floor, and x2 also raises the threshold for Fermi-shell states to $p_{\rm froz}$. ``Released'' counts the pool states below the ceiling that are not frozen, summed over the mesh. The spreads are $\Omega_I$ of the disentangled subspace and $\Omega_{\rm tot}$ after the Wannierization, and the band errors are index-paired values over the scoring windows of main-text Table~\ref{tab:lapw} (meV, windows relative to $E_{\mathrm{F}}$).}
\label{tab:sm-mixed}
\small
\begin{tabular*}{\textwidth}{@{\extracolsep{\fill}}l l r r r r r@{}}
\toprule
Cell & Variant & Released & $\Omega_I$ (\AA$^2$) & $\Omega_{\rm tot}$ (\AA$^2$) & Occupied & Conduction \\
\midrule
$\alpha$-Sn, 20 WF, $E_{\mathrm{F}}{+}6$ & release & 0 & 55.94 & 65.57 & 2.2 & 14.2 \\
 & x1 & 30 & 55.93 & 65.57 & 2.1 & 14.1 \\
 & x2 & 76 & 54.81 & 64.66 & 2.1 & 11.4 \\
Bi bilayer, 36 WF, $E_{\mathrm{F}}{+}4$ & release & 0 & 80.19 & 88.34 & 0.2 & 17.6 \\
 & x1 $=$ x2 & 222 & 58.45 & 65.42 & 0.2 & 1167.5 \\
\bottomrule
\end{tabular*}
\end{table}

\begin{figure}[htbp]
\centering
\includegraphics{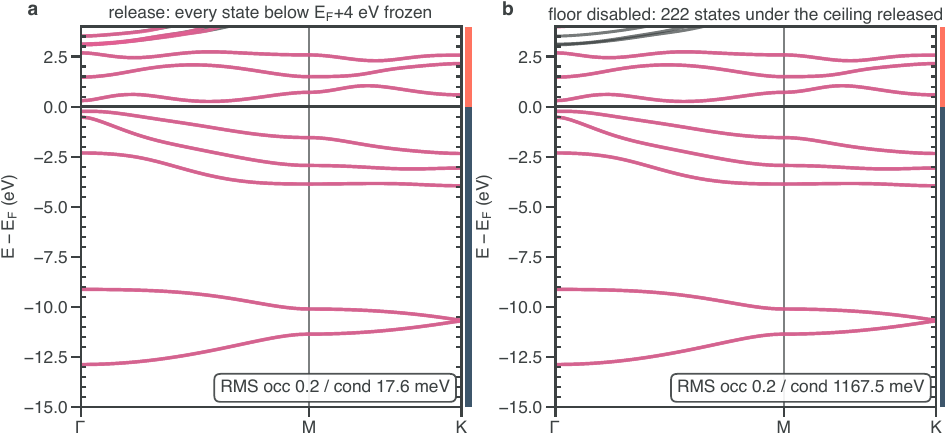}
\caption{The LAPW Bi bilayer with 36 Wannier functions, with the WIEN2k~\cite{Blaha2020} reference in gray. (a)~The released frozen set, which contains every pool state below $E_{\mathrm{F}}+4$~eV. (b)~The same run with the per-$\kk$ freezing floor disabled, in which the 222 states of low projectability under the ceiling are released and the band they belong to leaves the model. The boxes give the index-paired window error over $[-15,0]$ and $[0,+4]$~eV, which are marked by edge bars.}
\label{fig:sm-mixed}
\end{figure}

On $\alpha$-Sn the released states (30 in x1, all with $P<0.1$ just under the ceiling, and 76 in x2) lower $\Omega_{\rm tot}$ by at most 1.4\%. Variant x1 changes the band distance by at most 0.1~meV, and x2, which also releases Fermi-shell states with $P<p_{\rm froz}$, lowers it by 0.1 and 2.8~meV. On the Bi bilayer the two variants give the same mask, because the 222 released states, which lie between $+3.1$ and $+3.95$~eV with $P$ between 0.04 and 0.13, are already below the Fermi-shell threshold of 0.15. Releasing them lowers $\Omega_I$ from 80.19 to 58.45~\AA$^2$ and leaves the occupied window unchanged, but the model no longer contains the band to which these states belong (Fig.~\ref{fig:sm-mixed}). The index-paired conduction error rises from 17.6 to 1167.5~meV, while the nearest-band metric, which matches each DFT state to any Wannier band, reports a largest deviation of only 11~meV. The states of low projectability under the ceiling are the top of the target manifold, with their weight in regions that the projectors do not reach, and releasing them lets the $\Omega_I$ minimization replace them with states that give a lower spread. Both Bi disentanglements stopped at the cap of 30\,000 sweeps, with last changes of $|\Delta\Omega|=1.1\times10^{-7}$ and $8.5\times10^{-9}$~\AA$^2$, so both subspaces are reported at the cap and not at the tolerance. The x1 Wannierization met its criterion, and the model with the frozen set of the release, run to its criterion in one run, converges in 58\,500 iterations at the same 0.2 and 17.6~meV.

\paragraph{All five cells, both solvers.} The freezing floor freezes states on five of the eight LAPW cells, 1 on GaAs, 27 on SrVO$_3$, 30 on $\alpha$-Sn, 222 on the Bi bilayer and 2903 on Fe with spin--orbit coupling, whereas on MgB$_2$ and the two collinear Fe channels it freezes none, so that the trust mask and the window coincide there. Table~\ref{tab:sm-x1} compares variant x1, in which the trust mask is the frozen set, with the release on these five cells, for the Gauss--Seidel sweep and for conjugate gradients, with every Wannierization run to its criterion in one run. These are the LAPW rows of main-text Table~\ref{tab:extent}. The run files are in the directory of each cell in the \textsc{wien2wannier} test tree, the job scripts in \texttt{calculations/revision\_queue/job2.sh} and \texttt{calculations/c8\_cg}, and the scores, with a manifest of every file behind these rows and its checksum, in \texttt{calculations/revision\_queue/c2\_scores}.

On Fe with spin--orbit coupling the released states are spread over all 512 $\kk$-points. Releasing them lowers $\Omega_I$ from 11.22 to 10.82--10.84~\AA$^2$ and $\Omega_{\rm tot}$ from 12.73 to 11.48~\AA$^2$, but it raises the band errors from 4.5 and 5.5~meV to 156--174 and 227--301~meV. The weakest link of the hand-off stays near 0.5 (0.74 with the floor) and \textsc{Wannier90} meets its criterion, so neither the spread nor the continuity check identifies this model as poor. The Gauss--Seidel sweep of this variant stopped at its cap of 30\,000 sweeps with a flat tail ($\Omega_I$ moved by $3.7\times10^{-6}$ in relative terms over the last 3000 sweeps), and the conjugate-gradient solution converged to the same result. On SrVO$_3$ the two solvers reach the same subspace, and the 27 released states lower $\Omega_{\rm tot}$ from 23.47 to 20.92~\AA$^2$ and change the two errors by 0.15~meV in opposite directions. On GaAs the single released state changes $\Omega_I$ by at most $1\times10^{-4}$~\AA$^2$, and the two models differ by 0.6 and 2.2~meV with the Gauss--Seidel sweep and by 0.2 and 0.3~meV with conjugate gradients, which is less than the dependence of the Wannierization of this cell on the optimizer (Sec.~\ref{sm:method}). On $\alpha$-Sn and the Bi bilayer both solvers agree with the results above, and the conjugate-gradient x1 hand-off of the Bi bilayer, whose weakest link is 0.93, still lacks the target band.

\begin{table}[htbp]
\caption{The trust mask (variant x1) compared to the released window as the frozen set on the five LAPW cells in which the two differ, with the Gauss--Seidel sweep (GS) and with conjugate gradients (CG). The number after each cell is the count of pool states below the ceiling, summed over the mesh, that the trust mask leaves unfrozen. The weakest link is the smallest singular value of the hand-off overlaps, ``W90'' is the printed Wannierization iteration (every 500) at which the criterion was met. The spreads are $\Omega_I$ of the disentangled subspace and $\Omega_{\rm tot}$ after the Wannierization, and the band errors are index-paired values over the two scoring windows of main-text Table~\ref{tab:lapw} (meV, windows relative to $E_{\mathrm{F}}$).}
\label{tab:sm-x1}
\small\setlength{\tabcolsep}{3pt}
\begin{tabular*}{\textwidth}{@{\extracolsep{\fill}}l l l r r r r r r@{}}
\toprule
Cell (released) & Solver & Frozen set & $\Omega_I$ (\AA$^2$) & $\Omega_{\rm tot}$ (\AA$^2$) & Weakest link & W90 & Occupied & Conduction \\
\midrule
GaAs (1) & GS & window & 38.979 & 52.488 & $5.0\times10^{-4}$ & 59\,500 & 4.9 & 18.0 \\
 & & trust mask & 38.979 & 52.218 & $5.0\times10^{-4}$ & 223\,000 & 5.5 & 15.8 \\
 & CG & window & 38.957 & 51.403 & $2.8\times10^{-2}$ & 24\,500 & 5.6 & 17.7 \\
 & & trust mask & 38.957 & 51.408 & $2.8\times10^{-2}$ & 53\,500 & 5.4 & 17.4 \\
\addlinespace
Fe-SOC (2903) & GS & window & 11.224 & 12.726 & 0.74 & 64\,500 & 4.5 & 5.5 \\
 & & trust mask & 10.819$^{\rm a}$ & 11.475 & 0.47 & 17\,000 & 174.0 & 301.2 \\
 & CG & window & 11.224 & 12.728 & 0.74 & 12\,500 & 4.5 & 5.7 \\
 & & trust mask & 10.836 & 11.482 & 0.51 & 2\,500 & 155.9 & 226.9 \\
\addlinespace
SrVO$_3$ (27) & GS & window & 21.964 & 23.473 & 0.86 & 45\,500 & 0.7 & 1.6 \\
 & & trust mask & 20.581 & 20.916 & 0.95 & $<500$ & 0.5 & 1.8 \\
 & CG & window & 21.964 & 23.473 & 0.86 & 45\,500 & 0.7 & 1.6 \\
 & & trust mask & 20.581 & 20.916 & 0.95 & $<500$ & 0.5 & 1.8 \\
\addlinespace
$\alpha$-Sn (30) & GS & window & 55.944 & 65.597 & 0.49 & 18\,000 & 2.9 & 12.1 \\
 & & trust mask & 55.929 & 65.593 & 0.49 & 16\,500 & 2.9 & 12.0 \\
 & CG & window & 55.943 & 65.574 & 0.50 & 30\,500 & 2.2 & 14.2 \\
 & & trust mask & 55.929 & 65.570 & 0.50 & 34\,000 & 2.1 & 14.1 \\
\addlinespace
Bi bilayer (222) & GS & window & 80.192$^{\rm a}$ & 88.344 & 0.61 & 58\,500 & 0.2 & 17.6 \\
 & & trust mask & 58.446$^{\rm a}$ & 65.425 & --$^{\rm b}$ & 35\,000 & 0.2 & 1167.5 \\
 & CG & window & 80.499 & 88.388 & 0.62 & 68\,000 & 0.2 & 17.6 \\
 & & trust mask & 58.446 & 65.425 & 0.93 & 17\,000 & 0.2 & 1183.8 \\
\bottomrule
\end{tabular*}

\smallskip
\begin{minipage}{\textwidth}\footnotesize
$^{\rm a}$The sweep stopped at its 30\,000-sweep cap.\\
$^{\rm b}$Not recorded by the build that made this run.
\end{minipage}
\end{table}

\section{The admission floor}
\label{sm:floor}

The LCAO models admit a non-frozen pool state into the disentanglement only if $P_{n\kk}\ge p_{\rm floor}=0.10$ (main-text Eq.~\eqref{eq:masks}), whereas the LAPW models admit every pool state, and this section tests the floor on both frontends.

\begin{figure}[htbp]
\centering
\includegraphics{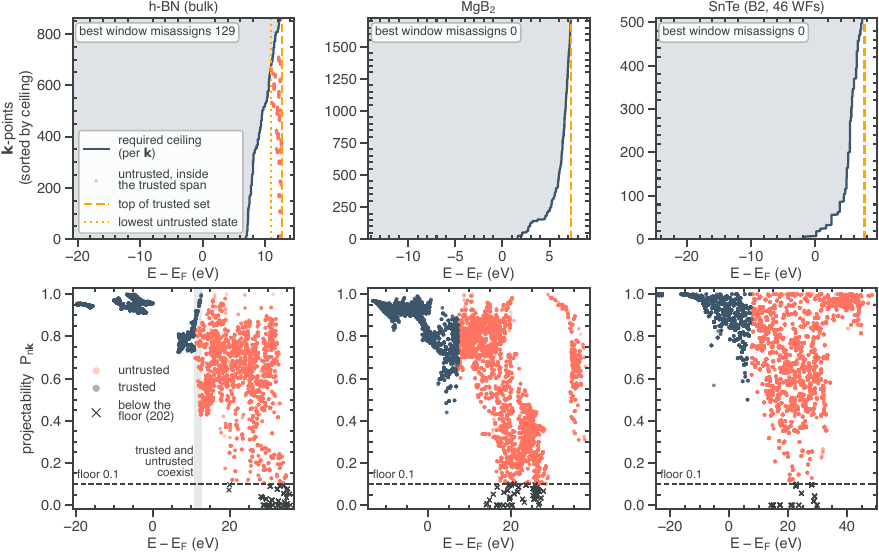}
\caption{Certificates of the LCAO trust masks of the bulk cells at the meshes of main-text Table~\ref{tab:defaults}, drawn with \texttt{scripts/plot\_window\_certificate.py}. SnTe is shown at CRYSTAL's Fermi energy with 46 functions, for which the trust mask is an energy window (main-text Sec.~\ref{sec:snte}). The top panels show the required ceiling at each $\kk$ (the top of the trusted set), the untrusted states inside the trusted span (coral), and the number of states the best global window misassigns. This window ends between the two amber lines, which mark the top of the trusted set (dashed) and the lowest untrusted state inside its span (dotted). The bottom panels show the projectability of every pool state against energy for trusted (slate) and untrusted (coral) states, and the gray band marks the energies at which both kinds coexist. The dashed line marks the admission floor $p_{\rm floor}=0.10$, and the untrusted states below it (crosses) are not admitted to the disentanglement. These are 202 states for h-BN ($+20$ to $+36$~eV), 406 for MgB$_2$ ($+14$ to $+28$~eV) and 188 for SnTe ($+14$ to $+30$~eV), all far above the frozen span. On SnTe the bands they belong to have average projectabilities of 0.55--0.68 and lose their projection only at some $\kk$-points, and the 64, 40 and 68 states with $P_{n\kk}<10^{-3}$ have no projection at all.}
\label{fig:sm-lcaocert-bulk}
\end{figure}

\paragraph{LCAO.} The hand-off at the $E_{\mathrm{F}}+8$~eV ceiling (CG disentanglement, and for SnTe 46 functions without the pocket skip, as in main-text Table~\ref{tab:routes}) was run with only $p_{\rm floor}$ varied and then Wannierized with the default optimizer of \textsc{Wannier90} to its criterion, which every run met (\texttt{calculations/floor\_scan/run.sh}). The floor of $10^{-3}$ excludes exactly the states without projection (Fig.~\ref{fig:sm-lcaocert-bulk}), which are the states below the largest gap of the $P$ distribution.

\begin{table}[!htbp]
\caption{Admission floor on the LCAO cells. ``Admitted$/\kk$'' gives the smallest and largest number of admitted pool
states per $\kk$-point, and the weakest link is the smallest singular value of $W^\dagger(\kk)M^{(\kk,\bb)}W(\kk+\bb)$ over
all links of the hand-off. ``W90'' gives the Wannierization iterations to the criterion, and the spreads are $\Omega_I$ of the disentangled subspace and $\Omega_{\rm tot}$ after the Wannierization. The band errors pair bands by index (meV) over the scoring windows of main-text Table~\ref{tab:defaults} (relative to $E_{\mathrm{F}}$), namely $[-20,0]$ and $[+5,+8]$~eV for h-BN,
$[-13,+7.2]$ and $[-2,+2]$~eV for MgB$_2$, and $[-12,0]$ and $[0,+6]$~eV for SnTe.}
\label{tab:sm-lcao-floor}
\small
\begin{tabular*}{\textwidth}{@{\extracolsep{\fill}}l l r r r r r r r@{}}
\toprule
Cell & $p_{\rm floor}$ & Admitted$/\kk$ & $\Omega_I$ (\AA$^2$) & Weakest link & W90 & $\Omega_{\rm tot}$ (\AA$^2$) & Window 1 & Window 2 \\
\midrule
h-BN & 0 & 24 & 9.7816 & 0.907 & 1\,430 & 11.885 & 0.56 & 5.50 \\
 & 0.001 & 20--24 & 9.7843 & 0.907 & 1\,610 & 11.882 & 0.57 & 5.23 \\
 & 0.02 & 20--24 & 9.7855 & 0.907 & 1\,910 & 11.881 & 0.56 & 5.23 \\
 & 0.05 & 20--24 & 9.7890 & 0.907 & 1\,180 & 11.882 & 0.56 & 5.22 \\
 & 0.10 & 20--24 & 9.7935 & 0.907 & 1\,400 & 11.883 & 0.56 & 5.15 \\
 & 0.2 & 19--24 & 9.8995 & 0.908 & 1\,650 & 11.985 & 0.57 & 4.99 \\
 & 0.3 & 19--24 & 11.0008 & 0.819 & 2\,520 & 13.149 & 0.87 & 4.47 \\
\addlinespace
MgB$_2$ & 0 & 18 & 11.2232 & 0.895 & 1\,320 & 13.138 & 5.75 & 3.43 \\
 & 0.001 & 16--18 & 11.3381 & 0.794 & 1\,080 & 13.289 & 5.52 & 3.68 \\
 & 0.02 & 16--18 & 11.4151 & 0.841 & 1\,020 & 13.470 & 4.87 & 3.49 \\
 & 0.05 & 16--18 & 11.5745 & 0.892 & 1\,380 & 14.084 & 3.64 & 3.35 \\
 & 0.10 & 16--18 & 11.8786 & 0.885 & 3\,040 & 15.405 & 3.62 & 3.38 \\
 & 0.2 & 13--18 & 13.7068 & 0.714 & 2\,100 & 17.741 & 3.72 & 3.47 \\
 & 0.3 & 13--18 & 16.7233 & 0.508 & 5\,860 & 21.087 & 4.17 & 3.07 \\
\addlinespace
SnTe & 0 & 69 & 45.1424 & 0.0012 & 4\,510 & 53.308 & 16.16 & 34.37 \\
 & 0.001 & 61--69 & 45.5664 & 0.224 & 10\,700 & 53.927 & 11.52 & 28.64 \\
 & 0.02 & 61--69 & 45.6011 & 0.225 & 11\,580 & 53.968 & 11.15 & 28.26 \\
 & 0.05 & 61--69 & 45.8181 & 0.221 & 6\,920 & 54.629 & 10.95 & 27.96 \\
 & 0.10 & 61--69 & 46.5191 & 0.205 & 1\,460 & 56.413 & 8.44 & 24.43 \\
 & 0.2 & 61--69 & 49.3340 & 0.071 & 4\,110 & 62.425 & 21.71 & 35.79 \\
 & 0.3 & 59--69 & 55.0710 & 0.194 & 6\,720 & 68.740 & 10.95 & 30.33 \\
\bottomrule
\end{tabular*}
\end{table}

On h-BN the floor has little effect, since floors from 0 to 0.2 give 0.56--0.57 and 4.99--5.50~meV and a floor of 0.3 raises the occupied error to 0.87~meV and lowers the conduction error to 4.47~meV. On MgB$_2$ every subspace stays continuous, and $\Omega_I$ and $\Omega_{\rm tot}$ fall with the floor ($\Omega_{\rm tot}$ from 15.40 to 13.14~\AA$^2$ between 0.10 and 0), while the error over the wide window rises from 3.6 to 5.8~meV. The added error lies on the $k_z=\pi$ plane (A--L--H), whereas the in-plane bands improve slightly. On SnTe the subspace loses its continuity at X (weakest link 0.0012) only when every state is admitted. Excluding the states without projection keeps it continuous, but from 0.001 to 0.05 the index-paired errors stay above those at 0.10 (11.0--11.5 and 28.0--28.6 vs.\ 8.4 and 24.4~meV), while the nearest-band conduction error falls (12.9--13.4 vs.\ 18.5~meV), because the lower floors reorder bands rather than shift them. Across the three cells a floor of 0.10 is within 0.2~meV of the best floor in every window except the h-BN conduction window and the MgB$_2$ Fermi window, where 0.3 is better (4.47 vs.\ 5.15 and 3.07 vs.\ 3.38~meV) but makes the other window of the cell worse (0.87 vs.\ 0.56~meV over the h-BN valence band and 4.17 vs.\ 3.62~meV over the MgB$_2$ wide window).
With the Gauss--Seidel solver of the extent studies, switching the floor off on h-BN (\texttt{ceiling\_study/hbn\_c8\_admitall}) gives a minimization that converges in 277 sweeps to $\Omega_I=9.966$~\AA$^2$ (9.969 with the floor) and a Wannierization that reaches $\Omega_{\rm tot}=12.088$~\AA$^2$ in 1740 iterations (12.071 with the floor), with band errors of 0.79 and 7.92~meV compared to 0.77 and 7.42~meV.

\paragraph{LAPW.} The LAPW models of main-text Table~\ref{tab:lapw} admit every pool state, and we tested the floor on the eight LAPW cells with every other input of the release protocol unchanged. Setting A admits every state, as the published models do, and rerunning it reproduces their hand-offs byte for byte. Setting Z excludes only the states of zero projectability, and on $\alpha$-Sn and the Bi bilayer, which have no such states, it places the floor at the largest gap of the $P$ distribution instead, at $3\times10^{-4}$ and $1.6\times10^{-3}$ (gaps of 1.7 and 0.34 decades). Setting F applies the LCAO floor of 0.10. The three settings correspond to the options \texttt{-\kern0pt-admit-all}, \texttt{-\kern0pt-p-floor 5e-7} and \texttt{-\kern0pt-p-floor 0.10}, and \textsc{Wannier90} runs the published input files with the iteration limit raised to 300\,000. The scripts are \texttt{tests/regress\_out/floor\_gap\_chain.sh}, \texttt{floor\_gap\_w90.sh} and, for the guided Bi input, \texttt{bi\_floor\_chain.sh}, and the scores come from \texttt{tools/band\_compare.py} with \texttt{band\_compare\_specs/floor\_gap.json} and \texttt{floor\_gap\_bi2d.json}.

\begin{table}[!htbp]
\caption{Admission settings on the LAPW cells. Setting A admits every pool state, as the published models do, setting Z excludes only the states of zero projectability (on $\alpha$-Sn and the Bi bilayer, which have none, it excludes the states below the largest gap of the $P$ distribution), and setting F applies the LCAO floor of 0.10. ``Excluded'' counts the non-frozen pool states that the admission mask leaves out, summed over the mesh, and the weakest link is the smallest singular value of $W^\dagger(\kk)M^{(\kk,\bb)}W(\kk+\bb)$ over all links of the hand-off. ``W90'' gives the Wannierization iterations, with ``met'' when the convergence criterion was met and ``cap'' when the run stopped at the iteration limit. The band errors are index-paired values over the two scoring windows of main-text Table~\ref{tab:lapw} (meV, windows relative to $E_{\mathrm{F}}$, with $[-13,+7.2]$ and $[-2,+2]$~eV for MgB$_2$), and $\Omega_I$ is the gauge-invariant spread of the disentangled subspace.}
\label{tab:sm-lapw-floor}
\small
\begin{tabular*}{\textwidth}{@{\extracolsep{\fill}}l l r r r r r r@{}}
\toprule
Cell & Setting & Excluded & $\Omega_I$ (\AA$^2$) & Weakest link & W90 & Occupied & Conduction \\
\midrule
GaAs & A & 0 & 38.979 & $5.0\times10^{-4}$ & 59\,500, met & 4.9 & 18.0 \\
 & Z & 3 & 39.344 & $7.3\times10^{-4}$ & 194\,000, met & 8.2 & 22.8 \\
 & F & 26 & 39.623 & $1.6\times10^{-3}$ & 300\,000 cap & 7.1 & 21.0 \\
\addlinespace
MgB$_2$ & A & 0 & 12.603 & 0.79 & 12\,000, met & 11.5 & 4.4 \\
 & Z & 59 & 12.818 & 0.73 & 3\,000, met & 5.7 & 2.3 \\
 & F & 2\,124 & 14.088 & 0.63 & 6\,000, met & 4.8 & 1.5 \\
\addlinespace
$\alpha$-Sn & A & 0 & 55.943 & 0.50 & 30\,500, met & 2.2 & 14.2 \\
 & Z & 2 & 55.943 & 0.50 & 40\,500, met & 2.6 & 13.9 \\
 & F & 557 & 63.702 & 0.24 & 1\,500, met & 4.2 & 17.5 \\
\addlinespace
Bi bilayer & A & 0 & 80.192 & 0.61 & 58\,500, met & 0.2 & 17.6 \\
 & Z & 4 & 80.193 & 0.61 & 56\,000, met & 0.2 & 17.6 \\
 & F & 7\,292 & 365.03 & $1.2\times10^{-3}$ & 300\,000 cap & 2.6 & 457.4 \\
\addlinespace
Fe$\uparrow$ & A & 0 & 5.876 & 0.65 & met & 8.0 & 15.1 \\
 & Z & 18 & 5.876 & 0.65 & 500, met & 8.2 & 14.9 \\
 & F & 1\,666 & 7.086 & 0.64 & 500, met & 14.5 & 19.0 \\
\addlinespace
Fe$\downarrow$ & A & 0 & 6.002 & 0.61 & 500, met & 11.8 & 7.0 \\
 & Z & 18 & 6.002 & 0.61 & met & 11.8 & 7.0 \\
 & F & 1\,420 & 6.941 & 0.50 & 500, met & 11.3 & 4.6 \\
\addlinespace
Fe-SOC & A & 0 & 11.224 & 0.74 & 64\,500, met & 4.5 & 5.5 \\
 & Z & 118 & 12.015 & $1.9\times10^{-5}$ & 300\,000 cap & 54.1 & 14.4 \\
 & F$^{\rm b}$ & 3\,242 & 17.380 & $3.2\times10^{-5}$ & 300\,000 cap & 125.0 & 38.2 \\
\addlinespace
SrVO$_3$ & A & 0 & 21.964 & 0.86 & 45\,500, met & 0.7 & 1.6 \\
 & Z & 6 & 21.967 & 0.86 & 41\,000, met & 0.7 & 1.6 \\
 & F$^{\rm a}$ & 3\,681 & 22.023 & 0.90 & met & 0.5 & 1.7 \\
\bottomrule
\end{tabular*}

\smallskip
\begin{minipage}{\textwidth}\footnotesize
$^{\rm a}$At the automatic ceiling ($+6.0$~eV) the subspace lost its continuity (weakest link $4.8\times10^{-19}$), and the
pocket skip lowered the ceiling to $+4.62$~eV, where the row was measured.\\
$^{\rm b}$The pocket skip tried $+6.90$~eV, did not restore the continuity, and kept $+8.0$~eV.
\end{minipage}
\end{table}

No setting is best on every cell, since each improves some cells and worsens others. The floor of 0.10 (setting F) improves MgB$_2$ from 11.5 and 4.4 to 4.8 and 1.5~meV, at an $\Omega_I$ higher by 1.5~\AA$^2$, and improves Fe$\downarrow$ slightly. It breaks the subspace continuity of the Bi bilayer and of Fe-SOC, whose Wannierizations then do not meet the criterion within 300\,000 iterations and whose band errors grow 7 to 28 times. On the Bi bilayer it excludes 45\% of the pool and raises $\Omega_I$ from 80.2 to 365.0~\AA$^2$ and the conduction error from 17.6 to 457~meV. It also worsens $\alpha$-Sn without breaking its continuity (weakest link 0.24, with errors rising from 2.2 and 14.2 to 4.2 and 17.5~meV), and worsens Fe$\uparrow$ and GaAs. On SrVO$_3$ it changes the band errors by at most 0.2~meV, although there it also breaks the continuity at the automatic ceiling, which the pocket skip restores by lowering the ceiling to $+4.62$~eV (Table~\ref{tab:sm-lapw-floor}, note~a).
Excluding only the states of zero projectability (setting Z) improves MgB$_2$, moves the two $\alpha$-Sn windows in opposite directions by at most 0.4~meV, leaves the Fe spin channels, SrVO$_3$ and the Bi bilayer unchanged, worsens GaAs, and breaks the continuity of Fe-SOC. On Fe-SOC the zero lies at H, on a band at $+9.86$~eV whose states at the neighboring $\kk$-points have $P\approx10^{-3}$ and stay admitted, so that the subspace loses the band at one $\kk$-point only (weakest link $1.9\times10^{-5}$ vs.\ 0.74). On the LCAO frontend the same exclusion keeps SnTe continuous at X (weakest link 0.22 vs.\ 0.0012 with every state admitted), so the effect of excluding a state at isolated $\kk$-points depends on the cell. Neither $\Omega_I$ nor the continuity check selects the most accurate setting, since on MgB$_2$ admitting every state (setting A) gives the lowest $\Omega_I$ and a continuous subspace but the worst interpolation. Admitting every state is, however, the only setting that keeps every tested subspace continuous, apart from that of GaAs, whose continuity is broken in all three settings. The released \textsc{wien2wannier} therefore admits every pool state, and the option \texttt{-\kern0pt-p-floor} sets a floor for a cell known to need one.

\section{Convergence, restarts, and execution paths}
\label{sm:convergence}

Table~\ref{tab:sm-conv} lists the convergence of every model reported in the main-text tables. The column ``dis'' refers to the constrained-$\Omega_I$ disentanglement and gives the gradient evaluations of the conjugate-gradient solver, the Gauss--Seidel sweeps to the relative tolerance of $10^{-10}$, or the cap and the last change. The column ``W90'' refers to the Wannierization in \textsc{Wannier90} and gives the iterations to the criterion, or the budget and the last per-iteration change. For the window route, ``dis'' gives the disentanglement performed by \textsc{Wannier90} itself.

\begin{table}[htbp]
\caption{Convergence of the reported models. ``Dis.'' gives the disentanglement steps (sweeps, CG gradient evaluations, or
\textsc{Wannier90} iterations for the window route), ``W90'' the Wannierization iterations, and $\Omega_{\rm tot}$ the total spread after the Wannierization. ``Met'' marks a run that stopped at the
convergence criterion, ``budget'' or ``cap'' a run that stopped there, and ``$N$ stopped'' a run stopped by hand after $N$ iterations,
with the terminal per-iteration $|\Delta\Omega|$ (\AA$^2$) in parentheses. Step counts are comparable only within one method, and a run stopped at a budget or cap has not met its criterion.}
\label{tab:sm-conv}
\small
\begin{tabular*}{\textwidth}{@{\extracolsep{\fill}}l l l r@{}}
\toprule
Model & Dis. & W90 & $\Omega_{\rm tot}$ (\AA$^2$) \\
\midrule
\multicolumn{4}{@{}l}{\emph{LCAO, released default (main-text Table~\ref{tab:defaults})}} \\
h-BN, $E_{\mathrm{F}}+8$ & CG, 149 evaluations & 1\,400, met & 11.883 \\
h-BN, trusted top $+12.65$ & CG, 144 evaluations & 8\,160, met & 14.411 \\
Bi bilayer, $E_{\mathrm{F}}+7$ & 2 sweeps (isolated) & 2\,230, met & 44.118 \\
MgB$_2$, $E_{\mathrm{F}}+8$ & CG, 655 evaluations & 3\,040, met & 15.405 \\
SnTe (46 WFs), default, $E_{\mathrm{F}}+6.18$ & CG, 627 + 4073$^{\rm a}$ & 5\,180, met & 53.445 \\
SnTe (46 WFs), $E_{\mathrm{F}}+8$ & CG, 627 evaluations & 1\,460, met & 56.413 \\
SnTe (46 WFs), $E_{\mathrm{F}}+8$ & 3230 sweeps & 1\,970, met & 56.414 \\
SnTe (36 WFs), $E_{\mathrm{F}}+8$ & CG, 1565 evaluations & 42\,130, met & 65.177 \\
Sc-COF $\alpha$, $E_{\mathrm{F}}+8$ & 14\,806 sweeps & $10^4$ budget ($2.7\times10^{-9}$) & 303.636 \\
Sc-COF $\beta$, $E_{\mathrm{F}}+8$ & 13\,881 sweeps & $10^4$ budget ($7.1\times10^{-6}$) & 306.358 \\
\addlinespace
\multicolumn{4}{@{}l}{\emph{LCAO, extent and ceiling studies (Gauss--Seidel)}} \\
h-BN, $E_{\mathrm{F}}+8$ & 186 sweeps & 2\,070, met & 12.071 \\
h-BN, window to $+12.65$ & 316 sweeps & 1\,900, met & 14.602 \\
MgB$_2$, $E_{\mathrm{F}}+8$ & 1510 sweeps & 5\,690, met & 16.679 \\
h-BN, $E_{\mathrm{F}}+8$, admission floor off & 277 sweeps & 1\,740, met & 12.088 \\
\addlinespace
\multicolumn{4}{@{}l}{\emph{LAPW (main-text Table~\ref{tab:lapw}): Gauss--Seidel in the colored order with over-relaxation$^{\rm c}$}} \\
GaAs & 120 sweeps & 59\,500, met & 52.488 \\
MgB$_2$ & 120 sweeps & 12\,000, met & 18.153 \\
$\alpha$-Sn & CG, 2\,883 evaluations & 30\,500, met & 65.574 \\
Bi bilayer (guided) & 30\,000 cap ($1.1\times10^{-7}$) & 58\,500, met & 88.344 \\
Fe$\uparrow$ / Fe$\downarrow$ & 110 / 110 sweeps & met / 500 & 7.476 / 7.583 \\
Fe-SOC & 8049 sweeps & 64\,500, met & 12.726 \\
SrVO$_3$ & 592 sweeps & 45\,500, met & 23.473 \\
\addlinespace
\multicolumn{4}{@{}l}{\emph{Window route at the released ceiling (main-text Table~\ref{tab:routes})}} \\
Bi bilayer, $E_{\mathrm{F}}+7$ & 6 iterations, met & 2\,170, met & 44.118 \\
h-BN, $E_{\mathrm{F}}+8$ & 1398 iterations, met & $10^4$ budget ($6.7\times10^{-4}$) & 16.861 \\
 & & one run: 18\,260, met & 11.849 \\
MgB$_2$, $E_{\mathrm{F}}+8$ & 8388 iterations, met & 1\,690, met & 13.028 \\
SnTe, $E_{\mathrm{F}}+8$, 69 bands & 10\,197 iterations, met & 29\,950, met$^{\rm b}$ & 53.656 \\
 & & two-stage$^{\rm b}$ 22\,830, met & 53.654 \\
GaAs, $E_{\mathrm{F}}+8$ (LAPW) & 1085 iterations, met & 66\,500, met & 50.049 \\
\bottomrule
\end{tabular*}

\smallskip
\begin{minipage}{\textwidth}\footnotesize
$^{\rm a}$First pass at the automatic $E_{\mathrm{F}}+8$~eV ceiling, followed by the pass of the pocket skip at $E_{\mathrm{F}}+6.18$~eV (main-text
Sec.~\ref{sec:masks}), whose subspace is the one handed off.\\
$^{\rm b}$Restarted from the converged disentanglement checkpoint ($N_b>N_w$, so the restart keeps the gauge). The default optimizer is that of main-text Table~\ref{tab:routes}. In the two-stage scheme the Stengel--Spaldin spread~\cite{StengelSpaldin2006} is minimized first, and from the per-$\kk$ SVD start it did not reach $10^{-6}$~\AA$^2$ within its 20\,000-iteration cap ($\Omega_{\rm tot}$ 2\,605~\AA$^2$). The Marzari--Vanderbilt spread~\cite{Marzari1997} was then minimized with \texttt{num\_cg\_steps}~$=200$ to the criterion (12.2 and 31.9~meV), and the 22\,830 iterations include both stages.\\
$^{\rm c}$See Sec.~\ref{sm:sweeporder}. The $\alpha$-Sn row uses conjugate gradients instead of the sweep. GaAs, Fe with spin--orbit coupling and the Bi bilayer were first Wannierized to a 50\,000-iteration cap, which gave the same values (5.44 vs.\ 5.47~meV over $[0,+4]$~eV for Fe with spin--orbit coupling).
\end{minipage}
\end{table}

\begin{figure}[htbp]
\centering
\includegraphics{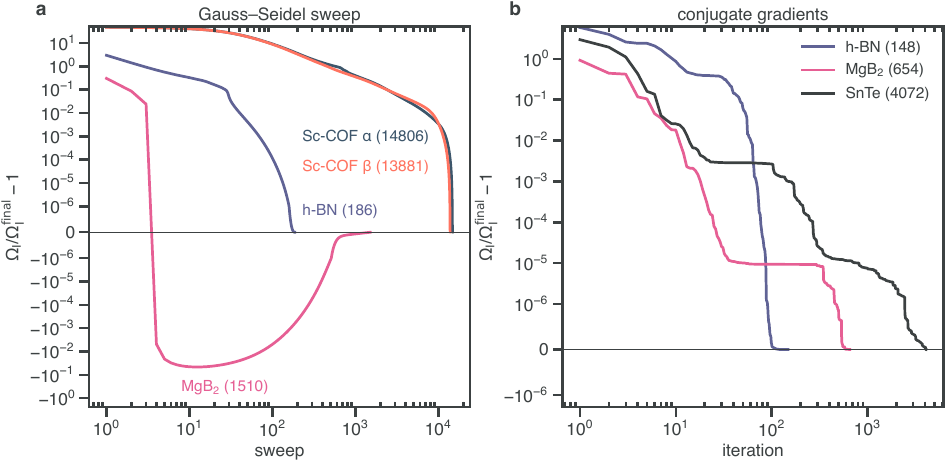}
\caption{Constrained-$\Omega_I$ disentanglement traces, given as the relative distance from the final value on a symmetric-log scale, with the number of steps in parentheses. (a)~Gauss--Seidel sweeps for h-BN and MgB$_2$ at $E_{\mathrm{F}}+8$~eV and for the two Sc-COF channels of main-text Table~\ref{tab:defaults}. (b)~Conjugate gradients for the h-BN, MgB$_2$ and SnTe (46 functions) models of main-text Table~\ref{tab:defaults}. The data are the \texttt{<seed>\_disentangle.log} files of the runs in \texttt{ceiling\_study/hbn\_c8}, \texttt{mgb2\_auto}, \texttt{table1\_cg} and \texttt{table1\_cg46}, plotted by \texttt{figures\_src/sm\_dis\_traces.py}.}
\label{fig:sm-traces}
\end{figure}

\paragraph{Disentanglement traces.} Figure~\ref{fig:sm-traces} shows $\Omega_I$ during the constrained disentanglement of the reported LCAO models. The conjugate-gradient runs decrease $\Omega_I$ at every iteration, as the backtracking line search requires, whereas the Gauss--Seidel sweep is a fixed-point iteration and not a descent method. The sweep descends at every step on h-BN and on both Sc-COF channels, but on MgB$_2$ $\Omega_I$ reaches 12.2772~\AA$^2$ at sweep 12 and then rises to the fixed point, 12.8724~\AA$^2$ at sweep 1510. The conjugate-gradient minimum on the same MgB$_2$ inputs is lower still, at 11.8786~\AA$^2$.

\paragraph{Checkpoint restarts.} In the \textsc{Wannier90} builds used here (the development build and the 4.0.x releases, Sec.~\ref{app:perf}), a run without disentanglement, which every $N_b{=}N_w$ hand-off is, must not be continued with \texttt{restart = wannierise}, because the restart discards the gauge, so such a model has to be Wannierized in one uninterrupted run, whereas runs with disentanglement ($N_b>N_w$) restart correctly. The defect shows in the identity that a checkpoint must satisfy, $M_{\rm chk}^{(\kk,\bb)}=U^\dagger(\kk)M_{\rm mmn}^{(\kk,\bb)}U(\kk+\bb)$, which holds for every continuous run and for every restart of a run with disentanglement ($\le1.5\times10^{-13}$, Table~\ref{tab:sm-restart}). For a run without disentanglement, \texttt{restart = wannierise} reads the checkpoint, sets \texttt{u\_matrix\_opt} to zero, then copies it into the gauge and orthonormalizes it again, so the gauge accumulated before the restart is discarded while the stored overlaps keep it. The optimizer continues correctly on the overlaps, so the spreads and centers appear continuous, but the stored $U$ no longer matches them, and $H(\mathbf R)$ and the path bands are wrong while the eigenvalues at the mesh points stay exact (Sec.~\ref{app:perf}). The apparent near-degenerate minima of the spread that appeared in earlier runs (0.77 vs.\ 51.9~meV on h-BN at the same $\Omega_{\rm tot}$, and 9.2 vs.\ 117~meV on SnTe) came from such restarts. Repairing the SnTe continuation ($U_{\rm true}=U_{\rm before}U_{\rm stored}$) gives 9.33 and 35.26~meV at the criterion, the value that an uninterrupted run reaches in 72\,740 iterations (both scored with the earlier Fermi reference and nearest-band matching). A restart of a run with disentanglement, in contrast, behaves correctly. The h-BN window route, continued from its checkpoint at 10\,000 iterations (per-iteration $|\Delta\Omega|=6.7\times10^{-4}$) to the criterion (17\,730 iterations in all, compared to 18\,260 for an uninterrupted run), lowers $\Omega_{\rm tot}$ from 16.86 to 11.85~\AA$^2$ and the band error from 258 and 11.7 to 0.40 and 5.18~meV over $[-20,0]$ and $[+5,+8]$~eV, which shows that a Wannierization stopped at its budget can be far from its minimum.

\begin{table}[htbp]
\caption{Checkpoint self-consistency (largest deviation of the stored overlaps from $U^\dagger M U$ over all
links) and band error on the path (nearest band, SnTe with the earlier Fermi reference).}
\label{tab:sm-restart}
\small
\begin{tabular*}{\textwidth}{@{\extracolsep{\fill}}l c c r@{}}
\toprule
Run & Disentangled & $\max|M_{\rm chk}-U^\dagger M U|$ & Band error (meV) \\
\midrule
h-BN hand-off, continuous (1 thread) & no & $5.2\times10^{-14}$ & 0.77 / 7.42 \\
h-BN hand-off, restart every 1000 & no & 1.99 & 51.9 / 10.1 \\
Bi hand-off, continuous (1 thread) & no & $6.0\times10^{-14}$ & 0.26 \\
Bi hand-off, restart every 1000 & no & 0.27 & 24.9 \\
Bi window route, restart every 1000 & yes & $5.8\times10^{-14}$ & 0.26 \\
h-BN window route, continued from 10k & yes & $1.5\times10^{-13}$ & 0.40 / 5.18 \\
SnTe hand-off, continued from 10k & no & 1.94 & 117 / 164 (as stored) \\
SnTe hand-off, same, gauge repaired & no & $6.2\times10^{-13}$ & 9.33 / 35.26 \\
\bottomrule
\end{tabular*}
\end{table}

\paragraph{Execution paths.} Each model of the main-text comparisons was rerun from its own \texttt{.amn}, \texttt{.mmn} and \texttt{.eig} files with only the execution path changed, to one BLAS thread, to two MPI ranks and, for the hand-offs, to guiding centers switched off (Table~\ref{tab:sm-paths}, with runs in \texttt{calculations/review\_checks/ens}). Every valid change leaves the h-BN hand-off, the MgB$_2$ hand-off and trust mask, and the h-BN window route within 0.03~meV of the reported values, and switching guiding centers off changes h-BN and MgB$_2$ by at most 0.01~meV. The h-BN window route converges in 2800 to 18\,260 iterations depending on the path, always to 0.37--0.40 and 5.18--5.21~meV. The MgB$_2$ window route agrees to 0.1~meV on one thread, but on two MPI ranks its Wannierization, started from the per-$\kk$ SVD projections, stalls. After 200\,000 iterations $\Omega_{\rm tot}$ remains at 23.06--23.19~\AA$^2$, compared to 13.03~\AA$^2$ on every other path at the same $\Omega_I$, and the band error is 7.7 and 6.5~meV. Started from the SCDM gauge of the hand-off, the same two-rank run converges in 1140 iterations to $\Omega_{\rm tot}=13.028$~\AA$^2$ and to 4.53 and 1.78~meV. On SnTe (scored with the earlier Fermi reference) two MPI ranks converge at 49\,960 iterations to 9.38 and 35.29~meV, 0.05~meV from the uninterrupted run, and with guiding centers off the run gives 18.6 and 39.1~meV at 49\,200 iterations, so guiding helps the occupied window of SnTe. The values in this paragraph and in Table~\ref{tab:sm-paths} use nearest-band matching, as these runs were scored.

\begin{table}[htbp]
\caption{Execution paths, with the total spread $\Omega_{\rm tot}$ after the Wannierization and the number of Wannierization iterations. The band error uses nearest-band matching (meV) over the scoring windows of the main text, relative to $E_{\mathrm{F}}$ ($[-20,0]$,
$[+5,+8]$ and $[+5,+12]$~eV for h-BN, $[-13,+7.2]$ and $[-2,+2]$~eV for MgB$_2$, and $[-12,0]$ and $[0,+6]$~eV for SnTe with the earlier
Fermi reference). Restarted hand-offs are omitted (Table~\ref{tab:sm-restart}).}
\label{tab:sm-paths}
\small
\begin{tabular*}{\textwidth}{@{\extracolsep{\fill}}l r r l@{}}
\toprule
Model and path & $\Omega_{\rm tot}$ (\AA$^2$) & Iterations & Band error (meV) \\
\midrule
\multicolumn{4}{@{}l}{\emph{h-BN hand-off, $E_{\mathrm{F}}+8$}} \\
\quad 1 thread & 12.070929 & 2\,070 & 0.77; 7.42; 1719 \\
\quad 2 MPI ranks & 12.070925 & 1\,880 & 0.78; 7.42; 1719 \\
\quad no guiding centers & 12.070923 & 2\,430 & 0.78; 7.41; 1719 \\
\addlinespace
\multicolumn{4}{@{}l}{\emph{h-BN hand-off, window to $+12.65$}} \\
\quad 1 thread & 14.601842 & 1\,900 & 1.48; 1.27; 4.84 \\
\quad 2 MPI ranks & 14.601834 & 1\,910 & 1.48; 1.27; 4.84 \\
\addlinespace
\multicolumn{4}{@{}l}{\emph{h-BN hand-off, trust mask}} \\
\quad 1 thread & 14.009848 & 1\,520 & 1.31; 1.91; 52.0 \\
\quad 2 MPI ranks & 14.009845 & 1\,450 & 1.31; 1.91; 51.9 \\
\addlinespace
\multicolumn{4}{@{}l}{\emph{h-BN window route, $E_{\mathrm{F}}+8$}} \\
\quad uninterrupted, 2 threads & 11.849247 & 18\,260 & 0.40; 5.19; 1705 \\
\quad 1 thread & 11.851301 & 2\,800 & 0.38; 5.21; 1705 \\
\quad 2 MPI ranks & 11.846610 & 4\,230 & 0.37; 5.21; 1705 \\
\addlinespace
\multicolumn{4}{@{}l}{\emph{MgB$_2$ hand-off, $E_{\mathrm{F}}+8$}} \\
\quad 1 thread & 16.678661 & 5\,690 & 7.04; 4.50 \\
\quad 2 MPI ranks & 16.678654 & 4\,540 & 7.03; 4.50 \\
\quad no guiding centers & 16.678691 & 4\,750 & 7.04; 4.50 \\
\addlinespace
\multicolumn{4}{@{}l}{\emph{MgB$_2$ hand-off, trust mask}} \\
\quad 1 thread & 16.452120 & 4\,520 & 24.1; 5.13 \\
\quad 2 MPI ranks & 16.452122 & 3\,760 & 24.1; 5.13 \\
\addlinespace
\multicolumn{4}{@{}l}{\emph{MgB$_2$ window route, $E_{\mathrm{F}}+8$}} \\
\quad 1 thread & 13.027769 & 2\,010 & 4.40; 1.50 \\
\quad 2 MPI ranks, SVD start & 23.072062 & 200\,000 (budget) & 7.70; 6.54 \\
\addlinespace
\multicolumn{4}{@{}l}{\emph{SnTe hand-off, $E_{\mathrm{F}}+8$}} \\
\quad uninterrupted, 2 threads & 70.733084 & 72\,740 & 9.33; 35.3 \\
\quad 1 thread & 70.733084 & 72\,740 & 9.33; 35.3 \\
\quad 2 MPI ranks & 70.733063 & 49\,960 & 9.38; 35.3 \\
\quad no guiding centers & 70.798459 & 49\,200 & 18.6; 39.1 \\
\bottomrule
\end{tabular*}
\end{table}

\paragraph{Sc-COF commands.} Each spin channel of the Sc-COF is one
command (\texttt{calculations/\allowbreak final\_default/\allowbreak job\_F.sh}):
\begin{center}
\verb|lcao2wannier --stage all --method hybrid --mmn-method analytic|\\
\verb|--input Sc_all_center_sp_f_101.out --seed sc_alpha|\\
\verb|--k-grid 6 6 1 --spin alpha --hybrid-dis-niter 20000|\\
\verb|--hybrid-dis-solver gs|
\end{center}
and the same with \verb|--seed sc_beta --spin beta|. The reported Sc-COF models were made with the code of 2026-09-14, before the conjugate-gradient default, and the last flag reproduces their Gauss--Seidel disentanglement with the release.

\paragraph{Cost by stage.} Table~\ref{tab:sm-route-cost} breaks down the times of main-text Table~\ref{tab:routes} into preparation, disentanglement and Wannierization.

\begin{table}[htbp]
\caption{Cost of the two routes of main-text Table~\ref{tab:routes} to a converged model, by stage, measured on one machine (Apple M5 Pro, two threads per LCAO run and one per LAPW \textsc{Wannier90} run, with the runs of different cells sharing the machine). ``Prep.'' covers Stages~1 and~2 from the CRYSTAL output, or \texttt{w2w\_hybrid} from the \textsc{wien2wannier} \texttt{.amn}, \texttt{.mmn} and \texttt{.eig} files, and excludes the disentanglement of the hand-off. The generation of these files is common to both routes and is not counted. ``Disentanglement'' is the constrained minimization of the hand-off (CG gradient evaluations, or Gauss--Seidel sweeps for GaAs) or the disentanglement in \textsc{Wannier90} (iterations). ``Wannierization'' is the default optimizer of \textsc{Wannier90} run to its criterion in one run, which for the window route is a restart from the converged disentanglement.}
\label{tab:sm-route-cost}
\small
\begin{tabular*}{\textwidth}{@{\extracolsep{\fill}}l l r r r r r r@{}}
\toprule
 & & Prep. & \multicolumn{2}{c}{Disentanglement} & \multicolumn{2}{c}{Wannierization} & Total \\
\cmidrule(lr){4-5}\cmidrule(lr){6-7}
Cell & Route & (s) & Steps & (s) & Iter. & (s) & (s) \\
\midrule
Bi bilayer & hand-off & 25 & --$^{\rm a}$ & 0 & 2\,230 & 27 & 52 \\
 & window & 26 & 3 & 0.5 & 2\,170 & 28 & 55 \\
\addlinespace
h-BN & hand-off & 169 & 149 & 7 & 1\,400 & 136 & 312 \\
 & window & 256 & 1\,398 & 386 & 18\,260 & 1\,877 & 2\,519 \\
\addlinespace
MgB$_2$ & hand-off & 145 & 655 & 49 & 3\,040 & 268 & 462 \\
 & window & 196 & 8\,388 & 2\,675 & 1\,690 & 170 & 3\,040 \\
\addlinespace
SnTe & hand-off & 204 & 627 & 127 & 1\,460 & 1\,008 & 1\,338 \\
 & window & 436 & 10\,197 & 7\,168 & 29\,950 & 20\,676$^{\rm c}$ & 28\,281 \\
\addlinespace
GaAs, LAPW & hand-off & 7.6 & 120$^{\rm b}$ & 1.9 & 59\,500 & 3\,536 & 3\,546 \\
 & window & 11 & 1\,085 & 43 & 66\,500 & 4\,040 & 4\,094 \\
\bottomrule
\end{tabular*}

\smallskip
\begin{minipage}{\textwidth}\footnotesize
$^{\rm a}$Isolated manifold, with nothing to disentangle.\\
$^{\rm b}$The Gauss--Seidel hand-off of Table~\ref{tab:lapw}.\\
$^{\rm c}$Timed in a separate run, restarted from the disentanglement checkpoint of the benchmark while other jobs ran on the machine. At 0.69~s per iteration it matches the Wannierization of the hand-off for the same 46 functions (0.69~s).
\end{minipage}
\end{table}

\section{Guiding centers}
\label{sm:guiding}

The two programs use different default protocols for the Wannierization in \textsc{Wannier90}. The CRYSTAL frontend, \textsc{lcao2wannier}, writes atom-centered projections for the target channels and enables \texttt{guiding\_centres} whenever real projection centers exist, disabling it only when the projection block falls back to \texttt{random}, because random guiding centers pull the functions apart. The released \textsc{wien2wannier} protocol starts from random projections without guiding centers, except on the Bi bilayer, where the unguided Wannierization oscillated between 220 and 310~\AA$^2$ and the reported model uses guiding centers on Bi $s$, $p$, $d$. To test whether the difference matters where both protocols converge, the $\alpha$-Sn hand-off was Wannierized a second time with guiding centers on Sn $s$, $p$, $d_{z^2}$ (spinor-doubled to 20 functions, \texttt{spinors = .true.}). It met the criterion after 26\,000 iterations, compared to 30\,500 without guiding, at $\Omega_{\rm tot}=65.574053$ compared to $65.574025$~\AA$^2$ and with the same band distance (2.2 and 14.2~meV) to 0.1~meV.

\begin{figure}[htbp]
\centering
\includegraphics{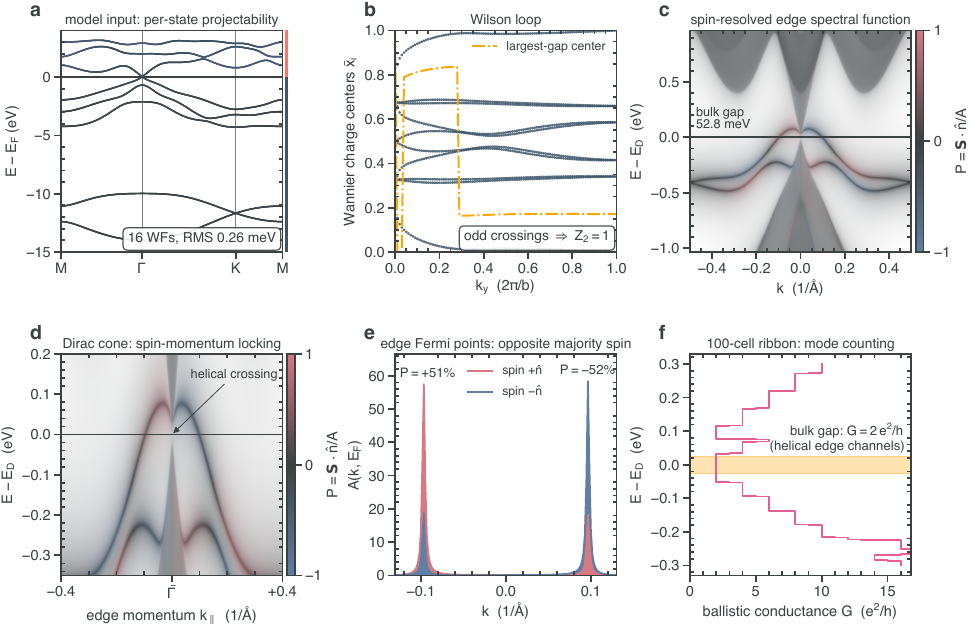}
\caption{The Bi(111) bilayer, a quantum spin Hall insulator~\cite{Murakami2006}, computed from the 16-function LCAO model. (a)~The model input, with the LCAO bands colored by their per-state projectability onto the 16-WF target (color scale of main-text Fig.~\ref{fig:concept}). The edge bar marks the $[-15,+4]$~eV scoring window, over which the RMS error is 0.26~meV with nearest-band matching on M--$\Gamma$--K--M and 0.27~meV with bands paired by index. (b)~Wilson loop of the Wannier charge centers~\cite{Soluyanov2011}, giving $Z_2=1$~\cite{KaneMele2005,FuKane2006}. (c)~Spin-resolved edge spectral function from the iterative surface Green's function~\cite{Sancho1985} in WannierTools~\cite{Wu2018}, with the intensity giving the spectral weight and the color the spin polarization $P$. (d)~The same data near the helical Dirac crossing. (e)~Spin-resolved $A(k,E_{\mathrm{F}})$, with two edge Fermi points of opposite majority spin ($P=+51\%$ and $-52\%$, partial because of the SOC admixture). (f)~Ballistic conductance from mode counting on a 100-cell ribbon, with a quantized $G=2e^2/h$ plateau across the $52.8$~meV gap. Panels (b) and (d) form main-text Fig.~\ref{fig:topo}, where the polarization $P$ is labeled $S_{\hat n}/A$.}
\label{fig:smtopo}\label{sm:topo}
\end{figure}

The topology of the Bi bilayer model of main-text Fig.~\ref{fig:topo} is shown in full in Fig.~\ref{fig:smtopo}.

\section{Certificates of the released LAPW masks}
\label{sm:lapwcert}

\begin{figure}[htbp]
\centering
\includegraphics{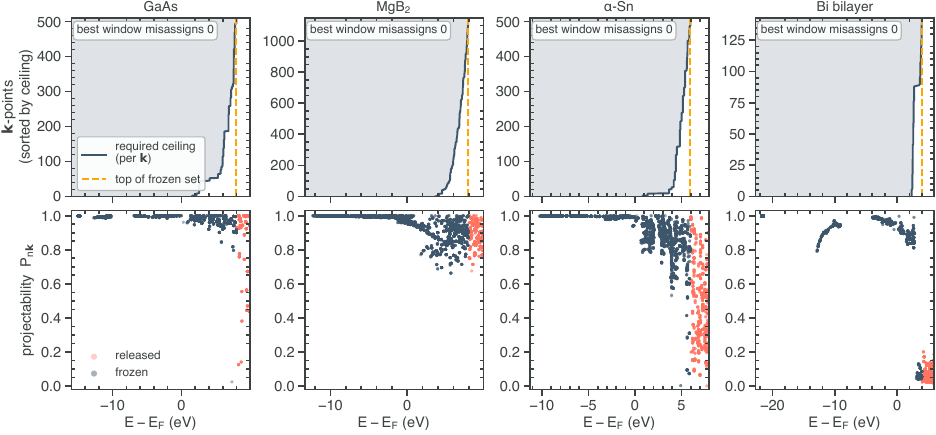}
\caption{Certificate of the released LAPW frozen sets for the first four cells. The top panels show the required ceiling at each $\kk$ and the top of the frozen set (dashed). No released state lies inside the frozen span, so in every cell the best global window misassigns no state. The bottom panels show the projectability against energy for frozen (slate) and released (coral) states.}
\label{fig:sm-lapwcert1}
\end{figure}

\begin{figure}[htbp]
\centering
\includegraphics{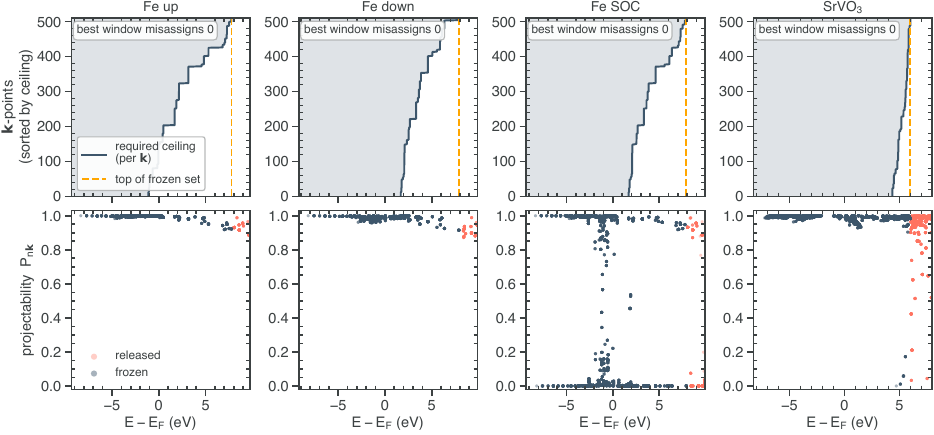}
\caption{The remaining four cells, drawn as in Fig.~\ref{fig:sm-lapwcert1}.}
\label{fig:sm-lapwcert2}
\end{figure}

With the release protocol every LAPW frozen set $F(\kk)$ is an energy window by construction, because the ceiling removes the trusted states above it and the freezing floor freezes every state below it. Applied to these frozen sets, the certificate of the main text therefore gives zero misassigned states in all eight cells (Figs.~\ref{fig:sm-lapwcert1} and \ref{fig:sm-lapwcert2}). This checks the construction of $F(\kk)$ and is not a statement about the trust masks $T(\kk)$, on which main-text Appendix~\ref{app:certificate} defines the certificate. The LAPW certificate of the trust masks built from muffin-tin projectability in the earlier porting runs is given in Sec.~\ref{sm:bridge}.

\begin{figure}[htbp]
\centering
\includegraphics{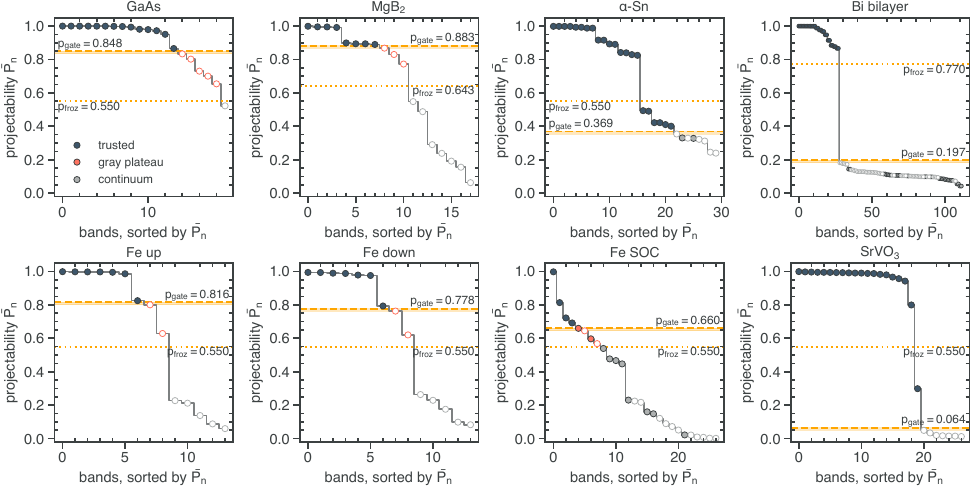}
\caption{Sorted band-averaged projectabilities $\bar P_n$ of the eight LAPW cells, with the derived gate $p_{\text{gate}}$ (dashed) and state threshold $p_{\text{froz}}$ (dotted). The markers show the trusted bands (slate), the gray plateau of intermediate projectability (coral) and the continuum (gray), and open markers are bands that never enter the fidelity range.}
\label{fig:sm-anatomy}
\end{figure}

\begin{figure}[htbp]
\centering
\includegraphics{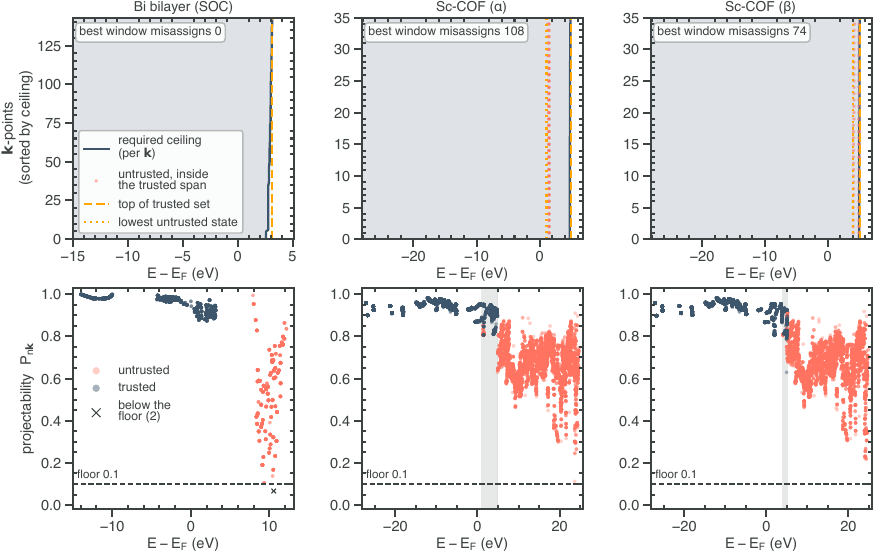}
\caption{Certificates of the LCAO trust masks of the two-dimensional cells, drawn as Fig.~\ref{fig:sm-lcaocert-bulk}. The best global window misassigns no state on the Bi bilayer, 108 states on the Sc-COF $\alpha$ channel and 74 on the $\beta$ channel. Two untrusted states lie below the admission floor on the Bi bilayer (at $+10.5$~eV), and none in either Sc-COF channel.}
\label{fig:sm-lcaocert-2d}
\end{figure}

\section{The earlier LAPW porting runs}
\label{sm:bridge}

This section describes the porting runs that preceded the released \textsc{wien2wannier}. The $\alpha$-Sn porting runs used an earlier self-consistent field of the cell, whose Fermi level encloses 0.1 more electrons than the cell's 28 valence electrons on its own bands. These runs compare constructions on the same Hamiltonian, and the released $\alpha$-Sn model uses the recomputed field of Table~\ref{tab:sm-wien2k}. Every number in this section belongs to that historical dataset unless it is labeled as the release protocol, and the release models are those of main-text Table~\ref{tab:lapw}.

Before the released \textsc{wien2wannier} performed Stage~2 itself, a set of porting runs fed \textsc{wien2wannier} matrix elements through a conversion layer into the reference implementation of the pipeline. Because part of the charge lies in the interstitial region, $P_{n\kk}$ computed inside the muffin-tin spheres alone is not a partition of unity. The conversion layer compensated for this by rescaling $P$ by its occupied-state percentile $q_{\mathrm{occ}}=99\%$, which leaves the threshold rule unaffected because the rule depends only on relative positions in the distribution. Over the windows of Table~\ref{tab:lapw} this construction gave 4.3~meV for GaAs, 23.7~meV for MgB$_2$ and 2.9~meV for $\alpha$-Sn. It was the reference case for the port and its regression tests, and its masks are the ones examined by the certificate analysis below.

\begin{figure}[htbp]
\centering
\includegraphics{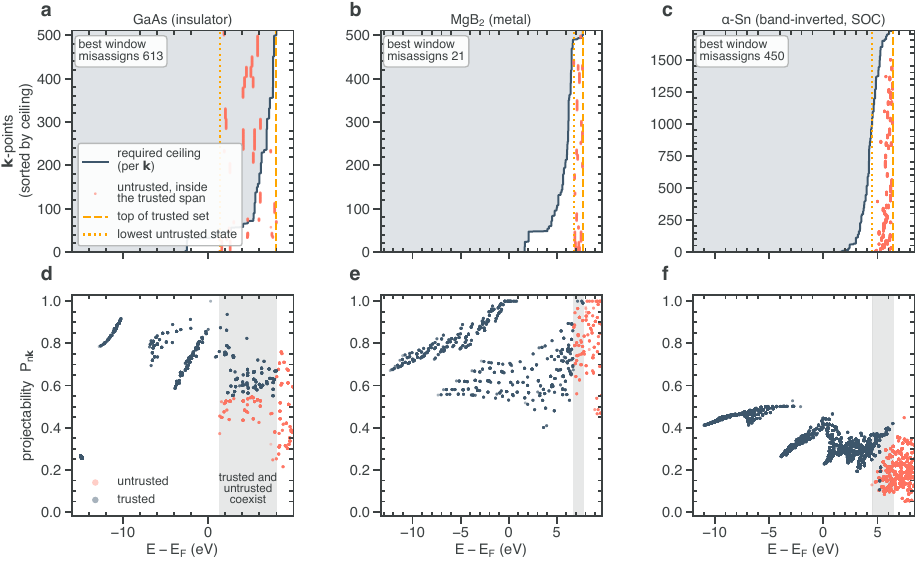}
\caption{Certificate of the LAPW trust masks built on the muffin-tin projectability in the earlier porting runs, drawn as main-text Fig.~\ref{fig:certificate}. The frozen sets of the release protocol are windows below the automatic ceiling by construction (Sec.~\ref{sm:lapwcert}). Each column is a material, and the two rows share the energy axis. (a)--(c)~The required ceiling at each $\kk$, which is the top of the trusted set (slate, $\kk$ sorted by ceiling), and the untrusted states inside the trusted span (coral). The best global window ends between the two amber lines. A window ending at the top of the trusted set (dashed) freezes all of these untrusted states, and a window ending just below the lowest of them (dotted) leaves trusted states out. The best global window misassigns 613 (GaAs), 21 (MgB$_2$) and 450 ($\alpha$-Sn) states, counting both wrongly frozen and wrongly omitted states. The count was computed up to several eV past each frozen manifold, so each minimum is bracketed (on GaAs it rises from 613 at $+7.8$~eV to about 3400 at $+15$~eV). (d)--(f)~The per-state projectability $P_{n\kk}$ on the same energy axis for trusted (slate) and untrusted (coral) states, and the gray band marks the conduction energies at which both kinds coexist. On GaAs, 330 of 512 $\kk$-points have trusted sets that are not intervals (coral inside the slate manifold), so condition (i) of main-text Appendix~\ref{app:certificate} already fails there.}
\label{fig:lapwcert}
\end{figure}

\emph{Certificate of the porting masks
(Fig.~\ref{fig:lapwcert}).} The construction of main-text Appendix~\ref{app:certificate} was applied to LAPW trust masks built from the muffin-tin projectability of the standard \textsc{wien2wannier} \texttt{.amn}, which is the route of the porting runs. On GaAs the frozen ceiling varies by 10.7~eV across the zone, the best global window misassigns 613 of 5\,960 frozen states, and at 330 of 512 $\kk$-points the frozen set is not an energy interval. On MgB$_2$ 21 states are misassigned, so the mask is nearly window-representable, and on $\alpha$-Sn 450 of 27\,648 are misassigned, all of which come from the band inversion at $\Gamma$ (unlike SnTe with the earlier count of 36 functions). A nonzero fraction of $\kk$-points with a non-interval set proves that no single energy window reproduces the mask exactly, but how a hand-set window then performs is an empirical question, which the porting runs answer. On GaAs the hand-set window fails (valence error 113~meV, gradient ${\sim}1$ after $10^4$ gauge iterations), on MgB$_2$ the two methods perform equally, and on $\alpha$-Sn a window run informed by the hybrid masks is slightly better (0.5 vs.\ 2.9~meV over the occupied bands) at about seven times the cost (below). The misassignment count does not predict this outcome, since the 450 states of $\alpha$-Sn exceed the 21 of MgB$_2$ and windows nevertheless do better on $\alpha$-Sn than on MgB$_2$. A large count can arise from a ceiling that only varies across the zone, which the window disentanglement tolerates. The outcome instead follows the fraction of $\kk$-points whose frozen set is not an energy interval (GaAs 64.5\%, MgB$_2$ 1.2\%, $\alpha$-Sn 0\%), that is, the fraction with states that no window edge can separate at a single $\kk$. For these three porting cases that fraction, which is computed from the masks in seconds, indicated before any minimization where a hand-set window fails (GaAs) and where it succeeds (MgB$_2$ and $\alpha$-Sn). It does not rule out the derived window, because at the release protocol the frozen set of every cell is the window below its automatic ceiling, and on GaAs the disentanglement of \textsc{Wannier90} inside that window reproduces the bands (5.5 and 17.1~meV) as the hand-off does (4.9 and 18.0~meV, Sec.~\ref{sec:two-routes}).

\emph{GaAs, the joint-model case
[Fig.~\ref{fig:lapwcert}(a,\,d)].} GaAs is the only one of the three LAPW systems with unfrozen states inside its own per-$\kk$ frozen span. At 330 of 512 $\kk$-points the frozen set is not an energy interval, so that a non-frozen state lies between frozen ones in energy at the same $\kk$, whereas this happens at 6 $\kk$-points on MgB$_2$ and at none on $\alpha$-Sn. The reason is the same as on h-BN, since GaAs is the only joint valence-and-conduction model of an insulator among the three, and the $sp$-antibonding character of its conduction states alternates in energy across the zone. At a representative $\kk$ the states at $+2.1$ and $+6.1$~eV (projectability $P\approx0.53$) are unfrozen while their neighbors at $+4.6$ and $+6.9$~eV ($P\approx0.61$--$0.68$) are frozen, an ordering that no \texttt{dis\_froz\_max} can separate. This is why GaAs has the largest certificate of the three, visible as the unfrozen states interleaved with the frozen ones in Fig.~\ref{fig:lapwcert}(a), and why it is the only case in which the hand-set window does not converge. Two features distinguish it from the LCAO h-BN cell. The deep Ga-$3d$ semicore is frozen by the occupied shell and not by its projectability ($P\approx0.26$ there), so it contributes no unfrozen states inside the frozen span. The conduction states also alternate at a small contrast in projectability (0.53 vs.\ 0.61, on either side of $p_{\text{froz}}=0.55$), whereas in h-BN a band of near-free-electron character ($P\approx0$) crosses atomic states ($P\approx1$). In GaAs the set is therefore not an interval because of a fine gradation of projectability, not because of a distinct free-electron band. The hand-off of the porting runs converged nevertheless ($\eta$ 4.3 and 19.3~meV), whereas the hand-set window did not. At the release protocol the frozen set is the window below $E_{\mathrm{F}}+8$~eV and both routes reproduce GaAs, so the obstruction excludes only the hand-set window.

\emph{Cost in the porting runs.} Both routes share the DFT vectors and the overlap generation of the interface, and the part of the interface that depends on the route is negligible. On $\alpha$-Sn the interface takes 26.0~s compared to 25.3~s of wall time per spin channel, because the larger anchor \texttt{.amn} and the dump of the per-state projectability cost about 3\% of a step dominated by the shared construction of $M^{(\kk,\bb)}$. The minimization stages differ much more. The timings were taken on one machine on the $\alpha$-Sn cell of the earlier porting runs ($12^3=1\,728$ $\kk$-points and 18 WFs, not the $8^3$, 20-function cell of Table~\ref{tab:lapw}). The hand-off route took about 12~min, of which about 4~min was the disentanglement in the reference implementation and 7.6~min the \textsc{Wannier90} Wannierization, which is limited to unitary gauge rotations. The window route took about 80~min, of which 73~min was the SMV~\cite{Souza2001} disentanglement of \textsc{Wannier90} and 7~min the Wannierization. In this setup the ordering follows from the sizes of the two problems, an $N_w$-dimensional rotation problem for \textsc{Wannier90} in the hand-off compared to the full $N_b$-dimensional self-consistency in the window route, and the difference grows with the frozen fraction, because the constrained sweep diagonalizes only the free block. This factor of about seven applies to the porting runs only, since at the release protocol the Wannierization dominates both routes and the difference shrinks to a factor of 1.2 (GaAs, both routes run to their criteria, Table~\ref{tab:sm-route-cost}).

The frozen-constrained $\Omega_I$ minimization takes more than 94\% of the run time of the reference implementation on every material. It is a Gauss--Seidel sweep in batched BLAS calls over the block of non-frozen states, with the convergence functional evaluated every few sweeps, so the remaining cost is interpreter overhead on the per-$\kk$ eigenproblems of small ($\le N_w$) matrices. A compiled Fortran version of the sweep reproduces it bit for bit, with identical $\Omega_I$ and sweep counts, but on one thread at production band counts it is not faster than the NumPy implementation, because NumPy calls Apple's Accelerate BLAS whereas the compiled version links OpenBLAS (\texttt{scripts/build\_disentangle\_fortran.sh}), and this difference offsets the interpreter overhead that the compiled version removes. A colored order of the $\kk$-points would allow a compiled version to run the sweep in parallel threads, but it changes the update order and with it, since $\Omega_I$ has several minima, the fixed point reached (Sec.~\ref{sec:dis}). The parallel solver whose result does not depend on the thread count is the Grassmann-CG solver, whose gradient is evaluated at one common point, but it is a different search and can stop at a different minimum from the serial sweep (Sec.~\ref{sec:dis}).

\section{Method details}
\label{sm:method}

This section collects the evidence and calibration behind main-text Secs.~\ref{sec:masks}--\ref{sec:auto} that the main text summarizes.

\paragraph{Fermi shell and the per-state mode.} For insulators the shell is the occupied manifold resolved per state, because rules at the level of whole bands freeze too much when a band changes character across the zone. On SnTe, in an earlier per-state configuration with a mis-set Fermi reference, freezing whole bands stalled the gauge optimization at $\Omega_{\text{OD}}/N_w=0.50$~\AA$^2$. Metals and insulators are distinguished by a test for bands that cross the Fermi level, because a criterion based on the sampled gap misclassifies metals on coarse meshes (MgB$_2$ on $8^3$ shows a spurious gap of $0.055$~eV when no $\kk$-point lies on the Fermi surface). The shell also has its own representability check, in which the projectability threshold $P\ge p_{\text{froz}}$ is imposed on shell states only when at most a fraction $f_{\mathrm{shell}}=0.01$ of them falls below it and is otherwise declined with an explicit warning, because the shell exists to fix the Fermi surface and removing a large part of it defeats that purpose (the shell-floor decision of main-text Sec.~\ref{sec:auto}). The fraction of shell states below $p_{\text{froz}}$ separates the material on which this override degrades the subspace from those on which it is harmless, with $16.45\%$ of the shell states below the threshold on MgB$_2$ at $12^3$ compared to at most $0.04\%$ on the other cells. In the per-state mode the frozen counts are capped at $N_w$ per $\kk$ (shell states first), and under the frozen ceiling the ceiling search enforces the same limit. In both cases the admitted set is filled up to at least $N_w$ states with the states of highest projectability, so the constrained problem is feasible at every $\kk$ by construction (the repair that the \texttt{dis\_proj\_min} option of \textsc{Wannier90} lacks, main-text Sec.~\ref{sec:masks}).

\paragraph{Band pool.} The preliminary selection uses a single contiguous pool of band indices whose size is at least $r_{\mathrm{pool}}N_w$, with $r_{\mathrm{pool}}=1.5$, and the pool grows past this size only when the index span of the seed manifold requires it. Contiguity matters because ghost bands (bands dominated by diffuse Gaussian functions, Sec.~\ref{app:lessons}, rule~1) at $+10$--$30$~eV project $\ge0.85$ onto compact target AOs, so any ranking by projectability alone pulls them in. Oversized pools also trap the minimization, since a 48-band h-BN pool converges to $\Omega_I/N_w=1.19$~\AA$^2$, compared to $0.54$~\AA$^2$ on the 24-band pool with identical masks.

\paragraph{Pocket skip on the validation set.} Among the 14 validation hand-offs, the first two conditions of the pocket skip hold on two. On SnTe the rule lowers the ceiling to $E_{\mathrm{F}}+6.18$~eV (Sec.~\ref{sec:snte}). On LAPW $\alpha$-Sn the weakest link (0.496) touches two Kramers-paired pool bands that are frozen at 8.8\% of the $\kk$-points, from $E_{\mathrm{F}}+4.93$~eV, with $\langle P\rangle=0.38$ there, and the rule leaves that pocket frozen because its character is below $p_{\rm mix}=0.5$. With $p_{\rm mix}$ lowered to 0.2 the rule acts, lowers the ceiling to $E_{\mathrm{F}}+4.83$~eV, raises the weakest link to 0.78, and improves both windows (from 2.2 and 14.2 to 2.0 and 6.3~meV over $[-12,0]$ and $[0,+4]$~eV), so on this cell the character threshold prevents an improvement. On the earlier $\alpha$-Sn field of the porting runs (Sec.~\ref{sm:bridge}) the corresponding pocket had $\langle P\rangle=0.25$, and lowering the ceiling under it by hand doubled the conduction error. The threshold of 0.5 separates SnTe from both of these $\alpha$-Sn cases. The GaAs hand-off has the weakest link of all ($5\times10^{-4}$), but on links that touch no pocket, so the rule does not apply there.

\paragraph{Solvers.} One solver is the serial Gauss--Seidel sweep of main-text Eq.~\eqref{eq:zmat}. On per-state frozen sets it reaches a lower $\Omega_I$ than the best window ($\Omega_I/N_w = 0.5362$ vs.\ $0.5371$~\AA$^2$ on h-BN at the $6^3$ mesh), but the lower $\Omega_I$ does not give the better model. Main-text Sec.~\ref{sec:extent} shows that the per-state frozen set gives larger band errors than the window at equal extent, and that on the LAPW Bi bilayer a model with a lower $\Omega_I$ has lost a target band. Because only the $D\times D$ subblock is diagonalized, the sweep runs as batched matrix products, $62\times$ faster than the naive contraction at production scale ($N_b=441$, $N_w=294$, 36 $\kk$-points, with hardware and timings in Sec.~\ref{app:perf}).

In \textsc{lcao2wannier} the sweep visits the $\kk$-points in serial order, and a compiled version reproduces it bit for bit (Sec.~\ref{app:perf}). The other solver, which is the default (Algorithm~\ref{alg:sm-cg}), minimizes $\Omega_I$ directly on the product Grassmannian by Polak--Ribi\`ere conjugate gradients with a backtracking line search and a QR retraction. Direct minimization of the entangled-band problem, in place of the alternating SMV iteration, was formulated in Ref.~\cite{DamleLevittLin2019}, and here it is applied to the frozen-constrained $\Omega_I$ alone. A CG step needs no per-$\kk$ eigendecomposition, and the gradient is evaluated at every $\kk$ from the \emph{same} iterate, so the result is bit-identical at every thread count. Every CG solve checks its gradient against finite differences before the first step (relative error $4\times10^{-9}$ to $9\times10^{-8}$ on h-BN, MgB$_2$ and SnTe) and stops when the projected gradient falls to $\max(10^{-10},10^{-6}|g_0|)$, where $g_0$ is the initial gradient. An earlier stopping rule, five consecutive small decrements of $\Omega_I$, stopped MgB$_2$ $1.1\times10^{-4}$~\AA$^2$ above its minimum while the subspace was still moving. On an isolated manifold used as a control (diamond) the two solvers agree to $2\times10^{-10}$ in $\Omega_I$.

On entangled cells the two solvers search a functional with inequivalent local minima and can stop in different ones. CG ends lower on h-BN (by 1.8\%, with 0.56 and 5.2~meV vs.\ 0.77 and 7.4~meV) and on MgB$_2$ (by 7.7\%, with 3.6 and 3.4~meV vs.\ 7.0 and 4.5~meV). On SnTe with the earlier count of 36 functions it ends 1.25\% higher, yet it is the solver that produces a usable model, because the Gauss--Seidel subspace jumps at the X points (main-text Sec.~\ref{sec:snte}). On the LAPW GaAs hand-off (Table~\ref{tab:lapw}) CG also ends lower ($\Omega_I$ 38.957 vs.\ 38.979~\AA$^2$). The model that \textsc{Wannier90} then produces depends on its optimizer as much as on the disentangler. On GaAs the CG hand-off, Wannierized with the default optimizer, converges at 5.6 and 17.7~meV, compared to 4.9 and 18.0~meV from the Gauss--Seidel one (Table~\ref{tab:sm-x1}), and a CG hand-off made with an earlier stopping rule, equal in $\Omega_I$ to six digits, ends between 5.2/14.0 and 12.1/36.5~meV depending on the optimizer (Table~\ref{tab:sm-optimizer}).

A colored Gauss--Seidel order of the $\kk$-points allows the sweep to run in parallel but changes the update order, and with it the fixed point, so that on SnTe $\Omega_I$ ends 1.6\% higher. In \textsc{lcao2wannier} the colored order is an option that no reported model uses, whereas the LAPW Gauss--Seidel models used it (Sec.~\ref{sm:sweeporder}). CG is also the cheaper solver when the free subspace is large, since its step costs matrix products, compared to the $n_{\text{dis}}^3$ eigendecompositions of the sweep, and the crossover per step lies near $n_{\text{dis}}\approx250$ free directions (on SnTe with one thread, 81 vs.\ 674~s). Starting the sweep from the CG result does not combine the two solvers, because on h-BN and MgB$_2$ the sweep returns to its own fixed point (after 323 and 2428 further sweeps), and on SnTe it stays at the CG minimum.

Neither solver comes with a guarantee of global convergence, since $\Omega_I$ is non-convex over the product of Grassmannians, and the self-consistent iteration of the original Souza--Marzari--Vanderbilt (SMV) scheme~\cite{Souza2001} has no such guarantee either. The SCDM construction for entangled bands~\cite{Damle2018} avoids the iteration but does not minimize $\Omega_I$. In our runs $\Omega_I$ decreases at every conjugate-gradient iteration, which the line search enforces, whereas the Gauss--Seidel sweep is a fixed-point iteration and need not descend. On MgB$_2$ the sweep passes $\Omega_I=12.277$~\AA$^2$ at sweep 12 and converges to its fixed point at 12.872~\AA$^2$, whereas on h-BN and the Sc-COF it descends at every sweep (traces in Sec.~\ref{sm:convergence}). On correctly sized pools the band distance does not depend on the seed (projection-SVD vs.\ random seeding at $r_{\mathrm{pool}}=1.5$), although the minimum reached can depend on the starting point, and started from the CG minimum the sweep returns to its own fixed point on h-BN and MgB$_2$ and stays at the CG minimum on SnTe. For the runs that reach the sweep cap before the tolerance, we report the final slope of $\Omega_I$ with the result.

\paragraph{Sweep options of the LAPW models.}\label{sm:sweeporder} The seven Gauss--Seidel hand-offs of main-text Table~\ref{tab:lapw} were made with three options of the \textsc{wien2wannier} sweep that are off by default, namely the colored update order (\texttt{-\kern0pt-colored}), scheduled over-relaxation starting at a factor of 1.9 (\texttt{-\kern0pt-relax -1.9}), and convergence checks at adaptive intervals starting from every 10 sweeps (\texttt{-\kern0pt-adaptive}, \texttt{-\kern0pt-check-every 10}). No LCAO model uses any of them, and conjugate gradients, the default of both releases and the solver of the $\alpha$-Sn row, ignores them. Over-relaxation and the adaptive checks change only the number of sweeps, and on MgB$_2$ the sweep needs 2\,211 sweeps without either option and 120 with both, with subspace eigenvalues that agree to 1.1~meV. The update order, in contrast, can change the result, and Table~\ref{tab:sm-sweeporder} therefore repeats each hand-off in the serial order, with the serial program of the release and every other option as published, and with the published options the same program reproduces the published hand-offs byte for byte (the Bi bilayer was not rerun). Below the frozen ceiling the two orders give the same subspace eigenvalues on every cell, because every state below the ceiling belongs to the subspace whatever the solver does, and on GaAs, both Fe spin channels and SrVO$_3$ they also give the same subspace above it, to within 0.9~meV. On Fe with spin--orbit coupling the two orders reach the same $\Omega_I$ to $8\times10^{-6}$~\AA$^2$ but different subspaces above $+8.2$~eV, just above the ceiling of $+8.0$~eV, where a subspace eigenvalue changes by up to 1.1~eV, and the serial-order hand-off of this cell was not Wannierized. On MgB$_2$ they stop at different minima, and conjugate gradients from the same inputs reaches the minimum of the colored order. Wannierized with the published protocol, the serial-order hand-off gives $\Omega_{\rm tot}=14.837$~\AA$^2$ and 7.0 and 1.5~meV over $[-13,+7.2]$ and $[-2,+2]$~eV, compared to 18.153~\AA$^2$ and 11.5 and 4.4~meV for the published model. The released driver at its defaults (conjugate gradients on the projections it chooses) also reaches the lower minimum on this cell, with 7.1 and 1.5~meV. On the Bi bilayer neither order converges within 30\,000 sweeps, and on $\alpha$-Sn the serial order does not, whereas the colored, relaxed sweep converges there in 164 sweeps.

\begin{table}[htbp]
\caption{The Gauss--Seidel hand-offs of main-text Table~\ref{tab:lapw} in the published colored order and in the serial order, every other option unchanged. The ``largest change'' is the largest change of a subspace eigenvalue at a mesh $\kk$-point between the two orders, and the ``lowest change'' is the lowest energy, relative to $E_{\mathrm{F}}$, at which an eigenvalue changes by more than 1~meV (--: none).}
\label{tab:sm-sweeporder}
\small
\begin{tabular*}{\textwidth}{@{\extracolsep{\fill}}l r r r r r r@{}}
\toprule
 & \multicolumn{2}{c}{Colored (published)} & \multicolumn{2}{c}{Serial} & & \\
\cmidrule(lr){2-3}\cmidrule(lr){4-5}
Cell & $\Omega_I$ (\AA$^2$) & Sweeps & $\Omega_I$ (\AA$^2$) & Sweeps & Largest change & Lowest change \\
\midrule
GaAs & 38.97932196 & 120 & 38.97932196 & 120 & 0.004~meV & -- \\
Fe$\uparrow$ & 5.87627656 & 110 & 5.87627656 & 110 & 0.002~meV & -- \\
Fe$\downarrow$ & 6.00223215 & 110 & 6.00223215 & 110 & 0.003~meV & -- \\
SrVO$_3$ & 21.96424225 & 592 & 21.96424224 & 1\,728 & 0.9~meV & -- \\
Fe-SOC & 11.22394220 & 8\,049 & 11.22393391 & 5\,239 & 1.1~eV & $+8.2$~eV \\
MgB$_2$ & 12.60297735 & 120 & 12.36906647 & 210 & 3.6~eV & $+8.0$~eV \\
Bi bilayer$^{\rm a}$ & 80.19172265$^{\rm b}$ & 30\,000 & 80.19139918$^{\rm b}$ & 30\,000 & 0.19~eV & $+4.3$~eV \\
$\alpha$-Sn$^{\rm c}$ & 55.94369571 & 164 & 55.94286476$^{\rm b}$ & 30\,000 & 0.31~eV & $+6.0$~eV \\
\bottomrule
\end{tabular*}

\smallskip
\begin{minipage}{\textwidth}\footnotesize
$^{\rm a}$Published values from the original run, not rerun with the published options.\\
$^{\rm b}$Stopped at the 30\,000-sweep cap.\\
$^{\rm c}$This row gives the sweep, whereas main-text Table~\ref{tab:lapw} reports the conjugate-gradient model ($\Omega_I=55.94273218$~\AA$^2$ in 2\,883 gradient evaluations).
\end{minipage}
\end{table}

\paragraph{Gauge.} Starting projections built from atomic-orbital trial functions can leave the Wannierization stalled. On h-BN at $6^3$ the window route had not met its criterion after 500\,000 iterations from the per-$\kk$ SVD of its projections, but it converged in 1\,870--2\,260 iterations from the SCDM gauge with the same subspace (main-text Sec.~\ref{sec:two-routes}). A gauge built from point rows, $B[\text{anchors}]^\dagger$, is exactly singular whenever a subspace state has zero amplitude on every anchor AO at some high-symmetry $\kk$. This occurs on h-BN, where $\sigma_{\min}$ rises from 0.000 to 0.103 when columns are used instead. Of the four candidates, a point-row candidate is the best conditioned on the Gauss--Seidel SnTe subspace ($\sigma_{\min}=1.8\times10^{-3}$), and a column candidate on the default conjugate-gradient subspace ($3.7\times10^{-3}$).

\paragraph{Thresholds.} In the per-state mode, where the trust mask is the frozen set, no fixed pair $(p_{\text{froz}},p_{\text{gate}})$ works for all materials, since the h-BN conduction band requires $p_{\text{froz}}\approx0.70$ and the Bi conduction band requires $p_{\text{froz}}\le0.87$. At $p_{\text{froz}}=0.95$, the freezing threshold recommended for PDWF~\cite{Qiao2023} and the fallback value of main-text Appendix~\ref{app:constants}, the Bi conduction band is left unfrozen and its error rises from 0.3 to 153~meV. The gray band of SnTe lies so high that a gate below $0.79$ admits continuum states again. The lower clip is itself calibrated, because setting the threshold from a trusted band whose $\min_\kk P$ dips to $0.43$ on SnTe (with the earlier count of 36 functions) freezes too many states and triples the error of the occupied manifold.

\paragraph{Channel scores.} Channel selection needs both the retention and the specificity. The retention $\Pi_c$ is needed because orthogonalization removes weight from diffuse channels and so lowers the projectability they can reach, and the projectability a channel achieves is therefore compared with $\Pi_c$ and not with one. The specificity is needed because the union of all channels is the full basis, for which $P\equiv1$, so the raw coverage rises whenever any channel is added. The h-BN cell shows what happens without it, because its polarization-$d$ continuum projects $P\approx1$, and with these channels in the target ($N_w=36$) the method fails, giving $\Omega_I/N_w = 7.5$~\AA$^2$ compared to $0.54$~\AA$^2$ for the chemical 16-function target (spreads are reported per Wannier function throughout) and a singular gauge. The selection itself is exact, and it was verified to be identical to an exhaustive enumeration of subsets over $5000$ randomized instances. The band error, unlike the selection objective, is not additive over channels (Table~\ref{tab:snte-channels}, earlier Fermi reference), since Sn-$d$ alone changes the conduction error from 1568 to 1557~meV and Te-$d$ alone gives 321~meV, while both together give 46.7~meV. In the count rule of main-text Eq.~\eqref{eq:decoupled}, the Te-$4d$ bands of SnTe at $E_{\mathrm{F}}-38.5$~eV lie below the pool and are not counted. Only admitted channels count, so the failure of the 36-function h-BN target described above does not recur, because the $d$ channels of h-BN have neither semicore bands nor specificity. An earlier version of the rule added the valence shell of a channel only while the frozen set did not fit, and on SnTe it stopped at 36 functions, with 32 frozen and 4 free directions per $\kk$-point (main-text Sec.~\ref{sec:snte}).

\paragraph{Checks.} The checks were built in the same way as the thresholds, since each criterion comes from a failure observed in a run, and the run either stops or reports the failure with the reason. Three checks stop the run, namely a gauge check, which prevents writing a rank-deficient SCDM \texttt{.amn}, the ceiling search, which stops when the frozen count exhausts $N_w$, and the hand-off acceptance test, which enforces Eq.~\eqref{eq:handoff-acceptance} on every link. The other checks only report, and they include a pool-coverage audit, which lists the states below the ceiling, at either end of the pool, that the band pool does not contain, a shell-floor decision, which reports when the per-state Fermi shell overrides the projectability threshold, a $\kk$-mesh audit, which lists the high-symmetry points that the mesh never samples, and a manifold check, which flags a band distance (Table~\ref{tab:defaults}) that nearest-band matching understates because a reference band has no Wannier partner.

Two further checks were added after failures on the metal SnTe. The subspace-continuity check reports the smallest singular value of the hand-off overlaps $W^\dagger(\kk)M^{(\kk,\bb)}W(\kk+\bb)$ and warns below $10^{-2}$. The value is 0.73--0.93 on h-BN, MgB$_2$ and Bi and 0.92 on SnTe (0.20 before the pocket skip of Sec.~\ref{sec:masks}), compared to $7\times10^{-4}$ on SnTe with the earlier count of 36 functions and the Gauss--Seidel solver, where the subspace jumps between neighboring $\kk$-points in a way that no gauge can smooth (Sec.~\ref{sec:snte}). A warning marks a jump to be inspected and not a failed model, since the LAPW GaAs hand-off of Table~\ref{tab:lapw} reaches $5\times10^{-4}$ and still interpolates at 4.9 and 18.0~meV once converged. The Fermi-level check keeps the electronic-structure code's $E_{\mathrm{F}}$ for a metal, where the band-filling estimate (the highest occupied eigenvalue over the mesh) lies above it and depends on the mesh (1.43~eV too high for SnTe on $8^3$). SnTe studies in the main text that were run with that estimate are marked as using the earlier Fermi reference.

The test of the frozen count against $N_w$ and the contractivity test follow from mathematical constraints, whereas the thresholds of the other checks had to be calibrated. The gauge check, for instance, must be scale-free, because the $\sigma_{\min}$ at the worst $\kk$ depends on the system and the Wannier count. The models of Table~\ref{tab:defaults} span $3.7\times10^{-3}$ on SnTe to $0.39$ on Bi, and the earlier per-state Sc-COF model lay below $5\times10^{-4}$, under the absolute floor of $2\times10^{-3}$ calibrated on the LAPW frontend. In our own runs the $\kk$-mesh audit found that the $8^3$ MgB$_2$ mesh of the earlier runs samples neither $K$ nor $H$ (Sec.~\ref{app:perf}), and the pool audit has raised no warning on the five materials, none of which has a pool defect. The certificate of main-text Appendix~\ref{app:certificate} is separate from these checks and measures, before any minimization, how far the trust mask departs from an energy window.

\section{Practical stability lessons}
\label{app:lessons}
Each of the following empirical rules was needed for at least one validated model, and they are collected here for anyone who reimplements the method.
\begin{enumerate}\setlength\itemsep{2pt}
\item A contiguous pool is better than a pool ranked by projectability, because diffuse-Gaussian ghost bands at $+10$--$30$~eV project $\ge0.85$ onto compact target AOs, and a ranking by $P$ alone admits and even freezes them ($\Omega_I/N_w$ 1.4--2.4 vs.\ 0.54~\AA$^2$ on h-BN). Ghost states at extreme energies ($P=1.0$ at $+1550$~eV in Bi) are rejected by the seed energy window.
\item A pool of about $1.5N_w$ bands is sufficient, because the optimum saturates there and oversized pools lead the Gauss--Seidel iteration to a worse result ($\Omega_I/N_w=1.19$ vs.\ 0.54~\AA$^2$ for a 48-band pool). On a pool of this size the seed did not change the band distance on h-BN or MgB$_2$, although it changes the minimum reached on SnTe (main-text Sec.~\ref{sec:dis}).
\item In the per-state mode, whole bands are frozen and the gray zone is gated, because whether the run converges is set by the band-average gate and not by the state threshold (the $2\times2$ experiment, Fig.~\ref{fig:hbn}), and the state threshold must not follow the minima of trusted bands below $P\approx0.55$ (the SnTe floor calibration, made with 36 functions).
\item Both SCDM trial types are formed and the better conditioned one is kept, because point-row trials can be exactly singular at high-symmetry $\kk$ (h-BN), whereas on the Gauss--Seidel SnTe subspace a point-row trial is the best conditioned (main-text Sec.~\ref{sec:scdm}).
\item The Fermi shell is resolved per state, because freezing whole bands freezes too much when a band changes character across the zone (on SnTe, in the per-state mode with the earlier Fermi reference, the run stalled at $\Omega_{\text{OD}}/N_w=0.50$~\AA$^2$ with a gauge $\sigma_{\min}$ of 0.005).
\item Interpolated bands are taken from the \texttt{band.dat} file of \textsc{Wannier90}, because a direct re-transform of \texttt{hr.dat} ignores the Wigner--Seitz \texttt{wsvec} corrections and is wrong on coarse grids.
\item Guiding centers are used only with real projections, because with the random-projection fallback they pull all WFs to random points (on the Sc-COF the mean spread is $5.7\times10^{4}$~\AA$^2$ with them and 11~\AA$^2$ without). The pipeline couples the flag to the success of the projections and derives PDWF centers from the valence configuration (MgB$_2$ $\Omega_{\text{tot}}/N_w$ from 3.7 to 0.68~\AA$^2$ with centers and the full iteration budget).
\end{enumerate}
The remaining rules come from fixes made to the released programs during this work.
\begin{enumerate}\setlength\itemsep{2pt}\setcounter{enumi}{7}
\item An $N_b{=}N_w$ Wannierization is never restarted from its checkpoint, because the \texttt{restart = wannierise} option of \textsc{Wannier90} discards the gauge of a run without disentanglement, so such a run is taken to the criterion in one run (Sec.~\ref{app:perf}). With $N_b>N_w$ the restart is safe, and the SnTe window-route runs restarted from their disentanglement checkpoints reproduce the spread of their first Wannierization to every printed digit (main-text Table~\ref{tab:routes}).
\item Subspace continuity is checked before the Wannierization, because a converged $\Omega_I$ can still belong to a subspace that jumps between neighboring $\kk$-points, and no gauge repairs such a jump (SnTe with 36 functions, main-text Sec.~\ref{sec:snte}).
\item The Fermi energy of the electronic-structure code is kept for a metal, because the highest occupied eigenvalue over the mesh lies above $E_{\mathrm{F}}$ and depends on the mesh.
\item The valence shell of a selected channel is counted, because an added channel whose semicore bands fill its Wannier functions leaves no function for the conduction character for which it was selected (SnTe, 36 to 46 functions, main-text Sec.~\ref{sec:snte}).
\item A pocket of mixed character at the ceiling is not frozen, because a band of half-target character frozen on an island of $\kk$-points makes the subspace jump at the edge of the island (SnTe, main-text Sec.~\ref{sec:masks}). Whether a pocket of lower character should be released depends on the cell (Sec.~\ref{sm:method}).
\end{enumerate}

\section{SnTe channel-set, Wannier-count, h-BN calibration and baseline tables}
\label{app:tables}

\begin{table}[htbp]
\caption{SnTe target channels and Wannier count ($8^3$ mesh, analytic overlaps). The upper block is the channel study in the per-state mode (trust mask frozen, no ceiling, fixed thresholds $p_{\text{froz}}=0.70$ and $p_{\text{gate}}=0.75$). Its bands are matched to the nearest reference band, and the RMS errors are taken over $[-12,0]$ (occupied) and $[0,+6]$~eV (conduction) relative to an earlier Fermi reference 1.43~eV above CRYSTAL's. Its last row is the automatic pipeline (main-text Sec.~\ref{sec:auto}). The lower block gives the automatic set $\{$Sn-$d$, Te-$d\}$ at the released defaults (CRYSTAL's $E_{\mathrm{F}}$, CG disentanglement) with the earlier count (36) and with the count rule of main-text Eq.~\eqref{eq:decoupled} (46), and the number after each label is the frozen ceiling $\varepsilon_{\rm c}-E_{\mathrm{F}}$ in eV. These rows pair bands by index, with nearest-band values in parentheses. $\Omega_{\text{tot}}/N_w$ is the total spread per Wannier function from \textsc{Wannier90}, the iterations are \textsc{Wannier90} iterations to convergence, and bold marks the lowest error in each block.}
\label{tab:snte-channels}
\small
\begin{tabular*}{\textwidth}{@{\extracolsep{\fill}}l r r r c r@{}}
\toprule
Target (spinor WFs) & $N_w$ & Occupied (meV) & Conduction (meV) & $\Omega_{\text{tot}}/N_w$ (\AA$^2$) & Iterations \\
\midrule
Sn($sp$)+Te($sp$)    & 16 & 21.0 & 1568 & 2.14 & -- \\
Sn($spd$)+Te($sp$)   & 26 & 24.3 & 1557 & 1.34 & -- \\
Sn($sp$)+Te($spd$)   & 26 & \textbf{6.1} & 321 & 2.18 & -- \\
Sn($spd$)+Te($spd$)  & 36 & 12.4 & 46.7 & 3.01 & -- \\
Automatic augmentation & 36 & 9.0 & \textbf{46.4} & 2.95 & -- \\
\midrule
Earlier count, $+8$ & 36 & 21.8 (12.3) & 47.1 (23.3) & 1.81 & 42\,130 \\
Earlier count, $+6.2$ & 36 & 5.7 (5.3) & 17.1 (16.7) & 1.57 & 24\,360 \\
Count rule, $+8$ & 46 & 8.4 (6.5) & 24.4 (18.5) & 1.23 & 1\,460 \\
Count rule, pocket skip, $+6.18$ & 46 & \textbf{3.9} (3.2) & \textbf{15.4} (13.0) & 1.16 & 5\,180 \\
\bottomrule
\end{tabular*}
\end{table}

In the SnTe channel study (upper block of Table~\ref{tab:snte-channels}) the channels act as sets rather than one at a time, since neither $d$ channel alone suffices, and Te-$spd$ alone gives the best occupied manifold but a conduction band seven times worse. The automatic pipeline selects Sn-$d$ and Te-$d$, rejects Te-$f$, whose specificity of $-0.24$ shows that its coverage gain comes only from the completeness of the basis, and clips the threshold at its lower bound. The study was run in the per-state mode with the earlier Fermi reference, and its count rule stopped at 36 functions.

The lower block measures the Wannier count for the same channel set at the released defaults (CRYSTAL's $E_{\mathrm{F}}$,
automatic ceiling, CG disentangler). The count rule of main-text Eq.~\eqref{eq:decoupled} adds the 10 Sn-$4d$ semicore bands
in the pool to the 36 functions of the earlier count, giving 46. At the automatic ceiling of $E_{\mathrm{F}}+8$~eV the 46
functions lower the occupied and conduction errors by factors of 2.6 and 1.9 and cut the Wannierization from
42\,130 to 1\,460 iterations (main-text Sec.~\ref{sec:snte}). With the ceiling just below the mixed-character pocket at X
(main-text Sec.~\ref{sec:snte}), set by hand to $E_{\mathrm{F}}+6.2$~eV for 36 functions and by the pocket skip to $E_{\mathrm{F}}+6.18$~eV for
46, the 46-function model is again the better one, 3.9 and 15.4~meV compared to 5.7 and 17.1~meV, and needs 5\,180
iterations instead of 24\,360. Lowering the ceiling therefore helps at both counts, and the larger count helps
at both ceilings.

\begin{figure}[htbp]
\centering
\includegraphics{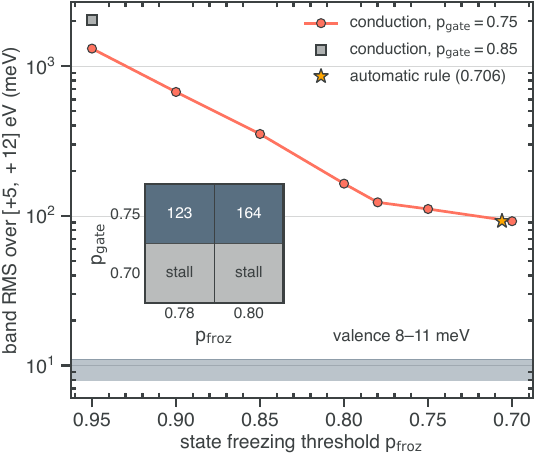}
\caption{Per-state mode on h-BN at $6^3$ (trust mask frozen, 16 WFs). The conduction RMS over $[+5,+12]$~eV (nearest band) is plotted against $p_{\text{froz}}$ at $p_{\text{gate}}=0.75$ (coral circles), with a gray square for the first run at $p_{\text{gate}}=0.85$ and a star for the thresholds of the automatic rule (92.2~meV). The slate band is the valence RMS (8--11~meV). The inset is the $2\times2$ gate experiment, giving the conduction RMS in meV for the converged runs (slate) or `stall' (gray). On the converged $12^2{\times}6$ mesh the same mode gives 52~meV (main-text Table~\ref{tab:extent}), and the difference from that value comes from the mesh.}
\label{fig:hbn}
\end{figure}

In the h-BN threshold calibration all runs converged, and the conduction RMS over $[+5,+12]$~eV changes with the pair $(p_{\text{froz}}, p_{\text{gate}})$ as
$(0.95,0.85)$~2038 $\to$ $(0.95,0.75)$~1310 $\to$ $(0.90,0.75)$~672 $\to$
$(0.85,0.75)$~352 $\to$ $(0.80,0.75)$~164 $\to$ $(0.78,0.75)$~123 $\to$
$(0.75,0.75)$~111 $\to$ $(0.70,0.75)$~$\mathbf{92}$~meV, while the valence error stays at
$8$--$11$~meV throughout. A gate of $0.70$, in contrast, fails for every state threshold tested (the $2\times2$ experiment, Fig.~\ref{fig:hbn}). The automatic rule
gives $(0.706, 0.740)$ on this $6^3$ mesh (the gate is 0.739 on the $12^2{\times}6$ mesh of main-text Table~\ref{tab:defaults}) and reproduces the hand-tuned reference at the same mesh, 92.2/10.0~meV nearest-band (52.2/1.3 paired by band index on the converged $12^2{\times}6$ mesh). These are per-state-mode values, whereas the released default of
main-text Table~\ref{tab:defaults} freezes the window below $E_{\mathrm{F}}+8$~eV instead.

\emph{Mechanism isolation on h-BN.} With the frozen set given by the trust mask (the per-state mode, \texttt{-\kern0pt-hybrid-froz-emax none}), two controlled experiments at the $6^3$ mesh establish what each threshold does (Fig.~\ref{fig:hbn}). (a)~\emph{The gate decides convergence.} A $2\times2$ scan $(p_{\text{froz}},p_{\text{gate}})\in\{0.80,0.78\}\times\{0.75,0.70\}$ shows both gate-$0.75$ runs converging ($164/123$~meV conduction) and both gate-$0.70$ runs stalling ($\Omega_{\rm tot}/N_w\approx1.6$--$1.9$~\AA$^2$, \textsc{Wannier90}'s total spread 26--31~\AA$^2$), because the $\bar P\approx0.72$ gray band freezes at scattered $\kk$-points whatever the state threshold. (b)~\emph{The threshold sets the conduction fidelity monotonically.} Lowering $p_{\text{froz}}$ at fixed gate takes the conduction RMS from 2038 to 92~meV, with the valence RMS at 8--11~meV and every run converged.

\emph{The baseline of main-text Fig.~\ref{fig:baseline}.} Table~\ref{tab:sm-baseline} lists the runs of main-text Fig.~\ref{fig:baseline}(a), which are standard \textsc{Wannier90} with four constraint settings and the two released routes, on identical h-BN inputs at the $6^3$ mesh.

\begin{table}[htbp]
\caption{The runs of main-text Fig.~\ref{fig:baseline} (h-BN, $6^3$, 16 functions, with the same band pool, analytic overlaps, L\"owdin target chemistry and iteration budgets). ``Frozen$/\kk$'' gives the frozen states per $\kk$-point, and $\Omega_{\text{OD}}/N_w$ is the off-diagonal spread per Wannier function at the end of the run. The per-state runs use PDWF's recommended thresholds $0.95/0.01$ (freezing/discard) or the derived freezing threshold $0.706$, without a frozen energy window. Band errors are given over the valence window $[-20,0]$~eV and the conduction window $[+5,+12]$~eV relative to $E_{\mathrm{F}}$, with bands paired by index. The \textsc{Wannier90} runs use the serial development build \texttt{ff955d1}, except the automatic-threshold run (v4.0.2), and -- marks values not recorded.}
\label{tab:sm-baseline}
\small
\begin{tabular*}{\textwidth}{@{\extracolsep{\fill}}l c c r r r@{}}
\toprule
Setting & Frozen$/\kk$ & $\Omega_{\text{OD}}/N_w$ & Valence & Conduction & Converged \\
 & & (\AA$^2$) & (meV) & (meV) & (iterations) \\
\midrule
Hand-set energy windows & -- & 2.3 & 437 & 516 & no \\
\texttt{dis\_froz\_proj}, $0.95/0.01$ & 1--7 & 3.4 & 609$^{\rm a}$ & 546$^{\rm a}$ & no \\
\texttt{dis\_froz\_proj}, $0.706$ & 8--14 & 3.9 & 679 & 662 & no \\
\texttt{dis\_proj\_auto}, $0.805/0.180$ & -- & -- & 498 & 803 & no \\
\addlinespace
Hand-off & all$^{\rm b}$ & -- & 10.1 & 69.3 & 1\,760 \\
Window route, SCDM start & all$^{\rm b}$ & -- & 10.1 & 69.1 & 1\,870 \\
Window route, SVD start & all$^{\rm b}$ & -- & 575 & 470 & no ($5\times10^5$) \\
\bottomrule
\end{tabular*}

\smallskip
\begin{minipage}{\textwidth}\footnotesize
$^{\rm a}$With the serial development build. The archived values, 595 and 605~meV, do not reproduce.\\
$^{\rm b}$Every pool state below the ceiling at the top of the trusted set, $+12.65$~eV.
\end{minipage}
\end{table}

\section{Higher-order midpoint overlaps}
\label{sm:midorder}

Main-text Appendix~\ref{app:contractivity} replaces each momentum-shifted overlap by the zeroth-order term of the expansion of $e^{-i\bb\cdot(\mathbf r-\mathbf c)}$ about the pair midpoint $\mathbf c=(\boldsymbol\tau_\mu+\boldsymbol\tau_\nu+\mathbf g)/2$. Keeping the terms through order $n$ gives
\begin{equation}
S^{(\bb,n)}_{\mu\nu}(\mathbf g)=e^{-i\bb\cdot\mathbf c}\sum_{j=0}^{n}\frac{(-i)^j}{j!}\,\langle\chi_\mu|[\bb\cdot(\mathbf r-\mathbf c)]^j|\chi_\nu\rangle ,
\label{eq:sm-midorder}
\end{equation}
which needs the moment integrals of the AO pairs about $\mathbf c$. We evaluate them with the Obara--Saika recursion, extended by a third index for the powers of $(x-c_x)$, over the same lattice sum and 16-bohr pair cutoff as the analytic overlaps. The $j=0$ term is taken from the printed $S(\RR)$, as in the pipeline (option \texttt{-\kern0pt-mmn-midpoint-order} of \textsc{lcao2wannier}~v1.0), and with $n=0$ the construction reproduces each run's own midpoint $M^{(\kk,\bb)}$ ($\sigma_{\max}$ to $2\times10^{-7}$). Table~\ref{tab:sm-midorder} lists the largest singular value and $\Omega_I(\mathds{1})$ for $n=0$ to 4 on the pools and meshes of main-text Fig.~\ref{fig:stability}, with h-BN $12{\times}12{\times}6$ added.

First order is worse than zeroth order, because $\Omega_I(\mathds{1})$ is then negative on every mesh. $\Omega_I$ is second order in $|\bb|$, so the expansion must be correct through second order to give even its sign. Second order gives a positive $\Omega_I(\mathds{1})$ on every mesh, but $\sigma_{\max}$ exceeds one on every h-BN mesh and on MgB$_2$ at $8^3$. At fourth order $\Omega_I(\mathds{1})$ lies within 0.23\% of the analytic value on h-BN at $3^3$ and within 0.03\% on the other meshes, and $\sigma_{\max}$ exceeds one only on h-BN at $3^3$. No finite order is contractive by construction, because the truncation error of order $n$ scales as $|\bb|^{n+1}$, while the margin $1-\sigma_{\max}$ of the analytic $M$ also shrinks as the mesh becomes denser ($2.2\times10^{-4}$ on h-BN at $3^3$, $2.4\times10^{-5}$ at $12{\times}12{\times}6$). Fifth order exceeds one again on h-BN at $3^3$ (1.00076) and $4^3$ (1.000011), and sixth order is contractive on every mesh measured. With the printed $S(\RR)$ as the zeroth-order term, the relative Frobenius distance to the analytic $M$ stops falling at $(0.8$--$3.5)\times10^{-4}$, the size of the difference between the printed and the rebuilt overlaps (main-text Appendix~\ref{app:contractivity}). With every term taken from the Gaussian moments, it falls to $1.5\times10^{-7}$ (h-BN $12{\times}12{\times}6$) and $8.8\times10^{-9}$ (MgB$_2$ $12^3$) at sixth order. The fourth-order moment integrals cost about three times the exact momentum-shifted integral per AO pair (355 vs.\ 108~$\mu$s for one $d$--$d$ pair), so the analytic construction remains the one we use.

\begin{table}[htbp]
\caption{Midpoint overlaps expanded to order $n$ (Eq.~\eqref{eq:sm-midorder}), with the zeroth-order term from the printed $S(\RR)$ as in the pipeline. $\sigma_{\max}$ is the largest singular value of $M^{(\kk,\bb)}$ over all $(\kk,\bb)$ blocks, and $\Omega_I(\mathds{1})/N_b$ is the spread of the pool per band with first-order Marzari--Vanderbilt weights, in \AA$^2$. The pools and meshes are those of the projectability runs of main-text Fig.~\ref{fig:stability}, h-BN with $N_b=16$ at $3^3$ and $4^3$ and 17 on the finer meshes, and MgB$_2$ with 14 (spin channel $\alpha$). The order $n=0$ is main-text Eq.~\eqref{eq:midpoint}. At $n=4$ the h-BN $3^3$ value exceeds one in 12 of 216 blocks.}
\label{tab:sm-midorder}
\centering\small
\begin{tabular*}{\textwidth}{@{\extracolsep{\fill}}l l r r r r r r r@{}}
\toprule
Cell & Mesh & $N_b$ & $n=0$ & $n=1$ & $n=2$ & $n=3$ & $n=4$ & Analytic \\
\midrule
\multicolumn{9}{@{}l}{$\sigma_{\max}$} \\
h-BN & $3^3$ & 16 & 3.050380 & 3.778089 & 1.284223 & 1.196223 & 1.000374 & 0.999776 \\
h-BN & $4^3$ & 16 & 3.179962 & 3.063765 & 1.156890 & 1.063645 & 0.999948 & 0.999823 \\
h-BN & $6^3$ & 17 & 3.229839 & 4.248738 & 1.082990 & 1.013240 & 0.999949 & 0.999949 \\
h-BN & $8^3$ & 17 & 2.049957 & 3.281741 & 1.020252 & 1.001897 & 0.999970 & 0.999970 \\
h-BN & $12{\times}12{\times}6$ & 17 & 1.332543 & 2.208522 & 1.003050 & 0.999947 & 0.999984 & 0.999976 \\
MgB$_2$ & $8^3$ & 14 & 1.146902 & 1.174009 & 1.000192 & 0.999901 & 0.999935 & 0.999936 \\
MgB$_2$ & $12^3$ & 14 & 1.063407 & 1.077500 & 0.999974 & 0.999965 & 0.999972 & 0.999972 \\
\midrule
\multicolumn{9}{@{}l}{$\Omega_I(\mathds{1})/N_b$ (\AA$^2$)} \\
h-BN & $3^3$ & 16 & $-0.3737$ & $-1.0148$ & 0.7393 & 0.8665 & 0.8175 & 0.8156 \\
h-BN & $4^3$ & 16 & $-0.0201$ & $-0.8621$ & 1.0656 & 1.1661 & 1.1064 & 1.1067 \\
h-BN & $6^3$ & 17 & 0.5376 & $-1.3006$ & 1.4147 & 1.5013 & 1.4477 & 1.4473 \\
h-BN & $8^3$ & 17 & 1.2841 & $-1.3478$ & 1.5916 & 1.6589 & 1.6167 & 1.6163 \\
h-BN & $12{\times}12{\times}6$ & 17 & 2.1119 & $-1.2711$ & 1.7449 & 1.7865 & 1.7613 & 1.7610 \\
MgB$_2$ & $8^3$ & 14 & 0.9553 & $-0.4692$ & 1.3706 & 1.4112 & 1.3941 & 1.3938 \\
MgB$_2$ & $12^3$ & 14 & 1.3048 & $-0.2218$ & 1.6558 & 1.6746 & 1.6666 & 1.6665 \\
\bottomrule
\end{tabular*}
\end{table}

\section{Structures and calculation parameters}
\label{sec:sm-structures}

Tables~\ref{tab:sm-structures}--\ref{tab:sm-wien2k} give the structure and the electronic-structure settings of every calculation in the paper, except the 89 atomic positions of the Sc-COF, which are not listed. The values are read from the files that the Wannier constructions used, namely the header of each CRYSTAL23 properties output, which repeats the summary of the converged self-consistent field, and the \texttt{case.struct} and \texttt{case.in1} of each WIEN2k case (script \nolinkurl{figures_src/gen_structure_tables.py}). Main-text Sec.~\ref{sec:results} and Appendix~\ref{app:lapw} give the functionals, basis sets and pseudopotentials with their references. The h-BN cell holds two BN layers with B above N and N above a hexagon center (space group $P\bar{6}m2$), not the AA$'$ stacking ($P6_3/mmc$) of the experimental crystal, and every comparison for this cell is against its own reference bands. The MgB$_2$ lattice constants of the two programs agree to 0.02\%, and the Bi bilayer has the same lattice constant and a buckling of $\pm0.8984$~\AA\ in both. The WIEN2k settings of Table~\ref{tab:sm-wien2k} come from each self-consistent field's own records (the options of \texttt{init\_lapw} and \texttt{run\_lapw} in its log, \texttt{case.in0}, \texttt{case.in2} and \texttt{case.scf}), with the irreducible $\kk$-point counts reproduced by \texttt{kgen}.

\emph{GaAs.} The record of the GaAs field was not kept, and the field was regenerated on the stored structure with
\begin{center}
\verb|init_lapw -b -prec 2 -vxc 5 -numk 600|;\quad \verb|run_lapw|
\end{center}
This reproduces the stored basis (\texttt{case.in1c}), $G_{\max}$ and interstitial plane-wave set exactly, the stored potential to $3.6\times10^{-4}$~Ry in its spherical part and $7.7\times10^{-5}$~Ry in its non-spherical part, within the scatter of two runs stopped at an energy change of $10^{-4}$~Ry, and the Fermi energy to 0.6~meV. The automatic $\kk$-mesh of \texttt{-prec 2} ($23^3$) reproduces the potential and the Fermi energy less closely.

\emph{$\alpha$-Sn.} The self-consistent field of the earlier porting runs (Sec.~\ref{sm:bridge}) was not kept, and its Fermi level enclosed 0.1 more electrons than the cell's 28 valence electrons on its own bands (28.11 on the $8^3$ Wannier mesh). The field was recomputed for this version on an unshifted $16^3$ mesh, which contains every high-symmetry point of the fcc zone, and converged in 14 iterations to a non-magnetic solution that encloses 28.04 electrons on the Wannier mesh, using
\begin{center}
\verb|init_lapw -b -prec 2 -sp|;\quad \verb|init_so_lapw| (default answers, magnetization axis [001]);\quad \verb|runsp_lapw -so -ec 0.0001 -cc 0.0001|
\end{center}

\begin{table}[htbp]
\caption{Structures of the twelve calculations. Lattice parameters are in \AA, and positions are in fractions of the conventional lattice vectors of the program's input, except the out-of-plane coordinate of the two CRYSTAL slabs, which is a Cartesian $z$ in \AA. The space group is that of the structure, although the two-component CRYSTAL calculations (SnTe and Bi) run in $P1$. $R_{\rm MT}$ is the WIEN2k muffin-tin radius in bohr. The positions of the Sc-COF (89 atoms) are not listed.}
\label{tab:sm-structures}
\centering\small
\begin{tabular}{@{}l l l l l@{}}
\toprule
Cell & Space group & Lattice (\AA) & Atom: position & $R_{\rm MT}$ \\
\midrule
\multicolumn{5}{@{}l}{\emph{CRYSTAL23 (LCAO)}} \\
h-BN & $P\bar{6}m2$ (187) & $a=2.50000$, $c=6.77000$ & B: $(0,\,0,\,0)$ & -- \\
 &  &  & N: $(\tfrac{1}{3},\,\tfrac{2}{3},\,0)$ & -- \\
 &  &  & B: $(\tfrac{1}{3},\,\tfrac{2}{3},\,\tfrac{1}{2})$ & -- \\
 &  &  & N: $(\tfrac{2}{3},\,\tfrac{1}{3},\,\tfrac{1}{2})$ & -- \\
MgB$_2$ & $P6/mmm$ (191) & $a=3.08458$, $c=3.52048$ & Mg: $(0,\,0,\,0)$ & -- \\
 &  &  & B: $(\tfrac{1}{3},\,\tfrac{2}{3},\,\tfrac{1}{2})$ & -- \\
 &  &  & B: $(\tfrac{2}{3},\,\tfrac{1}{3},\,\tfrac{1}{2})$ & -- \\
SnTe (B2) & $Pm\bar{3}m$ (221) & $a=3.46800$ & Te: $(0,\,0,\,0)$ & -- \\
 &  &  & Sn: $(\tfrac{1}{2},\,\tfrac{1}{2},\,\tfrac{1}{2})$ & -- \\
Bi bilayer & $P\bar{3}m1$ (164) & $a=4.29363$ & Bi: $(\tfrac{1}{3},\,\tfrac{1}{6},\,+0.89836\,\text{\AA})$ & -- \\
 &  &  & Bi: $(\tfrac{2}{3},\,\tfrac{5}{6},\,-0.89836\,\text{\AA})$ & -- \\
Sc-COF & $P\bar{6}m2$ (187) & $a=15.17639$ & 89 atoms, C$_{36}$H$_{24}$B$_{12}$O$_{12}$Sc$_5$ & -- \\
\addlinespace
\multicolumn{5}{@{}l}{\emph{WIEN2k (LAPW)}} \\
GaAs & $F\bar{4}3m$ (216) & $a=5.6532$ & Ga: $(0,\,0,\,0)$ & 2.23 \\
 &  &  & As: $(\tfrac{1}{4},\,\tfrac{1}{4},\,\tfrac{1}{4})$ & 2.23 \\
MgB$_2$ & $P6/mmm$ (191) & $a=3.0849$, $c=3.5211$ & Mg: $(0,\,0,\,0)$ & 1.85 \\
 &  &  & B: $(\tfrac{1}{3},\,\tfrac{2}{3},\,\tfrac{1}{2})$ & 1.35 \\
 &  &  & B: $(\tfrac{2}{3},\,\tfrac{1}{3},\,\tfrac{1}{2})$ & 1.35 \\
$\alpha$-Sn & $Fd\bar{3}m$ (227) & $a=6.6375$ & Sn: $(\tfrac{7}{8},\,\tfrac{3}{8},\,\tfrac{3}{8})$ & 2.35 \\
 &  &  & Sn: $(\tfrac{1}{8},\,\tfrac{1}{8},\,\tfrac{1}{8})$ & 2.35 \\
Bi bilayer & $P\bar{3}m1$ (164) & $a=4.2936$, $c=25.0000$ & Bi: $(\tfrac{1}{3},\,\tfrac{1}{6},\,0.035934)$ & 2.50 \\
 &  &  & Bi: $(\tfrac{2}{3},\,\tfrac{5}{6},\,0.964066)$ & 2.50 \\
bcc Fe & $Im\bar{3}m$ (229) & $a=2.8665$ & Fe: $(0,\,0,\,0)$ & 1.98 \\
SrVO$_3$ & $Pm\bar{3}m$ (221) & $a=3.8410$ & V: $(0,\,0,\,0)$ & 1.89 \\
 &  &  & Sr: $(\tfrac{1}{2},\,\tfrac{1}{2},\,\tfrac{1}{2})$ & 2.50 \\
 &  &  & O: $(\tfrac{1}{2},\,0,\,0)$ & 1.71 \\
 &  &  & O: $(0,\,\tfrac{1}{2},\,0)$ & 1.71 \\
 &  &  & O: $(0,\,0,\,\tfrac{1}{2})$ & 1.71 \\
\bottomrule
\end{tabular}
\end{table}

\begin{table}[htbp]
\caption{CRYSTAL23 calculations, read from the header of each properties output. PBE$+$10\% is PBE with 10\% Fock exchange, and two-comp.\ is a two-component (spin--orbit) calculation. $N_{\rm AO}$ is the number of spatial atomic orbitals per cell (a two-component Hamiltonian has twice this dimension), and $N_e$ is the number of electrons per cell outside the pseudopotential cores. Net gives the Pack--Monkhorst shrinking factors, with the Gilat net used for $E_{\mathrm{F}}$ in parentheses, and $N_{\kk}$ is the number of $\kk$-points in the irreducible zone. \texttt{TOLINTEG} gives the five truncation thresholds as $-\log_{10}$, and $|\Delta E|$ is the energy change in the last SCF cycle. $E_{\mathrm{F}}$ is as printed (Hartree), and for h-BN it is the valence-band maximum.}
\label{tab:sm-crystal}
\centering\footnotesize\setlength{\tabcolsep}{2.8pt}
\begin{tabular}{@{}l l l r r l r l r l r@{}}
\toprule
Cell & Hamiltonian & Spin & $N_{\rm AO}$ & $N_e$ & Net & $N_{\kk}$ & \texttt{TOLINTEG} & Cycles & $|\Delta E|$ (Ha) & $E_{\mathrm{F}}$ (Ha) \\
\midrule
h-BN & PBE0 & closed shell & 72 & 24 & $18^3$ (18) & 370 & 8,8,8,10,34 & 8 & $6.9{\times}10^{-10}$ & $-0.240$ \\
MgB$_2$ & PBE$+$10\% & open shell, $m=0$ & 55 & 22 & $18^3$ (18) & 370 & 9,9,9,11,38 & 10 & $2.2{\times}10^{-10}$ & $-0.164$ \\
SnTe (B2) & PBE$+$10\% & two-comp. & 71 & 46 & $12^3$ (12) & 1728 & 9,9,9,11,38 & 17 & $3.9{\times}10^{-10}$ & $-0.0627$ \\
Bi bilayer & PBE$+$10\% & two-comp. & 56 & 46 & $15^2{\times}1$ (15) & 225 & 8,8,8,8,20 & 45 & $8.5{\times}10^{-9}$ & $-0.137$ \\
Sc-COF & HSE06 & collinear, $5\,\mu_B$ & 1424 & 501 & $3^2{\times}1$ (6) & 3 & 9,9,9,9,18 & 77 & $4.5{\times}10^{-7}$ & $-0.104$ \\
\bottomrule
\end{tabular}
\end{table}

\begin{table}[htbp]
\caption{WIEN2k calculations. $R_{\rm MT}^{\min}K_{\max}$ is as in \texttt{case.in1}, with $\ell_{\max}=10$ in every case, local orbitals for the semicore states, and for GaAs and $\alpha$-Sn a $p$ local orbital from the second energy derivative. $G_{\max}$ is in bohr$^{-1}$. ``SCF mesh'' is the $\kk$-mesh of the self-consistent field, with its irreducible points in parentheses, and ``Criteria'' are the changes of total energy (Ry) and charge (e) between the last two iterations at which \texttt{run\_lapw} stops (\texttt{-ec}/\texttt{-cc}). In the Spin column, none is spin-unpolarized, pol.\ is collinear spin-polarized, and SOC is spin--orbit coupling by second variation, self-consistent for $\alpha$-Sn and the Bi bilayer and added in one step for Fe on the converged potential of the polarized calculation. $\alpha$-Sn was run spin-polarized and converged to zero moment. In the $E_{\mathrm{F}}$ column, tetra.\ is the tetrahedron method and FD Fermi--Dirac smearing. ``Wannier mesh'' is the $\Gamma$-centered full-zone mesh of the reference eigenvectors (main-text Table~\ref{tab:lapw}). For GaAs the settings are those of the regenerated self-consistent field (text).}
\label{tab:sm-wien2k}
\centering\footnotesize\setlength{\tabcolsep}{3pt}
\begin{tabular*}{\textwidth}{@{\extracolsep{\fill}}l l l r r l l l l@{}}
\toprule
Cell & Functional & Spin & $R_{\rm MT}^{\min}K_{\max}$ & $G_{\max}$ & SCF mesh & Criteria & $E_{\mathrm{F}}$ & Wannier mesh \\
\midrule
GaAs & LDA & none & 7.5 & 20 & $8^3$ (29) & $10^{-4}$/-- & tetra. & $8^3$ \\
MgB$_2$ & PBE & none & 6.04 & 16 & $8^3$ (50) & $10^{-4}$/-- & tetra. & $12^2{\times}8$ \\
$\alpha$-Sn & PBE & pol.$+$SOC & 7.5 & 20 & $16^3$ (621) & $10^{-4}$/$10^{-4}$ & tetra. & $8^3$ \\
Bi bilayer & PBE & SOC & 7.0 & 16 & $12^2{\times}1$ (144) & $10^{-4}$/$10^{-3}$ & FD, 2~mRy & $12^2{\times}1$ \\
bcc Fe & PBE & pol. & 8.0 & 16 & $17^3$ (165) & $10^{-4}$/$10^{-4}$ & tetra. & $8^3$ \\
bcc Fe (SOC) & PBE & pol.$+$SOC & 8.0 & 16 & $17^3$ (165) & $10^{-4}$/$10^{-4}$ & tetra. & $8^3$ \\
SrVO$_3$ & PBE & none & 7.0 & 16 & $10^3$ (35) & $10^{-4}$/$10^{-3}$ & tetra. & $8^3$ \\
\bottomrule
\end{tabular*}
\end{table}

\section{Performance, $\kk$-mesh construction, and reproducibility}
\label{app:perf}
\emph{Scale.} The production scale is set by the Sc-COF (89 atoms, 1424 AOs, $N_w=294$, a pool of 441 bands $=1.5N_w$, collinear $\alpha/\beta$), whereas the mesh series of main-text Table~\ref{tab:mmn} ran the projectability method, whose band count for this cell is 318. The costs per spin channel on an Apple M5 Pro (18 CPU cores, 48~GB) are listed below.
\begin{itemize}\setlength\itemsep{0pt}
\item Parsing the 530~MB CRYSTAL output takes 4.7~s, or 0.06~s from the parse cache.
\item The analytic $S^{(\bb)}(\mathbf g)$ precompute runs in parallel, with one $(\bb,\mathbf g)$ task per worker.
\item The constrained-$\Omega_I$ sweep uses batched dis-row matrix products for the $\kk$-local update, which are $62\times$ faster than the naive contraction ($20\times$ at SnTe scale). An exact frozen-block split (the frozen columns are a per-$\kk$ constant) and dis-column restriction give a further $1.5$--$2.7\times$ on the $\kk$-loop at bit-identical subspaces. A sweep takes about 3~s at $N_w=294$, measured with other jobs running, so 2000 sweeps take about 1.7~h.
\item Opt-in over-relaxation of the $\kk$-local update (mixing against the previous effective operator) needs $4.8$--$7.4\times$ fewer sweeps at identical $\Omega_I$ (h-BN $278\to58$, MgB$_2$ $1105\to149$, SnTe from the sweep cap to convergence at 646). It remains an option because the mixed iteration does not always preserve the fixed point, and on SnTe with the shell floor active it settles $0.006$~\AA$^2$ per function above the unmixed $\Omega_I$. No \textsc{lcao2wannier} model uses it, whereas the LAPW Gauss--Seidel models used the scheduled over-relaxation of \textsc{wien2wannier} (Sec.~\ref{sm:sweeporder}).
\item The parallel \texttt{.mmn} writer reduces Stage~2 from 115 to 39~s on the Sc-COF.
\item Peak memory is about 5~GB, dominated by the 2.3~GB $S^{(\bb)}$ store and the 0.9~GB pool-overlap tensor.
\item For large bases the eigensolve is restricted to the lowest $N$ bands (LAPACK subset drivers, $2.5\times$ faster at $N\approx35\%$ of the 1424-orbital basis), with a check that aborts if the band pool reaches the solve ceiling.
\end{itemize}

\emph{$\kk$-mesh construction.} The pipeline runs in one invocation and
one eigensolve per spin channel, and the \texttt{wannier90.x -pp} round trip
is not needed. \textsc{Wannier90} never reads the \texttt{.nnkp} back (it regenerates its neighbor list at Wannierization time) and matches each \texttt{.mmn} block to that list by its $(\kk,\kk{+}\bb,\mathbf G)$ label, so the only requirement is that the \texttt{.mmn} contains exactly the neighbors and $\mathbf G$ vectors that this internal construction derives. We reimplement that construction exactly
(shell-distance search, reproducible-degeneracy supercell sort, automatic
shell selection with SVD-solved completeness weights, and the acceptance
loop order), validated against 82 \texttt{-pp}-generated neighbor files
across every material and mesh, including pseudo-hexagonal
cells whose shell distances split at the sort's $10^{-8}$ degeneracy
tolerance. Two practical rules follow from this construction and from the sampling of the zone. The mesh must be built from the same full-precision lattice and $\kk$-points that are written to the \texttt{.win}, because inputs rounded to seven decimals reorder near-degenerate $\bb$-vectors. The mesh must also be commensurate with the high-symmetry points it should contain, since a $\Gamma$-centered uniform $N_1{\times}N_2{\times}N_3$ mesh (the Monkhorst--Pack set~\cite{Monkhorst1976} for odd $N_i$, whereas for even $N_i$ that set is offset from $\Gamma$ by half a step) samples a high-symmetry point only when its reduced coordinates are commensurate with the divisions (a hexagonal cell needs $3\,|\,N$ in-plane to contain $K$ and $H$), and an undersampled zone corner is a loss of fidelity that standard workflows do not flag.
The $\kk$-mesh
audit of main-text Sec.~\ref{sec:auto} reports every named point that the requested mesh misses. On the hexagonal cells of main-text Table~\ref{tab:defaults} the effect goes in both directions. With the per-state frozen set, h-BN improved from $10.0/92.2$ to $1.3/52.0$~meV (nearest band, $6^3\to12^2{\times}6$) once the mesh converged with $K$ sampled at both meshes, while MgB$_2$ rose from 18.6 to 24.1~meV ($8^3\to12^3$), the lower value being an artifact of an $8^3$ mesh that never samples $K$ or $H$. The derived thresholds are mesh-stable in both cases (h-BN $p_{\text{froz}}=0.706$ at both meshes, MgB$_2$ $0.766$), so those shifts came from the sampling of the zone and not from the calibration of the thresholds.

\emph{Reproducibility.} Every result in main-text Table~\ref{tab:defaults} is
produced by one command per material plus a \textsc{Wannier90} Wannierization (the development build \texttt{ff955d1}, which v4.0.2 reproduces bitwise, run serially with \texttt{OMP\_NUM\_THREADS=2}), with no hand-set thresholds. The numbers of threads and MPI ranks change the trajectory of a Wannierization at the level of $10^{-5}$~\AA$^2$ and, on the rerun models, its band errors by at most 0.03~meV, and the reported window-route state of MgB$_2$ reproduces only with two threads. The command is
\begin{center}
\verb|lcao2wannier --stage all --method hybrid|\\
\verb|--mmn-method analytic --input <material>.out|\\
\verb|--seed <seed> --k-grid <n1 n2 n3>|
\end{center}
The software is \textsc{lcao2wannier}~v1.0~\cite{lcao2wannierZenodo} with Python 3.11.15, NumPy 2.4.6 and SciPy 1.17.1, \textsc{wien2wannier} v3.0~\cite{wien2wannierZenodo} with WIEN2k 24.1, CRYSTAL23 v1.0.1, and the \textsc{Wannier90} development build \texttt{ff955d1}, run on macOS 26.5 on an Apple M5 Pro (18 CPU cores, 48~GB) with at most six MPI ranks. The public release of \textsc{lcao2wannier} reproduces the hand-offs of main-text Table~\ref{tab:defaults} byte for byte (checked on h-BN, MgB$_2$ and SnTe in \texttt{calculations/revision\_queue/release\_check}), whereas the Sc-COF models were made with the code of 2026-09-14 (Sec.~\ref{sm:convergence}). CRYSTAL truncates its direct-lattice sums by overlap, penetration and pseudo-overlap thresholds~\cite{Pisani1980,Pisani1988}, set by \texttt{TOLINTEG}~\cite{CrystalManual} to $10^{-n}$ with $n=(8,8,8,10,34)$ for h-BN, $(9,9,9,11,38)$ for MgB$_2$ and SnTe, $(8,8,8,8,20)$ for the Bi bilayer and $(9,9,9,9,18)$ for the Sc-COF. The analytic overlaps run over the cells this truncation retains (main-text Appendix~\ref{app:contractivity}).

\emph{Convergence and restarts.} Every constrained-$\Omega_I$ disentanglement behind a reported model converged to the tolerances of main-text Appendix~\ref{app:constants} within its sweep cap (20\,000, or 30\,000 in \textsc{wien2wannier}), with the sweep count reported. The exceptions are three LAPW Gauss--Seidel runs that stopped at the 30\,000-sweep cap, namely the Bi bilayer of main-text Table~\ref{tab:lapw} (last change $1.1\times10^{-7}$~\AA$^2$, Table~\ref{tab:sm-conv}) and the trust-mask models of the Bi bilayer and of Fe with spin--orbit coupling in main-text Table~\ref{tab:extent} (Table~\ref{tab:sm-x1}, note~a), whose conjugate-gradient counterparts converged. Every \textsc{Wannier90} disentanglement in a route comparison converged to its own criterion.
Each Wannierization is reported with its iteration count and either the criterion met or the terminal per-iteration
$|\Delta\Omega|$, and route comparisons use Wannierizations run to their criterion on both sides. The criterion is \textsc{Wannier90}'s own test with \texttt{conv\_tol}~$=10^{-10}$~\AA$^2$, which requires $|\Delta\Omega|<10^{-10}$~\AA$^2$ in each of 3 (LCAO) or 5 (LAPW) consecutive iterations. This tolerance is $10N_{\rm atom}$ times tighter than that of the most precise protocol of the AiiDA \textsc{Wannier90} workflows ($10^{-9}$~\AA$^2$ per atom over 6 iterations)~\cite{AiidaW90}. Applied to the default optimizer's trajectories, their moderate tolerance ($10^{-7}$~\AA$^2$ per atom over 3 iterations) stops on plateaus short of the minimum, for example the h-BN window route at $\Omega_{\rm tot}=16.76$~\AA$^2$ (11.85 at our criterion) and the GaAs hand-off after 6\,500 iterations at 5.5 compared to 4.9~meV. The reported iteration counts are those of \textsc{Wannier90}'s default optimizer, and the two-stage scheme below reaches the default runs' minima, or lower ones, in 1\,550 to 6\,420 iterations. A Wannierization stopped at its budget can be far from its minimum, as the h-BN window route shows, which interpolates at 258~meV over the valence window after 10\,000 iterations ($\Omega_{\rm tot}=16.8$~\AA$^2$) and at 0.4~meV at its criterion (18\,260 iterations in one run, 11.85~\AA$^2$).

Continuations of runs are treated in Sec.~\ref{sm:convergence}, where a run without disentanglement, which every $N_b{=}N_w$ hand-off is, is taken to the criterion in one run, because \texttt{restart = wannierise} discards its gauge (Table~\ref{tab:sm-restart}), whereas runs with disentanglement restart correctly.

Valid changes of the optimizer path (one thread, two MPI ranks, guiding centers off) move the h-BN band errors by at
most 0.01~meV, and rerunning a Wannierization from the same \texttt{.amn}, \texttt{.mmn}, and \texttt{.eig}
reproduces it exactly (the Gauss--Seidel h-BN hand-off: 2070 iterations, $\Omega_{\rm tot}=12.070929$~\AA$^2$, 0.766 and 7.418~meV). Band plots
are generated by a \texttt{restart = plot} pass on a copy of the run, so the Wannierization record is never
overwritten.

\emph{Wannierization optimizer.} The iteration counts reported in this document are those of \textsc{Wannier90}'s default optimizer, conjugate gradients reset to steepest descent every five steps (\texttt{num\_cg\_steps}~$=5$), run to $|\Delta\Omega|<10^{-10}$~\AA$^2$ on three consecutive iterations (five for GaAs). Table~\ref{tab:sm-optimizer} compares it with \texttt{num\_cg\_steps}~$=200$ and with a two-stage scheme, on the Gauss--Seidel hand-offs and on CG hand-offs made with an earlier stopping rule (five consecutive small decrements of $\Omega_I$, since replaced by the gradient criterion of main-text Sec.~\ref{sec:dis}). With \texttt{num\_cg\_steps}~$=200$ most hand-offs meet the same criterion sooner and with the same bands, but converging in fewer iterations does not always give the same model, since the CG GaAs hand-off then converges at 57.57 instead of 50.12~\AA$^2$, with band errors more than twice as large. On that plateau and in that minimum one Wannier function has diagonal overlaps $|M_{nn}^{(\kk,\bb)}|$ of 0.003--0.02 on some links, compared to at least 0.31 everywhere in the 50.12~\AA$^2$ minimum. The Marzari--Vanderbilt spread contains $\mathrm{Im}\ln M_{nn}^{(\kk,\bb)}$ for every link, which is undefined where $M_{nn}^{(\kk,\bb)}=0$, so the functional is not smooth there. Minima trapped by inconsistent branch choices were noted in Ref.~\cite{Marzari1997}, and Ref.~\cite{Li2023} suggests that suboptimal minima met by \textsc{Wannier90} may be non-smooth points of this spread and reports that its gauge-continuous spread typically needs 10 to 70 times fewer iterations. The Stengel--Spaldin spread~\cite{StengelSpaldin2006} (\texttt{use\_ss\_functional}, available in the \textsc{Wannier90} development build used here and in v4.0.1 and later) takes the phase of the $\kk$-averaged overlap $N_k^{-1}\sum_\kk M_{nn}^{(\kk,\bb)}$ instead, so a single small overlap does not make it singular, and on GaAs its gauges keep $|M_{nn}^{(\kk,\bb)}|\ge0.43$. The two-stage scheme minimizes it first, to $|\Delta\Omega|<10^{-6}$~\AA$^2$, and then runs the Marzari--Vanderbilt Wannierization afresh from its gauge, written as the \texttt{.amn}, with \texttt{num\_cg\_steps}~$=200$. It reaches the default runs' minima, or lower ones, in 1\,550 to 6\,420 iterations, except on the 36-function SnTe Gauss--Seidel hand-off, whose subspace itself jumps at X (main-text Sec.~\ref{sec:snte}).

\begin{table}[htbp]
\caption{The optimizer of \textsc{Wannier90} on the same hand-offs, with the default setting (\texttt{num\_cg\_steps}~$=5$), with 200 steps (\texttt{num\_cg\_steps}~$=200$), and with the two-stage scheme (Stengel--Spaldin spread, then Marzari--Vanderbilt with 200 steps). Every run meets the criterion, the CG hand-offs were made with the earlier stopping rule, and $\Omega_{\rm tot}$ is the total spread after the Wannierization. Band errors are over the scoring windows of the main-text tables (relative to $E_{\mathrm{F}}$), with bands paired by index unless marked, ``$\pm x$'' means within $x$~meV of the default run, and -- marks values not recorded.}
\label{tab:sm-optimizer}
\small\setlength{\tabcolsep}{2.5pt}
\begin{tabular*}{\textwidth}{@{\extracolsep{\fill}}l l r r l@{}}
\toprule
Hand-off & Optimizer & Iterations & $\Omega_{\rm tot}$ (\AA$^2$) & Band errors (meV) \\
\midrule
h-BN, Gauss--Seidel & default & 2\,070 & -- & -- \\
 & 200 steps & 890 & -- & $\pm0.05$ \\
h-BN, CG & default & 1\,340 & -- & -- \\
 & 200 steps & 460 & -- & $\pm0.05$ \\
MgB$_2$, Gauss--Seidel & default & 5\,690 & -- & -- \\
 & 200 steps & 820 & -- & $\pm0.05$ \\
MgB$_2$, CG & default & 5\,790 & -- & -- \\
 & 200 steps & 1\,680 & -- & $\pm0.05$ \\
h-BN and MgB$_2$, all four & two-stage & 1\,550--2\,290 & default minima & $\pm0.12$ \\
\addlinespace
GaAs (LAPW), Gauss--Seidel & default & 59\,500 & 52.488 & 4.9; 18.0 \\
 & 200 steps & 4\,000 & 52.466 & 4.8; 16.3 \\
GaAs (LAPW), CG & default & 104\,500$^{\rm a}$ & 50.12 & 5.2; 14.0 \\
 & 200 steps, or preconditioner & -- & 57.57 & 12.1; 36.5 \\
 & two-stage & $<4\,000$ & 50.115 & 5.3; 14.1 \\
\addlinespace
SnTe, 36 WFs, CG & default & 59\,960$^{\rm b}$ & 65.177 & 11.1; 23.8$^{\rm c}$ \\
 & two-stage & 6\,420 & 64.916 & 10.7; 22.1$^{\rm c}$ \\
SnTe, 36 WFs, Gauss--Seidel & two-stage & 127\,130$^{\rm d}$ & -- & 88.8; 88.0$^{\rm c}$ \\
\bottomrule
\end{tabular*}

\smallskip
\begin{minipage}{\textwidth}\footnotesize
$^{\rm a}$After 80\,000 iterations between 59.2 and 57.5~\AA$^2$.\\
$^{\rm b}$A separate run. The run of main-text Sec.~\ref{sec:snte} took 42\,130 iterations to the same spread.\\
$^{\rm c}$Nearest band.\\
$^{\rm d}$The Marzari--Vanderbilt stage, in which one Wannier function's overlaps fall to $4\times10^{-4}$ because the subspace jumps at X.
\end{minipage}
\end{table}

The baseline comparisons of main-text Fig.~\ref{fig:baseline} need per-state freezing (\texttt{dis\_froz\_proj}), released in the \textsc{Wannier90} 4.0.x series (checked in v4.0.2~\cite{W90v402}). Every \textsc{Wannier90} run of this paper used the development build \texttt{ff955d1} (the PDWF merge), whose results v4.0.2 reproduces bitwise, except the automatic-threshold run of main-text Fig.~\ref{fig:baseline}, run with v4.0.2. The hand-off needs no disentanglement feature, but with guiding centers, which the LCAO runs use, the result depends on the build, and the Bi bilayer interpolates at 0.27~meV with this build and with v4.0.2 but at 0.74~meV with release 3.1.0.

\section{Stage-2 procedures}
\label{sm:procedures}

Main-text Algorithm~\ref{alg:stage2} lists the steps of the hand-off route in words, and this section gives the five procedures behind those steps in the detail needed to reimplement them, each preceded by a description. They cover the addition of channels to the target (step~3), the frozen ceiling (steps~6 and~8), the constrained $\Omega_I$ minimization (step~7), the AO-SCDM gauge (step~9), and the certificate of main-text Appendix~\ref{app:certificate}. The constants are those of the released \textsc{lcao2wannier}~v1.0. The released \textsc{wien2wannier}~v3.0 implements Algorithms~\ref{alg:sm-ceiling} and~\ref{alg:sm-cg} with the same constants, and Algorithm~\ref{alg:sm-certificate} is also the computation performed by its \texttt{tools/plot\_certificate.py}. Energies are relative to $E_{\mathrm{F}}$, $\bar\varepsilon_n$ and $\bar P_n$ are $\kk$-averages over band $n$, and $\langle P\rangle_L$ is the mean of $P_{n\kk}$ over the states of the bands in a set $L$.

\paragraph{Channel augmentation.} The procedure determines which orbital channels the low conduction bands are made of and adds those channels to the target. For every channel $c=(\text{atom},\ell)$ of the basis outside the valence configuration it measures three quantities. The coverage $C_c$ is the mean projectability of the four lowest conduction bands onto $c$. The specificity $s_c$ is the coverage minus the channel's projectability onto twelve bands of the continuum above the fidelity range, which removes channels whose coverage only reflects the completeness of the basis. The semicore bands $\mathcal{S}_c$ are deep bands that project almost entirely onto $c$. A channel is admitted if its L\"owdin retention $\Pi_c$ (main-text Eq.~\eqref{eq:ceiling}) is at least $\Pi_{\min}=0.40$, below which it is nearly linearly dependent on the other channels, and if it has semicore bands or a specificity of at least $s_{\min}=0.05$. Channels are disjoint sets of orbitals, so the coverage of a set of channels is the sum of their coverages, and the smallest set whose coverage lies within $\delta_C=0.02$ of that of all admitted channels is the leading run of the admitted channels sorted by coverage. The set is accepted only if it raises the coverage by at least $\Delta C=0.05$. Each selected channel adds one valence shell of Wannier functions, and one function for each of its semicore bands inside the pool window.

\begin{breakablealgorithm}
\caption{Channel augmentation of the target (LCAO frontend, main-text Sec.~\ref{sec:auto}). Energies are relative to $E_{\mathrm{F}}$, and the constants are $\Pi_{\min}=0.40$, $s_{\min}=0.05$, $\delta_C=0.02$ and $\Delta C=0.05$.}
\label{alg:sm-augment}
\renewcommand{\baselinestretch}{1.1}\selectfont
\begin{algorithmic}[1]
\Statex \textbf{Symbols}
\Statex \begin{tabular}{@{}p{0.2\linewidth}@{\quad}p{0.74\linewidth}@{}}
$\mathcal{T}$, $P_{n\kk}$ & the target AO set, and the projectability onto it\\
$c$ & a candidate channel $(\text{atom},\ell)$ of the basis outside the valence configuration\\
$P^c_{n\kk}$, $\Pi_c$ & the projectability onto channel $c$, and its L\"owdin retention, main-text Eq.~\eqref{eq:ceiling}\\
$\bar\varepsilon_n$, $\bar P^c_n$ & the $\kk$-averages of $\varepsilon_{n\kk}$ and of $P^c_{n\kk}$ over band $n$\\
$\langle P\rangle_{\mathcal{L}}$ & the mean of $P_{n\kk}$ over every state of the bands in a set $\mathcal{L}$\\
$C_c$, $s_c$ & the coverage of the low conduction by channel $c$, and its specificity\\
$\mathcal{S}_c$, $n_c$ & the semicore bands of $c$, and the number of them inside the pool window\\
$\mathcal{C}_{\rm adm}$, $\mathcal{C}_{\rm sel}$ & the admitted channels, and the selected ones\\
$g$ & 2 for spinors, 1 otherwise\\
\end{tabular}
\Statex
\Require $\mathcal{T}$, $P_{n\kk}$, the overlap matrices $S(\kk)$, the channels $c$
\Ensure an enlarged $\mathcal{T}$, $P$ and $N_w$, or no change
\Statex
\LineComment{Two reference sets of bands}
\State $\mathcal{L}\gets$ the 4 lowest bands with $0<\bar\varepsilon_n<8$~eV \Comment{the low conduction}
\State $\mathcal{J}\gets$ the 12 lowest bands with $\min_\kk\varepsilon_{n\kk}>8$~eV and $\bar\varepsilon_n<40$~eV \Comment{the continuum above the fidelity range}
\Statex
\LineComment{Score every candidate channel}
\For{each channel $c$}
\State $P^c_{n\kk}\gets$ the L\"owdin projectability onto $c$; \quad $\Pi_c\gets$ its retention
\State $C_c\gets\langle P^c\rangle_{\mathcal{L}}$ \Comment{coverage of the low conduction}
\State $s_c\gets C_c-\langle P^c\rangle_{\mathcal{J}}$ \Comment{coverage beyond what basis completion gives}
\State $\mathcal{S}_c\gets\{n:\bar\varepsilon_n\le-2~\text{eV and }\bar P^c_n\ge0.85\}$ \Comment{deep bands made of $c$}
\State $n_c\gets|\{n\in\mathcal{S}_c:\bar\varepsilon_n\ge-30~\text{eV}\}|$ \Comment{those inside the pool window}
\EndFor
\Statex
\LineComment{Admit channels, then select the smallest set that is nearly as good as all of them}
\State $\mathcal{C}_{\rm adm}\gets\{c:\Pi_c\ge\Pi_{\min}\ \text{and}\ (\mathcal{S}_c\ne\emptyset\ \text{or}\ s_c\ge s_{\min})\}$
\If{$\mathcal{C}_{\rm adm}=\emptyset$}
\State \Return no change
\EndIf
\State $C_0\gets\langle P\rangle_{\mathcal{L}}$; \quad $C_{\max}\gets C_0+\sum_{c\in\mathcal{C}_{\rm adm}}C_c$ \Comment{coverage adds over disjoint channels}
\State $\mathcal{C}_{\rm sel}\gets\emptyset$; \quad $C\gets C_0$ \Comment{$C$: coverage of the selected set}
\For{$c\in\mathcal{C}_{\rm adm}$ in order of decreasing $C_c$}
\If{$\mathcal{C}_{\rm sel}\ne\emptyset$ and $C\ge C_{\max}-\delta_C$}
\State \textbf{break} \Comment{within $\delta_C$ of all admitted channels}
\EndIf
\State $\mathcal{C}_{\rm sel}\gets\mathcal{C}_{\rm sel}\cup\{c\}$; \quad $C\gets C+C_c$
\EndFor
\If{$C-C_0<\Delta C$}
\State \Return no change \Comment{the gain is too small}
\EndIf
\State $\mathcal{T}\gets\mathcal{T}\cup\mathcal{C}_{\rm sel}$; \quad $P\gets P+\sum_{c\in\mathcal{C}_{\rm sel}}P^c$
\State $N_w\gets N_w+\sum_{c\in\mathcal{C}_{\rm sel}}\big[g(2\ell_c+1)+n_c\big]$ \Comment{a valence shell per channel, plus its pooled semicore bands, main-text Eq.~\eqref{eq:decoupled}}
\end{algorithmic}
\end{breakablealgorithm}

\paragraph{Frozen ceiling.} The ceiling sets the extent of the frozen manifold, and its search starts at the fidelity limit $E_{\text{fid}}=E_{\mathrm{F}}+8$~eV and lowers the ceiling in steps of $\Delta E_{\rm step}=1$~eV until no $\kk$-point has more pool states below the ceiling than there are Wannier functions. If even $E_{\mathrm{F}}+1$~eV is too high, the run stops and reports the highest ceiling that would fit, $\min_\kk\varepsilon_{N_w+1,\kk}$. Once the hand-off has been made, the pocket skip inspects it for bands that are frozen only on islands of $\kk$-points. A band that lies below the ceiling at fewer than $f_{\rm p}=25\%$ of the $\kk$-points is frozen on that island and free outside it, and when the band is of mixed character (mean projectability $p_{\rm mix}=0.5$ or more on its frozen states) the disentangled subspace can jump at the edge of the island. The pocket skip acts when the weakest link of the hand-off, the smallest singular value of $W^\dagger(\kk)M^{(\kk,\bb)}W(\kk+\bb)$, is below $\sigma_{\rm p}=0.5$ and touches such a band. It lowers the ceiling $\delta_{\rm p}=0.1$~eV below the lowest such band, disentangles again, and keeps the new result only if its weakest link is at least $\sigma_{\rm p}$, doing so at most three times and never below $E_{\mathrm{F}}+1$~eV. A ceiling given explicitly is used as given, without the pocket skip.

\begin{breakablealgorithm}
\caption{The frozen ceiling, with the ceiling search of main-text Eq.~\eqref{eq:walk} and the pocket skip of main-text Sec.~\ref{sec:masks}. The pocket skip runs only for an automatic ceiling, and an explicit ceiling is used as given.}
\label{alg:sm-ceiling}
\renewcommand{\baselinestretch}{1.1}\selectfont
\begin{algorithmic}[1]
\Statex \textbf{Symbols}
\Statex \begin{tabular}{@{}p{0.2\linewidth}@{\quad}p{0.74\linewidth}@{}}
$\varepsilon_{n\kk}$ & the energy of pool band $n$ at $\kk$\\
$\varepsilon_{\rm c}$ & the frozen ceiling, at or below which every pool state is frozen\\
$A(\kk)$ & the admitted states of main-text Eq.~\eqref{eq:masks}\\
$W(\kk)$ & the hand-off frames, that is, the disentangled states in the eigenbasis of the projected Hamiltonian (main-text Eq.~\eqref{eq:handoff})\\
link $(\kk,\bb)$ & a pair of neighboring mesh points $\kk$ and $\kk+\bb$\\
$\sigma_{\kk\bb}$ & the strength of link $(\kk,\bb)$, the smallest singular value of $W^\dagger(\kk)M^{(\kk,\bb)}W(\kk+\bb)$, which is near 1 where the subspace varies smoothly and near 0 where it jumps\\
$f_n$, $q_n$ & the fraction of $\kk$-points at which band $n$ lies below $\varepsilon_{\rm c}$, and the mean $P_{n\kk}$ over those states\\
$\mathcal{B}$, $\mathcal{K}$ & the pocket bands, and the $\kk$-points that weak links touch\\
$\sigma_{\rm p}$, $f_{\rm p}$ & the weak-link threshold, 0.5, and the largest pocket fraction, 0.25\\
$p_{\rm mix}$, $\delta_{\rm p}$ & the mixed-character threshold, 0.5, and the margin under a pocket, 0.1~eV\\
\end{tabular}
\Statex
\Function{LowerCeiling}{\,}
\For{$\varepsilon=E_{\mathrm{F}}+8,\ E_{\mathrm{F}}+7,\ \dots,\ E_{\mathrm{F}}+1$~eV}
\State $d\gets\max_\kk|\{n\in\text{pool}:\varepsilon_{n\kk}\le\varepsilon\}|$ \Comment{the frozen count at this ceiling}
\If{$d\le N_w$}
\State \Return $\varepsilon$ \Comment{the frozen states fit the Wannier count}
\EndIf
\EndFor
\State stop, reporting $\min_\kk\varepsilon_{N_w+1,\kk}$, the highest ceiling that would fit
\EndFunction
\Statex
\Function{Disentangle}{$\varepsilon$}
\State $F(\kk)\gets\{n\in\text{pool}:\varepsilon_{n\kk}\le\varepsilon\}$ \Comment{the frozen window}
\State $V\gets$ minimize $\Omega_I$ over $A(\kk)\cup F(\kk)$ with $F(\kk)$ frozen \Comment{Algorithm~\ref{alg:sm-cg}}
\State $W(\kk)\gets V(\kk)R(\kk)$ \Comment{rotation of main-text Eq.~\eqref{eq:handoff}}
\State \Return $W$
\EndFunction
\Statex
\Function{PocketSkip}{$\varepsilon_{\rm c}$, $W$}
\For{at most three attempts}
\State $\sigma_{\kk\bb}\gets\sigma_{\min}\big[W^\dagger(\kk)M^{(\kk,\bb)}W(\kk+\bb)\big]$ for every link
\State $(\kk_0,\bb_0)\gets$ the link with the smallest $\sigma_{\kk\bb}$
\If{$\sigma_{\kk_0\bb_0}\ge\sigma_{\rm p}$}
\State \Return $\varepsilon_{\rm c}$, $W$ \Comment{no weak link}
\EndIf
\LineComment{Pockets: bands below the ceiling at only a few $\kk$-points, of mixed character there}
\State $\mathcal{B}\gets$ the pool bands $n$ with $0<f_n<f_{\rm p}$ and $q_n\ge p_{\rm mix}$
\If{no band of $\mathcal{B}$ lies below $\varepsilon_{\rm c}$ at $\kk_0$ or at $\kk_0+\bb_0$}
\State \Return $\varepsilon_{\rm c}$, $W$ \Comment{the weakest link is not at a pocket}
\EndIf
\State $\mathcal{K}\gets$ the $\kk$-points of every link with $\sigma_{\kk\bb}<\sigma_{\rm p}$
\State $\varepsilon'\gets\min\{\min_\kk\varepsilon_{n\kk}: n\in\mathcal{B}$ below $\varepsilon_{\rm c}$ somewhere in $\mathcal{K}\}-\delta_{\rm p}$ \Comment{just under those bands}
\If{$\varepsilon'<E_{\mathrm{F}}+1$~eV}
\State \Return $\varepsilon_{\rm c}$, $W$ \Comment{the ceiling may not drop further}
\EndIf
\State $W'\gets$ \Call{Disentangle}{$\varepsilon'$}
\If{the smallest link strength of $W'$ is below $\sigma_{\rm p}$}
\State \Return $\varepsilon_{\rm c}$, $W$ \Comment{lowering did not restore the link, so keep the first result}
\EndIf
\State $\varepsilon_{\rm c}\gets\varepsilon'$; \quad $W\gets W'$ \Comment{accept, and inspect the new hand-off}
\EndFor
\State \Return $\varepsilon_{\rm c}$, $W$
\EndFunction
\end{algorithmic}
\end{breakablealgorithm}

\paragraph{Constrained $\Omega_I$ minimization.} The minimization keeps the frozen states fixed and moves only the free columns. At each $\kk$ the frozen states are unit columns of $V(\kk)$, and the $n_\kk=N_w-|F(\kk)|$ free columns form an orthonormal frame $X_\kk$ on the admitted non-frozen states $D(\kk)$, a point on a Grassmannian. Only the span of the free columns matters for $\Omega_I$, because rotating them among themselves leaves the subspace unchanged, which is why the search runs over Grassmannians. Splitting $V$ into its frozen and free columns in main-text Eq.~\eqref{eq:omegai} gives
\begin{equation}
\Omega_I = N_w\sum_b w_b - \frac{1}{N_k}\Big(c_0 + \sum_{\kk}\operatorname{tr}X_\kk^\dagger Q_\kk X_\kk
+ \sum_{\kk,\bb}\big\|X_\kk^\dagger \tilde M_{\kk\bb} X_{\kk+\bb}\big\|_F^2\Big),
\label{eq:sm-omegai-split}
\end{equation}
with $\tilde M_{\kk\bb}=w_b^{1/2}M^{(\kk,\bb)}[D(\kk),D(\kk+\bb)]$. The constant $c_0=\sum_{\kk,\bb}w_b\|M^{(\kk,\bb)}[F(\kk),F(\kk+\bb)]\|_F^2$ collects the frozen--frozen blocks, and the constant Hermitian matrix
\begin{equation}
Q_\kk = \sum_{\bb} w_b\,M^{(\kk,\bb)}[D,F]\,M^{(\kk,\bb)}[D,F]^\dagger
+ \sum_{\kk'+\bb'=\kk} w_{b'}\,M^{(\kk',\bb')}[F,D]^\dagger M^{(\kk',\bb')}[F,D]
\label{eq:sm-Qk}
\end{equation}
the frozen--free blocks, with the rows of $M^{(\kk,\bb)}[D,F]$ in $D(\kk)$ and its columns in $F(\kk+\bb)$, and the rows of $M^{(\kk',\bb')}[F,D]$ in $F(\kk')$ and its columns in $D(\kk)$. We write $f$ for the bracket in Eq.~\eqref{eq:sm-omegai-split}, so that maximizing $f$ minimizes $\Omega_I$, and the gradient of $f$ with respect to $X_\kk^*$ is $Z_\kk X_\kk$, with
\begin{equation}
Z_\kk = Q_\kk + \sum_{\bb}\tilde M_{\kk\bb}X_{\kk+\bb}X_{\kk+\bb}^\dagger\tilde M_{\kk\bb}^\dagger
+ \sum_{\kk'+\bb'=\kk}\tilde M_{\kk'\bb'}^\dagger X_{\kk'}X_{\kk'}^\dagger\tilde M_{\kk'\bb'},
\label{eq:sm-Zk}
\end{equation}
where the last sum runs over the links on which $\kk$ is the neighbor. Its projection $\nabla f_\kk=(1-X_\kk X_\kk^\dagger)Z_\kk X_\kk$ is the gradient on the Grassmannian, and $\langle A,B\rangle=\mathrm{Re}\sum_\kk\operatorname{tr}A_\kk^\dagger B_\kk$. For a tangent vector $Y$, the retraction $R(X,Y)$ takes at every $\kk$ the $Q$ factor of the QR decomposition of $X_\kk+Y_\kk$, with the diagonal of the $R$ factor made non-negative. The solver is a conjugate-gradient ascent of $f$ on the product of these Grassmannians. It starts from the projections of the states $D(\kk)$ onto the target and checks the analytic gradient against finite differences before the first step, stopping if they disagree. It then steps along a Polak--Ribi\`ere direction with a backtracking line search that accepts a step only if $\Omega_I$ decreases, restarts from the steepest direction after a failed line search (and stops after three failures in a row), and ends when the projected gradient has fallen to $\max(10^{-10},10^{-6}g_0)$, where $g_0$ is its initial norm. The search direction is carried from one tangent space to the next without transport, and is restarted along the gradient whenever it no longer raises $f$. Every quantity is evaluated at all $\kk$ from the same iterate, so the result does not depend on the thread count. A stall or the iteration cap with $\|\nabla f\|$ already below the threshold also counts as converged, and the run log records which exit ended the solve.

\begin{breakablealgorithm}
\caption{Frozen-constrained $\Omega_I$ minimization by conjugate gradients (CG) on the product of Grassmannians, the default solver of both programs. It raises $f$, the bracket of Eq.~\eqref{eq:sm-omegai-split}, which lowers $\Omega_I$. Every operation on $X$, $\mathit{grad}$ and $\mathit{dir}$ acts at all $\kk$ at once. A solve that stalls or reaches the cap with its gradient already below the convergence test is reported as converged.}
\label{alg:sm-cg}
\renewcommand{\baselinestretch}{1.1}\selectfont
\begin{algorithmic}[1]
\Statex \textbf{Symbols}
\Statex \begin{tabular}{@{}p{0.2\linewidth}@{\quad}p{0.74\linewidth}@{}}
$F(\kk)$, $D(\kk)$ & the frozen pool states at $\kk$, and the admitted pool states that are not frozen\\
$n_\kk$ & $N_w-|F(\kk)|$, the number of free columns at $\kk$\\
$X$ & the free columns, forming at each $\kk$ an orthonormal frame $X_\kk$ of $n_\kk$ columns on the states $D(\kk)$\\
$f$, $\nabla f$ & the objective of Eq.~\eqref{eq:sm-omegai-split}, and its gradient on the Grassmannian, $\nabla f_\kk=(1-X_\kk X_\kk^\dagger)Z_\kk X_\kk$ with $Z_\kk$ of Eq.~\eqref{eq:sm-Zk}\\
$\langle A,B\rangle$, $\|A\|$ & $\mathrm{Re}\sum_\kk\operatorname{tr}A_\kk^\dagger B_\kk$, and $\langle A,A\rangle^{1/2}$\\
$R(X,Y)$ & the retraction, which takes at every $\kk$ the $Q$ factor of the QR decomposition of $X_\kk+Y_\kk$, with the diagonal of the $R$ factor made non-negative\\
$\mathit{grad}$, $\mathit{dir}$ & the current gradient and search direction, both tangent vectors at $X$\\
$t$, $n_{\rm fail}$ & the step length, and the number of consecutive failed line searches\\
\end{tabular}
\Statex
\Require $F(\kk)$ and $D(\kk)$, the constants $c_0$ and $Q_\kk$, the scaled overlaps $\tilde M_{\kk\bb}$ of Eq.~\eqref{eq:sm-omegai-split}, and the iteration cap $N^{\max}=20\,000$
\Ensure $X$ at a stationary point of $\Omega_I$, or a report that the solve stalled or reached the cap
\Statex
\LineComment{Starting frames}
\State $X_\kk\gets$ the $n_\kk$ leading left singular vectors of the projections of the states $D(\kk)$ onto the target $\mathcal{T}$
\Statex
\LineComment{Check the analytic gradient against finite differences once, before the first step}
\State $\Theta\gets$ a random tangent vector at $X$; \quad $h\gets10^{-6}$
\State $s_{\rm FD}\gets\big[f(R(X,h\Theta))-f(R(X,-h\Theta))\big]/2h$ \Comment{slope of $f$ along $\Theta$}
\If{$|s_{\rm FD}-2\langle\nabla f(X),\Theta\rangle|>10^{-4}\,|s_{\rm FD}|$}
\State stop \Comment{a wrong gradient still descends, to a worse minimum}
\EndIf
\Statex
\LineComment{Conjugate-gradient ascent of $f$}
\State $\mathit{grad}\gets\nabla f(X)$; \quad $\mathit{dir}\gets\mathit{grad}$; \quad $g_0\gets\|\mathit{grad}\|$ \Comment{first step along the gradient}
\State $t_0\gets1/\max_\kk\max_{ab}|(Q_\kk)_{ab}|$; \quad $t\gets t_0$; \quad $n_{\rm fail}\gets0$ \Comment{initial step from the scale of $Q$}
\For{$i=1,\dots,N^{\max}$}
\State $X'\gets R(X,t\,\mathit{dir})$, halving $t$ up to 20 times until $\Omega_I(X')<\Omega_I(X)$ \Comment{backtracking line search}
\If{no trial step lowered $\Omega_I$}
\If{$n_{\rm fail}=3$}
\State \Return $X$ \Comment{stalled}
\EndIf
\State $n_{\rm fail}\gets n_{\rm fail}+1$; \quad $\mathit{dir}\gets\mathit{grad}$; \quad $t\gets t_0\,10^{-n_{\rm fail}}$ \Comment{restart along the gradient, shorter step}
\State \textbf{continue}
\EndIf
\State $n_{\rm fail}\gets0$; \quad $X\gets X'$; \quad $t\gets1.8\,t$ \Comment{accept, and try a longer step next}
\State $\mathit{grad}'\gets\nabla f(X)$
\State $\beta\gets\max\big(0,\ \langle\mathit{grad}',\mathit{grad}'-\mathit{grad}\rangle/\langle\mathit{grad},\mathit{grad}\rangle\big)$ \Comment{Polak--Ribi\`ere, never negative}
\State $\mathit{dir}\gets\mathit{grad}'+\beta\,\mathit{dir}$ \Comment{$\mathit{dir}$ is reused at the new $X$ without transport}
\If{$\langle\mathit{grad}',\mathit{dir}\rangle\le0$}
\State $\mathit{dir}\gets\mathit{grad}'$ \Comment{$\mathit{dir}$ no longer raises $f$: restart}
\EndIf
\State $\mathit{grad}\gets\mathit{grad}'$
\If{$\|\mathit{grad}\|\le\max(10^{-10},\,10^{-6}g_0)$}
\State \Return $X$ \Comment{converged}
\EndIf
\EndFor
\State \Return $X$ \Comment{iteration cap reached}
\end{algorithmic}
\end{breakablealgorithm}

\paragraph{AO-SCDM gauge.} The gauge is built from the density matrix of the hand-off subspace in the atomic-orbital basis. QR with column pivoting (QRCP) selects $N_w$ anchor orbitals, either from the $\kk$-averaged density matrix $\rho$ or from the states at $\Gamma$. Each anchor set gives two candidate gauges, a column candidate and a row candidate. The column candidate consists of the density-matrix columns at the anchors, projected onto the states, which lie in the span of the manifold. The row candidate consists of the amplitudes of the states on the anchor orbitals, and it is exactly singular when a state vanishes on every anchor at some $\kk$. Of the four candidates the one whose worst $\kk$ is best conditioned is kept, and the run stops if that gauge is still rank-deficient (condition number above $10^{10}$).

\begin{breakablealgorithm}
\caption{The AO-SCDM gauge (LCAO frontend, main-text Sec.~\ref{sec:scdm}). The LAPW frontend selects the anchors on its channel index and, in place of the four candidates and the choice among them, keeps the SCDM columns at well-conditioned $\kk$-points and transports the gauge to the others (main-text Appendix~\ref{app:lapw}).}
\label{alg:sm-scdm}
\renewcommand{\baselinestretch}{1.1}\selectfont
\begin{algorithmic}[1]
\Statex \textbf{Symbols}
\Statex \begin{tabular}{@{}p{0.2\linewidth}@{\quad}p{0.74\linewidth}@{}}
$B(\kk)$ & the AO coefficients of the $N_w$ hand-off states at $\kk$, one column per state, orthonormal in the analytic metric\\
$\rho$ & the $\kk$-averaged density matrix of the hand-off states in the AO basis\\
$a$ & an anchor set of $N_w$ AO indices\\
$\rho[:,a]$, $B(\kk)[a,:]$ & the columns of $\rho$ and the rows of $B(\kk)$ at the anchors\\
$G(\kk)$ & the $N_w\times N_w$ gauge matrix written to the \texttt{.amn} file\\
$\sigma_{\min}$, $\sigma_{\max}$ & the smallest and largest singular values\\
\end{tabular}
\Statex
\Require $B(\kk)$ at every $\kk$
\Ensure $G(\kk)$ at every $\kk$, or a stop with a report
\Statex
\State $\rho\gets N_k^{-1}\sum_\kk B(\kk)B^\dagger(\kk)$
\Statex
\LineComment{Two anchor sets, each chosen by QR with column pivoting (QRCP)}
\State $a_\rho\gets$ the first $N_w$ pivots of QRCP$(\rho)$
\State $a_\Gamma\gets$ the first $N_w$ pivots of QRCP$\big(B^\dagger(\Gamma)\big)$ \Comment{the anchors of SCDM-$\kk$}
\Statex
\LineComment{Two gauge forms for each anchor set, giving four candidates}
\For{$a\in\{a_\Gamma,a_\rho\}$}
\State column candidate: $G(\kk)\gets B^\dagger(\kk)\,\rho[:,a]$ \Comment{density-matrix columns, in the span of the manifold}
\State row candidate: $G(\kk)\gets\big(B(\kk)[a,:]\big)^\dagger$ \Comment{singular where a state vanishes on every anchor}
\EndFor
\State keep the candidate with the largest $\min_\kk\sigma_{\min}[G(\kk)]$ \Comment{the best-conditioned worst $\kk$}
\If{$\max_\kk\sigma_{\max}[G(\kk)]\,/\,\min_\kk\sigma_{\min}[G(\kk)]>10^{10}$}
\State stop \Comment{rank-deficient gauge}
\EndIf
\end{algorithmic}
\end{breakablealgorithm}

\paragraph{Certificate.} The certificate counts how many states the best single energy window would assign differently from the trust mask. The window starts at the lowest trusted state $E_{\rm lo}$ and ends at a ceiling $E$. It wrongly freezes the untrusted states below $E$ inside the trusted span and leaves out the trusted states above $E$. The count changes only at state energies, so evaluating it at every one of those energies (and once below $E_{\rm lo}$) gives the exact minimum. A second number records at how many $\kk$-points the trust mask is not an energy interval at all, which no window can represent.

\begin{breakablealgorithm}
\caption{The certificate of main-text Appendix~\ref{app:certificate}, which is the smallest number of states that a single energy window $[\varepsilon_{\rm lo},\varepsilon]$, starting at the lowest trusted state, assigns differently from the trust mask.}
\label{alg:sm-certificate}
\renewcommand{\baselinestretch}{1.1}\selectfont
\begin{algorithmic}[1]
\Statex \textbf{Symbols}
\Statex \begin{tabular}{@{}p{0.2\linewidth}@{\quad}p{0.74\linewidth}@{}}
$T(\kk)$ & the trust mask of main-text Eq.~\eqref{eq:masks}, over the pool, whose states are the trusted states\\
$\varepsilon_{\rm lo}$, $\varepsilon_{\rm hi}$ & the lowest and highest energy of a trusted state\\
$U$ & the untrusted pool states with energies in $[\varepsilon_{\rm lo},\varepsilon_{\rm hi}]$\\
$m(\varepsilon)$ & the number of states the window $[\varepsilon_{\rm lo},\varepsilon]$ misassigns\\
\end{tabular}
\Statex
\Require $T(\kk)$ and the energies $\varepsilon_{n\kk}$ of the pool
\Ensure the certificate $\min_\varepsilon m(\varepsilon)$, and the number of $\kk$ at which $T(\kk)$ is not an energy interval
\Statex
\State $\varepsilon_{\rm lo},\varepsilon_{\rm hi}\gets$ the lowest and highest energy of a trusted state
\State $U\gets$ the untrusted pool states with $\varepsilon_{\rm lo}\le\varepsilon_{n\kk}\le\varepsilon_{\rm hi}$
\Statex
\LineComment{The window $[\varepsilon_{\rm lo},\varepsilon]$ wrongly freezes the states of $U$ below $\varepsilon$ and abandons the trusted states above it}
\For{$\varepsilon\in\{\varepsilon_{\rm lo}-1~\text{eV}\}\cup\{\text{energies of }U\}\cup\{\text{energies of the trusted states}\}$} \Comment{$m$ changes only at these}
\State $m(\varepsilon)\gets|\{u\in U:\varepsilon_u\le\varepsilon\}|+|\{t\ \text{trusted}:\varepsilon_t>\varepsilon\}|$
\EndFor
\State \Return $\min_\varepsilon m(\varepsilon)$, and the number of $\kk$ at which $T(\kk)$ is not an energy interval
\end{algorithmic}
\end{breakablealgorithm}

\end{document}